\documentclass[11 pt]{article}
\pdfoutput=1 
\usepackage{jheppub} 
\usepackage{amsmath}
\usepackage{feynman}
\allowdisplaybreaks
\usepackage{subcaption,longtable,stmaryrd}
\usepackage{slashed}
\usepackage{bigints}
\usepackage{cancel}
\usepackage{multicol}
\usepackage{blindtext}
\usepackage{tikz}
\usetikzlibrary{decorations.markings}
\usepackage{enumitem}
\usepackage[customcolors,shade]{hf-tikz}
\usepackage{graphicx}
\usepackage{cancel}
\DeclareSymbolFont{usualmathcal}{OMS}{cmsy}{m}{n}
\DeclareMathAlphabet\mathbfcal{OMS}{cmsy}{b}{n}
\DeclareSymbolFontAlphabet{\mathcal}{usualmathcal}
\DeclareSymbolFont{rmlargesymbols}{OMX}{mdbch}{m}{n}
\DeclareMathSymbol{\rmintop}{\mathop}{rmlargesymbols}{82}
\DeclareMathSymbol{\rmointop}{\mathop}{rmlargesymbols}{72}
\newcommand{\rmint}{\rmintop\nolimits}
\DeclareMathSymbol{\rmintop}{\mathop}{rmlargesymbols}{82}
\definecolor{mygray}{gray}{0.5}
\title{\boldmath{Worldline effective field theory of inspiralling black hole binaries in presence of dark photon and axionic dark matter}}

\author[]{Arpan Bhattacharyya,}
\author[]{Saptaswa Ghosh,}
\author[]{Sounak Pal}
\affiliation[]{\it Indian Institute of Technology, Gandhinagar, Gujarat-382055, India}
\emailAdd{abhattacharyya@iitgn.ac.in}
\emailAdd{saptaswaghosh@iitgn.ac.in}
\emailAdd{palsounak@iitgn.ac.in}
\abstract{We investigate the correction to the potential that gives rise to the bound orbits and radiation from non-spinning inspiralling binary black holes in a dark matter environment consisting of \textit{axion-like particles} and \textit{dark photons} using the techniques of \textit{Worldline Effective Field Theory}. We compute the conservative dynamics up to $1$PN order for gravitational, electromagnetic, and Proca fields and up to $2$PN order for the scalar field. The effect of axion-electromagnetic coupling ($g_{a\gamma\gamma}$) arises in the conservative dynamics at $2.5$PN order and the kinetic mixing constant ($\gamma$) at $1$PN order. Furthermore, we calculate the radiation due to the various fields present in our theory. We find that the contribution of $g_{a\gamma\gamma}$ to the gravitational radiation appears at $N^{(7)}LO$ and to the scalar radiation appears at $N^{(5)}LO$. We also find that these radiative corrections due to the coupling $g_{a\gamma\gamma}$ vanish for any orbit confined to a plane because of the existence of a \textit{binormal} like term in the effective radiative action, but give rise to non-zero contributions for any orbit that lies in \textit{three dimensions}. Last but not the least, $\gamma$  contributes to the gravitational radiation at $N^{(2)}LO$ and $N^{(4)}LO$.}
\makeatletter
\begin{document}
\maketitle
\flushbottom
\section{Introduction} 
Detection of gravitational waves (GW) \cite{LIGOScientific:2016aoc, LIGOScientific:2016sjg, LIGOScientific:2016vlm, LIGOScientific:2017bnn, Kokeyama:2020dkg, LIGOScientific:2019hgc} opens a new pathway to test different theories of gravity. Gravitational waves can be generated mostly from binary systems consisting of Black holes and Neutron stars. These remarkable achievements in detecting gravitational waves suggest probing the accurate structure of the source. There are three phases of the evolution of these binaries: the \textit{inspiral}, \textit{merger}, and \textit{ringdown} phase. To probe the merger phase where the gravity is strong, one requires non-perturbative methods \cite{Lehner:2014asa, LIGOScientific:2014oec}, based on numerical simulations. However, one can use perturbative (analytic) techniques to probe the inspiral phase, assuming the gravity is weak. \textcolor{black}{The wave signal detected from the ground-based detector has a greater scope of detection due to the large time duration} of the existence of the inspiral phase, where the orbital velocity of the binaries is small compared to the velocity of light, i.e. $\frac{v}{c}\ll 1$. So, this binary inspiral problem can be solved using the Post-Newtonian approximation technique \cite{Blanchet:2013haa}. 
\par
The gravitational wave radiation from non-spinning binaries till 3PN order has been computed in \cite{Tagoshi:2000zg, Faye:2006gx, Blanchet:2006gy, Blanchet:2004ek, Damour:2000ni, Itoh:2003fy,Boetzel:2019nfw,Faye:2012we,Mishra:2013rna} using traditional methods by solving Green's functions. Recently this computation has been extended upto 4.5PN in \cite{Fujita:2010xj,Faye:2014fra, Blanchet:2023sbv,Blanchet:2023bwj} \footnote{Interestingly, both UV and IR divergences occur during the computation. The UV divergences appear because the binaries are modeled by a point-like particle without any internal structures. On the other hand, IR divergences occur due to the insertion of a typically divergent  (at long distance) PN expansion in the near zone while defining the source moments. Interested readers are referred to  \cite{Larrouturou:2021gqo, Blanchet:2004bb,Blanchet:2023soy} for a detailed discussion regarding the origin of these divergences as well as how to regularize them.}. Using these results of the gravitational waveform, parameter estimation based on the Fisher information matrix has been done in \cite{Arun:2004hn}. Later these studies have been extended by taking into account the effect of binary's spin at leading PN order in \cite{PhysRevD.12.329, PhysRevD.2.1428, Kidder:1992fr}. After that, several works have been done to extend this analysis to higher PN order \cite{Cho:2022syn}. Also, Hamiltonians generating the dynamics of the binary, including the binary's spin upto 3PN order, were derived in \cite{Steinhoff:2007mb, Steinhoff:2008ji, Hergt:2008jn, Hergt:2010pa,Porto:2012as} by using ADM formalism. Also, the computation of the gravitational wave radiation is being done for eccentric orbits \cite{Enoki:2006kj, Favata:2011qi, Munna:2019fjz}. Apart from general relativity (GR), these computations are being extended for certain alternative theories of gravity \cite{Zhang:2017srh,AbhishekChowdhuri:2022ora, Zhang:2018prg,Saffer:2018jmx, Lin:2018ken,Li:2022grj,Shiralilou:2021mfl,Julie:2019sab}.
\par
In parallel with the traditional method, perhaps a more efficient, systematic methodology based on \textit{\textcolor{black}{Worldline} Effective Field Theory} (WEFT) to compute various PN corrections emerged. This is sometimes called ``Non-Relativistic General Relativity'' (NRGR) \cite{Goldberger:2004jt}. It has been used to compute the gravitational wave radiation upto LO in \cite{Goldberger:2004jt}. \textcolor{black}{ One can decouple and treat the conservative and radiative degrees of freedom separately in this approach. Integrating out the potential modes (heavy modes), one can find the effective action of radiative modes (light modes). From the real part of this effective action, one can find the effective potential from which one can derive the binary dynamics, and from the imaginary part of this action, one can find the expression for the total power radiation using an optical theorem.}\footnote{Like the traditional method, various divergences occur while computing various multipole moments of the source using the EFT method. Interested readers are referred to \cite{Porto:2017dgs,Porto:2007pw} for more details about how to handle these divergences (in particular, the IR ones). Also, the renormalizability of EFTs of binary systems and the necessity of a time-dependent mass counterterm have been discussed in \cite{Goldberger:2012kf,Galley:2015kus}.} Subsequently, the method is developed further to include effects of binary's spin (upto quadratic order) in the  radiative multipole moments upto \cite{Porto:2010zg} 
\textcolor{black}{3PN order}, and then it has been extended to higher PN  orders in \cite{Levi:2011eq, Levi:2015uxa, Levi:2015ixa, Levi:2014sba}. Apart from these, this EFT based approach has been used to study various phenomena like self-force effect on the curved spacetime, gravitational tail effects and also inspecting conservative and radiative dynamics of the binary inspiral problem (via computing effective Hamiltonian/Lagrangian) \cite{Maia:2017gxn,Maia:2017yok,Foffa:2019yfl,Kalin:2022hph,Porto:2008jj,Porto:2008tb,Porto:2007px,Mandal:2022nty,Mandal:2022ufb,Goldberger:2009qd,Ross:2012fc,Goldberger:2020fot,Goldberger:2017ogt,Goldberger:2006bd,Goldberger:2005cd}
\footnote{The list is by no means exhaustive. Interested readers are referred to the following reviews \cite{Goldberger:2022rqf,Goldberger:2022ebt} and references therein for more details.}.
 Last but not least, the EFT approach has been extended to compute the conservative dynamics as well as power radiation due to a massive (as well as massless) scalar field minimally coupled with GR upto 1PN order \cite{Kuntz:2019zef,Huang:2018pbu} and to study the conservative dynamics of electromagnetic field minimally coupled with GR in \cite{Patil:2020dme,Gupta:2022spq}.
\par
Ultra-light bosonic fields are promising candidates for dark-matter \cite{GrillidiCortona:2015jxo, Sanchis-Gual:2022ooi, Goldstein:2022pxu, Schutz:2020jox} and  dynamical dark-energy \cite{Kamionkowski:2014zda}
\footnote{The list is by no means exhaustive.  Interested readers are referred to this Snowmass review \cite{Adams:2022pbo} and references therein for more details.}.  One famous example of such ultra-light bosonic fields is the QCD axion.  It stems from the Peccei–Quinn mechanism \cite{Fukuda:2021drn,CAPP:2020utb,Sakhelashvili:2021eid, Peccei:2006as, RevModPhys.82.557,PhysRevLett.40.223,PhysRevLett.40.279,Preskill:1982cy,Gorghetto:2018ocs,DiLuzio:2021pxd,Berezhiani:2000gh,Fukuda:2015ana,Dimopoulos:2016lvn,Gherghetta:2016fhp,Kim:1998va,DiLuzio:2020wdo,Conlon:2006tq,Chakraborty:2021fkp,Harigaya:2019qnl} proposed as one of the models for solving the strong CP problem originally proposed as one of the models to solve the strong CP problem \cite{PhysRevLett.38.1440,PhysRevD.16.1791,PhysRevLett.40.279,Clowe:2006eq,Hsu:2004mf,Agrawal:2017ksf,Gupta:2020vxb,PhysRevLett.43.103}.  Experimental findings for neutron electric dipole moment expects the strong CP angle to be much smaller than $10^{-10}$, while CP angle in CKM matrix has been found to be $\mathcal{O}(1)$ \cite{Baker:2006ts}.  The solution to this paradox can be resolved by introducing the axionic coupling \cite{Huang:2018pbu, Zhang:2021mks}.$$\frac{a}{f_a}\frac{g_{a\gamma\gamma}^2}{32\pi^2}F^{\mu\nu}\tilde{F}^{\mu\nu},$$
where $g_{a\gamma\gamma}$ is the strong coupling constant, $\tilde{F}^{\mu\nu}$ is the dual field strength tensor, and $f_a$ is the axionic decay constant. QCD Axions have several constraints.  For example, the ADMX experiment showed that the first constraint on the QCD axion parameter space in the $\mu$eV mass range \cite{PhysRevLett.120.151301}.  Also, it was suggested that one could possibly obtain constraints on the parameter space for QCD axions by studying its effects on the stellar configuration of a white dwarf \cite{Balkin:2022qer}.  There are other ultra-light bosons with similar properties to the QCD axion, with only the difference that their mass is not related to the decay constant.  They are often termed as \textit{Axion-like particles (ALP)}. They can arise from different considerations, e.g. from string compactifications  \cite{Cicoli:2013ana,Svrcek:2006hf,Hiramatsu:2010yu}.  There exist phenomenological constraints on the mass of these ultra-light bosons stemming from dark-matter phenomenology, black hole superradiance  \cite{Arias:2012az, DiLuzio:2021pxd, QuilezLasanta:2021yzt, Ishii:2022lwc, 
 Cardoso:2018tly, Brito:2015oca,Ghosh:2023tyz} as well as on its anomalous coupling with electromagnetic field ($g_{a\gamma\gamma}$) by using the induced oscillating electric dipole moment of the electron as advocated in \cite{Hill:2015vma}.
Besides axions (ultra-light boson fields), \textit{ultra-light vector fields} (ULV) are also possible candidates for dark-matter \cite{Cardoso:2018tly}.  One example of ultra-light vector fields is the so-called \textit{dark photon} \cite{Flambaum:2019cih}.  This can often be thought of as a \textit{hidden U(1)} massive vector field which can interact with the photon \cite{Cardoso:2018tly}.
\par
Both the ultra-light boson and vector fields can couple with gravity, e.g., axions can couple with gravity minimally and non-minimally.  The effect of axions on
the detected waveform is negligible until the binaries are separated by roughly a Compton wavelength of the axion \cite{Zhang:2019eid}. As the orbital length scale decays to the Compton wavelength, scalar radiation may become an important source
of orbital energy loss \cite{Huang:2018pbu}, especially for large Compton wavelengths.  Scalar radiation is detectable for both NS-NS and NS-BH binaries \cite{Dar:2018dra}.  Furthermore, one can write down Chern-Simons type terms both in the electromagnetic and gravitational sectors by which the massive scalar (axion) can couple with both electromagnetic and gravitational fields.  This is natural in the context when we consider the axion to play the role of dark matter \cite{Yoshida:2017cjl}.  If the axions interact with gravity via the Chern-Simons term, GW waves induce axion decays into gravitons \cite{Yoshida:2017cjl}.  Apart from this, in \cite{Machado:2018nqk, Machado:2019xuc}, it was shown that if axions couple with the dark photon with an unbroken $U(1)$ symmetry, there is a possibility of the generation of a stochastic gravitational wave when it experiences a tachyonic instability.  The resulting signal can be detected by ground-based GW detectors  \cite{Machado:2018nqk, Machado:2019xuc,10.21468/SciPostPhys.12.5.171}.  Last but not least, as advocated in \cite{ Ejlli:2022zah, Nagano:2021kwx}, polarimetry experiments on GW signals may provide a novel way to probe the axion-like particles.
\par 
Previously, there have been works that have taken steps towards constraining \textcolor{black}{nuclear coupling of ALPs} using GW observations \cite{Huang:2018pbu, Zhang:2021mks}.  This is done by considering a massive scalar field minimally coupled with GR and then computing the additional contribution to the total power radiation (at the leading PN order) and the correction to the phase of the emitted GW due to the presence of the scalar field.  In this paper, we will consider a more generic scenario.  Motivated by the dark matter model discussed in \cite{Cardoso:2018tly}, we will consider a model where we will have both the ALP (in the form of a massive scalar field which has an anomalous coupling with EM field via a Chern-Simons type term) and the ULV (in the form of a massive vector field (Proca field) which also couples with EM field by kinetic mixing term).  For simplicity, we neglect the coupling of ALP via the gravitational Chern-Simons term.  Our bigger goal is to inspect, apart from the masses of the scalar and Proca field, whether we can put constraints on the axion-photon coupling parameters $g_{a\gamma\gamma}$ as well as the Proca-electromagnetic coupling constant $\gamma$ (coming from the kinetic mixing term), also using GW observations.  To do that, we take a primary step forward and initiate a computation of the power radiation from a binary system in this dark matter environment, which is the key to computing the phase of the gravitational waveform using the WEFT approach, and eventually comment on the possibility of detecting these couplings.
\par
This paper is organized as follows: In Sec.~(\ref{Sec2}), we briefly review the general machinery of the EFT approach. We also discuss various length scales associated with the EFT of inspiralling binaries, which is crucial for separating various degrees of freedom. In Sec.~(\ref{Sec3}) we discuss how the binaries can be modeled by a point particle action. We review the procedure of integrating out the potential modes to get effective action for Einstein's gravity. We also give the EFT power counting for the bound sector. In Sec.~(\ref{Sec4}), we introduce our model, which consists of a scalar field (axion), electromagnetic (photon), and Proca field (dark photon) minimally coupled with Einstein gravity and an interaction term between photon and dark photon as well between axion and photon. Then we focus on the bound sector, which gives rise to the corrections to the effective potential. \textit{Interestingly, we observe that the correction due to the axion-photon coupling ($g_{a\gamma\gamma}$) enters at the 2.5PN order.} In Sec.~(\ref{Sec5}), we discuss the dissipative sector. We used the familiar optical theorem to calculate the power radiation. We again discuss the EFT power counting for the radiative sector. We describe the corrections to the gravitational radiation coming from different field vertices. We reproduce the well-known expression for the power radiation coming from the pure gravitational and scalar sectors at the leading order. \par 
We extend the computation to the $N^{(4)}LO$ and also showed that the $g_{a\gamma\gamma}$ appears at $N^{(5)}LO$ for the  scalar sector. Due to the presence of other fields, we obtain new contributions to the power radiation from the gravitational sector at different orders of perturbations. We observe that the contribution due to the axion-photon coupling ($g_{a\gamma\gamma}$) enters at the $N^{(7)}LO$ in the expression for the power radiation from the gravitational sector. Furthermore, we also compute the contribution of electromagnetic and Proca fields to the total power radiation at the leading order.  We also comment on the possibility of detecting axion coupling ($g_{a\gamma\gamma}$) through the gravitational flux. Lastly, the kinetic mixing constant $\gamma$ starts contributing from $1PN$ in conservative sector and appears in radiative sector at $N^{(2)}LO$ and $N^{(4)}LO$ through gravitational radiation.\par
Finally, in Sec.~(\ref{disc}), we summarize our main findings and conclude with some future directions. Some details regarding the computation of a few integrals are given in Appendices ~(\ref{app2}), (\ref{App1}), (\ref{app3}) and (\ref{ch1:app:E}), while comments on the middle vertex contribution to radiation are collected in Appendix~ (\ref{D}).
\subsection*{\textbf{ \textit{Notations and conventions:}}}

\begin{multicols}{2}
\begin{itemize}
    \item Metric signature: (-,+,+,+).
    \item $\rmint \mathcal{D}\hat\xi \,e^{iS_{\text{quad}}}\rightarrow \rmint \Bar{\mathcal{D}}\hat\xi$.
    \item $\rmint \frac{d^3\Vec{k}}{(2\pi)^3}\rightarrow \rmint_{\Vec{k}}.$
    \item ${\Vec{x}}_0=\frac{\Vec{x}_1-\Vec{x}_2}{2}.$
    \item Reduced mass: $\mu=\frac{m_1m_2}{M}$\,\,\text{with},\,\,$M=m_1+m_2$.
    \item Mass ratio: $\nu=\frac{\mu}{m_1+m_2}$.
    \item Planck mass: $m_p:=\frac{1}{\sqrt{8\,\pi\, G}},\textstyle{with\, (\hbar,\,c)=1.}$
\end{itemize}
\end{multicols}
\begin{itemize}
\item Levi-Civita symbol convention: $\epsilon^{\mu\nu\rho\sigma}\equiv\frac{\Hat{\epsilon}^{\,\mu\nu\rho\sigma}}{\sqrt{-g}}$, with $\Hat{\epsilon}^{\,0123}=1$.
\item $nPN$ in conservative sector symbolizes $\sim \mathcal{O}(Lv^{2n})$, where $L$ denotes angular momentum.
 $LO$ (Leading Order) symbolizes $\sim\mathcal{O}$($L^{1/2}v^{1/2}$) in radiation sector. Then the subsequent orders are denoted by $N^{(n)}LO\sim \mathcal{O}(L^{1/2}v^{n+1/2}).$

\end{itemize}

\section{Brief review of machinery of EFTs}\label{Sec2}
Consider a QFT with a general tensor field $\xi_{\alpha\beta..}$ in 3+1 dimensions described by the action $S[\xi]$. In order to compute the Effective action, one could first write down the generating functional and then take the logarithm. Usually, in QFT, one writes down the generating functions corresponding to the connected diagrams and, by Legendre transformation, can systematically evaluate the effective action as,
\begin{eqnarray}
        \mathcal{Z}[J]&=&\rmint \mathcal{D}\xi_{\alpha\beta..} \, e^{iS[\xi]+\rmint J\cdot \xi},\nonumber\\
      e^{i W[J]}&=& \mathcal{Z}[J],\\
      S_{\text{eff}}[\xi_{\text{cl}}]&=& W[J]-\rmint d^4x J\,\xi_{\text{cl}}.\nonumber \label{2.3m}
\end{eqnarray}
Here $W[J]$ have contributions only from connected diagrams and $\xi_{\text{cl}}$ is defined as $\xi_{\text{cl}}=\frac{\delta W[J]}{\delta J}$. 
 Effective action in (\ref{2.3m}) is complex in general. \textcolor{black}{The real part gives the conservative dynamics of the system, and the imaginary part gives the radiative dynamics of the system. Hence, for a binary inspiral problem, the real part gives the equation of motion, i.e., the orbit equation of the binary system, and the imaginary part gives the power radiation from the system.} The perturbative computation in $\hbar$ also leads to effective action at the classical and quantum levels.
 \begin{align}
     \begin{split}
S_{\text{eff}}^{\text{tot}}=\sum \underbrace{\text{Tree-level diagrams}}_{\mathcal{O}(\hbar^0)(\text{classical contribution})}+ \sum \underbrace{\text{loop level diagram}}_{\mathcal{O}(\hbar^n)(\text{quantum corrections})}. \label{Stotal}
     \end{split}
 \end{align}
 \par
 At this point, we assume that the theory has two different energy scales, a high energy scale ($\Lambda$) and a low energy scale ($\epsilon$). As we are interested in the low energy dynamics of the quantum field theory, it is convenient to decompose the field in those two modes: $\xi_{\alpha\beta...}=\hat{\xi}_{\alpha\beta..}+\Bar{\xi}_{\alpha\beta..}$ such that,
\begin{itemize}
    \item $\Bar{\xi}_{\alpha\beta..}$ are the light modes with energy scales $\epsilon$.
    \item  $\hat{\xi}_{\alpha\beta..}$ are the heavy modes with energy scale $\Lambda>>\epsilon$.
\end{itemize}
Then the effective theory of the light modes $\Bar{\xi}_{\alpha\beta..}$ can then be determined integrating out the heavy modes $\hat{\xi}_{\alpha\beta..}$. This can be done using  path integral in the following way,
\begin{align}
    \begin{split}
    \mathcal{Z}[\Bar{\xi}_{\alpha\beta..}]:=    e^{i\mathcal{S}_{\text{eff}}[\Bar{\xi}_{\alpha\beta..}]}=\rmint \mathcal{D}\hat{\xi}_{\alpha\beta..}\,e^{i\,\mathcal{S}^{\mathcal{O}(\hbar^0)}_{\text{eff}}[\Bar{\xi}_{\alpha\beta..},\hat{\xi}_{\alpha\beta..}]}\,,
        \end{split} \label{stotal1}
\end{align}
where $S^{\mathcal{O}(\hbar^0)}_{eff}$ is the classical part of the action mentioned in (\ref{Stotal}).
The path integral eventually gives the effective action for the light modes as follows,
\begin{align}
    \begin{split}
       \mathcal{S}_{\text{eff}}[\Bar{\xi}_{\alpha\beta..}]= \rmint d^4x\Big[\frac{1}{2}\partial_{\mu}\Bar{\xi}_{\alpha\beta..}\partial^{\mu}\Bar{\xi}^{\alpha\beta..}+
        \sum_{n}C_{n}\mathcal{O}_{n}(\Bar{\xi}_{\alpha\beta..})
\Big]\,.
    \end{split}\label{2.4m}
\end{align}
The coefficients $C_{n}$ are sometimes called the Wilson Coefficients, which contain information about the UV sector of the theory. This approach is called the \textit{top-down} approach. One can go the other way around. In that approach, one needs first to write down an effective action with $C_{n}$ completely unfixed, and then we need to fix them by comparing the experimental data or the computation from the \textit{top-down} approach. This approach is called the \textit{bottom-up} approach. We have to perturbatively compute the effective action in the small parameter of $\frac{\epsilon}{\Lambda}$. So we need to know how many $\mathcal{O}_n$ are in a given order. This can be fixed by EFT's power counting rules. We must understand how the quantities scale with $\frac{\epsilon}{\Lambda}$. It is systematically discussed in Sec.~(\ref{Sec4}) for our case. \par 
\par
 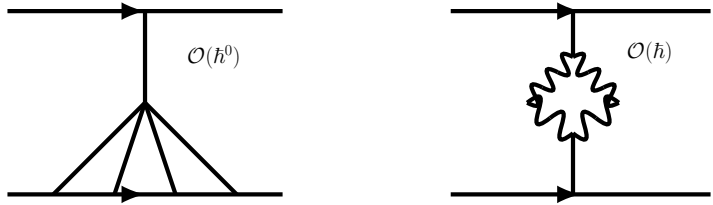
\begin{figure}
     \centering
    \scalebox{0.4}{\begin{feynman}
    \fermion[lineWidth=4, showArrow=false]{5.40, 4.00}{5.80, 5.20}
    \fermion[showArrow=false, lineWidth=4]{5.80, 5.20}{7.00, 4.00}
    \fermion[lineWidth=4, showArrow=false]{11.40, 5.80}{11.40, 6.40}
    \electroweak[flip=true, lineWidth=4]{10.80, 5.20}{11.40, 5.80}
    \fermion[lineWidth=4]{9.80, 6.40}{13.20, 6.40}
    \fermion[showArrow=false, lineWidth=4, label=$\mathcal{O}(\hbar^0)$]{5.80, 5.20}{5.80, 6.40}
    \fermion[lineWidth=4]{4.00, 4.00}{7.60, 4.00}
    \fermion[showArrow=false, lineWidth=4]{4.60, 4.00}{5.80, 5.20}
    \fermion[lineWidth=4, showArrow=false]{6.20, 4.00}{5.80, 5.20}
    \fermion[lineWidth=4]{9.80, 4.00}{13.20, 4.00}
    \electroweak[lineWidth=4]{11.40, 4.80}{10.80, 5.20}
    \fermion[lineWidth=4]{4.00, 6.40}{7.60, 6.40}
    \electroweak[label=$\mathcal{O}(\hbar)$, lineWidth=4]{11.40, 5.80}{12.00, 5.20}
    \electroweak[flip=true, lineWidth=4]{11.40, 4.80}{12.00, 5.20}
    \fermion[showArrow=false, lineWidth=4]{11.40, 4.00}{11.40, 4.80}
\end{feynman}
}
     \caption{Tree level diagram (left) vs. loop level (one loop) (right) diagram in WEFT}
     \label{mfig}
 \end{figure}
At this point, we are not interested in the quantum effective action as the quantum corrections are very much sub-leading. Hence we have neglected $\mathcal{O}(\hbar^n)$ terms in right hand side of (\ref{stotal1}) and focused only on the tree-level diagrams, i.e. $\mathcal{O}(\hbar^0)$ diagrams as shown in Fig.~(\ref{mfig}). Furthermore, we expand the classical effective action in the powers of $(v/c).$ This expansion is known as the Post-Newtonian (PN) expansion, and the n-PN order is of $\mathcal{O}(\frac{v}{c})^{2n}$. It is interesting to notice that to compute the tree-level, i.e., $\mathcal{O}(\hbar^0)$ PN diagrams, we need to compute several Feynman loop integrals. These tree-level PN diagrams are sometimes called ``Classical Loops''. For more details, interested readers are suggested to consult \cite{Burgess:2020tbq,Porto:2016pyg}. As we explained before, EFT approaches are very useful when there is a clear separation of scales, which is the case for the binary inspiral problem. The different length scales for this case are:
\begin{itemize}
    \item The size of the astrophysical object $R\sim G\,M$.
    \item The orbital radius r.
    \item The wavelength of radiation: $\lambda$.
\end{itemize}
These three parameters are not independent but depend on each other by the relative velocity of the binaries $v$ by, $\frac{R}{r}\sim v^2$. Keeping in mind the sensitivity of the gravitational wave detector, one can roughly assume: $\frac{1}{\lambda}\sim \omega_{GW}\sim \frac{v}{r}=\Omega_{orbit}$. Due to these connections, one can convince that the different orders in PN expansion have the information of different physical length scales with the following hierarchy: $R\ll r \ll\lambda$. Hence, in EFT language, the two field modes: the heavy and light ones, are decoupled, and their dynamics can be treated separately.  
\par
\begin{figure}
    \centering
    \includegraphics[scale=0.20]{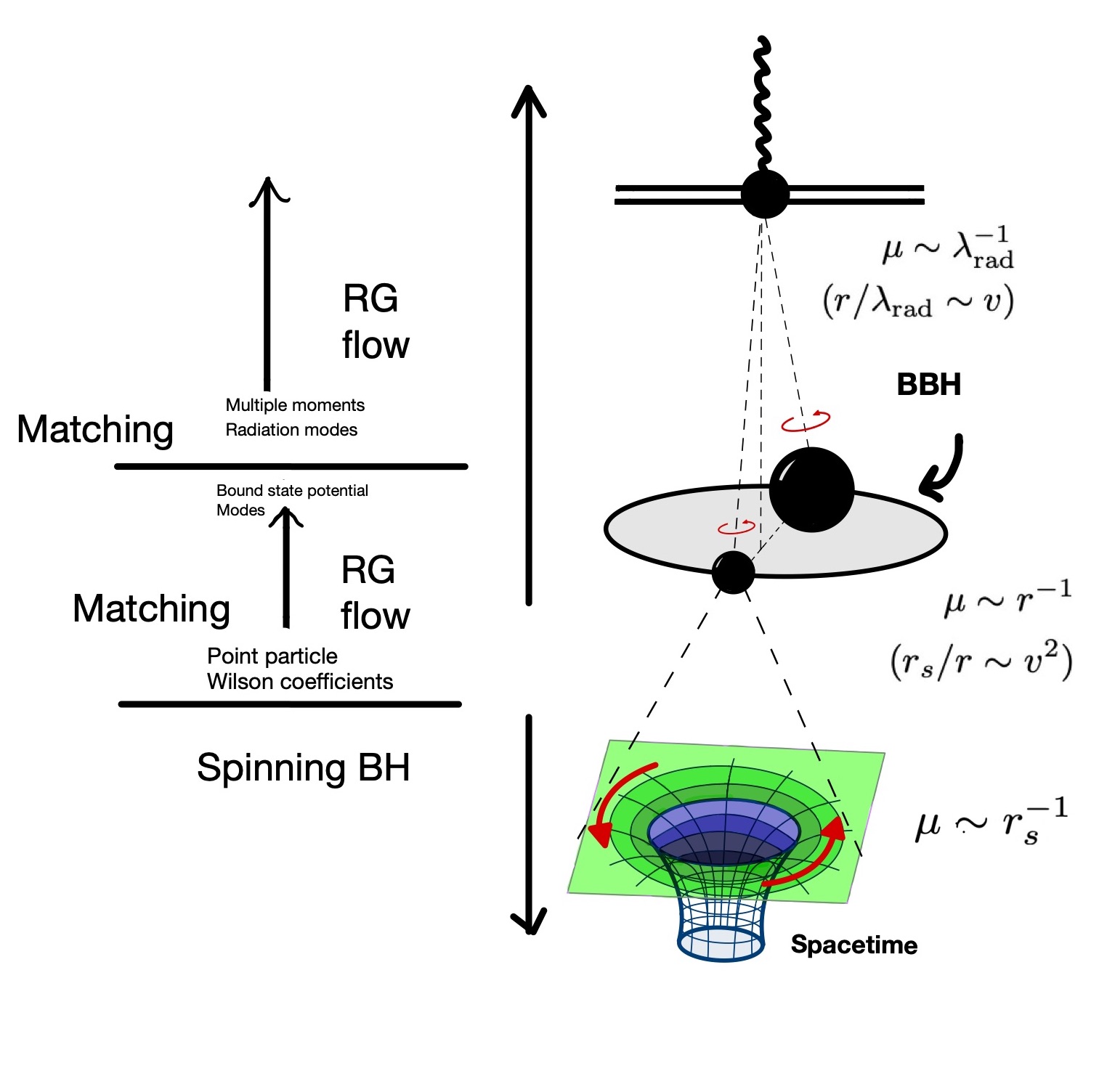}
    \caption{Different scales in the binary inspiral problem. We reproduce the figure from \cite{Porto:2016pyg}.}
    \label{newfig}
\end{figure}

\textcolor{black}{In the Worldline Effective Field Theory (WEFT) approach used to study the binary inspiral problem in this paper, apart from field degrees of freedom we do have worldline degrees of freedom $\{x^{\mu}_{a}\}$. We treat the worldline degrees of freedom as non-dynamical DOFs. Then integrating out the heavy modes, we get the total effective action: $\mathcal{S}_{\text{eff}}^{\text{tot}}[\Vec{x}_a,\Bar{\xi}]$ which has the following form,
\begin{align}
    \begin{split}
        \mathcal{S}_{\text{eff}}^{\text{tot}}[\Vec{x}_a,\Bar{\xi}]=-i\,\log\,\mathcal{Z}[\Vec{x}_a,\Bar{\xi}]=\underbrace{\mathcal{S}_{\text{eff}}^{\text{cons}}[\Vec{x}_a]}_{\text{real}}+\underbrace{\mathcal{S}_{\text{eff}}^{\text{rad}}[\Vec{x}_a,\bar\xi]}_{\text{complex}}.
    \end{split}
\end{align}
Extremizing $\mathcal{S}_{\text{eff}}^{\text{cons}}[\Vec{x}_a]$ we get the conservative dynamics of the binary system:
\begin{align}
    \begin{split}
     &    \frac{\delta}{\delta \Vec{x}_a(t)}\,\mathcal{S}_{\text{eff}}^{\text{cons}}[\Vec{x}_a]\Big |_{\mathcal{O}(\hbar^0)}=0 \implies \text{Orbit equation of the binary system.}
    \end{split}
\end{align}
and, $ \mathcal{S}_{\text{eff}}^{\text{rad}}[\Vec{x}_a,\bar\xi]$ is in general complex and related to the radiated power from the system via optical theorem as,
\begin{align}
    \begin{split}
        \,\text{Im}[\gamma_{\text{eff}}[\Vec{x}_a]]\equiv \text{Im}[-i\log\rmint \mathcal{D}\bar\xi \,e^{i\mathcal{S}_{\text{eff}}^{\text{rad}}[\Vec{x}_a,\bar\xi]}]=\frac{\mathcal{T}}{2}\rmint dE\,d\Omega \frac{d^2\Gamma}{dEd\Omega},\,\text{with radiated power},\, dP=E\,d\Gamma.
    \end{split}
\end{align}}
\\
The hierarchy of EFT for the binary inspiral problem is shown in Fig.~(\ref{newfig}).  
\section{Review of worldline effective field theory for a point particle in GR}\label{Sec3}
In this section, we briefly discuss the action of an inspiralling non-spinning black hole binary in Einstein theory. The background metric is chosen to be in the Kaluza-Klein form, consisting of a non-relativistic gravitational field. 
We start with the following action, 
\begin{equation}
    S=S_{\text{EH}}+S_{\text{pp}} \label{2.1}
\end{equation}
where the Einstein-Hilbert (EH) action is given by,
\begin{eqnarray}
    S_{EH}=\frac{1}{16\pi G}\rmint d^4x \sqrt{-g}\,R
\end{eqnarray}
and describes the dynamics of the graviton. $S_{\text{pp}}$ models the binary (determines the dynamics of the two-body system (black holes or any other compact objects)) and takes the following form, 
\begin{eqnarray} \label{eqnew}
    S_{pp}=\sum_{a=1}^2 m_a \rmint d\tau_{a}+\sum_{a=1}^2 C^{\text{R}}_a \rmint d\tau_{a} R({x_a})+\sum_{a=1}^2 C_{a}^{\text{V}}\rmint d\tau_a R_{\mu\nu}(x_a)\,\dot{x}_a^\mu \,\dot{x}_\nu^a+.....\,.
\end{eqnarray}

Here, $\tau_a$ denotes the proper time along the particle's world line and  $C^R$, $C^V$ are the Wilson coefficients. In this paper, we will only consider non-spinning binaries; we can ignore the finite size effects. Henceforth, we will not consider the proportional to $C^R$, $C^V$ from (\ref{eqnew}).  \par
A gravitational wave detector such as LIGO measures the gravitational waveform. The first step towards computing such waveforms is to calculate the radiated energy flux from the binary system. Given the action in (\ref{2.1}) we can calculate this quantity by considering the perturbation around a flat metric, i.e., $g_{\mu\nu}=\eta_{\mu\nu}+h_{\mu\nu}$. Integrating out the graviton field, one can obtain the effective action for the point particles,
\begin{eqnarray}
    e^{i\,\mathcal{S}_{\text{eff}}}=\rmint \mathcal{D}h_{\mu\nu} \exp{[iS_{EH}+iS_{pp}]}\,.
\end{eqnarray}\label{2.4}
The real part of the effective action describes the dynamics of the two-body system, and the imaginary part measures the graviton emission from the binary system over a large time. In principle, one can directly evaluate the full relativistic path integral using the Feynman rules for the Einstein-Hilbert action, which involves tree-level as well as loop-level diagrams. However, after calculating the amplitudes, it is not easy to take the non-relativistic limit. In order to overcome the problem, following \cite{Kol:2007bc}, one can decompose the metric in terms of the non-relativistic gravitational fields and then perform the calculations. We will only focus on the tree-level diagrams relevant to classical results. The methodology is discussed in detail in the subsequent sections. \par

\subsubsection*{Non relativistic general relativity}
In this section, we will briefly discuss how one can arrive at the effective action in a non-relativistic regime \cite{Goldberger:2004jt,Levi:2018nxp,Porto:2016pyg}. We expand the point particle action in powers of the particle's three-velocities.
\begin{eqnarray}
    S_{pp}=\sum_{a=1}^2\Big[\frac{1}{2}v_a^2-\frac{1}{2}h_{00}-h_{0i}\,v^a_{i}-\frac{1}{2}h_{00}v_{a}^2-\frac{1}{2}h_{ij}v^{i}_a\,v^{j}_a+\frac{1}{8}v_a^4+.......\Big]\,.
\end{eqnarray}
Here we should remember that all components of the metric perturbations $h_{\mu\nu}$ are evaluated at the particular particle world line.\par 
However, at this point, the propagator for $h_{\mu\nu}$ is fully relativistic. Hence, we can not distinguish between the potential gravitons (heavy modes) and the long-wavelength radiation gravitons (light modes). We are not interested in potential gravitons as the gravitational wave detector can not detect them. So due to our interest in finding out the effective action in the non-relativistic regime, we integrate over the potential gravitons. Now, to connect with the discussion in Sec.~(\ref{Sec2}) about integrating out the heavy modes, we decompose the metric perturbation as follows,
\begin{eqnarray}
    h_{\mu\nu}=\Bar{h}_{\mu\nu}+
\mathcal{H}_{\mu\nu}\,,\label{3.6n}
\end{eqnarray}
where $\mathcal{H}_{\mu\nu}$ is the potential gravitons with the properties $\partial_{i}\mathcal{H}_{\mu\nu}\sim\frac{\mathcal{H}_{\mu\nu}}{r},\partial_{0}\mathcal{H}_{\mu\nu}\sim{\mathcal{H}_{\mu\nu}}\frac{v}{r}$. And $\Bar{h}_{\mu\nu}$ is the radiation gravitons with $\partial_{\beta}\bar{h}_{\mu\nu}\sim{\bar{h}_{\mu\nu}}\frac{v}{r}$. It is more convenient to express $\mathcal{H}_{\mu\nu}$ as,
\begin{eqnarray}
    \mathcal{H}_{\mu\nu}(x^\mu)=\rmint\frac{d^3\Vec{k}}{(2\pi)^3}\,e^{i\Vec{k}.\Vec{x}}\,\mathcal{H}^{k}_{\mu\nu}(x^0)\,.
\end{eqnarray}
Now, the conservative (or radiative) effective action for the binaries $S_{\text{NRGR}}$ can be computed by integrating over the non-relativistic potential gravitons,
\begin{eqnarray}
    e^{i\mathcal{S}_{\text{eff}}[x_a,\Bar{h}_{\mu\nu}]}=\rmint \mathcal{D}\mathcal{H}_{\mu\nu}\,\exp{[iS[\Bar{h}+\mathcal{H},x_a]+iS_{GF}]}\label{2.8}
\end{eqnarray}
where, $S_{GF}$ is the gauge fixing term. While evaluating the path integral, ghost terms arise in the effective action, but at the classical level, they do not contribute. Now, to preserve the gauge invariance, we choose the gauge fixing term as \cite{Goldberger:2004jt},
\begin{eqnarray}
    S_{GF}=\rmint d^4x\sqrt{\Bar{g}}\,\Gamma_\mu \Gamma^\mu
\end{eqnarray}
which is invariant under the general coordinate transformation of the background metric $\Bar{g}_{\mu\nu}=\eta_{\mu\nu}+\Bar{h}_{\mu\nu}$ and $\Gamma_\mu=\nabla_{\alpha}\mathcal{H}^{\alpha}_{\mu}-\frac{1}{2}\nabla_{\mu}\mathcal{H}_{\alpha}^{\alpha}.$ From the EH Lagrangian (upto $\mathcal{O}(\mathcal{H}^2)$) in momentum space one can write down the \textcolor{black}{orbital} graviton propagator as \cite{Goldberger:2004jt},
\begin{eqnarray}
    \Big\langle\mathcal{H}^{k}_{\mu\nu}(x^0)\mathcal{H}^{q}_{\alpha\beta}(0)\Big\rangle=-(2\pi)^3 \delta^{(3)}(\Vec{k}+\Vec{q})\delta(x^0)\frac{i}{\Vec{k}^2}P_{\mu\nu,\alpha\beta}\label{2.10}
\end{eqnarray}
with, $P_{\mu\nu,\alpha\beta}=\frac{1}{2}[\eta_{\mu\alpha}\eta_{\nu\beta}+\eta_{\mu\beta}\eta_{\nu\alpha}-\frac{2}{d-2}\eta_{\mu\nu}\eta_{\alpha\beta}]$. So in order to find $S_{NRGR}$ one have to compute the functional integral in (\ref{2.8}) using the (\ref{2.10}). To do this, one has to sum over the relevant Feynman diagram, which has the following topological properties \cite{Goldberger:2022ebt, Goldberger:2004jt},
\begin{itemize}
    \item All diagrams must be connected with stripped-off particle worldlines.
    \item Diagram can contain internal graviton potential modes ($\mathcal{H}_{\mu\nu}$) but can not have external potential mode.
    \item Radiative graviton modes ($\Bar{h}_{\mu\nu}$) are only appear in the external lines.
\end{itemize}
As there is a gauge redundancy, it is better to choose a particular gauge from the beginning and use the proper decomposition of the non-relativistic graviton field \cite{Kol:2007bc}. This helps us to identify the correct diagrams at each PN order and avoid any confusion arising due to the mixing of various components of $h_{\mu\nu}$. Hence, following \cite{Kol:2007bc}, we choose the standard Kaluza-Klein (KK) decomposition,
\begin{align}
    \begin{split}
        ds^2=g_{\mu\nu}dx^{\mu}dx^{\nu}=-e^{2\psi}(dt-\tilde{\mathcal{A}_i}dx^{i})^2+e^{-2{\psi}}\gamma_{ij}dx^{i}dx^{j}\,.
    \end{split} \label{KK}
\end{align}
In the component form:
$$g_{\mu\nu}=
\begin{pmatrix}
-e^{2\psi} & e^{2\psi} \tilde{\mathcal{A}}_{j}\\
e^{2\psi} \tilde{\mathcal{A}}_{i} & \,\,e^{-2\psi}\gamma_{ij}-e^{2\psi}\tilde{\mathcal{A}}_{i}\tilde{\mathcal{A}}_{j}\\
\end{pmatrix}$$
where $\psi,\tilde{\mathcal{A}_i},\gamma_{ij}=\delta_{ij}+\sigma_{ij}$ are the set of \textit{non-relativistic graviton field} (NRG), sometimes called as Kol-Smolkin variables, are Newtonian potential, gravitomagnetic potential, and metric 3-tensor, respectively \cite{Kol:2007bc}.
For our future computations to make all NRG fields have mass-dimension one, one can rescale the fields as follows,
\begin{align}
    \begin{split}
        \psi\rightarrow\frac{\psi}{m_p},\,\,\tilde{\mathcal{A}}_{i}\rightarrow\frac{\tilde{\mathcal{A}}_{i}}{m_p},\,\,\sigma_{ij}\rightarrow \frac{\sigma_{ij}}{m_p}.
    \end{split}
\end{align}
\section{Conservative dynamics of binary black holes in a theory of ALP and ULV }\label{Sec4}
In this section, we describe the action for inspiralling binary black holes in a theory of \textit{Axion-like particles and Ultra-light vectors}. Apart from the usual EH term, it contains a Maxwell field, a Proca field, and a massive scalar field.  We have the following action for this theory \cite{Cardoso:2018tly}:
\begin{align}
\begin{split}
S_{\text{field}}=\rmint d^4x\sqrt{-g}&\Big(\frac{m_{p}^2}{2}R-\frac{1}{4}F_{\mu\nu}F^{\mu\nu}-\frac{1}{2}\partial_\mu\phi\partial^\mu\phi-\frac{1}{2}m^2\phi^2+\frac{g_{a\gamma\gamma}}{4\,m_{p}}\phi F_{\mu\nu}^*F^{\mu\nu}-\frac{1}{4}B_{\mu\nu}B^{\mu\nu}\\ &
+\frac{\gamma}{2}F_{\mu\nu}B^{\mu\nu}
-\frac{1}{2}\mu_{\gamma}^2B_{\mu}B^{\mu}\Big)\,.
\label{2.11}
\end{split}
\end{align}
The action in (\ref{2.11}) describes the Axion-like-particles (ALPs) (scalar), massless photon, and massive (often known as \textit{dark photon}) gauge field (Proca) minimally coupled with gravity. In (\ref{2.11}) $R,\, \phi $ are the Ricci scalar and ALP. Also, 
$$F_{\mu\nu}=\partial_{\mu} {A}_{\nu}-\partial_{\nu} {A}_{\mu},\, B_{\mu\nu}=\partial_{\mu} {B}_{\nu}-\partial_{\nu} {B}_{\mu},$$ where ${A}_{\mu}$ and ${B}_{\mu}$ are Electromagnetic and Proca field respectively. $\gamma$ is the coupling constant between photon and dark photon. Furthermore, $g_{a\gamma\gamma}$ is the coupling constant for axion-photon, and it has a negative mass dimension, implying the theory is not renormalizable, and the term $\phi F^{*} F$ is sometimes called the \textit{Theta term}. However, that is not a problem for our case as we finally intend to write down a classical effective action for the light modes in powers of $(v/c)$ as mentioned in Sec.~(\ref{Sec2}). A good EFT does not need to be renormalized. We are only concerned about the tree-level diagrams as we want classical results. Eventually, we will encounter loop integrals while computing the effective action, and we call them classical loops as they are proportional to $\hbar ^0$. \par

Next we compute the effective action at $\mathcal{O}{(\hbar^0)}$ and perturbatively at $\mathcal{O}(\frac{v}{c})^n$. To do that,  we work in the non-relativistic regime and choose the background metric in KK form as mentioned in (\ref{KK}). We work in the non-relativistic regime of the KK metric and consider the non-trivial worldline coupling of the Non-relativistic gravitational (NRG) Fields. We compute the effective action for the binaries \textit{upto 1PN in gravitational, electromagnetic, Proca and upto 2PN for scalar interaction.} Also, we will show that the leading order correction from the  CP-violating term ($ g_{a\gamma \gamma}$ term) in  (\ref{2.11}) to the effective action comes in such a way that it affects the conservative dynamics at \textit{2.5 PN}. \par

 We use the following prescription to calculate the contribution of the following term in the Lagrangian. Our goal is to calculate the effective action in the given theory. The EFT of radiation fields, in KK parametrization, can be obtained by decomposing fields in the potential and radiation sectors as, $\psi=\Tilde{\psi}+\Bar{\psi},\Tilde{\mathcal{A}}_i=\Hat{\mathcal{A}}_i+\Bar{\mathcal{A}}_i,\sigma_{ij}=\zeta_{ij}+\Bar{\sigma}_{ij}\,,\phi=\varphi+\Bar{\phi}, A_{\mu}=\mathcal{A}_{\mu}+\Bar{a}_{\mu}$ and $B_{\mu}=\mathcal{B}_{\mu}+\Bar{b}_{\mu}$ , and integrating over the potential modes. The partition function has the following form:
\begin{align}
\begin{split}
  \mathcal{Z}[\Vec{x}_a,\Bar{\psi},\Bar{{\mathcal{A}}}_i,\Bar{\sigma}_{ij},\Bar{a}_{\mu},\Bar{b}_{\mu}]  &:=e^{i\mathcal{S}_{\text{eff}}[\Vec{x}_a,\Bar{\psi},\Bar{{\mathcal{A}}}_i,\Bar{\sigma}_{ij},\Bar{a}_{\mu},\Bar{b}_{\mu}]}\,,\\ &
  =\rmint \underbrace{\mathcal{D}\tilde{\psi} \,\mathcal{D}\hat{\mathcal{A}}_{i}\,\mathcal{D}\varphi \,\mathcal{D}\mathcal{A}_{\mu}\,\mathcal{D}\mathcal{B}_{\nu}\,
  \mathcal{D}\boldsymbol{\zeta}_{ij}\,}_{\text{integrating over potential modes}\,(\mathcal{D}\hat{\xi})}\,e^{iS_{\text{field}}+iS_{\text{pp}}}\,.\label{2.13}
\end{split}
\end{align}
We can compute the effective action in (\ref{2.13}) by evaluating the  amplitudes of the suitable Feynman diagrams by taking the logarithm as,
\begin{align}
    \begin{split}
        i\mathcal{S}_{\text{eff}}[\Vec{x}_a,\Bar{\xi}]&=\log\Big[{\rmint\mathcal{D}\hat\xi\,e^{iS_{\text{field}}+iS_{\text{pp}}}}\Big]\,.\label{4.3mm}
    \end{split}
\end{align}
One can compute (\ref{4.3mm}) perturbatively by evaluating the connected Feynman diagrams. We need two kinds of vertices, worldline vertices and field vertices. To get the worldline vertices, we need to consider the point particle action, including all the fields.
Now we can write down the field action by, $S_{\text{field}}=S_{\text{quad}}+S_{\text{int}}$ and use,
\begin{align}
    \begin{split}
        \Big\langle\mathcal{O}_{1}(\Vec{x}_1(t_1))...\mathcal{O}_{n}(\Vec{x}_n(t_n))\Big\rangle=\rmint\mathcal{D}\hat\xi \,e^{iS_{\text{quad}}}\,\underbrace{\mathcal{O}_{1}(\Vec{x}_1(t_1),\hat\xi)...\mathcal{O}_{n}(\Vec{x}_n(t_n),\hat\xi)}_{\text{got expanding \,$e^{iS_{\text{pp}}+iS_{\text{int}}}$}}
    \end{split}
\end{align}
and the wick contraction to reduce the higher point correlation function into the two-point functions to compute the effective action.
\par
In order to evaluate suitable Feynman diagrams, one needs to identify the Feynman propagators (two-point functions) as discussed earlier. To do this, we first write down the action for different field configurations and then identify the corresponding propagators. Next, we will give the details. We write down the action in the KK parametrization and separate the terms containing the spatial and temporal derivatives. Then we can identify the propagators. \\

\textbf{Propagators of different fields:} We list the two-point functions, i.e. the propagators of different fields in our theory. We also do the conventional non-relativistic expansion of the propagator, assuming $|k_0|\ll |\Vec{k}|$. In this limit, we expand the relativistic Feynman propagator, which contains an infinite sum of the delta function and its time derivatives (even). We only consider the non-relativistic contribution and the first-order relativistic corrections.  \\\\
\underline{\textit{Scalar propagator:}}\\
 In KK parametrization, the scalar action has the following form:
 \begin{align}
    \begin{split}
        S_{\phi}&=\rmint d^4x \sqrt{-g}\Big[-\frac{1}{2}\partial_{\mu}\phi\partial^{\mu}\phi-\frac{1}{2}m^2 \phi^2\Big]\,,\\ &
        =\rmint d^3x \,dt \sqrt{\gamma}\Big[\frac{e^{-4\psi}}{2}\partial_{0}\phi\partial_{0}\phi-2\Hat{\mathcal{A}}^{i}\partial_{i}\phi\partial_{0}\phi-\frac{1}{2}\gamma^{ij}\partial_{i}\phi\partial_{j}\phi-\frac{1}{2}e^{-2\psi}m^2\phi^2 \nonumber
        \Big] \,,\end{split}
\end{align}
\begin{align}
    \begin{split}
\,\,\,\,\,\,\,\,\,\,\,\,\,\,\,\,\,\,\,\,\,\,\,\,\,\,\,=\rmint d^3x dt \,\Big(1+\frac{\text{Tr}(\sigma)}{2}\Big)&\Big[\frac{1}{2}\partial_{0}\phi\partial_{0}\phi-2\psi \partial_{0}\phi\partial_{0}\phi-2\Hat{\mathcal{A}}^{i}\partial_{i}\phi\partial_{0}\phi-\frac{1}{2}\delta^{ij}\partial_{i}\phi\partial_{j}\phi \\&+ \frac{1}{2}\sigma^{ij}\partial_{i}\phi\partial_{j}\phi-\frac{1}{2}(1-2\psi)m^2\phi^2\Big]\,.\label{24m}
    \end{split}
\end{align}
From the action (\ref{24m}), one can identify the scalar propagator as follows,
 \begin{align}
    \begin{split}
\Big\langle\varphi(x_1)\varphi(x_2)\Big\rangle &=\rmint \frac{d^4 k}{(2\pi)^4} \frac{e^{ik\cdot(x_1-x_2)}}{-k^2-m^2}\\&=\rmint \frac{dk_0}{2\pi}\,e^{-ik_0 (t_1-t_2)}\rmint \frac{d^3k}{(2\pi)^3}\frac{e^{i\Vec{k}\cdot(\Vec{x}_1-\Vec{x}_2)}}{\Vec{k}^2+m^2}\Big[1+\frac{k_0^2}{\Vec{k}^2+m^2}+...\Big]
        \\ &
        =\delta(t_1-t_2)\rmint \frac{d^3k}{(2\pi)^3}\frac{e^{i\Vec{k}\cdot\Vec{r}}}{\Vec{k}^2+m^2}+\partial_{t_1}\partial_{t_2}\delta(t_1-t_2)\rmint \frac{d^3k}{(2\pi)^3}\frac{e^{i\Vec{k}\cdot\Vec{r}}}{(\Vec{k}^2+m^2)^2}+...\,. 
    \end{split}
\end{align}
\underline{\textit{ Electromagnetic propagator:}}\\
The action for free Maxwell theory is given by,
\begin{align}
    \begin{split}
        S_{\text{EM}}=-\frac{1}{4}\rmint d^3x dt \sqrt{\gamma}&\Big[-F_{0i}F_{0i}+e^{2\psi}\gamma^{ij}\gamma^{kl}F_{ik}F_{jl}+e^{2\psi}\Hat{\mathcal{A}}^{i}\Hat{\mathcal{A}}^{j}F_{0j}F_{i0}+e^{2\psi}\Hat{\mathcal{A}}^{i}\gamma^{jk}F_{0j}F_{ik}\Big]\\ &
        +(\text{higher order in fields})\,.\label{25m}
    \end{split}
\end{align}
From the quadratic part of the action (\ref{25m}), one can identify the propagator of the electromagnetic field as follows,
\begin{align}
    \begin{split}
&\Big\langle\mathcal{A}_{0}(x_1)\mathcal{A}_0(x_2)\Big\rangle=\textcolor{black}{-}\Big[\delta(t_1-t_2)\rmint \frac{d^3k}{(2\pi)^3}\frac{e^{i\Vec{k}\cdot\Vec{r}}}{\Vec{k}^2}+\partial_{t_1}\partial_{t_2}\delta(t_1-t_2)\rmint \frac{d^3k}{(2\pi)^3}\frac{e^{i\Vec{k}\cdot\Vec{r}}}{\Vec{k}^4}...\Big]\,,\\ &
        \Big\langle\mathcal{A}_{i}(x_1)\mathcal{A}_k(x_2)\Big\rangle=\Big[\delta(t_1-t_2)\rmint \frac{d^3k}{(2\pi)^3}\frac{e^{i\Vec{k}\cdot\Vec{r}}}{\Vec{k}^2}+\partial_{t_1}\partial_{t_2}\delta(t_1-t_2)\rmint \frac{d^3k}{(2\pi)^3}\frac{e^{i\Vec{k}\cdot\Vec{r}}}{\Vec{k}^4}...\Big]\delta_{ik}\,.\\ &
    \end{split}
\end{align}
\underline{\textit{ Proca propagator:}}

The action for the Proca field is the same as (\ref{25m}) with a mass term, hence the propagator has the following form:
\begin{align}
    \begin{split}
    &\Big\langle\mathcal{B}_{0}(x_1)\mathcal{B}_0(x_2)\Big\rangle=\textcolor{black}{-}\Big[\delta(t_1-t_2)\rmint \frac{d^3k}{(2\pi)^3}\frac{e^{i\boldsymbol{k}\cdot\boldsymbol{r}}}{\boldsymbol{k}^2+\mu_{\gamma}^2}+\partial_{t_1}\partial_{t_2}\delta(t_1-t_2)\rmint \frac{d^3k}{(2\pi)^3}\frac{e^{i\boldsymbol{k}\cdot\boldsymbol{r}}}{(\boldsymbol{k}^2+\mu_{\gamma}^2)^2}+.....\Big]\,.\\ &
\Big\langle\mathcal{B}_{i}(x_1)\mathcal{B}_k(x_2)\Big\rangle=\Big[\delta(t_1-t_2)\rmint \frac{d^3k}{(2\pi)^3}\frac{e^{i\boldsymbol{k}\cdot\boldsymbol{r}}}{\boldsymbol{k}^2+\mu_{\gamma}^2}+\partial_{t_1}\partial_{t_2}\delta(t_1-t_2)\rmint \frac{d^3k}{(2\pi)^3}\frac{e^{i\boldsymbol{k}\cdot\boldsymbol{r}}}{(\boldsymbol{k}^2+\mu_{\gamma}^2)^2}...\Big]\delta_{ik}\,.\label{2.28jjj}
    \end{split}
\end{align}
The propagator given in \eqref{2.28jjj} is not the full expression; in general, it also contains a correction term of the form \(\frac{k^\mu k^\nu}{\mu_{\gamma}^2}\). However, this additional piece contributes only at subleading order to both the potential and the waveform. Since, for the massive modes, our interest is restricted to the leading contribution, we neglect these extra terms.\\\\
\underline{\textit{Graviton propagators:}}\\
Lastly, the gravitational action takes the following form in KK parametrization:
\begin{align}
    \begin{split}
        S_{\text{G}}=\frac{m_{p}^2}{2}\rmint d^3x dt \,\sqrt{\gamma}&\Big[R[\gamma]-2\gamma^{ij}\partial_{i}\psi\partial_{j}\psi+\frac{e^{4\psi}}{4}\Hat{F}_{ij}\Hat{F}_{kl}\gamma^{ik}\gamma^{jl}+\frac{e^{-4\psi}}{4}[\Dot{\gamma}_{ij}\Dot{\gamma}_{kl}\gamma^{ik}\gamma^{jl}-(\gamma^{ij}\Dot{\gamma}_{ij})^2]\\ &
-4\gamma^{ij}\Dot{\Hat{\mathcal{A}}}\,\partial_{j}\psi
        +e^{-4\psi}(2\Dot{\psi}\gamma^{ij}\Dot{\gamma}_{ij}-6\Dot{\psi}^2)\Big]\,.\label{23m}
    \end{split}
\end{align}
For the action in (\ref{23m}), one can identify the static graviton propagator and its relativistic time corrections as:
\begin{align}
  \begin{split}
&\Big\langle\tilde{\psi}(x_1)\tilde{\psi}(x_2)\Big\rangle=\frac{1}{2}\Big[\delta(t_1-t_2)\rmint \frac{d^3k}{(2\pi)^3}\frac{e^{i\Vec{k}\cdot\Vec{r}}}{\Vec{k}^2}+\partial_{t_1}\partial_{t_2}\delta(t_1-t_2)\rmint \frac{d^3k}{(2\pi)^3}\frac{e^{i\Vec{k}\cdot\Vec{r}}}{\Vec{k}^4}...\Big]\,,\\ &
        \Big\langle\Hat{\mathcal{A}}_{i}(x_1)\Hat{\mathcal{A}}_k(x_2)\Big\rangle=-2\Big[\delta(t_1-t_2)\rmint \frac{d^3k}{(2\pi)^3}\frac{e^{i\Vec{k}\cdot\Vec{r}}}{\Vec{k}^2}+\partial_{t_1}\partial_{t_2}\delta(t_1-t_2)\rmint \frac{d^3k}{(2\pi)^3}\frac{e^{i\Vec{k}\cdot\Vec{r}}}{\Vec{k}^4}...\Big]\delta_{ik}\,,\\ &
        \Big\langle\boldsymbol{\zeta}_{ij}(x_1)\boldsymbol{\zeta}_{kl}(x_2)\Big\rangle=4\Big[\delta(t_1-t_2)\rmint \frac{d^3k}{(2\pi)^3}\frac{e^{i\Vec{k}\cdot\Vec{r}}}{\Vec{k}^2}+\partial_{t_1}\partial_{t_2}\delta(t_1-t_2)\rmint \frac{d^3k}{(2\pi)^3}\frac{e^{i\Vec{k}\cdot\Vec{r}}}{\Vec{k}^4}...\Big]P_{ij,kl} \,.
    \end{split}
\end{align}
We denote the relativistic time correction to the non-relativistic propagator as $\Big\langle....\Big\rangle_{\varoslash}$.\\\\
\textbf{Point Particle action:} In our analysis, we model the binaries as charged point particles, ignoring any internal structure of the astrophysical objects under consideration. Our first task is to identify the worldline operators. For that, we first write down the point particle worldline action. As we have a massive scalar field in the theory, we expect the objects' masses to depend on the scalar field \cite{Dyadina:2018ryl}. We also assume that the binary possesses electromagnetic and Proca charges (dark charges). So we include the coupling between the charges (electromagnetic (EM) and Proca) with the relevant fields. Here, we ignore the finite-size effects terms. 
The action is therefore given (at leading order in gauge fields) by,
\begin{align}
    \begin{split}
        S_{pp}=&-\sum_{a}\rmint m_a(\phi)d\tau_{a}+\sum_a\rmint\,Q_a \,d\tau_a \textcolor{black}{A^\mu}\,v_\mu+\sum_a\rmint\,Q_a'\, d\tau_a B^\mu\,v_\mu \,,\\ &
        =\sum_{a}\, m_a\rmint dt\Big[-1+\frac{1}{2}v_a^2-\frac{\psi}{m_{p}}-\frac{\hat{\mathcal{A}}_{i}}{m_{p}}\,v^i_{a}-\frac{\psi^2}{2m_{p}^2}-\frac{3}{2}\frac{\psi}{m_{p}} v_{a}^2\textcolor{black}{+}\frac{1}{2}\frac{\sigma_{ij}}{m_{p}}v^{i}_a\,v^{j}_a+\frac{1}{8}v_a^4+\mathcal{O}(v^6)\Big]\Tilde{\mathcal{L}} \\ &
        +\sum_a \rmint dt \Big[Q_a\,A_{0}(1-\frac{\psi}{m_{p}}-\frac{\psi^2}{2m_{p}^2}) +Q_a\,v^{i}_a A_{i}(1-\frac{\psi}{m_{p}}-\frac{\psi^2}{2m_p^2})+\cdots\Big]\\ &
        \sum_a \rmint dt \Big[Q_a'\,B_{0}(1-\frac{\psi}{m_{p}}-\frac{\psi^2}{2m_{p}^2}) +Q_a'\,v^{i}_a B_{i}(1-\frac{\psi}{m_{p}}-\frac{\psi^2}{2m_p^2})+\cdots\Big]\,, \label{4.10}
    \end{split}
 \end{align}
 where, $\Tilde{\mathcal{L}}=\Big[1+s_a\,\frac{\phi}{m_{p}}+g_a\,\frac{\phi^2}{m_{p}^2}\Big].$ Also, $a=1,2$ denotes the two objects the binary system consists of. We have expanded $m_a(\phi)$ in Taylor series upto second order and  $s_a=\frac{1}{m_a}\frac{\partial m_a(\phi)}{\partial \phi}|_{\phi=0}, \,g_a=\frac{1}{m_a}\frac{\partial^2 m_a(\phi)}{\partial \phi^2}|_{\phi=0} $. \textit{ $Q_1'$ and $Q_2'$ denote two dark charges. They are not associated with an electric or magnetic field. They are sometimes also referred to as `hidden' charges because they would not interact with ordinary matter in the same way that electric and magnetic charges do.}
\footnote{Terms with time derivatives in fields will contribute as corrections to the non-relativistic propagator.  One can show that taking the 2-point self-interacting temporal vertices produces the same corrections to the non-relativistic propagator. 
}\\\\

\begin{table}[htb!]
\centering
\scalebox{0.70}{
\setlength{\arrayrulewidth}{0.3mm}
\setlength{\tabcolsep}{20pt}
\renewcommand{\arraystretch}{2.2}
\begin{tabular}{|p{3cm}|p{3cm}|p{3cm}|p{3cm}|p{3cm}|}
\hline
\multicolumn{4}{|c|}{EFT PN COUNTING} \\
\hline
Fields & Worldline-Coupling & Diagram &order \\
\hline
$\hat{\mathcal{A}}_i$& -$\frac{m_a}{m_{p}} \rmint dt\,\Hat{\mathcal{A}}^i\,v_i^a$ & \scalebox{0.3}{\begin{feynman}
    \gluon[lineWidth=4,color=fcb900,label=$\Hat{\mathcal{A}}^i$]{4.00, 5.20}{5.80, 5.20}
    \fermion[lineWidth=4, showArrow=true, flip=false]{4.00, 4.00}{4.00, 6.40}
   
\end{feynman}}& $\sim v$\\
\hline
$ \mathcal{A}_i$& $Q\rmint dt \mathcal{A}_i v^i$& \scalebox{0.3}{\begin{feynman}
    \fermion[lineWidth=4, showArrow=false]{4.00, 4.00}{4.00, 6.20}
    \electroweak[color=9900ef, lineWidth=4, label=$\mathcal{A}_i$]{4.00, 5.00}{5.40, 5.00}
\end{feynman}}& $\sim v$\\
\hline
$ \mathcal{B}_i$& $Q'\rmint dt\, \mathcal{B}_i \,v^i$& \scalebox{0.2}{\begin{feynman}
    \gluon[lineWidth=4, flip=true, endcaps=false, color=eb144c, label=$\mathcal{B}_i$]{4.00, 5.20}{5.60, 5.20}
    \fermion[lineWidth=4, showArrow=true, flip=false]{4.00, 4.00}{4.00, 6.40}
    
\end{feynman}}& $\sim v$\\
\hline
$ \mathcal{A}_0$& $Q\rmint dt\,  \mathcal{A}_0$& \scalebox{0.3}{\begin{feynman}
    \dashed[lineWidth=4, showArrow=false, label=$\mathcal{A}_0$,flip=true]{4.00, 5.00}{6.00, 5.00}
    \fermion[lineWidth=4]{4.00, 4.00}{4.00, 6.00}
    
\end{feynman}}& $\sim v^0$\\
\hline
$\tilde{\psi}$& $\frac{-m_a}{m_{p}}\rmint dt\, \,\tilde{\psi}$& \scalebox{0.3}{\begin{feynman}
    \electroweak[lineWidth=4, label=$\boldsymbol{\tilde{\psi}}$]{4.00, 5.20}{6.00, 5.20}
    \fermion[lineWidth=4]{4.00, 4.00}{4.00, 6.20}
\end{feynman}
}& $\sim v^0$\\
\hline
$ \mathcal{B}_0$& $Q'\,\rmint dt\,  \mathcal{B}_0$ & \scalebox{0.3}{\begin{feynman}
    \dashed[lineWidth=4, showArrow=false, color=9900ef, label=$\mathcal{B}_0$]{4.00, 5.20}{5.40, 5.20}
    \fermion[lineWidth=4, showArrow=true, flip=false]{4.00, 4.00}{4.00, 6.40}
   
\end{feynman}}& $\sim v^0$\\
\hline
$\varphi$& -$s_a\, \frac{m_a}{m_{p}} \rmint dt\, \varphi $ & \scalebox{0.3}{\begin{feynman}
    \electroweak[lineWidth=4, color=0693e3, label=$\boldsymbol{\varphi}$, flip=true]{4.00, 5.00}{5.40, 5.00}
    \fermion[lineWidth=4]{4.00, 4.00}{4.00, 6.20}
\end{feynman}}& $\sim v^0$\\
\hline
\end{tabular}
}
\caption{Table showing the different worldline couplings and the order at which they contribute.}
\label{table1m}
\end{table}

\textbf{EFT counting for different fields:}
As discussed in Sec.~(\ref{Sec2}), we will write the effective action for light modes upto certain PN order. This results in EFT power counting. Hence, we need to know how different field couplings with the worldline scale with $v\,.$ We use Table~(\ref{table1m}) to find out how the different terms in the conservative effective action scale with the velocities of the black holes. 
The contribution of effective action corresponding to every diagram has the following form $Lv^{2n}$, which we call the $n PN$ diagram.
\subsection{The Gravitational bound sector}
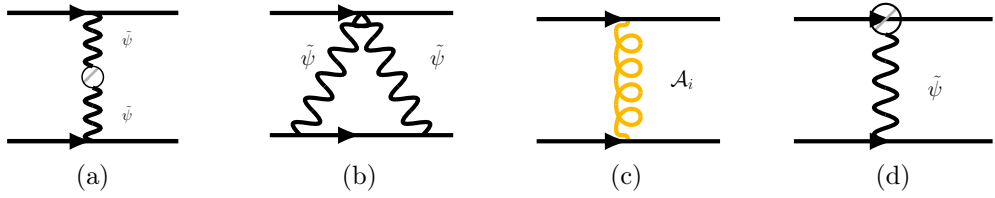
\begin{figure}
    \centering
    \begin{subfigure}{0.22\textwidth}
        \centering
\scalebox{0.28}{\begin{feynman}
    \electroweak[lineWidth=6, label=$\tilde{\psi}$]{5.60, 4.00}{5.60, 5.00}
    \fermion[lineWidth=6]{4.00, 6.40}{7.20, 6.40}
    \electroweak[label=$\tilde{\psi}$, lineWidth=6]{5.60, 5.40}{5.60, 6.40}
    \fermion[lineWidth=6]{4.00, 4.00}{7.20, 4.00}
    \parton{5.60,5.20}{0.20}
\end{feynman}
}
        \caption{}
        \label{fig1a}
    \end{subfigure}
    \begin{subfigure}{0.22\textwidth}
        \centering
\scalebox{0.4}{\begin{feynman}
    \electroweak[label=$\Tilde{\psi}$, lineWidth=4]{4.40, 4.00}{5.20, 5.60}
    \fermion[lineWidth=4]{4.00, 5.60}{6.40, 5.60}
    \electroweak[flip=true, label=$\Tilde{\psi}$, lineWidth=4]{6.00, 4.00}{5.20, 5.60}
    \fermion[lineWidth=4]{4.00, 4.00}{6.40, 4.00}
\end{feynman}}
        \caption{}
        \label{fig1b}
    \end{subfigure}
    \begin{subfigure}{0.22\textwidth}
        \centering
\scalebox{0.4}{\begin{feynman}
    \gluon[label=$\mathcal{A}_i$, color=fcb900, lineWidth=4]{5.20, 4.00}{5.20, 5.60}
    \fermion[lineWidth=4]{4.00, 5.60}{6.40, 5.60}
    \fermion[lineWidth=4]{4.00, 4.00}{6.40, 4.00}
\end{feynman}}
        \caption{}
        \label{fig1c}
    \end{subfigure}
    \begin{subfigure}{0.22\textwidth}
        \centering
\scalebox{0.4}{\begin{feynman}
    \fermion[lineWidth=4]{4.00, 5.60}{6.60, 5.60}
    \electroweak[lineWidth=4, label=$\Tilde{\psi}$]{5.20, 4.00}{5.20, 5.40}
    \fermion[lineWidth=4]{4.00, 4.00}{6.60, 4.00}
    \parton{5.20,5.60}{0.20}
\end{feynman}
}
        \caption{}
        \label{fig1d}
    \end{subfigure}
    \caption{Diagrams contributing to the gravitational bound sector at 1PN. }
    \label{fig:my_labela}
\end{figure}

Now, after identifying the propagators and given the power counting rules, we are ready to compute the effective action. As discussed in Sec.~(\ref{Sec2}), the effective action, in general, is a complex one. From the real part of the action, we get the equation of the orbit of the binary. We first focus on this bound sector. This will enable us to find the corrections to the usual Kepler's orbit upto some PN order. We will spell out the computation sector by sector. First, we will focus on the gravitational bound sector. 
Though its results are well known in the literature, we give a brief review of it for the sake of completeness \cite{Kuntz:2019zef}. Next, compute the effective action upto 1PN. In Fig.~(\ref{fig:my_labela}) we show all the diagrams that contribute to this order. Below, we calculate the amplitudes of these diagrams.\footnote{To do the integrals, we take the help of the master integrals given in Appendix~(\ref{ch1:app:E}).}
\begin{itemize}
\item The amplitude corresponds to the diagram in Fig.~(\ref{fig1a}) with worldline vertex $\frac{-m_a}{m_p}\rmint dt \Tilde{\psi} $:
\begin{align}
    \begin{split}
   \mathcal{S}_{\text{eff}}\Big|_{\text{fig}(\ref{fig1a})}& =\frac{m_1 m_2}{m_p^2}\rmint \Bar{\mathcal{D}}\hat\xi\rmint dt_1\,\tilde{\psi}(\vec{x_1}
    (t_1))\rmint dt_2\,\tilde{\psi}(\vec{x_2}(t_2))
  \\ & =\frac{m_1m_2}{m_{p}^2}\rmint dt_1dt_2\Big{\langle}\tilde{\psi}(\vec{x_1}
    (t_1))\tilde{\psi}(\vec{x_2}(t_2))\Big{\rangle}_{\varoslash}\,.\\ &
    =\frac{ m_1m_2}{16 \pi m_p^2}\rmint dt\, \frac{(\Vec{v}_1\cdot \Vec{v}_2)-(\Vec{v}_1\cdot \Vec{r})(\Vec{v}_2\cdot \Vec{r})}{r}\sim \mathcal{O}(Lv^2)\,. \label{4.13a}
    \end{split} 
\end{align}
Here, we use the Wick contraction between the fields to evaluate the diagrams going from the first to the second line. We will use this same strategy to evaluate all the subsequent terms throughout this paper. 
\item The amplitude corresponds to the diagram in Fig.~(\ref{fig1b}) with worldline vertex $\frac{-m_a}{2m_p^2}\rmint dt \Tilde{\psi}^2 $:
\begin{align}
    \begin{split}
    \mathcal{S}_{\text{eff}}\Big |_{\text{fig}(\ref{fig1b})}&=-\frac{m_1m_2^2}{2m_{p}^4}\rmint\Bar{\mathcal{D}}\Hat{\xi}\rmint dt_1dt_2dt_3\tilde{\psi}^2(\vec{x_1}
    (t_1))\tilde{\psi}(\vec{x_2}(t_2))\tilde{\psi}(\vec{x_2}(t_3))\,,\\ &
      =-\frac{m_1m_2^2}{2m_{p}^4}\rmint dt_1dt_2dt_3\Big{\langle}\tilde{\psi}(\vec{x_1}
    (t_1))\tilde{\psi}(\vec{x}_2(t_2))\Big{\rangle}\Big{\langle}\tilde{\psi}(\vec{x}_1(t_1))\tilde{\psi}(\vec{x_2}(t_3))\Big{\rangle}\\ &
    =-\frac{m_1m_2^2}{128 \pi^4 m_p^4}\rmint \frac{ dt}{r^2}+(1\leftrightarrow 2)\,\sim \mathcal{O}(Lv^2).
    \end{split}
\end{align}
\item The amplitude corresponds to the diagram in Fig.~(\ref{fig1c}) with worldline vertex $\frac{-m_a}{m_p}\rmint dt\, \hat{\mathcal{A}}_{i}v^{i}$:
\begin{align}
    \begin{split}
    \mathcal{S}_{\text{eff}}\Big |_{\text{fig}(\ref{fig1c})}
    &=\frac{m_1 m _2}{m_p^2}\rmint\Bar{\mathcal{D}}\Hat{\xi}\rmint dt_1 v_1^i(t_1)\Hat{\mathcal{A}}_i(\vec{x}_1(t_1))\rmint dt_2\,v_1^i(t_1)\Hat{\mathcal{A}}_i(\vec{x}_2(t_2))\\ &
    =\frac{m_1m_2}{m_{p}^2} \rmint dt_1dt_2\,v_1^i\,v_2^j\Big{\langle}\Hat{\mathcal{A}}_i(\vec{x_1}(t_1))\,\Hat{\mathcal{A}}_j(\vec{x_2}(t_2))\Big{\rangle}\\ &
    =-\frac{m_1m_2}{2\pi m_p^2}\rmint dt \,\frac{(\Vec{v}_1\cdot \Vec{v}_2)}{r}+(1\leftrightarrow 2)\,\sim \mathcal{O}(Lv^2).
    \end{split}
\end{align}
\item The amplitude corresponds to the diagram in Fig.~(\ref{fig1d}) with worldline vertex $\frac{-m_a}{m_p}\rmint dt \, \tilde{\psi}$ and $\frac{-3}{2m_p}\rmint dt\, v^2\,\Tilde{\psi}$:
\begin{align}
    \begin{split}
    \mathcal{S}_{\text{eff}}\Big |_{\text{fig}(\ref{fig1d})}
    &=\frac{3m_1m_2}{2m_{p}^2}\rmint\Bar{\mathcal{D}}\Hat{\xi}\rmint dt_1\tilde{\psi}(\vec{x_1}(t_1))v_1^2(t_1)\rmint dt_2\,\tilde{\psi}(\vec{x_2}(t_2)) \\ &
    =\frac{3m_1m_2}{2m_{p}^2}\rmint dt_1dt_2\Big{\langle}{\tilde{\psi}(\vec{x_1}(t_1))}\tilde{\psi}(\vec{x_2}(t_2)) \Big{\rangle}v_{1}^2+(1\leftrightarrow2)\\ &
    =-\frac{3  m_1m_2}{16\pi m_p^2}\rmint dt\, \frac{(v_1^2+v_2^2)}{r}\,\sim \mathcal{O}(Lv^2)\,.
    \end{split}
\end{align}
\end{itemize}
\subsection{The Electromagnetic (EM) bound sector} \label{EM section}
Now, the EM and Proca sectors are left to be investigated. The EM bound sector consists of the following terms :
\begin{align}
    \begin{split}
  F_{\mu\nu}F^{\mu\nu}&=(\nabla_{\mu} A_{\nu}-\nabla_{\nu} A_{\mu})(\nabla^{\mu} A^{\nu}-\nabla^{\nu} A^{\mu})\\ &
=(\partial_0 A_i\partial_0 A_i-\partial_0 A_i\partial_iA_0+\partial_lA_i\partial_l A_i-\partial_l A_i\partial_iA_l-\partial_k A_0\partial_k A_0+\partial_kA_0\partial_0 A_k)\,.
      \end{split}
\end{align}

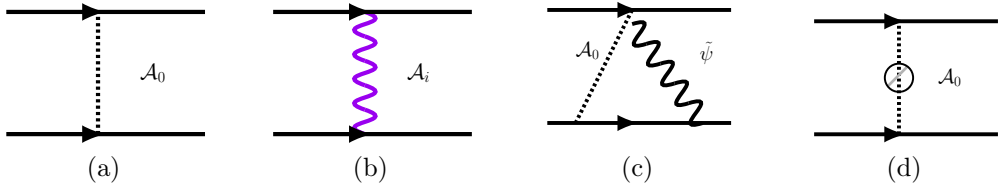
\begin{figure}
    \centering
    \begin{subfigure}{0.22\textwidth}
        \centering
\scalebox{0.4}{\begin{feynman}
    \dashed[lineWidth=4, showArrow=false, label=$\mathcal{A}_{0}$]{5.20, 4.00}{5.20, 5.60}
    \fermion[lineWidth=4]{4.00, 4.00}{6.60, 4.00}
    \fermion[lineWidth=4]{4.00, 5.60}{6.60, 5.60}
\end{feynman}}
        \caption{}
        \label{mfig4a}
    \end{subfigure}
      \begin{subfigure}{0.22\textwidth}
        \centering
\scalebox{0.4}{\begin{feynman}
    \electroweak[lineWidth=4, label=$\mathcal{A}_{i}$, color=9900ef]{5.20, 4.00}{5.20, 5.60}
    \fermion[lineWidth=4]{4.00, 4.00}{6.60, 4.00}
    \fermion[lineWidth=4]{4.00, 5.60}{6.60, 5.60}
\end{feynman}}
        \caption{}
        \label{mfig4b}
    \end{subfigure}
      \begin{subfigure}{0.22\textwidth}
        \centering
\scalebox{0.37}{\begin{feynman}
    \dashed[showArrow=false, lineWidth=4, label=$\mathcal{A}_{0}$]{4.40, 4.00}{5.20, 5.60}
    \electroweak[lineWidth=4, label=$\Tilde{\psi}$]{6.20, 4.00}{5.20, 5.40}
    \fermion[lineWidth=4]{4.00, 4.00}{6.60, 4.00}
    \fermion[lineWidth=4]{4.00, 5.60}{6.60, 5.60}
\end{feynman}}
        \caption{}
        \label{mfig4c}
    \end{subfigure}
    \begin{subfigure}{0.22\textwidth}
        \centering
\scalebox{0.37}{\begin{feynman}
    \dashed[showArrow=false, lineWidth=4, label=$\mathcal{A}_{0}$]{5.20, 4.00}{5.20, 5.60}
    \fermion[lineWidth=4]{4.00, 4.00}{6.60, 4.00}
    \fermion[lineWidth=4]{4.00, 5.60}{6.60, 5.60}
    \parton{5.20,4.80}{0.20}
\end{feynman}}
        \caption{}
        \label{mfig4d}
    \end{subfigure}
    \caption{Diagrams contributing to electromagnetic bound sector upto 1PN.}
    \label{fig:my_label}
\end{figure}
Remember that the terms that are quadratic in fields contribute to the propagator. So they do not contribute as separate interaction vertices. Furthermore, we also have a gauge fixing term $(\partial_\mu A^\mu)^2$. However, this term does not contribute to the effective action at  1PN order. But there will be some bound sector diagrams for EM and Proca fields. We show the effective action for the electromagnetic field (coupled with gravity) upto 1PN order as follows. There are total \textcolor{black}{four} contributing diagrams.
\begin{itemize}
    \item The amplitude corresponding to the diagram in Fig.~(\ref{mfig4a}) with worldline vertex\,\,
 $Q_a\rmint dt\,\mathcal{A}_0$ has the following form,
    \begin{align}
        \begin{split}
         \mathcal{S}_{\text{eff}}\Big|_{\text{fig}(\ref{mfig4a})}&=Q_1Q_2\rmint\Bar{\mathcal{D}}\Hat{\xi}\rmint dt_1\,\mathcal{A}_{0}(\Vec{x}_1(t_1))\rmint dt_2\,\mathcal{A}_{0}(\Vec{x}_2(t_2))\,,\\ &
         =Q_1Q_2\rmint dt_1 dt_2\,\Big\langle\mathcal{A}_{0}(\Vec{x}_1(t_1))\mathcal{A}_{0}(\Vec{x}_2(t_2))\Big\rangle\,,\\ &
         =-\frac{Q_1Q_2}{4\pi}\rmint \frac{dt}{r}+(1\leftrightarrow 2)\sim \mathcal{O}(Lv^0)\,.
        \end{split}
    \end{align}
    \item The amplitude corresponding to the diagram in Fig.~(\ref{mfig4b}) with worldline coupling \,\,$Q_a\rmint dt\,v_a^{i}\mathcal{A}_{i}$ has the following form,
    \begin{align}
        \begin{split}
            \mathcal{S}_{\text{eff}}\Big|_{\text{fig}(\ref{mfig4b})}&=Q_1Q_2\rmint\Bar{\mathcal{D}}\Hat{\xi}\rmint dt_1 dt_2\,v_1^{i}\mathcal{A}_{i}(\Vec{x}_1(t_1))\rmint dt_2\,v_2^{j}(t_2)\mathcal{A}_{j}(\Vec{x}_2(t_2))\,,\\ &
            =Q_1 Q_2 \rmint dt_1 dt_2\,v_1^{i}(t_1)v_2^{j}(t_2)\Big\langle\mathcal{A}_{i}(\Vec{x}_1(t_1))\mathcal{A}_{j}(\Vec{x}_2(t_2))\Big\rangle\,,\\ &
            =\frac{Q_1 Q_2}{4\pi}\rmint dt\, \frac{\Vec{v}_1\cdot \Vec{v}_2}{r}+(1\leftrightarrow 2)\sim \mathcal{O}(Lv^2)\,.
        \end{split}
    \end{align}
    \item The amplitude corresponding to the diagram in Fig.~(\ref{mfig4c}) with two different worldline couplings $Q_a\rmint dt\,\mathcal{A}_0$ and $-\rmint \frac{m_b}{m_p}\Tilde{\psi}$ has the following form,
    \begin{align}
        \begin{split}
        \mathcal{S}_{\text{eff}}\Big|_{\text{fig}(\ref{mfig4c})}&=\frac{m_2 Q_1Q_2}{m_p^2}\rmint\Bar{\mathcal{D}}\Hat{\xi}\rmint dt_1 Q_1 \mathcal{A}_{0}(\Vec{x}_1(t_1))\Tilde{\psi}(\Vec{x}_1(t_1))\rmint dt_2 \mathcal{A}_{0}(\Vec{x}_2(t_2))\rmint dt_3\,\Tilde{\psi}(\Vec{x}_2(t_3))\,,\\ &
        =\frac{m_2 Q_1Q_2}{m_p^2}\rmint dt_1 dt_2 dt_3 \Big\langle\mathcal{A}_{0}(\Vec{x}_1(t_1))\mathcal{A}_{0}(\Vec{x}_2(t_2))\Big\rangle \,\Big\langle\Tilde{\psi}(\Vec{x}_1(t_1))\Tilde{\psi}(\Vec{x}_2(t_3))\Big\rangle\,,\\ &
        =-\frac{m_2Q_1Q_2}{32\pi^2 m_p^2}\rmint \frac{dt}{r^2}+(1\leftrightarrow 2)\,. \sim \mathcal{O}(Lv^2)\,.
        \end{split}
    \end{align}
    \item The last term coming from the modified propagator of $\mathcal{A}_0\,\mathcal{A}_0$ as shown in the Fig.~(\ref{mfig4d}), with the following amplitude with worldline vertices $\rmint dt Q \mathcal{A}_{0}$, 
    \begin{align}
        \begin{split}
            \mathcal{S}_{\text{eff}}\Big|_{\text{fig}(\ref{mfig4d})}
            &=Q_1 Q_2 \rmint\Bar{\mathcal{D}}\Hat{\xi}\rmint  dt_1\mathcal{A}_{0}(\Vec{x}_1(t_1))\rmint dt_2\mathcal{A}_{0}(\Vec{x}_2(t_2))\\ &
            =Q_1Q_2\rmint dt_1 dt_2\,\Big\langle\mathcal{A}_{0}(\Vec{x}_1(t_1))\mathcal{A}_{0}(\Vec{x}_2(t_2))\Big\rangle_{\varoslash }\,,\\ &
            =-Q_1 Q_2\rmint dt_1 dt_2\,\delta(t_1-t_2)\partial_{t_1}\partial_{t_2}\Big\{\frac{-|\Vec{x}_1(t_1)-\Vec{x}_2(t_2)|}{8\pi}\Big\}\,,\\ &
            =\frac{Q_1Q_2}{8\pi}\rmint dt\,\Big[\frac{\Vec{v}_1\cdot \Vec{v}_2}{r}-\frac{(\Vec{v}_1\cdot \Hat{n})(\Vec{v}_2\cdot \Hat{n})}{r}\Big]\sim \mathcal{O}(Lv^2)\,.
        \end{split}
    \end{align}
\end{itemize}
\subsection{The Proca  bound sector}
The Proca bound sector consists of the following term :
\begin{align}
    \begin{split}
  B_{\mu\nu}B^{\mu\nu}&=\frac{1}{2}(\nabla_\mu B_\nu-\nabla_\nu B_\mu)(\nabla^\mu B^\nu-\nabla^\nu B^\mu)\\ &
=(\partial_0B_i\partial_0B_i-\partial_0B_i\partial_iB_0+\partial_lB_i\partial_lB_i-\partial_lB_i\partial_iB_l-\partial_kB_0\partial_kB_0+\partial_kB_0\partial_0B_k)\,.
      \end{split}
\end{align}

We show the effective action for the Proca field (coupled with gravity) upto 1PN order as follows. There are total of four contributing diagrams. The computation is similar to that of Sec.~(\ref{EM section}).
\begin{figure}
    \centering
    \begin{subfigure}{0.22\textwidth}
        \centering
\scalebox{0.38}{\begin{feynman}
    \dashed[showArrow=false, lineWidth=4, label=$\mathcal{B}_{0}$, color=9900ef]{5.20, 4.00}{5.20, 5.60}
    \fermion[lineWidth=4]{4.00, 4.00}{6.60, 4.00}
    \fermion[lineWidth=4]{4.00, 5.60}{6.60, 5.60}
\end{feynman}}
        \caption{}
        \label{mfig5a}
    \end{subfigure}
         \begin{subfigure}{0.22\textwidth}
        \centering
\scalebox{0.38}{\begin{feynman}
    \fermion[lineWidth=4]{4.00, 4.00}{6.60, 4.00}
    \gluon[lineWidth=4, color=eb144c, label=$\mathcal{B}_i$]{5.20, 4.00}{5.20, 5.60}
    \fermion[lineWidth=4]{4.00, 5.60}{6.60, 5.60}
\end{feynman}}
        \caption{}
        \label{mfig5b}
    \end{subfigure}
      \begin{subfigure}{0.20\textwidth}
        \centering
\scalebox{0.39}{\begin{feynman}
    \electroweak[lineWidth=4, label=$\Tilde{\psi}$]{6.20, 4.00}{5.20, 5.60}
    \fermion[lineWidth=4]{4.00, 4.00}{6.60, 4.00}
    \dashed[lineWidth=4, showArrow=false, color=9900ef, label=$\mathcal{B}_{0}$]{4.40, 4.00}{5.20, 5.60}
    \fermion[lineWidth=4]{4.00, 5.60}{6.60, 5.60}
\end{feynman}}
        \caption{}
        \label{mfig5c}
    \end{subfigure}
      \begin{subfigure}{0.20\textwidth}
        \centering
\scalebox{0.38}{\begin{feynman}
    \fermion[lineWidth=4]{4.00, 4.00}{6.60, 4.00}
    \dashed[lineWidth=4, showArrow=false, color=9900ef, label=$\mathcal{B}_{0}$]{5.20, 4.00}{5.20, 5.60}
    \fermion[lineWidth=4]{4.00, 5.60}{6.60, 5.60}
    \parton{5.20,4.80}{0.20}
\end{feynman}}
        \caption{}
        \label{mfig5d}
    \end{subfigure}
    \caption{Diagrams contributing to the proca bound sector upto 1PN.}
    \label{fig:my_label}
\end{figure}
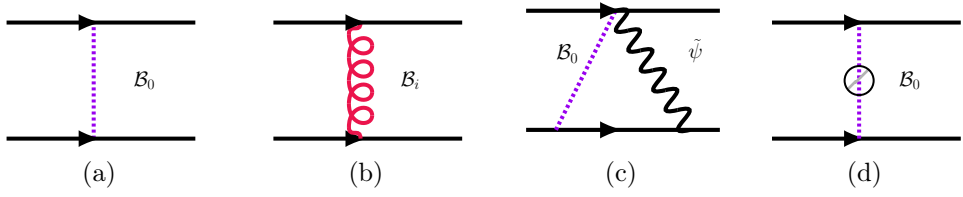
\begin{itemize}
    \item The amplitude corresponding the diagram in Fig.~(\ref{mfig5a}) with worldline vertex $Q_a'\rmint dt\,\mathcal{B}_0$ has the following form,
    \begin{align}
        \begin{split}
         \mathcal{S}_{\text{eff}}\Big|_{\text{fig}(\ref{mfig5a})}&=Q_1'Q_2'\rmint \Bar{\mathcal{D}}\hat{\xi}\rmint dt_1\,\mathcal{B}_{0}(\Vec{x}_1(t_1))\rmint dt_2\,\mathcal{B}_{0}(\Vec{x}_2(t_2))\,,\\ &
         =Q_1'Q_2'\rmint dt_1 dt_2\,\Big\langle\mathcal{B}_{0}(\Vec{x}_1(t_1))\mathcal{B}_{0}(\Vec{x}_2(t_2))\Big\rangle\,,\\ &
         =-\frac{Q_1'Q_2'}{4\pi}\rmint dt\,\frac{e^{-\mu_{\gamma}r}}{r}+(1\leftrightarrow 2)\,.\sim \mathcal{O}(Lv^0)\,.
        \end{split}
    \end{align}
    \item The amplitude corresponds to the diagram in Fig.~(\ref{mfig5b}) with worldline coupling $Q_a'\rmint dt\,v_a^{i}\mathcal{B}_{i}$ has the following form,
    \begin{align}
        \begin{split}
            \mathcal{S}_{\text{eff}}\Big|_{\text{fig}(\ref{mfig5b})}&=Q_1'Q_2'\rmint \Bar{\mathcal{D}}\hat{\xi}\rmint dt_1 dt_2\,v_1^{i}\mathcal{B}_{i}(\Vec{x}_1(t_1))\rmint dt_2\,v_2^{j}(t_2)\mathcal{B}_{j}(\Vec{x}_2(t_2))\,,\\ &
            =Q_1' Q_2' \rmint dt_1 dt_2\,v_1^{i}(t_1)v_2^{j}(t_2)\Big\langle\mathcal{B}_{i}(\Vec{x}_1(t_1))\mathcal{B}_{j}(\Vec{x}_2(t_2))\Big\rangle\,,\\ &
            =\frac{Q_1' Q_2'}{4\pi}\rmint dt\, \frac{(\Vec{v}_1\cdot \Vec{v}_2)e^{-\mu_{\gamma}r}}{ r}+(1\leftrightarrow 2)\,.\sim \mathcal{O}(Lv^2)\,.
        \end{split}
    \end{align}
    \item The amplitude corresponds to the diagram in Fig.~(\ref{mfig5c}) with two different worldline couplings $Q_a\rmint dt\,\mathcal{B}_0$ and $-\rmint \frac{m_b}{m_p}\Tilde{\psi}$ has the following form,
    \begin{align}
        \begin{split}
        \mathcal{S}_{\text{eff}}\Big|_{\text{fig}(\ref{mfig5c})}&=\frac{m_2 Q_1'Q_2'}{m_p^2}\rmint \Bar{\mathcal{D}}\hat{\xi}\rmint dt_1  \mathcal{B}_{0}(\Vec{x}_1(t_1))\Tilde{\psi}(\Vec{x}_1(t_1))\rmint dt_2 \mathcal{B}_{0}(\Vec{x}_2(t_2))\rmint dt_3\,\Tilde{\psi}(\Vec{x}_2(t_3))\,,\\ &
        =\frac{m_2 Q_1'Q_2'}{m_p^2}\rmint dt_1 dt_2 dt_3 \Big\langle\mathcal{B}_{0}(\Vec{x}_1(t_1))\mathcal{B}_{0}(\Vec{x}_2(t_2))\Big\rangle \,\Big\langle\Tilde{\psi}(\Vec{x}_1(t_1))\Tilde{\psi}(\Vec{x}_2(t_3))\Big\rangle\,,\\ &
        =-\frac{m_2Q_1'Q_2'}{32\pi^2 m_p^2}\rmint dt \,\frac{e^{-\mu_{\gamma}r}}{r^2}+(1\leftrightarrow 2)\,. \sim \mathcal{O}(Lv^2)\,.
        \end{split}
    \end{align}
    \item The last term coming from the modified propagator of $\mathcal{B}_0\,\mathcal{B}_0$ as shown in the Fig.~(\ref{mfig5d}), with the following amplitude, 
    \begin{align}
        \begin{split}
            \mathcal{S}_{\text{eff}}\Big|_{\text{fig}(\ref{mfig5d})}&
            Q_1' Q_2' \rmint\Bar{\mathcal{D}}\Hat{\xi}\rmint  dt_1\mathcal{B}_{0}(\Vec{x}_1(t_1))\rmint dt_2\mathcal{B}_{0}(\Vec{x}_2(t_2))\\ &
            =Q_1'Q_2'\rmint dt_1 dt_2\,\Big\langle\mathcal{B}_{0}(\Vec{x}_1(t_1))\mathcal{B}_{0}(\Vec{x}_2(t_2))\Big\rangle_{\varoslash}\,,\\ &
    =-\frac{Q_1'Q_2'}{8\pi}\rmint dt \,e^{-\mu_{\gamma}r}\Big[\frac{\Vec{v}_1\cdot\Vec{v}_2}{r}-\frac{(\Vec{v}_1\cdot\Hat{n})(\Vec{v}_2\cdot\Hat{n})}{r}(1+\mu_{\gamma}r)\Big]+(1\leftrightarrow 2)\,.\sim\mathcal{O}(Lv^2)\,.     
        \end{split}
    \end{align}
\end{itemize}

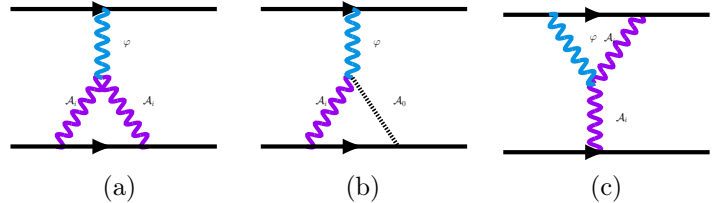
\begin{figure}[htb!]
    \centering 
    \begin{subfigure}{0.2\textwidth}
   \scalebox{0.2}{ \begin{feynman}
    \electroweak[flip=true, lineWidth=8, label=$\mathcal{A}_{i}$, color=9900ef]{5.20, 4.00}{6.40, 5.80}
    \electroweak[lineWidth=8, label=$\mathcal{A}_{i}$, color=9900ef]{6.40, 5.80}{7.60, 4.00}
    \fermion[lineWidth=8]{4.00, 7.60}{9.40, 7.60}
    \electroweak[lineWidth=8, label=$\varphi$, color=0693e3]{6.40, 5.80}{6.40, 7.60}
    \fermion[lineWidth=8]{4.00, 4.00}{9.40, 4.00}
\end{feynman}}
\caption{}
\label{fig4an}
\end{subfigure}
\begin{subfigure}{0.2\textwidth}
    \centering
     \scalebox{0.2}{\begin{feynman}
    \electroweak[flip=true, lineWidth=8, label=$\mathcal{A}_{i}$, color=9900ef]{5.20, 4.00}{6.40, 5.80}
    \fermion[lineWidth=8]{4.00, 7.60}{9.40, 7.60}
    \electroweak[lineWidth=8, label=$\varphi$, color=0693e3]{6.40, 5.80}{6.40, 7.60}
    \fermion[lineWidth=8]{4.00, 4.00}{9.40, 4.00}
    \dashed[showArrow=false, lineWidth=8, label=$\mathcal{A}_0$]{6.40, 5.80}{7.60, 4.00}
\end{feynman}}
    \caption{}
    \label{fig4cn}
\end{subfigure}
\begin{subfigure}{0.2\textwidth}
    \centering 
\scalebox{0.2}{\begin{feynman}
    \electroweak[flip=true, lineWidth=8, label=$\mathcal{A}_{i}$, color=9900ef]{6.40, 4.00}{6.40, 5.80}
    \electroweak[lineWidth=8, label=$\mathcal{A}_i$, color=9900ef]{6.40, 5.80}{7.60, 7.60}
    \fermion[lineWidth=8]{4.00, 7.60}{9.40, 7.60}
    \electroweak[lineWidth=8, label=$\varphi$, color=0693e3]{6.40, 5.80}{5.20, 7.60}
    \fermion[lineWidth=8]{4.00, 4.00}{9.40, 4.00}
\end{feynman}
}
\caption{}
\label{fig4bn}
\end{subfigure}
    \caption{Scalar-Electromagnetic interaction diagrams which contribute to the bound sector.}
    \label{figtheta1}
\end{figure}

\subsection{The EM-Axion bound sector}\label{Sec4.2}
In this section, we mainly focus on the \textit{Theta term} as we want to investigate the effect of $g_{a\gamma\gamma}$ into the bound potential as well as the power radiation as discussed in Sec.~(\ref{sec5.5}). \textcolor{black}{The bulk interaction vertex which is relevant for this is the following,
\begin{equation}
    S_{int}^{\text{Theta}}=
\frac{g_{a\gamma\gamma}}{4m_p}\rmint d^4x \sqrt{-g}\,\phi\,F_{\mu\nu}\,\Tilde{F}^{\mu\nu}. 
\end{equation}
}
The contribution comes from this term at 2.5 PN order and has the following form\footnote{From the subsequent sections we replace $\hat\epsilon\rightarrow \epsilon$.},
\begin{equation}
    S_{int}^{\text{Theta}}\Big |_{2.5PN}= \frac{g_{a\gamma\gamma}}{m_p}\rmint d^4x [\hat\epsilon^{\,0ikm}\phi\,\partial_{0}\mathcal{A}_{i}\partial_{k}\mathcal{A}_{m}+ \hat\epsilon^{\,i0km}\phi\,\partial_{i}\mathcal{A}_{0}\partial_{k}\mathcal{A}_{m}]\,.
\end{equation}
 Three possible diagrams will contribute to this specific interaction. The contributing diagrams have the two-photon line and one dynamical scalar line as shown in Fig.~(\ref{figtheta1}).
\begin{itemize}
\item The first diagram that contributes has two electromagnetic propagators from the same worldline as shown in Fig.~(\ref{fig4an}) and Fig.~(\ref{fig4cn}) \textcolor{black}{with vertex factors $\phi\,\partial_{0}\mathcal{A}_{i}\partial_{k}\mathcal{A}_{m}$ and $\phi\,\partial_{i}\mathcal{A}_{0}\partial_{k}\mathcal{A}_{m}$ respectively}. The amplitude is given by, $  \mathcal{S}_{\text{eff}}=\frac{g_{a\gamma\gamma}}{m_p}Q_2^2\,m_1\,s_1\epsilon^{i0km}(  \mathcal{S}^{\text{eff}}_{ikm}+\Bar{  \mathcal{S}}^{\text{eff}}_{ikm}).$ The first and second terms are defined in (\ref{2.15}) and (\ref{4.18n})\,.
\begin{align}
   \begin{split}
      \mathcal{S}_{\text{eff}}^{ikm}\Big|_{\text{fig}(\ref{fig4an})}&=\rmint \Bar{\mathcal{D}}\hat{\xi}\rmint dt_2  \mathcal{A}_{j}(\Vec{x}_2(t_2))v_{2}^{j}(t_2)\rmint dt_{3}  \mathcal{A}_{l}(\Vec{x}_2(t_3))v_2^l(t_3)\rmint dt_1  \frac{\varphi(\Vec{x}_1(t_1))}{m_p}\\ &
      \hspace{1.2 cm}\rmint d^4x\,  \varphi\,\partial_0  \mathcal{A}_{i}\partial_{k}\mathcal{A}_{m}(x)\,, \\ &
    =\frac{1}{m_p}\rmint \prod_{i=1}^{3}dt_i\,  dt\, v_{2}^{j}(t_2) v_2^l(t_3) \rmint d^3x\, \partial_{0}\Big\langle \mathcal{A}_{j}(\Vec{x}_2(t_2))\mathcal{A}_{i}(x)\Big\rangle\,\partial_{k}\boldsymbol{\Big\langle}\mathcal{A}_{l}(\Vec{x}_2(t_3))\mathcal{A}_{m}(x)\Big\rangle\,\\&\hspace{6.6cm}\Big\langle\varphi(\Vec{x}_1(t_1))\varphi({x})\Big\rangle,\\ &
     =-\frac{i}{m_p}\rmint dt \,v_2^{m}(t)\Big[a_{2}^{i}(t)\rmint_{k_1,k_2}\frac{k_2^{k}}{\Vec{k}_1^2\Vec{k}_2^2[(\Vec{k}_1+\Vec{k}_2)^2+m^2]}e^{i(\Vec{k}_1+\Vec{k}_2)\cdot \Vec{r}}\\ &
   \,\,\,\,\,\,\,\,\,\,\,\,\,\,  +v_2^{i}(t)v_2^{a}(t)\rmint_{k_1,k_2}\frac{i\,k_2^k\,k_1^a\,e^{i(\Vec{k}_1+\Vec{k}_2)\cdot \Vec{r}}}{\Vec{k}_1^2\Vec{k}_2^2[(\Vec{k}_1+\Vec{k}_2)^2+m^2]}\Big]
   +(1\leftrightarrow 2)
   \label{2.15}
    \end{split}
\end{align}
and
\begin{align}
    \begin{split}
        \mathcal{ \Bar{S}_{\text{eff}}}^{ikm}\Big |_{\text{fig}(\ref{fig4cn})}&=-\rmint\Bar{\mathcal{D}}\hat{\xi}\rmint dt_2 \mathcal{A}_{0}(\Vec{x}_2(t_2))\rmint dt_3 \mathcal{A}_{j}(\Vec{x}_2(t_3))v^{j}(t_3)\rmint dt_1\,\varphi(\Vec{x}_1(t_1))\\ &
        \hspace{4.3 cm}\rmint d^4x \,\varphi(x)\partial_{i}\mathcal{A}_{0}\partial_{k}\mathcal{A}_{m}\,,\\ &
        =-\rmint dt\, dt_1 dt_2 dt_3 \,v^{j}(t_3)\rmint d^3 x\,\partial_{i}\Big\langle \mathcal{A}_{0}(\Vec{x}_2(t_2))\mathcal{A}_{0}(x)\Big\rangle \partial_{k}\Big\langle \mathcal{A}_{j}(\Vec{x}_2(t_3))\mathcal{A}_{m}(x) \Big\rangle\\ &
       \hspace{5.8 cm} \Big\langle 
\varphi(\Vec{x}_1(t_1))\varphi(x)\Big\rangle\,,\\&
        =\rmint dt \,v^{m}(t)\rmint_{\Vec{k_1},\Vec{k_2}}\frac{k_1^i\,k_2^k\,e^{i(\Vec{k_1}+\Vec{k_2})\cdot\Vec{r}(t)}}{\Vec{k_1}^2\Vec{k_2}^2[(\Vec{k_1}+\Vec{k_2})^2+m^2]}\,,\\ &
        =-\frac{1}{64}\rmint dt\, v^m(t)\Big[f(r)n^in^k+g(r)\delta^{ik}\Big]\,.\label{4.18n}
    \end{split}
\end{align}
Now due to the symmetric nature of $n^in^k$ and $\delta^{ik},$  $\mathcal{ \Bar{S}_{\text{eff}}}^{ikm}\Big |_{\text{eff}(\ref{fig4cn})}$ does not contribute to the amplitude after getting contracted with the $\epsilon_{i0km}.$
Therefore only $ \mathcal{S}_{\text{eff}}^{ikm}\Big|_{\text{fig}(\ref{fig4an})}$ contributes to the amplitude. The details of the computations are given in Appendix~(\ref{app2}).
\begin{align}
    \begin{split}
          \mathcal{S}_{\text{eff}}\Big|_{\text{fig}(\ref{fig4an})}&=\frac{g_{a\gamma\gamma}}{m_{p}^2}\,Q_2^2 \,m_1\,s_1\,\epsilon^{i0km}\,  \mathcal{S}^{\text{eff}}_{ikm}\,,\\ &
= \frac{\pi^{3/2}\, g_{a\gamma\gamma}Q_2^2m_1s_1 }{8m_p^2}\Bigg[\rmint dt\, \Vec{a_2}\cdot(\hat{n}\times\Vec{v_2})\Bigg\{\frac{1}{2} \, m \,G_{1,3}^{2,1}\left(\frac{m^2 r^2}{4}\Big|
\begin{array}{c}
 -\frac{1}{2} \\
 -\frac{1}{2},-\frac{1}{2},0 \\
\end{array}
\right)\\&\hspace{5.3cm}-\frac{\, G_{1,3}^{2,1}\left(\frac{m^2 r^2}{4}\Big|
\begin{array}{c}
 \frac{1}{2} \\
 \frac{1}{2},\frac{1}{2},0 \\
\end{array}
\right)}{m\, r^2}\Bigg\}\Bigg] +1\leftrightarrow 2
\sim 
 \mathcal{O}(Lv^5).
\
\label{4.31}
\end{split}
\end{align}
Also we can easily check that, when one takes $m\rightarrow 0$ limit, (\ref{4.31}) smoothly goes to the following, 

\begin{align}
    \begin{split}
          \mathcal{S}_{\text{eff}}\Big|_{\text{fig}(\ref{fig4an})}&
        = \frac{ g_{a\gamma\gamma}\,Q_2^2\,m_1\,s_1\,}{32\pi^2\,m_p^2}\rmint dt\,\frac{\Vec{a_2}.(\hat{n}\times\Vec{v_2})}{r}+(1\leftrightarrow2)\sim \mathcal{O}(Lv^5)\,.
    \end{split}
\end{align}
\par

\item  There is another diagram that consists of two electromagnetic fields coupled with two different worldline vertices  $\rmint dt Q_a v^{i}_{a}\mathcal{A}_{i}.$ The corresponding \textcolor{black}{amplitude} is given by,
\begin{align}
    \begin{split}
\mathcal{S}_{\text{eff}}^{ikm}\Big |_{\text{fig}(\ref{fig4bn})}&=\frac{g_{a\gamma\gamma Q_1Q_2m_1s_1}}{m_p^2}{\rmint\Bar{\mathcal{D}}\Hat{\xi}\rmint dt_1\,v_1^{j}(t_1) \mathcal{A}_{j}(\Vec{x}_1(t_1))\rmint dt_2\,\varphi(\Vec{x}_1(t_2))\rmint dt_3 v_2^{l}(t_1)\mathcal{A}_{l}(\Vec{x}_2(t_3))}\\ & \hspace{7 cm}
\rmint d^4x \,\varphi \partial_{0}  \mathcal{A}_{i}\partial_{k}\mathcal{A}_{m}(x)\,,\\ &
        =\frac{g_{a\gamma\gamma}}{m_p^2}\rmint dt\prod_{i=1}^{3}dt_{i}\,v_1^{j}(t_1)v_2^{l}(t_3)\rmint d^3x \partial_{t}\Big\langle \mathcal{A}_{j}(\Vec{x}_1(t_1))\mathcal{A}_{i}(x)\Big\rangle\partial_{k}\Big\langle \mathcal{A}_{l}(\Vec{x}_2(t_3))\mathcal{A}_{m}(x)\Big\rangle \\& \hspace{5cm} \Big\langle\varphi(\Vec{x}_1(t_2))\varphi(x,t)\Big\rangle\,,\\ &
        =\frac{2ig_{a\gamma\gamma}Q_1Q_2m_1s_1}{m_p^2}\rmint dt v_2^{m}(t)a_1^{i}(t)\rmint_{k}\frac{k^k\,e^{i\Vec{k}\cdot\Vec{r}}}{\Vec{k}^2}\textcolor{black}{\rmint_{k_3}\frac{1}{(\Vec{k}-\Vec{k_3})^2(\Vec{k_3}^2+m^2)}}\,,\\ &
  =\frac{ g_{a\gamma\gamma}Q_1Q_2m_1s_1}{2\pi m_p^2}\rmint dt \, v_2^m (t)a_1^i(t)\,\partial_r\alpha(m;r)n^k\,,\label{4.33mm}
    \end{split}
\end{align}
Therefore, the contracted amplitude contributing to the effective action is given by,
\begin{align}
    \begin{split}
        \mathcal{S}_{\text{eff}}\Big|_{\text{fig}(\ref{fig4bn})}=\frac{ g_{a\gamma\gamma}Q_1Q_2m_1s_1}{2\pi m_p^2}\rmint dt \,(\Vec{a}_1 \times \Hat{n})\cdot \Vec{v}_2\, \partial_{r}\alpha(m;r)\sim \mathcal{O}(Lv^5).
    \end{split}
\end{align}
\end{itemize}
\textcolor{black}{where the function $\alpha(m;r)$ is defined in (\ref{C3m}) of Appendix~(\ref{app3}).}
\subsection{The pure scalar sector}\label{Sec4.3}
\textcolor{black}{The contributing terms in this case are simply given by expanding the 3rd and 4th terms of the Lagrangian mentioned in (\ref{2.11}).}
The non-relativistic decomposition of the action is given by,
\begin{align}
    \begin{split}
         S[{\varphi}]_{\text{int}}&
        =\frac{1}{m_p}\rmint d^3x dt \,[-2\psi \partial_{0}\varphi\partial_{0}\varphi-2\Hat{\mathcal{A}}^{i}\partial_{i}\varphi\partial_{0}\varphi-\frac{1}{4}\sigma^{lm}\delta_{lm}\partial_{i}\varphi\partial_{i}\phi+\frac{1}{2}\sigma^{ij}\partial_{i}\varphi\partial_{j}\varphi+\psi m^2 \varphi^2]\,.\label{4.24}
    \end{split}
\end{align}
Note that all the interaction terms except the last one will contribute to the effective action at the order of $L\,v^4$, i.e., of 2PN order. The last term in (\ref{4.24}) corresponds to scalar-graviton interaction at 1PN. There are two diagrams contributing to the 1PN effective action that comes from the 3-point vertex, as shown in Fig.~(\ref{fig:my_label}). Next, we give the details of the amplitudes corresponding to these diagrams.
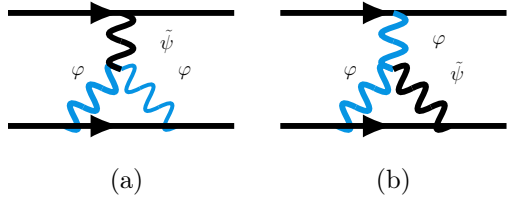
\begin{figure}[hbt!]
    \centering
    \begin{subfigure}{0.22\textwidth}
\scalebox{0.37}{\begin{feynman}
    \electroweak[flip=true, label=$\varphi$, lineWidth=6, color=0693e3]{4.80, 4.00}{5.60, 4.80}
    \fermion[lineWidth=6]{4.00, 5.60}{7.20, 5.60}
    \electroweak[label=$\varphi$, lineWidth=4, color=0693e3]{5.60, 4.80}{6.40, 4.00}
    \electroweak[lineWidth=6, label=$\Tilde{\psi}$]{5.60, 4.80}{5.60, 5.60}
    \fermion[lineWidth=6]{4.00, 4.00}{7.20, 4.00}
\end{feynman}}
        \caption{}
        \label{fig5aa}
    \end{subfigure}
    \begin{subfigure}{0.22\textwidth}
        \centering
\scalebox{0.37}{\begin{feynman}
    \electroweak[flip=true, label=$\varphi$, lineWidth=6, color=0695e3]{4.80, 4.00}{5.60, 4.80}
    \fermion[lineWidth=6]{4.00, 5.60}{7.20, 5.60}
    \electroweak[lineWidth=6, label=$\varphi$, color=0693e3]{5.60, 4.80}{5.60, 5.60}
    \electroweak[lineWidth=6, label=$\Tilde{\psi}$]{5.60, 4.80}{6.40, 4.00}
    \fermion[lineWidth=6]{4.00, 4.00}{7.20, 4.00}
\end{feynman}
}
        \caption{}
        \label{fig5bb}
    \end{subfigure}
    \caption{\textcolor{black}{1PN diagrams for the scalar sector with 3-point vertex coming from scalar-graviton interaction.}}
    \label{fig:my_label}
\end{figure}
\begin{itemize}
    \item The amplitude corresponds to the diagram in Fig.~(\ref{fig5aa}) with the bulk interaction vertex $\rmint d^4 x \,\Tilde{\psi}\varphi\varphi(x)$ where two scalar fields are contracted with the same worldline:
    \begin{align}
        \begin{split}
\mathcal{S}_{\text{eff}}\Big |_{\text{fig}(\ref{fig5aa})}&=\frac{m^2m_1 m_2^2 s_2^2}{m_{p}^4}\rmint\Bar{\mathcal{D}}\Hat{\xi}\rmint dt_1 \,\Tilde{\psi}(\Vec{x}_1(t_1))\rmint dt_2 \,\varphi(\Vec{x}_2(t_2))\rmint dt_3 \varphi(\Vec{x}_2(t_3))\rmint d^4 x \,\Tilde{\psi}\varphi\varphi(x)\,,\\ &
            =\frac{m^2 m_1 m_2^2 s_2^2}{m_p^4} \rmint dt\prod_{i=1}^3 dt_i \Big\langle\Tilde{\psi}(\Vec{x}_1(t_1))\Tilde{\psi}(x)\Big\rangle \Big\langle\varphi(\Vec{x}_2(t_2))\varphi(x) \Big\rangle\Big\langle\varphi(\Vec{x}_2(t_3))\varphi(x) \Big\rangle\,,\\ &
            =\frac{m^2 m_1 m_2^2 s_2^2}{2m_p^4} \rmint dt \rmint_{\Vec{k},\Vec{k}_1}\,\frac{e^{i\Vec{k}\cdot \Vec{r}}}{\Vec{k}^2(\Vec{k}_1^2+m^2)[(\Vec{k}-\Vec{k}_1)^2+m^2]}\,,\\ &
            =\frac{m^2 m_1 m_2^2 s_2^2}{8\pi m_p^4}\rmint dt \rmint \frac{e^{i\Vec{k}\cdot \Vec{r}}}{\Vec{k}^3}\arctan(|\Vec{k}|/2m)\,,\\&
            =\frac{m_1 m_2^2 s_2^2}{64\pi^2 m_p^4}\rmint dt \,\frac{m-2m^2r\,\text{Ei}(-2mr)-m\,e^{-2mr}}{r}+(1\leftrightarrow 2)\sim\mathcal{O}(Lv^2)\,.
            \label{4.25}
        \end{split}
    \end{align}
    In (\ref{4.25}), $\text{Ei}(z)=-\rmint_{-z}^{\infty}dt\,\frac{e^{-t}}{t}$ is the exponential integral function.
 \item The amplitude corresponds to the diagram in Fig.~(\ref{fig5bb}) with the bulk interaction vertex $\rmint d^4 x \,\Tilde{\psi}\varphi\varphi(x)$ where two scalar fields are contracted with two different worldlines:
    \begin{align}
        \begin{split}
            \mathcal{S}_{\text{eff}}\Big |_{\text{fig}(\ref{fig5bb})}&=\frac{m^2m_1 m_2^2 s_2^2}{m_{p}^4}\rmint\Bar{\mathcal{D}}\Hat{\xi}\rmint dt_1 \,\varphi(\Vec{x}_1(t_1))\rmint dt_2 \,\Tilde{\psi}(\Vec{x}_2(t_2))\rmint dt_3 \varphi(\Vec{x}_2(t_3))\rmint d^4 x \,\Tilde{\psi}\varphi\varphi(x)\,,\\ &=\frac{m^2m_1s_1 m_2^2 s_2}{m_p^4} \rmint d^3x\,dt\prod_{i=1}^3 dt_i \Big\langle\Tilde{\psi}(\Vec{x}_2(t_2))\Tilde{\psi}(x)\Big\rangle \Big\langle\varphi(\Vec{x}_1(t_1))\varphi(x) \Big\rangle\Big\langle\varphi(\Vec{x}_2(t_3))\varphi(x) \Big\rangle\,,\\ &
            =\frac{m^2m_1s_1 m_2^2 s_2}{m_p^4}\rmint_{k}\frac{e^{i\Vec{k}\cdot \Vec{r}}}{\Vec{k}^2+m^2}\rmint_{k_1}\frac{1}{\Vec{k}_1^2[(\Vec{k}_1-\Vec{k})^2+m^2]}\,,\\ &
           = \frac{m^2 m_1s_1 m_2^2 s_2}{4\pi m_p^4}\rmint dt \,\beta(m;r)+(1\leftrightarrow 2)\,\sim \mathcal{O}(Lv^2).\label{4.26}
        \end{split}
    \end{align}
    In (\ref{4.26}), $\beta(m;r)$ has no closed form expression. It's  nature is shown in Fig.~(\ref{lastfig}) in Appendix~(\ref{App1}).
\end{itemize}


   \begin{figure}[t!]
    \centering
    \begin{subfigure}{0.25\textwidth}
       \centering
\scalebox{0.3}{\begin{feynman}
    \electroweak[label=$\varphi$, lineWidth=6, color=0693e3]{5.60, 4.00}{5.60, 5.00}
    \electroweak[label=$\varphi$, lineWidth=6, color=0693e3]{5.60, 6.40}{5.60, 5.40}
    \fermion[lineWidth=6]{4.00, 6.40}{7.20, 6.40}
    \fermion[lineWidth=6]{4.00, 4.00}{7.20, 4.00}
    \parton{5.60,5.20}{0.20}
\end{feynman}}
        \caption{}
        \label{fig6ma}
    \end{subfigure}
    \begin{subfigure}{0.25\textwidth}
        \centering
\scalebox{0.3}{\begin{feynman}
    \electroweak[lineWidth=6, color=0693e3, label=$\varphi$]{5.60, 6.00}{6.80, 4.00}
    \fermion[lineWidth=6]{4.00, 4.00}{7.20, 4.00}
    \electroweak[flip=true, lineWidth=6, label=$\Tilde{\psi}$]{4.40, 4.00}{5.60, 6.00}
    \fermion[lineWidth=6]{4.00, 6.00}{7.20, 6.00}
\end{feynman}}
        \caption{}
        \label{fig6mb}
    \end{subfigure}
    \begin{subfigure}{0.25\textwidth}
        \centering
\scalebox{0.3}{\begin{feynman}
    \electroweak[lineWidth=6, color=0693e3, label=$\varphi$]{5.60, 6.00}{6.80, 4.00}
    \fermion[lineWidth=6]{4.00, 4.00}{7.20, 4.00}
    \electroweak[flip=true, lineWidth=6, label=$\varphi$, color=0693e3]{4.40, 4.00}{5.60, 6.00}
    \fermion[lineWidth=6]{4.00, 6.00}{7.20, 6.00}
\end{feynman}}
        \caption{}
        \label{fig6mc}
    \end{subfigure}
    \caption{1PN scalar diagrams that contribute to the bound sector.}
    \label{fig6m}
\end{figure}
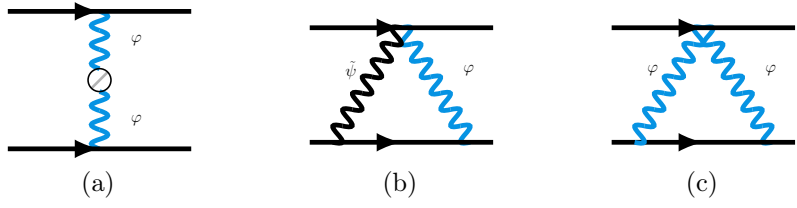
There are three more diagrams contributing to 1PN coming purely from the worldline coupling operators, one is from the correction to the scalar propagator, and another is from the $\Tilde{\psi}\varphi$ coupling at any of the two worldlines. 
\begin{itemize}
\item The amplitude corresponds to the diagram in Fig.~(\ref{fig6ma}) with worldline vertices $\rmint dt\,\varphi$ and with the corrected scalar propagator:
\begin{align}
\begin{split}
    \mathcal{S}_{\text{eff}}\Big|_{\text{fig}(\ref{fig6ma})}&=\frac{m_1s_1m_2s_2}{m_{p}^2}\rmint\Bar{\mathcal{D}}\Hat{\xi}\rmint dt_1 \,\varphi(\Vec{x}_1(t_1)) \rmint dt_2 \,\varphi(\Vec{x}_2(t_2)),\\ &
    =\frac{m_1s_1m_2s_2}{m_p^2}\rmint dt_1 dt_2\,\Big\langle\varphi(\Vec{x}_1(t_1))\varphi(\Vec{x}_2(t_2))\Big\rangle_{\varoslash}\,,\\ &
    =\frac{m_1s_1m_2s_2}{m_p^2}\rmint dt_1 dt_2\, \partial_{t_1}\partial_{t_2}\delta(t_1-t_2)\rmint_{k}\frac{e^{i\Vec{k}\cdot(\Vec{x}_1-\Vec{x}_{2})}}{(\Vec{k}^2+m^2)^2}\,,\\ &
    =\frac{m_1s_1m_2s_2}{8\pi m_p^2}\rmint dt \,e^{-mr}\Big[\frac{\Vec{v}_1\cdot\Vec{v}_2}{r}-\frac{(\Vec{v}_1\cdot\Hat{n})(\Vec{v}_2\cdot\Hat{n})}{r}(1+mr)\Big]\,\sim \mathcal{O}(Lv^2).
    \end{split}
\end{align}
\newpage
\item The amplitude corresponds to the diagram as shown in Fig.~(\ref{fig6mb}) with the worldline vertices $\rmint dt \,\Tilde{\psi}\varphi,$  $\rmint dt \,\Tilde{\psi}$ and  $\rmint dt \,\varphi$ : 
\begin{align}
    \begin{split}
        \mathcal{S}_{\text{eff}}\Big |_{\text{fig}(\ref{fig6mb})}&=\frac{m_1s_1m_2^2s_2}{m_{p}^4}\rmint\Bar{\mathcal{D}}\Hat{\xi}\rmint dt_1 \,\varphi(\Vec{x}_1(t_1))\Tilde{\psi}(\Vec{x}_1(t_1)) \rmint dt_2 \,\varphi(\Vec{x}_2(t_2))\rmint dt_3 \Tilde{\psi}(\Vec{x}_2(t_3)) \,,\\ &=\frac{m_1 s_1 m_2^2s_2}{m_p^4}\rmint dt_1 dt_2 dt_3 \Big\langle\varphi(\Vec{x}_1(t_1))\varphi(\Vec{x}_2(t_2))\Big\rangle \Big\langle\Tilde{\psi}(\Vec{x}_1(t_1))\Tilde{\psi}(\Vec{x}_2(t_3))\Big\rangle\,,\\ &
        =\frac{m_1 s_1 m_2^2s_2}{2 m_p^4}\rmint dt \rmint_{k_1}\frac{e^{i\Vec{k}_1\cdot\Vec{r}}}{\Vec{k}_1^2+m^2}\rmint_{k_2}\frac{e^{i\Vec{k}_2\cdot\Vec{r}}}{\Vec{k}_2^2}\,,\\ &
        =\frac{m_1 s_1 m_2^2s_2}{32 \pi^2 m_p^4}\rmint dt \frac{e^{-mr}}{r^2}+(1\leftrightarrow 2)\,\,\sim \mathcal{O}(Lv^2).
    \end{split}
\end{align}
\item The amplitude corresponds to the diagram in Fig.~(\ref{fig6mc}) with the worldline vertices $\rmint dt\, \varphi\varphi$ and $\rmint dt \,\varphi$ : 
\begin{align}
    \begin{split}
\mathcal{S}_{\text{eff}}\Big |_{\text{fig}(\ref{fig6mc})}&=\frac{m_1g_1m_2^2s_2^2}{m_{p}^4}\rmint\Bar{\mathcal{D}}\Hat{\xi}\rmint dt_1 \,\varphi(\Vec{x}_1(t_1))\varphi(\Vec{x}_1(t_1)) \rmint dt_2 \,\varphi(\Vec{x}_2(t_2))\rmint dt_3 \varphi(\Vec{x}_2(t_3)) \,,\\ & =\frac{m_1 g_1 m_2^2 s_2^2}{m_p^4}\rmint dt_1 dt_2 dt_3 \Big\langle\varphi(\Vec{x}_1(t_1))\varphi(\Vec{x}_2(t_2))\Big\rangle \Big\langle\varphi(\Vec{x}_1(t_1))\varphi(\Vec{x}_2(t_3))\Big\rangle\,,\\ &
         =\frac{m_1 g_1 m_2^2 s_2^2}{16\pi^2 m_p^4}\rmint dt \frac{e^{-2mr}}{r^2}+(1\leftrightarrow 2)\,\,\sim \mathcal{O}(Lv^2).
    \end{split}
\end{align}
\end{itemize}
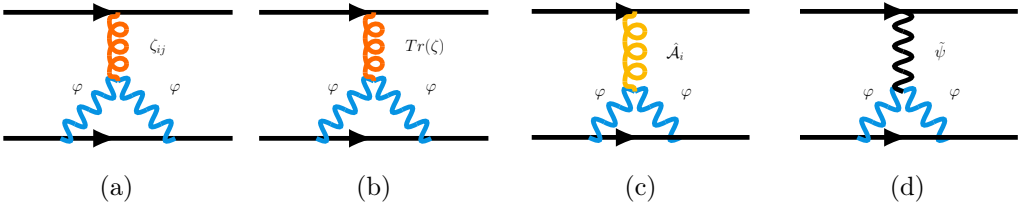
\begin{figure}
\centering
\begin{subfigure}{.22\textwidth}
  \centering
  \scalebox{0.3}{\begin{feynman}
    \fermion[lineWidth=6]{4.00, 6.20}{8.00, 6.20}
    \fermion[lineWidth=6]{4.00, 4.00}{8.00, 4.00}
    \electroweak[lineWidth=6, color=0693e3, label=$\varphi$]{6.00, 5.00}{7.00, 4.00}
    \electroweak[flip=true, lineWidth=6, color=0693e3, label=$\varphi$]{5.00, 4.00}{6.00, 5.00}
    \gluon[lineWidth=6, color=ff6900, label=$\zeta_{ij}$]{6.00, 5.00}{6.00, 6.20}
\end{feynman}}
  \caption{}
  \label{fig3a}
\end{subfigure}%
\begin{subfigure}{.22\textwidth}
  \centering
   \scalebox{0.3}{\begin{feynman}
    \fermion[lineWidth=6]{4.00, 6.20}{8.00, 6.20}
    \fermion[lineWidth=6]{4.00, 4.00}{8.00, 4.00}
    \electroweak[lineWidth=6, color=0693e3, label=$\varphi$]{6.00, 5.00}{7.00, 4.00}
    \electroweak[flip=true, lineWidth=6, color=0693e3, label=$\varphi$]{5.00, 4.00}{6.00, 5.00}
    \gluon[lineWidth=6, color=ff6900, label=$Tr(\zeta)$]{6.00, 5.00}{6.00, 6.20}
\end{feynman}
}
  \caption{}
  \label{fig3b}
\end{subfigure}
\begin{subfigure}{.22\textwidth}
\centering
\scalebox{0.3}{\begin{feynman}
    \electroweak[lineWidth=6, color=0693e3, label=$\varphi$]{5.80, 4.80}{6.60, 4.00}
    \fermion[lineWidth=6]{4.00, 6.20}{7.80, 6.20}
    \fermion[lineWidth=6]{4.00, 4.00}{7.80, 4.00}
    \electroweak[flip=true, lineWidth=6, color=0693e3, label=$\varphi$]{5.00, 4.00}{5.80, 4.80}
    \gluon[lineWidth=6, label=$\Hat{\mathcal{A}}_{i}$, color=fcb900]{5.80, 4.80}{5.80, 6.20}
\end{feynman}
}
    \caption{}
  \label{fig3c}
\end{subfigure}
\begin{subfigure}{0.22\textwidth}
  \centering
  \scalebox{0.3}{\begin{feynman}
    \electroweak[lineWidth=6, color=0693e3, label=$\varphi$]{5.80, 4.80}{6.60, 4.00}
    \fermion[lineWidth=6]{4.00, 6.20}{7.80, 6.20}
    \fermion[lineWidth=6]{4.00, 4.00}{7.80, 4.00}
    \electroweak[flip=true, lineWidth=6, color=0693e3, label=$\varphi$]{5.00, 4.00}{5.80, 4.80}
    \electroweak[lineWidth=6, label=$\Tilde{\psi}$]{5.80, 4.80}{5.80, 6.20}
\end{feynman}}
\caption{}
\label{fig3d}
\end{subfigure}
\caption{\textcolor{black}{Scalar diagrams coming from the 3-point vertices that contribute at 2 PN order in scalar effective action.}}
\label{fig3}
\end{figure}
\textcolor{black}{For the scalar sector, we find the effective action upto 2PN order. At 2PN order, corrections appear due to the scalar-graviton interaction vertices as shown in Fig.~(\ref{fig3}).} We now list out their contributions.
\newpage
\begin{itemize}
\item The amplitude corresponds to the diagram in Fig.~(\ref{fig3a}) with the following bulk interaction vertex $\rmint d^4 x\frac{1}{2}\boldsymbol{\zeta}^{lm}\partial_{l}\varphi\partial_{m}\varphi(x)$:
\begin{align}
    \begin{split}
        \mathcal{S}_{\text{eff}}\Big|_{\text{fig}(\ref{fig3a})}&=-\frac{m_1}{2m_p^4}\rmint\bar {\mathcal{D}}{\hat\xi}\,\rmint dt_1 \boldsymbol{\zeta}_{ij}(\Vec{x}_1(t_1))v_1^i v_1^j\rmint dt_2 m_2 s_2 \varphi(\Vec{x}_2(t_2))\rmint dt_2 m_2 s_2 \varphi(\Vec{x}_2(t_3))\\&\hspace{5cm}\rmint d^3x \,dt\frac{1}{2}\boldsymbol{\zeta}^{lm}\partial_{l}\varphi\partial_{m}\varphi(x)\,,\\ &
        =-\frac{ m_1 m_2^2 s_2^2}{4m_p^4}\rmint \prod_{i=1}^3 dt_i dt \,v_1^{i}(t_1)v_1^{j}(t_1)\rmint d^3x \,\partial_{l}\Big\langle\varphi(\Vec{x}_2(t_2))\varphi(x)\Big\rangle \partial_{m}\Big\langle\varphi(\Vec{x}_2(t_3))\varphi(x)\Big\rangle \\& \hspace{8.5cm}
        \Big\langle\boldsymbol{\zeta}_{ij}(\Vec{x}_1(t_1))\boldsymbol{\zeta}^{lm}(x)\Big\rangle\,,\\ &
        =-\frac{2 m_1m_2^2 s_2^2}{m_p^4}\rmint dt\, v_1^{i}(t)v_1^{j}(t)\rmint_{\Vec{k},\Vec{k}_1}\frac{e^{i\Vec{k}\cdot \Vec{r}}k_{1l}(k_{1m}-k_m)}{(\Vec{k}_1^2+m^2)[(\Vec{k}-\Vec{k}_1)^2+m^2]\Vec{k}^2}\\&\,\,\,\,\,\,\,\,\,\,\,\,\,\,\,\,\,\,\,\,\,\,\,\,\,\,\,\,\,\,\,\,\,\,\,\,\,\,\,\,\,\,\,\,\,\,\,\,\,\,\,\,\,\,\,\,\,\,\,\,\,\,\,\,\, \,\,\,\,\,\,\,\,\,\,\,\,\,\,\,\,\,\,\,\,\,\,\,\,\,\,\,\,\,\,\,\,\,\,\,\,\,\,\,\,\,\,\,\,\,\,\,\,\,\,\,\,\,\,\,\,\,\,\,\,\,\,\,\,\,\,\,\,\,\,\,\,\,\,\,\,(\delta_{i}^{l}\delta_j^m+\delta_{i}^m\delta_j^l-2\delta_{ij}\delta^{lm})\,,\\ &
        =-\frac{4m_1 m_2^2 s_2^2}{m_p^4}\rmint dt \Big[(-2\lambda_1+2\chi_1-\lambda_2+\chi_2)v_1^2+(\lambda_2-\chi_2)(\Vec{v}_1\cdot\Hat{n})^2\Big]\\&\,\,\,\,\,\,\,\,\,\,\,\,\,\,\,\,\,\,\,\,\,\,\,\,\,\,\,\,\,\,\,\,\,\,\,\,\,\ +(1\leftrightarrow2)\sim \mathcal{O}(Lv^4)\,.
        \label{4.38}
    \end{split}
\end{align}
\vspace{-0.4cm}
The calculation and the functions $\lambda_{1,2}(r)$, $\chi_{1,2}(r)$ are explicitly given in Appendix~ (\ref{App1}).

\vspace{0.4cm}
\item The amplitude corresponds to the diagram in Fig.~(\ref{fig3b}) with the following bulk interaction vertex $\rmint d^4x\,\frac{1}{4}\boldsymbol{\zeta}^{lm}\delta_{lm}\partial_{k}\varphi\partial_{k}\varphi(x)$:
\begin{align}
    \begin{split}
          \mathcal{S}_{\text{eff}}\Big |_{\text{fig}(\ref{fig3b})}&=-\frac{m_1}{2m_p^4}\rmint\Bar{\mathcal{D}}\Hat{\xi}\rmint dt_1 \boldsymbol{\zeta}_{ij}(\Vec{x}_1(t_1))v_1^i v_1^j\rmint dt_2 m_2 s_2 \varphi(\Vec{x}_2(t_2))\rmint dt_2 m_2 s_2 \varphi(\Vec{x}_2(t_3))\\&\hspace{5cm}\rmint d^3x \,dt\frac{1}{4}\boldsymbol{\zeta}^{lm}\delta_{lm}\partial_{k}\varphi\partial_{k}\varphi(x)\,,\\ &
        =-\frac{m_1 m_2^2 s_2^2}{8 m_p^4}\rmint dt_1 dt_2 dt_3 dt v_{1}^i(t_1)v_1^j(t_1) \rmint d^3x \,\partial_{k}\Big\langle\varphi(\Vec{x}_2(t_2))\varphi(x)\Big\rangle \partial_{k}\Big\langle\varphi(\Vec{x}_2(t_3))\varphi(x)\Big\rangle
       \\&\hspace{8.5cm} \Big\langle\boldsymbol{\zeta}_{ij}(\Vec{x}_1(t_1))\boldsymbol{\zeta}^{lm}(x)\Big\rangle \delta_{lm}\,,\\ &
        =-\frac{m_1 m_2^2 s_2^2}{ m_p^4}\rmint dt v_1^{i}(t)v_1^j(t)\rmint_{k,k_1}\frac{e^{i\Vec{k}\cdot \Vec{r}}(\Vec{k}_1^2-\Vec{k}\cdot \Vec{k}_1)}{(\Vec{k}_1^2+m^2)[(\Vec{k}_1-\Vec{k})^2+m^2]\Vec{k}^2}\delta_{ij}\,,\\ &
        =-\frac{m_1 m_2^2 s_2^2}{m_p^4}\rmint dt\, v_1^2\Big(3(\lambda_1-\chi_1)+(\lambda_2-\chi_2)\Big)+(1\leftrightarrow 2)\sim \mathcal{O}(Lv^4)\,.
    \end{split}
\end{align}
Again all the details of the integrals are given in Appendix~(\ref{App1}). The functions $\lambda_{1,2},\chi_{1,2}$ are defined in  (\ref{B.5m}) of Appendix~(\ref{App1}).
\item The amplitude corresponds to the diagram in Fig.~(\ref{fig3c}) with the bulk interaction vertex $\rmint d^4x\,\Hat{\mathcal{A}}^{k}\partial_{k}\varphi\partial_{0}\varphi(x)$:
\begin{align}
    \begin{split}
          \mathcal{S}_{\text{eff}}\Big|_{\text{fig}(\ref{fig3c})}&=2m_1\rmint\Bar{\mathcal{D}}\Hat{\xi}\rmint dt_1\, \frac{\Hat{\mathcal{A}}_{i}(\Vec{x}_1(t_1))}{m_{p}}v_1^{i}(t_1)\rmint dt_2\,\frac{m_2 s_2}{m_p}\varphi(\Vec{x}_2(t_2))\rmint dt_3\,\frac{m_2 s_2}{m_p}\varphi(\Vec{x}_2(t_3))\\&\hspace{5cm}\rmint d^3x dt \,\frac{\Hat{\mathcal{A}}^{k}}{m_p}\partial_{k}\varphi\partial_{0}\varphi(x)\,,\\ &
        =\frac{2m_1 m_2^2 s_2^2}{m_p^4}\rmint dt dt_1 dt_2 dt_3 v_1^i(t_1)\rmint d^3x \Big\langle\Hat{\mathcal{A}}_{i}(\Vec{x}_1(t_1))\mathcal{\Hat{A}}^{k}(x)\Big\rangle \partial_{k}\Big\langle\varphi(\Vec{x}_2(t_2))\varphi(x)\Big\rangle \\&\hspace{9.5cm}\partial_{0}\Big\langle\varphi(\Vec{x}_2(t_3))\varphi(x  )\Big\rangle\,,\\ &
        =-\frac{4m_2 m_1^2 s_1^2}{m_p^4}\rmint dt v_1^i(t)v_2^{l}(t)\rmint_{k_2,k_3}\frac{k_2^i\,k_3^l\,e^{-i(\Vec{k}_2+\Vec{k}_3)\cdot \Vec{r}}}{(\Vec{k}_2+\Vec{k}_3)^2(\Vec{k}_2^2+m^2)(\Vec{k}_3^2+m^2)}\,,\\ &
        =-\frac{4m_2 m_1^2s_1^2}{m_p^4}\rmint dt[(\lambda_1+\chi_1)\Vec{v}_1\cdot\Vec{v}_2+(\lambda_2+\chi_2)(\Vec{v}_1\cdot \Hat{n})(\Vec{v}_2\cdot \Hat{n})]+(1\leftrightarrow 2)\sim \mathcal{O}(L v^4)\,.
    \end{split}
\end{align}
Again all the  details of the computation and the functions $\lambda_{1,2},\chi_{1,2}$ are given in Appendix~(\ref{App1}).
\item The amplitude corresponds to the diagram in Fig.~(\ref{fig3d}) with the bulk interaction vertex $\rmint d^4x\, \frac{\Tilde{\psi}}{m_p}\partial_{0}\varphi\partial_{0}\varphi$:
\begin{align}
    \begin{split}
          \mathcal{S}_{\text{eff}}\Big |_{\text{fig}(\ref{fig3d})}&=2m_1\rmint\Bar{\mathcal{D}}\Hat{\xi}\rmint dt_1 \frac{\Tilde{\psi}(\Vec{x}_1(t_1))}{m_p}\rmint dt_2 \frac{m_2 s_2}{m_p}\varphi(\Vec{x}_2(t_2))\rmint dt_3 \frac{m_2 s_2}{m_p}\varphi(\Vec{x}_2(t_3))\\& \,\,\,\,\,\,\,\,\,\,\,\,\,\,\,\,\,\,\,\,\,\,\,\,\,\,\,\,\,\,\,\,\,\,\,\,\,\,\,\,\,\,\,\,\,\,\,\,\,\,\,\,\,\,\,\,\,\,\,\,\,\,\,\,\,\,\,\,\,\,\,\,\,\,\,\,\,\,\,\,\,\,\,\,\,\,\,\,\,\,\,\,\,\,\,\,\,\,\,\,\,\,\,\,\,\,\,\,\,\,\,\,\,\,\,\,\,\,\,\,\,\,\,\, \rmint d^3x\,dt\, \frac{\Tilde{\psi}}{m_p}\partial_{0}\varphi\partial_{0}\varphi\,.\\ &
        =\textstyle{\frac{2m_1 m_2^2 s_2^2}{m_p^4}\rmint dt_1 dt_2 dt_3 dt\rmint d^3x \Big\langle\Tilde{\psi}(\Vec{x}_1(t_1))\Tilde{\psi}(x)\Big\rangle\partial_{0}\Big\langle\varphi(\Vec{x}_2(t_2))\varphi(x)\Big\rangle \partial_{0}\Big\langle\varphi(\Vec{x}_2(t_3))\varphi(x  )\Big\rangle}\,,\\ &
        =\frac{m_1 m_2^2 s_2^2}{m_p^4}\rmint dt \, v_2^a(t)v_2^b(t)\rmint_{k,k_1}\frac{k_{1a}(k_{1b}-k_b)e^{i\Vec{k}\cdot \Vec{r}}}{\Vec{k}^2(\Vec{k}_1^2+m^2)[(\Vec{k}-\Vec{k}_1)^2+m^2]}\,,\\ &
        =\frac{m_1 m_2^2 s_2^2}{m_p^4}\rmint dt \, [v_2^2(\lambda_1-\chi_1)+(\Vec{v}_2\cdot \Hat{n})^2(\lambda_2-\chi_2)]+(1\leftrightarrow 2)\sim \mathcal{O}(Lv^4)\,.
    \end{split}
\end{align}
\end{itemize}
At 2PN there are several more terms that are of the order $\mathcal{O}(Gv^4)$. We again list the contribution of those diagrams to the effective action. \newpage
\begin{figure}[t!]
    \centering
    \begin{subfigure}{0.25\textwidth}
        \centering
        \scalebox{0.3}{\begin{feynman}
    \electroweak[lineWidth=4, label=$\varphi$, color=0693e3]{5.80, 4.00}{5.80, 6.20}
    \fermion[lineWidth=4, label=$v^4$]{4.00, 6.20}{7.60, 6.20}
    \fermion[lineWidth=4, label=$v^0$]{4.00, 4.00}{7.60, 4.00}
\end{feynman}}
\caption{}
\label{fig4a}
    \end{subfigure}
    \begin{subfigure}{0.25\textwidth}
        \centering
        \scalebox{0.3}{\begin{feynman}
    \electroweak[lineWidth=4, label=$\varphi$, color=0693e3]{5.80, 4.00}{5.80, 6.20}
    \fermion[lineWidth=4, label=$v^2$]{4.00, 6.20}{7.60, 6.20}
    \fermion[lineWidth=4, label=$v^2$]{4.00, 4.00}{7.60, 4.00}
\end{feynman}}
\caption{}
\label{fig4b}
    \end{subfigure}
    \caption{Diagrams contributing to scalar bound sector at 2 PN order coming from worldline vertices. }
    \label{fig4}
\end{figure}
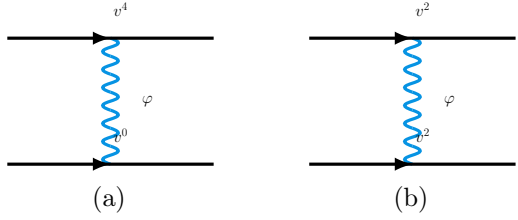
\begin{itemize}
    \item One term that contributes consists of the worldline vertex $\frac{1}{8}m_1 s_1\rmint dt \,v^4\frac{\varphi}{m_p}$ as shown in the Fig.~(\ref{fig4a}). The corresponding amplitude is given by,
    \begin{align}
        \begin{split}
              \mathcal{S}_{\text{eff}}\Big|_{\text{fig}(\ref{fig4a})}&=-\frac{m_1 s_1 m_2 s_2 }{8m_p^2}\rmint\Bar{\mathcal{D}}\Hat{\xi}\rmint dt_1\,v_1^4(t_1) \varphi(\Vec{x}_1(t_1))\rmint dt_2 \varphi(\Vec{x}_2(t_2))\,,\\ &
            =-\frac{m_1 s_1 m_2 s_2 }{8m_p^2}\rmint dt_1 \,dt_2\,v_1^4(t_1)\Big\langle\varphi(\Vec{x}_1(t_1))\varphi(\Vec{x}_2(t_2))\Big\rangle\,,\\ &
            =-\frac{m_1 s_1 m_2 s_2}{32\pi m_p^2}\rmint dt \,v_1^4(t)\,\frac{e^{-mr}}{r}+(1\leftrightarrow2)\sim \mathcal{O}(Lv^4)\,.
        \end{split}
    \end{align}
    \item Another term that contributes at this order consists of the worldline vertex vertex:$\rmint dt \frac{1}{2m_p}v^2\,m_a s_a \varphi$ as shown in Fig.~(\ref{fig4b}). The corresponding amplitude is given by,
    \begin{align}
        \begin{split}
              \mathcal{S}_{\text{eff}}\Big|_{\text{fig}(\ref{fig4b})}&=\frac{m_1 s_1 m_2 s_2}{4m_p^2}\rmint\Bar{\mathcal{D}}\Hat{\xi}\rmint dt_1 v_1^2(t_1)\varphi(\Vec{x}_1(t_1))\rmint dt_2 v_2^2(t_2)\varphi(\Vec{x}_2(t_2))\,,\\ &
            =\frac{m_1 s_1 m_2 s_2}{4m_p^2}\rmint dt_1 \, dt_2 v_1^2(t_1)v_2^2(t_2)\Big\langle\varphi(\Vec{x}_1(t_1))\varphi(\Vec{x}_2(t_2))\Big\rangle\,,\\ &
            =\frac{m_1 s_1 m_2 s_2}{16 \pi m_p^2}\rmint dt\,v_1^2(t)v_2^2(t)\frac{e^{-mr}}{r}+(1\leftrightarrow 2)\,\sim \mathcal{O}(Lv^4)\,.
        \end{split}
    \end{align}
\end{itemize} 
\subsection{Proca-electromagnetic interaction sector}
Now we focus on the Proca-electromagnetic interaction sector. This has an interaction term like $F_{\mu\nu}B^{\mu\nu}$. This term can be expanded \textcolor{black}{upto 1PN order }in the following fashion.
\begin{align}
    \begin{split}
  \frac{\gamma}{2} F_{\mu\nu}B^{\mu\nu}&=\frac{\gamma}{2}(\nabla_\mu A_\nu-\nabla_\nu A_\mu)(\nabla^\mu B^\nu-\nabla^\nu B^\mu)\\ &
=\gamma(\partial_0A_i\partial_0B_i-\partial_0A_i\partial_iB_0+\partial_lA_i\partial_lB_i-\partial_lA_i\partial_iB_l-\partial_kA_0\partial_kB_0+\partial_kA_0\partial_0B_k)\,.\label{4.35}
      \end{split}
\end{align}
In the action mentioned in (\ref{4.35}), the two fields are different (although they are quadratic), and they provide us with two-point interaction vertices. 
\begin{figure}[t!]
    \centering
    \begin{subfigure}{0.22\textwidth}
    \centering
    \scalebox{0.3}{ \begin{feynman}
    \electroweak[lineWidth=6, color=9900ef, label=$\mathcal{A}_i$]{6.00, 5.00}{6.00, 6.00}
    \dashed[showArrow=false, color=9900ef, lineWidth=6, label=$\mathcal{B}_0$]{6.00, 4.00}{6.00, 5.00}
    \fermion[lineWidth=6]{4.00, 4.00}{8.00, 4.00}
    \fermion[lineWidth=6]{4.00, 6.00}{8.00, 6.00}
\end{feynman}}
        \caption{}
        \label{fig9am}
    \end{subfigure}
    \begin{subfigure}{0.22\textwidth}
        \centering
        \scalebox{0.3}{\begin{feynman}
    \electroweak[lineWidth=6, color=9900ef, label=$\mathcal{A}_i$]{6.00, 5.00}{6.00, 6.00}
    \gluon[flip=true, lineWidth=6, color=eb144c, label=$\mathcal{B}_i$]{6.00, 4.00}{6.00, 5.00}
    \fermion[lineWidth=6]{4.00, 4.00}{8.00, 4.00}
    \fermion[lineWidth=6]{4.00, 6.00}{8.00, 6.00}
\end{feynman}
}
\caption{}
\label{fig9bm}
    \end{subfigure}
    \begin{subfigure}{0.22\textwidth}
        \centering
        \scalebox{0.3}{\begin{feynman}
    \electroweak[lineWidth=6, color=9900ef, label=$\mathcal{A}_i$]{6.00, 5.00}{6.00, 6.00}
    \gluon[flip=true, lineWidth=6, color=eb144c, label=$\mathcal{B}_l$]{6.00, 4.00}{6.00, 5.00}
    \fermion[lineWidth=6]{4.00, 4.00}{8.00, 4.00}
    \fermion[lineWidth=6]{4.00, 6.00}{8.00, 6.00}
\end{feynman}
}
\caption{}
\label{fig9cm}
    \end{subfigure}
    \begin{subfigure}{0.22\textwidth}
        \centering
        \scalebox{0.3}{\begin{feynman}
    \dashed[showArrow=false, lineWidth=6, color=9900ef, label=$\mathcal{B}_0$]{6.00, 4.00}{6.00, 5.00}
    \fermion[lineWidth=6]{4.00, 4.00}{8.00, 4.00}
    \fermion[lineWidth=6]{4.00, 6.00}{8.00, 6.00}
    \dashed[showArrow=false, lineWidth=6, label=$\mathcal{A}_0$]{6.00, 5.00}{6.00, 6.00}
\end{feynman}}
\caption{}
\label{fig9dm}
    \end{subfigure}
    \begin{subfigure}{0.22\textwidth}
        \centering
        \scalebox{0.3}{\begin{feynman}
    \gluon[flip=true, color=eb144c, lineWidth=6, label=$\mathcal{B}_k$]{6.00, 4.00}{6.00, 5.00}
    \fermion[lineWidth=6]{4.00, 4.00}{8.00, 4.00}
    \fermion[lineWidth=6]{4.00, 6.00}{8.00, 6.00}
    \dashed[showArrow=false, lineWidth=6, label=$\mathcal{A}_0$]{6.00, 5.00}{6.00, 6.00}
\end{feynman}}
\caption{}
\label{fig9em}
    \end{subfigure}
    \caption{Diagrams contributing to the bound sector at 1PN order due to the Proca-electromagnetic interaction term.}
    \label{fig:my_label2}
\end{figure}
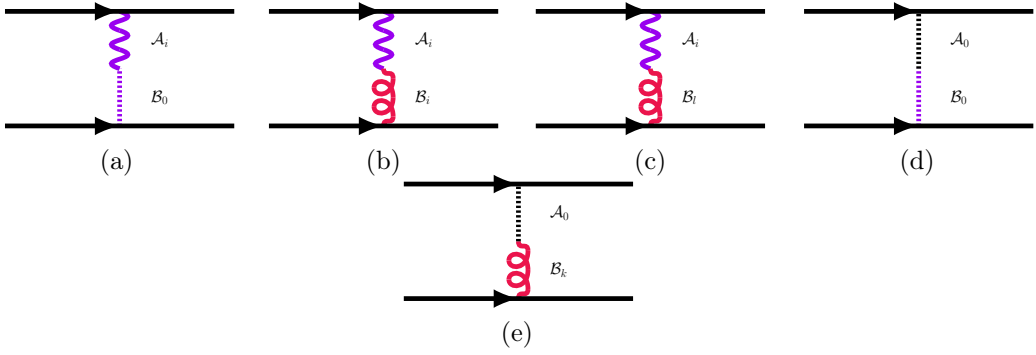
Now we need to calculate the amplitude of all the diagrams shown in Fig.~(\ref{fig:my_label}) that contribute to 1PN order in the effective action. Below we list all such contributions. 
\begin{itemize}
\item The amplitude corresponding to the diagram in Fig.~(\ref{fig9am}) with bulk interaction vertex $\rmint d^4 x\,\partial_0\mathcal{A}_i\partial_i\mathcal{B}_0$:
\begin{align}
    \begin{split}
   \mathcal{S}_{\text{eff}}\Big |_{{\text{fig}(\ref{fig9am})}}&=\gamma \,Q_1 Q_2'\rmint\Bar{\mathcal{D}}\Hat{\xi}\rmint dt_1v_1^j(t_1)\mathcal{A}_j(\Vec{x}_1(t_1))\rmint dt_2\, \mathcal{B}_0(\Vec{x}_2(t_2))\rmint d^4x\,\partial_0\mathcal{A}_i(x)\partial_i\mathcal{B}_0(x),\\ &=\gamma \,Q_1 Q_2'\rmint dt_1dt_2v_1^j(t_1)\rmint d^4x\,\partial_0\Big{\langle}\mathcal{A}_j(\Vec{x}_1(t_1))\mathcal{A}_i(x)\Big{\rangle}\partial_i\Big{\langle}\mathcal{B}_0(\Vec{x}_2(t_2))\mathcal{B}_0(x)\Big{\rangle},\\ &
=i\gamma\,Q_1 Q_2'\rmint dt \Bigg[a^{i}_{1}(t)\rmint_{\Vec{k}}\frac{k^i\,e^{-i\,\Vec{k}\cdot\Vec{r}}}{\Vec{k}^2(\Vec{k}^2+\mu_{\gamma}^2)}+v_1^{i}(t)v_{1}^{a}(t)\frac{i\,k^a\,k^i\,e^{-i\Vec{k}\cdot\Vec{r}}}{\Vec{k}^2(\Vec{k}^2+\mu_{\gamma}^2)}\Bigg]\,,\\ &
=-\frac{\gamma\,Q_1 Q_2'}{2\pi}\rmint dt \Bigg[(\Vec{a}_1\cdot \Hat{n})\Big\{\frac{  e^{-\mu_{\gamma} r}}{2 \mu_{\gamma} \,r}-\frac{1 -  e^{-\mu_{\gamma} r}}{2\, \mu_{\gamma}^2\, r^2}\Big\}+\Vec{v}_1^2 \,\frac{e^{-r \mu _{\gamma }} \left(-r \mu _{\gamma }+e^{r \mu _{\gamma }}-1\right)}{  r^3 \mu _{\gamma }^2}\\ &
\,\,\,\,\,\,\,\,\,\,\,\,\,\,\,\,\,\,\,\,\,\,\,\,\,\,\,\,\,
+(\Vec{v}_1\cdot \Hat{n})^2\,\frac{e^{-r \mu _{\gamma }} \Big(r \mu _{\gamma } \left(r \mu _{\gamma }+3\right)-3 e^{r \mu _{\gamma }}+3\Big)}{  r^3 \mu _{\gamma }^2}
\Bigg]+(1\leftrightarrow 2)\,\sim \mathcal{O}(L v^2)\,.
\end{split}
\end{align}
\vspace{-0.5cm}
\item The amplitude corresponding to the diagram in Fig.~(\ref{fig9bm}) with bulk interaction vertex $\rmint d^4x\,\partial_l\mathcal{A}_i\partial_l\mathcal{B}_i$:
\begin{align}
    \begin{split}
   \mathcal{S}_{\text{eff}}\Big|_{\text{fig}(\ref{fig9bm})}&=\gamma \,Q_1 Q_2'\rmint\Bar{\mathcal{D}}\Hat{\xi}\rmint dt_1v_1^j(t_1)\mathcal{A}_j(\Vec{x}_1(t_1))\rmint dt_2\, \mathcal{B}_m(\Vec{x}_2(t_2))v_2^m(\Vec{x}_2(t_2))\\ & \hspace{6.5 cm}\rmint d^4x\,\partial_l\mathcal{A}_i(x)\partial_l\mathcal{B}_i(x),\\&=\gamma Q_1 Q_2'\rmint dt_1dt_2v_1^j(t_1)v_2^m(t_2)\rmint d^4x\partial_l\Big{\langle}\mathcal{A}_j(\Vec{x}_1(t_1))\mathcal{A}_i(x)\Big{\rangle}\partial_l\Big{\langle}\mathcal{B}_m(\Vec{x}_2(t_2))\mathcal{B}_i(x)\Big{\rangle}\,,\\ &
=\frac{\gamma Q_1 Q_2'}{4\pi}\rmint dt\, (\vec{v}_1\cdot \vec{v}_2) \frac{e^{-\mu_{\gamma}r}}{r}+(1\leftrightarrow 2)\,\sim \mathcal{O}(L v^2)\,.
\end{split}
\end{align}
\newpage
\item The amplitude corresponding to the diagram in Fig.~(\ref{fig9cm}) with bulk interaction vertex $\rmint d^4x\,\partial_l\mathcal{A}_i\partial_i\mathcal{B}_l$:
\begin{align}
    \begin{split}
   \mathcal{S}_{\text{eff}}\Big|_{\text{fig}(\ref{fig9cm})}&=\gamma \,Q_1 Q_2'\rmint\Bar{\mathcal{D}}\Hat{\xi}\rmint dt_1v_1^j(t_1)\mathcal{A}_j(\Vec{x}_1(t_1))\rmint dt_2\, \mathcal{B}_m(\Vec{x}_2(t_2))v_2^m(\Vec{x}_2(t_2))\\ &\hspace{6.5 cm}\rmint d^4x\,\partial_l\mathcal{A}_i(x)\partial_i\mathcal{B}_l(x),\\&=\gamma Q_1 Q_2'\rmint dt_1dt_2v_1^j(t_1)v_2^m(t_2)\rmint d^4x\partial_l\Big{\langle}\mathcal{A}_j(\Vec{x}_1(t_1))\mathcal{A}_i(x)\Big{\rangle}\partial_i\Big{\langle}\mathcal{B}_m(\Vec{x}_2(t_2))\mathcal{B}_l(x)\Big{\rangle}\,,\\ &
=\gamma Q_1Q_2'\rmint dt\, \,\Big[v_1^i(t) v_2^l(t)\
\rmint \frac{k_i k_l\,e^{-i\vec k\cdot \vec{r}}}{k^2(\vec{k}^2+\mu^2)}\Bigg]\,,                 \\ &
=\gamma Q_1 Q_2'\rmint dt \, \Big[(\Vec{v_1}\cdot \Vec{v}_2)\frac{e^{-r \mu _{\gamma }} \left(-r \mu _{\gamma }+e^{r \mu _{\gamma }}-1\right)}{2 \pi  r^3 \mu _{\gamma }^2}\\ &
\,\,\,\,\,\,\,\,\,\,\,\,\,\,\,\,\,\,\,\,\,\,\,\,\,\,\,\,\,\,\,\,\,+(\Vec{v}_1\cdot \Hat{n})(\Vec{v}_2\cdot \Hat{n})\frac{e^{-r \mu _{\gamma }} \Big(r \mu _{\gamma } \left(r \mu _{\gamma }+3\right)-3 e^{r \mu _{\gamma }}+3\Big)}{2 \pi  r^3 \mu _{\gamma }^2}\Big]+(1\leftrightarrow 2) \sim \mathcal{O}(L v^2)\,.
\end{split}
\end{align}
\item The amplitude corresponding to the diagram in Fig.~(\ref{fig9dm}) with bulk interaction vertex $\rmint d^4x\,\partial_k \mathcal{A}_0\partial_k \mathcal{B}_0$:
\begin{align}
    \begin{split}
       \mathcal{S}_{\text{eff}}\Big|_{\text{fig}(\ref{fig9dm})}&=\gamma \,Q_1 Q_2'\rmint\Bar{\mathcal{D}}\Hat{\xi}\rmint dt_1 \mathcal{A}_0(\Vec{x}_1(t_1))\rmint dt_2\, \mathcal{B}_0(\Vec{x}_2(t_2))\rmint d^4x\,\partial_k\mathcal{A}_0(x)\partial_k\mathcal{B}_0(x),\\&
       =\gamma Q_1 Q_2'\rmint dt_1dt_2\rmint d^4x\partial_k\Big{\langle}\mathcal{A}_0(\Vec{x}_1(t_1))\mathcal{A}_0(x)\Big{\rangle}\partial_k\Big{\langle}\mathcal{B}_0(\Vec{x}_2(t_2))\mathcal{B}_0(x)\Big{\rangle}\,,\\ &
=\gamma Q_1Q_2'\rmint dt\, \,\Big[\rmint \frac{e^{-i\vec k\cdot \vec{r}}}{(\vec{k}^2+\mu^2)}\Bigg]    \,,             \\ &
=\gamma Q_1Q_2'\rmint dt\,\frac{e^{-\mu_{\gamma}r}}{4\pi r}+(1\leftrightarrow 2)\,\sim \mathcal{O}(L v^2)\,.
    \end{split}
\end{align}
\item The amplitude corresponding to the diagram in Fig.~(\ref{fig9em}) with the bulk interaction vertex $\rmint d^4x \,\partial_k\mathcal{A}_0\partial_0 \mathcal{B}_k$:
\begin{align}
    \begin{split}
   \mathcal{S}_{\text{eff}}\Big|_{\text{fig}(\ref{fig9em})}&=\gamma \,Q_1 Q_2'\rmint\Bar{\mathcal{D}}\Hat{\xi}\rmint dt_1\mathcal{A}_0(\Vec{x}_1(t_1))\rmint dt_2\, \mathcal{B}_j(\Vec{x}_2(t_2))v_2^j(\Vec{x}_2(t_2))\rmint d^4x\,\partial_k\mathcal{A}_0(x)\partial_0\mathcal{B}_k(x),\\&=\gamma Q_1 Q_2'\rmint dt_1dt_2v_2^j(t_2)\rmint d^4x\partial_k\Big{\langle}\mathcal{A}_0(\Vec{x}_1(t_1))\mathcal{A}_0(x)\Big{\rangle}\partial_0\Big{\langle}\mathcal{B}_j(\Vec{x}_2(t_2))\mathcal{B}_k(x) \Big{\rangle}\,,\\ &
=i\gamma Q_1 Q_2'\rmint dt \Bigg[a^{k}_{2}(t)\rmint_{\Vec{k}}\frac{k^k\,e^{-i\,\Vec{k}\cdot\Vec{r}}}{\Vec{k}^2(\Vec{k}^2+\mu^2)}-v_2^{k}(t)v_{2}^{a}(t)\frac{i\,k^a\,k^k\,e^{-i\Vec{k}\cdot\Vec{r}}}{\Vec{k}^2(\Vec{k}^2+\mu^2)}\Bigg]\,,\\ &
=-\frac{\gamma Q_1 Q_2'}{2\pi}\rmint dt \Bigg[(\Vec{a}_2\cdot \Hat{n})\Big\{\frac{  e^{-\mu_{\gamma} r}}{2 \mu_{\gamma} \,r}-\frac{1 -  e^{-\mu_{\gamma} r}}{2\, \mu_{\gamma}^2\, r^2}\Big\}+\Vec{v}_2^2 \,\frac{e^{-r \mu _{\gamma }} \left(-r \mu _{\gamma }+e^{r \mu _{\gamma }}-1\right)}{  r^3 \mu _{\gamma }^2}\\ &
\,\,\,\,\,\,\,\,\,\,\,\,\,\,\,\,\,\,\,\,\,\,\,\,\,\,\,\,\,
+(\Vec{v}_2\cdot \Hat{n})^2\,\frac{e^{-r \mu _{\gamma }} \Big(r \mu _{\gamma } \left(r \mu _{\gamma }+3\right)-3 e^{r \mu _{\gamma }}+3\Big)}{  r^3 \mu _{\gamma }^2}
\Bigg]+(1\leftrightarrow 2)\,\sim \mathcal{O}(L v^2)\, \label{4.52}
\end{split}
\end{align}
\end{itemize}

Before we conclude this section, we summarize our results in Table~(\ref{table2mm}). We show the number of diagrams responsible for the corrections to the conservative dynamics due to the various fields and their interactions, apart from the pure gravity contributions. 

\begin{table}[b!]
 \centering
 \scalebox{0.91}{\begin{tabular}{|c|c|c|c|c|}
 \hline
 & $0 PN$ & $1PN$ &  $2PN$ &  $2.5PN$ \\
  \hline
 \textit {scalar} & - & 5 & 6 &-\\
  \hline
 \textit{electromagnetic} & 1 & 3 & - &-\\
 \hline 
 \textit{proca} & 1 & 3 & - &- \\
 \hline 
 \textit{gravitational} & - & 4 & - &-\\
 \hline
 \textit{scalar-em interaction}
 & - & - & - &2($\sim g_{a\gamma\gamma}$)\\
 \hline
 \textit{proca-em interaction}&- & 4 $(\sim \gamma)$ & - &-\\
 \hline
 \end{tabular}}
 \caption{Table showing the number of diagrams contributing to a specific PN order for bound sector. Also, the orders at which terms to due the axion-photon coupling ($g_{a\gamma\gamma}$) and photon-dark photon kinetic mixing term ($\gamma$).}
 \label{table2mm}
 \end{table}
\vspace{0.5cm}
\textcolor{black}{{\bf Corrections to the bound orbit}: \textcolor{black}{Now we can add all the contributions mentioned in 
 (\ref{4.13a})-(\ref{4.52}) and that gives us the real part of the total effective action. To find out the orbit equation, one needs to extremize this.}
\begin{eqnarray}
  &  \frac{\delta}{\delta \Vec{x}_a(t)}\,\text{Re}\,\mathcal{S}_{\text{eff}}^{\text{tot}}[\Vec{x}_a]\Big |_{\mathcal{O}(\hbar^0)}=0\,.\label{4.53m}
\end{eqnarray}}
The leading order contribution to effective action is Newtonian potential. That gives rise to the Keplerian orbit as the solution of (\ref{4.53m}). This we use to study the radiative dynamics in Sec.~(\ref{Sec5}). Note that we will get corrections to the usual Keplerian orbit solution due to other terms in the effective action at higher-order PN. Although one should consider these corrections and find out the corrected orbit equation and then use it to study the radiative dynamics, in this paper, we will neglect such corrections and work with the leading order solution, i.e. Keplerian orbit solution (in fact, the circular orbit).

\section{The radiative dynamics}\label{Sec5}
\textcolor{black}{ Till now, we have discussed the conservative dynamics of the binary black holes where we do not consider the Feynman diagrams with the external radiative line. In this section, we will discuss the radiative sector of the different fields of the theory. To convince consider a field $\xi$ which can be decoupled into potential modes and radiation modes as, $\xi=\Hat{\xi}+\Bar{\xi}$, described by the action $S[\xi]$. Now first integrating over the potential modes (\ref{2.13}) we get the effective action for the radiative modes which has the following form: \begin{equation}\mathcal{S}_{\text{eff}}^{\Bar{\xi}}[\Vec{x}_a,\Bar{\xi}]=S_{\text{free}}^{\Bar{\xi}}+\rmint d^4 x\,J(x)\Bar{\xi}(x)\,,
\end{equation}
with the source term $J(x)$. \textcolor{black}{By comparing with (\ref{2.4m}) one can identify the source terms as Wilson coefficient $C_{1}$}. Now, we can expand the effective actions around $\bar\xi=0$ as,
\begin{align}
    \begin{split}
       & \mathcal{S}_{\text{eff}}^{\Bar{\xi}}[\Vec{x}_a,\Bar{\xi}]= \mathcal{S}_{0}[\Vec{x}_a]+\mathcal{S}_{1}[\Vec{x}_a,\Bar{\xi}]+\mathcal{S}_{2}[\Vec{x}_a,\Bar{\xi}]+\mathcal{O}(\bar\xi^3).\label{5.1m}
    \end{split}
\end{align}
The first term in (\ref{5.1m}) does not depend upon the radiative field and gives the conservative dynamics. The second term corresponds to the linear radiating field. The third term provides a kinetic term for the radiative fields and the higher-order terms give higher-order field coupling.\\ 
Now we discuss how we can calculate power radiation by evaluating the effective action for the source term. The effective action for the source term can be obtained by integrating out the fields as,
\begin{align}
\begin{split}
   \mathcal{Z}[J]:= e^{i{\gamma}_{\text{eff}}[J]}&=\rmint \mathcal{D}\Bar{\xi}\rmint \mathcal{D}\Hat{\xi}\,e^{iS[{\xi}]}\,,\\ &
   =\rmint \mathcal{D}\Bar{\xi}\, e^{i\mathcal{S}_{\text{eff}}^{\Bar{\xi}}[\Bar{\xi},J]}\,.
   \label{4.3}
   \end{split}
\end{align}
Again, in the in-out formalism $\mathcal{Z}[J]$ is the overlap between in and out states,
\begin{eqnarray}
    \mathcal{Z}[J]=\Big\langle0_+|0_-\Big\rangle.\label{4.4}
\end{eqnarray}
using the two relations (\ref{4.3}) and (\ref{4.4}) we get \cite{nastase_2019},
\begin{eqnarray}
   | \langle0_+|0_-\rangle|^2=e^{-2\text{Im}[\gamma_{\text{eff}}]}\,.
\end{eqnarray}
Expanding in small $\text{Im}[\gamma_{\text{eff}}]$ we will get,
\begin{align}
    \begin{split}
        2\,\text{Im}[\gamma_{\text{eff}}]=\mathcal{T}\rmint dE\,d\Omega \frac{d^2\Gamma}{dEd\Omega}\label{4.6}
    \end{split}
\end{align}
and the radiated power is given by,
\begin{align}
    \begin{split}
        P=\rmint dE\,d\Omega\,E\,\frac{d^2\Gamma}{dEd\Omega}\label{5.6m}
    \end{split}
\end{align}
where, $\mathcal{T}$ is the interaction time, $d\Gamma$ is the rate of particle emission and $P$ is the power radiation. We will use (\ref{4.6}) to calculate the power radiated for different fields.}
Recipe for calculating the effective action for the radiative field:
\begin{itemize}
    \item Draw the Feynman diagrams with external radiation field lines that do not contribute to the propagators.
    \item Calculate the amplitudes that give the effective action of the radiation field.
    \item Identify the source term and use equation (\ref{5.6m}) to compute the power radiation.
\end{itemize}
\subsection{Radiative sector PN counting} \label{EFTcounting}
 Now to compute the radiated power upto certain \textcolor{black}{$N^{(n)}LO$ } we will need to systematically associate EFT power counting rules to the radiative fields. To make things more compact we depict the scaling of the radiation fields: $\bar{\xi}\equiv\{\bar\psi,\bar{{\mathcal{A}}}_i,\bar\sigma_{ij},\bar{\phi},\Bar{b}_{\mu}, \Bar{a}_{\mu}\}$ in Table~(\ref{table3mm}). One should note that for the potential fields, one should perform a partial Fourier transform. Therefore,
$$\hat\xi(x^0,\vec{x})=\rmint_k e^{i\vec{k}\cdot \vec{x}}\,\hat\xi_k(x^0)$$
where $\xi_k\equiv \{\psi_k,\Hat{\mathcal{A}}_{{\Vec{k}}i},\zeta_{{\Vec{k}}ij}, \mathcal{A}_{\Vec{k}i},\mathcal{B}_{\Vec{k}i},\varphi_{\Vec{k}}\}$ are the potential fields. Here in this sector, we define the diagrams as $N^{(n)}LO$ i.e. leading order, next to leading order and so on. The specification of scaling for the radiative sector is denoted in terms of these notations as scaling rules are a bit subtle and different in the radiative sector. 
\begin{table}[ht]
\begin{center}
\begin{tabular}{|c|c|}
    \hline
   \multicolumn{2}{|c|}{EFT scaling} \\
\hline
Fields & scaling\\
\hline
t &  $\sim\frac{r}{v}$\\
\hline
  $\delta x$ &$\sim r$\\
  \hline
  $\frac{m}{M_{p}}$ & $\sim L^\frac{1}{2}v^\frac{1}{2}$\\
  \hline
  $Q^2$ & $ \sim L\,v$\\
  \hline
  $\xi_k$ & $\sim r^2v^\frac{1}{2}$\\
  \hline
$\bar{\Xi}$  & $\sim\frac{v}{r}$\\
\hline
$\partial_0\xi_k$ & $\sim\frac{v}{r}\xi_k$\\
\hline
  $\partial_{\mu}\bar{\Xi}$ & $\sim\frac{v}{r}\bar{\Xi}$\\
  \hline 
\end{tabular}
\end{center}
\caption{EFT PN counting for radiative sector}
\label{table3mm}
\end{table} 
\par  \textcolor{black}{Now expanding the metric at the linear level the  relation between NRG fields and the linearized metric components in the radiative sector is given by:$$
\bar g_{\mu\nu}=\eta_{\mu\nu}+\frac{1}{m_p}
\begin{pmatrix}
-2\bar\psi & \bar{\mathcal{A}_j}\\
\bar{\mathcal{A}_i} & \,\,\bar{\sigma_{ij}}-2\delta_{ij}\bar\psi\\
\end{pmatrix}\equiv\eta_{\mu\nu}+\frac{1}{m_p}
\begin{pmatrix}
\bar h_{00} & \bar h_{0j}\\
\bar h_{i0} & \bar h_{ij}\\
\end{pmatrix}
$$
The gravitational radiation formula we used in the later section in (\ref{5.65}) is basically considering the $\bar\psi$ radiation. One subtle point to note here is that when we calculate radiation, usually we consider the TT gauge along with the harmonic gauge condition. But from the beginning, we can't set $\bar h_{00}=0$ in the effective Lagrangian for radiation demanded by \textit{Transverse Traceless (TT)} gauge condition.}

\subsection{Scalar interaction and scalar radiation}\label{subsec:radiated_scalars}

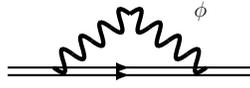
\begin{figure}
    \centering
\scalebox{0.4}{\begin{feynman}
    \electroweak[flip=true, lineWidth=4]{4.60, 4.10}{5.60, 4.80}
    \fermion[]{4.00, 4.00}{7.20, 4.00}
    \electroweak[lineWidth=4, label=$\bar\phi$]{5.60, 4.80}{6.60, 4.10}
    \fermion[]{4.00, 4.10}{7.20, 4.10}
\end{feynman}
}
    \caption{Self-energy diagram contributing to the effective action for the source term $J$.}
    \label{mfig12}
\end{figure}

Following a similar strategy used in Sec.~(\ref{Sec4}), we will calculate the power radiation sector by sector. First, we will start with the scalar sector. We will discuss the necessary details in this section, which will be useful for the subsequent computations of the power radiation in other sectors. For the scalar sector, we will compute the power radiation formula upto 2PN order. To begin with, we integrate the potential scalar modes and get the effective action for the radiating scalars.  
\begin{equation}
\mathcal{S}_{\text{eff}}^{\Bar{\phi}}=  \rmint d^4x  \left(\underbrace{ -\frac{1}{2} \eta^{\mu \nu}  \partial_\mu \bar{\phi} \partial_\nu \bar{\phi} 
-\frac{1}{2}m^2\bar{\phi}^2 }_{\mathcal{S}_{\rm free}^{\bar\phi}} + \underbrace{\frac{1}{m_p} J   \bar{\phi}}_{\mathcal{S}_{\text{int}}^{\bar\phi}} \right)\;, 
\end{equation}
which leads to the following equation of motion 
\begin{equation}
\textcolor{black}{(\square-m^2 )\bar{\phi} = \frac{J}{m_p} \;, \qquad \square \equiv \eta^{\mu \nu}  \partial_\mu  \partial_\nu \;.}
\end{equation}
We will now calculate the imaginary part of the effective action $\gamma_{\text{eff}}^{\bar\phi}$ for the source obtained by integrating out the radiation scalars as discussed in Eq.~(\ref{4.3}) 
and then using the optical theorem, we obtain the power radiation for the scalar sector. We will work out the power radiation essentially for circular orbits. Now,
\begin{equation}
\begin{split}
\mathcal{S}_{\rm int}^{\bar\phi} = \rmint d^4 x \frac{J \bar \phi }{m_p}= \rmint dt \, \rmint d^3 x \frac{J (t, \Vec{x})}{m_p}  \bigg( \bar{\phi}(t, \Vec{0}) + x^i \partial_i \bar{\phi}(t,\Vec{0}) + \frac{1}{2} x^i x^j \partial_i \partial_j \bar{\phi}(t, \Vec{0})  \\
+ \frac{1}{3!} x^i x^j x^k \partial_i \partial_j \partial_k \bar{\phi}(t, \Vec{0}) + \ldots \bigg) \;,
\label{5.10}
\end{split}
\end{equation}
From which we can write,
\begin{align}
\begin{split}
S_{\rm int}^{\bar\phi} = \frac{1}{m_p} \rmint dt \left( I_\phi \bar{\phi} + I_\phi^i \partial_i \bar{\phi} + \frac{1}{2} I_\phi^{ij} \partial_i \partial_j \bar{\phi}  + \ldots \right) \;,
\end{split}
\label{eq:multipole_expansion_scalar}
\end{align}
where 
\begin{equation}
\label{5.12mn}
\textcolor{black}{I_\phi  \equiv  \rmint d^3 x \left( J + \frac{1}{6}  (\partial_t^2+m^2) J  x^2 \right) \;, \quad I_\phi^i  \equiv \rmint d^3x \, x^i \left(J + \frac{1}{10} (\partial_t^2+m^2) J x^2 \right) \;, \quad  I_\phi^{ij} \equiv \rmint d^3x J  Q^{ij} \;}
\end{equation}
are respectively the scalar monopole, dipole, and quadrupole.
\textcolor{black}{ To get the irreducible forms of the multipole moments, instead of writing $x^ix^j$ one should use,
\begin{align}
    \begin{split}
        {Q}^{ij}=x^ix^j-\frac{1}{3}x^2\delta^{ij}\,.
    \end{split}
\end{align}}
Then we integrate over the scalar radiation field and get the effective action for the source, which is eventually equivalent to calculating the self-energy diagram Fig.~(\ref{mfig12}).
 \begin{align}
 \begin{split}
 i\gamma_{\text{eff}}^{\bar\phi}[J]&\sim \log\, \rmint \mathcal{D
        }\Bar{\phi}\,e^{i \mathcal{S}_{\rm free}^{\bar\phi}+i\mathcal{S}_{\rm int}^{\bar\phi}},\\ &
        =\log\,\rmint \mathcal{D}\Bar{\phi}\,e^{i \mathcal{S}_{\rm free}^{\bar\phi}}\Big[1+i \mathcal{S}_{\rm int}^{\bar\phi}+\frac{i^2}{2}(\mathcal{S}_{\rm int}^{\bar\phi})^{2}+..\Big],\\ &
        \sim \frac{-1}{2m_p^2}\rmint \mathcal{D}\Bar{\phi}\,e^{i \mathcal{S}_{\rm free}^{\bar\phi}} \rmint dt_1 dt_2 \Big[I_{\phi}(t_1)I_{\phi}(t_1) \Bar{\phi}(t_1,\Vec{0})\Bar{\phi}(t_2,\Vec{0})+I_{\phi}^{i}(t_1)I_{\phi}^{j}(t_1) \partial_{i}\Bar{\phi}(t_1,\Vec{0})\partial_{j}\Bar{\phi}(t_2,\Vec{0})+...\Big],\\ &
        =-\frac{1}{2m_p^2}\rmint dt_1 dt_2 \Big[I_{\phi}(t_1)I_{\phi}(t_1) \Big\langle\Bar{\phi}(t_1,\Vec{0})\Bar{\phi}(t_2,\Vec{0})\Big\rangle +I_{\phi}^{i}(t_1)I_{\phi}^{j}(t_1) \Big\langle\partial_{i}\Bar{\phi}(t_1,\Vec{0})\partial_{j}\Bar{\phi}(t_2,\Vec{0})\Big\rangle\\ &\,\,\,\,\,\,\,\,\,\,\,\,\,\,\,\,\,\,\,\,\,\,\,\,\,\,\,\,\,\,\,\,\,\,\,\,\,\,\,+I_{\phi}^{ij}(t_1)I_{\phi}^{kl}(t_2) \Big\langle\partial_{i}\partial_{j}\Bar{\phi}(t_1,\Vec{0})\partial_{k}\partial_{l}\Bar{\phi}(t_2,\Vec{0})\Big\rangle+...\Big]\\ &
      \,\,\,\,\,\,\,\,\,\,\,\,\,\,\,\,\,\,\,\,\,\,\,\,\,\,\,\,\,\,\,\,\,\,\,\,\,\,\,  +(\text{correlators involving even field operators})\,.\label{5.14n}
        \end{split}
        \end{align}
Here we have assumed the odd-point correlation functions are zero i.e. $\Big\langle\underbrace{\Bar{\phi}\Bar{\phi}...\Bar{\phi}}_{2n+1}\Big\rangle=0.\footnote{Note that here, we are linearizing the fields against the Minkowski background, and we chose the background value of the scalar field to be zero. It will be interesting to perform this computation for the case where the fields will have a non-zero vacuum expectation value i.e, $\langle 0 |\bar\phi|0\rangle\ne 0$.}$  
Then, using the following expression for the Feynman propagator of the scalar field,
\begin{align}
    \begin{split}
        \Big\langle T \Bar{\phi}(t_1,\Vec{x_1})\Bar{\phi}(t_2,\Vec{x_2})\Big\rangle=\rmint \frac{d^4k}{(2\pi)^4}\frac{i}{-k^2-m^2+i\epsilon}e^{-ik\cdot(x_1-x_2)}\,,\label{4.13}
    \end{split}
\end{align}
we get the following, 
\begin{align}
    \begin{split}
        \Big\langle T \partial_i \Bar{\phi}(t_1,\Vec{0})\partial_j \Bar{\phi}(t_2,\Vec{0})\Big\rangle&=\rmint \frac{d^4k}{(2\pi)^4}\frac{i}{-k^2-m^2+i\epsilon}e^{ik_0(t_1-t_2)}[\partial_{i}^{(x_1)}e^{-i\Vec{k}.\Vec{x_1}}][\partial_{j}^{(x_2)}e^{i\Vec{k}\cdot\Vec{x_2}}]\big|_{\Vec{x_1},\Vec{x_2}=0}\\ &
        =\rmint \frac{d^4k}{(2\pi)^4}\frac{i}{-k^2-m^2+i\epsilon}e^{ik_0(t_1-t_2)}\langle\langle k_{i}\,k_{j}\rangle\rangle\,,\\ &
        =\frac{1}{3}\rmint \frac{d^4k}{(2\pi)^4}\frac{i}{-k^2-m^2+i\epsilon}e^{ik_0(t_1-t_2)}|\Vec{k}|^2\delta_{ij}\,,\label{4.14}
    \end{split}
\end{align}
\begin{align}
    \begin{split}
          \Big\langle T \,\partial_i\partial_j \Bar{\phi}(t_1,\Vec{0})\,\partial_k\partial_l \Bar{\phi}(t_2,\Vec{0})\Big\rangle&=\rmint \frac{d^4k}{(2\pi)^4}\frac{i}{-k^2-m^2+i\epsilon}e^{ik_0(t_1-t_2)}[\partial_{i}\partial_j^{(x_1)}e^{-i\Vec{k}.\Vec{x_1}}][\partial_{k}\partial_l^{(x_2)}e^{i\Vec{k}\cdot\Vec{x_2}}]\big|_{\Vec{x_1},\Vec{x_2}=0}\,,\\ &
          =\rmint \frac{d^4k}{(2\pi)^4}\frac{i}{-k^2-m^2+i\epsilon}e^{ik_0(t_1-t_2)}\langle\langle k_ik_jk_kk_l\rangle\rangle\,,\\ &
          =\rmint \frac{d^4k}{(2\pi)^4}\frac{i}{-k^2-m^2+i\epsilon}e^{ik_0(t_1-t_2)}|\Vec{k}|^4 T_{ij,kl}\,,\label{4.15}
    \end{split}
\end{align}
where $T_{ij,kl}$ is constructed from $\langle\langle n_in_jn_kn_l\rangle\rangle$ and $\langle\langle n_in_j \rangle\rangle$ which is symmetric in $ij$ and $kl$ indices \cite{poissonwill}. Then finally we get, 
\begin{equation}
  \gamma_{\text{eff}}^{(\bar\phi)} = \frac{1}{2 m_p^2} \rmint \frac{d^4 k}{(2 \pi)^4} \frac{1}{-k^2-m^2 +i \epsilon} \bigg( | \mathcal{I}_\phi (k_0)|^2 + \frac{1}{3} |\Vec{k}|^2   | \mathcal{I}_\phi^i (k_0)|^2 + \frac{1}{30} |\Vec{k}|^4   | \mathcal{I}_\phi^{ij} (k_0)|^2  \bigg) \;, \label{5.18n}
\end{equation}
where, 
\begin{equation}
\mathcal{I}_\phi(k_0) = \rmint dt \,\mathcal{I}_\phi (t) e^{i k_0 t} \;, \mathcal{I}_{\phi}^{i}(k_0) = \rmint dt \,\mathcal{I}_\phi^i (t) e^{i k_0 t} \;, \mathcal{I}_{\phi}^{ij}(k_0) = \rmint dt \,\mathcal{I}_\phi^{ij} (t) e^{i k_0 t} \;.
\end{equation}
To extract the imaginary part, we need to find  the principal value of the propagator, i.e.,
\begin{align}
    \begin{split}
        \frac{1}{-k^2-m^2+i\epsilon}=PV\Big(\frac{1}{-k^2-m^2}\Big)-i\pi\,\delta(-k^2-m^2)\,.
    \end{split}
\end{align}

Therefore, besides the quadrupole,  the monopole and the dipole also contribute to the scalar radiation. 
But for the massive case, we will have some changes in the quadrupole formula compared to \cite{Kuntz:2019zef}.
\begin{align}
    \begin{split}
        P_{\phi}^m=\frac{1}{4\pi^2 m_p^2\mathcal{T}}\sum_{l}\frac{1}{l!(2l+1)!!}\rmint_0^\infty d\omega\,\omega\,(\omega^2-m^2)^{l+1/2}|\mathcal{I}^{m}_{(L)}(\omega)|^2 .
        \label{5.21}
    \end{split}
\end{align}

We will truncate our computation up to $l=1$ i.e. we will compute only the dipole radiation. Here in (\ref{5.21}) $`L$' denotes the \textit{symmetric trace free (STF)} indices $i,j,k,\cdots.$

\subsubsection*{Computing the source term $J_{\phi}$:}
 Now we explicitly write down the effective action for the source for the scalar sector upto $N^{(4)}LO$. We use the Feynman rules as spelled out in the previous section. The diagrams are depicted in Fig.~(\ref{Figure10}).
\begin{itemize}
\item The $LO$ contribution comes from the worldline vertex $\rmint dt \,m_a s_a \Bar{\phi}.$ The amplitude corresponding to the diagram in Fig.~(\ref{10a}):
\begin{align}
    \begin{split}
\mathcal{S}_{\text{eff}}\Big|_{\text{fig}(\ref{10a})}=-\sum_{a=1}^{2}\rmint dt\,\frac{m_a\,s_a}{m_p}\,\Bar{\phi}(\Vec{x}_a(t))\,\sim\mathcal{O}(L^{1/2}v^{1/2}).
    \end{split}
\end{align}
\begin{figure}
    \centering
  \begin{subfigure}{0.22\textwidth}
\scalebox{0.3}{\begin{feynman}
    \electroweak[lineWidth=6, color=0693e3, label=$\Bar{\phi}$]{6.00, 6.00}{7.00, 7.00}
    \fermion[lineWidth=6]{4.00, 6.00}{8.00, 6.00}
    \fermion[lineWidth=6]{4.00, 4.00}{8.00, 4.00}
\end{feynman}
}
      \caption{}
      \label{10a}
  \end{subfigure}
   \begin{subfigure}{0.22\textwidth}
 \scalebox{0.3}{\begin{feynman}
    \electroweak[lineWidth=6, color=0693e3, label=$\varphi$]{5.60, 4.00}{5.60, 6.00}
    \electroweak[lineWidth=6, label=$\Bar{\phi}$, color=0693e3]{5.60, 6.00}{6.80, 6.80}
    \fermion[lineWidth=6]{4.00, 6.00}{7.60, 6.00}
    \fermion[lineWidth=6]{4.00, 4.00}{7.60, 4.00}
\end{feynman}}
      \caption{}
      \label{10b}
  \end{subfigure}
   \begin{subfigure}{0.20\textwidth}
 \scalebox{0.25}{\begin{feynman}
    \fermion[lineWidth=6]{4.00, 4.00}{7.80, 4.00}
    \fermion[lineWidth=6]{4.00, 6.40}{7.80, 6.40}
    \electroweak[lineWidth=6, label=$\tilde{\psi}$]{5.80, 6.40}{5.80, 4.00}
    \electroweak[lineWidth=6, label=$\Bar{\phi}$, color=0693e3]{5.80, 6.40}{7.60, 7.60}
\end{feynman}
}
      \caption{}
      \label{10c}
  \end{subfigure}
  \begin{subfigure}{0.22\textwidth} \centering
        \scalebox{0.3}{
\begin{feynman}
    \electroweak[label=$\Bar{\phi}$, lineWidth=6, color=0693e3]{5.40, 5.00}{7.60, 5.00}
    \fermion[lineWidth=6]{4.00, 4.00}{6.80, 4.00}
    \electroweak[label=$\varphi$, lineWidth=6, color=0693e3]{5.40, 6.00}{5.40, 5.00}
    \electroweak[label=$\Tilde{\psi}$, lineWidth=6]{5.40, 5.00}{5.40, 4.00}
    \fermion[lineWidth=6]{4.00, 6.00}{6.80, 6.00}
\end{feynman}}
\caption{}
\label{10h}
\end{subfigure}
  \begin{subfigure}{0.20\textwidth}
 \scalebox{0.25}{\begin{feynman}
    \electroweak[label=$\varphi$, lineWidth=6, color=0693e3]{5.60, 4.00}{5.60, 5.00}
    \electroweak[label=$\bar{\phi}$, color=0693e3, lineWidth=6]{5.60, 6.40}{7.40, 7.80}
    \electroweak[label=$\varphi$, lineWidth=6, color=0693e3]{5.60, 6.40}{5.60, 5.40}
    \fermion[lineWidth=6]{4.00, 6.40}{7.20, 6.40}
    \fermion[lineWidth=6]{4.00, 4.00}{7.20, 4.00}
    \parton{5.60,5.20}{0.20}
\end{feynman}
}
      \caption{}
      \label{10d}
  \end{subfigure}
     \begin{subfigure}{0.22\textwidth}
        \centering
        \scalebox{0.25}{
\begin{feynman}
    \electroweak[color=0693e3, lineWidth=6, label=$\phi$]{5.60, 5.20}{5.60, 4.00}
    \fermion[lineWidth=6, showArrow=false]{4.00, 6.40}{7.20, 6.40}
    \electroweak[color=0693e3, lineWidth=6, label=$\bar{\phi}$]{5.60, 5.20}{8.20, 5.20}
    \electroweak[lineWidth=6, label=$\psi$]{5.60, 6.40}{5.60, 5.20}
    \fermion[lineWidth=6, showArrow=false]{4.00, 4.00}{7.20, 4.00}
\end{feynman}
}
\caption{}
\label{10e}
\end{subfigure}
    \begin{subfigure}{0.22\textwidth}
        \centering
        \scalebox{0.25}{\begin{feynman}
    \electroweak[color=0693e3, lineWidth=6, label=$\varphi$]{5.60, 5.20}{5.60, 4.00}
    \fermion[lineWidth=6, showArrow=false]{4.00, 6.40}{7.20, 6.40}
    \gluon[lineWidth=6, label=$\hat{\mathcal{A}_i}$, color=fcb900]{5.60, 6.40}{5.60, 5.20}
    \electroweak[color=0693e3, lineWidth=6, label=$\bar{\phi}$]{5.60, 5.20}{8.20, 5.20}
    \fermion[lineWidth=6, showArrow=false]{4.00, 4.00}{7.20, 4.00}
\end{feynman}
}
\caption{}
\label{10f}
\end{subfigure}
\begin{subfigure}{0.22\textwidth} \centering
        \scalebox{0.3}{
 \begin{feynman}
    \electroweak[label=$A_m$, color=9900ef, lineWidth=6]{5.40, 5.00}{5.40, 4.00}
    \electroweak[label=$A_i$, color=9900ef, lineWidth=6]{5.40, 6.00}{5.40, 5.00}
    \fermion[lineWidth=6]{4.00, 6.00}{7.00, 6.00}
    \fermion[lineWidth=6]{4.00, 4.00}{7.00, 4.00}
    \electroweak[label=$\bar{\phi}$, lineWidth=6, color=0693e3]{5.40, 5.00}{7.60, 5.00}
\end{feynman}}
\caption{}
\label{10g}
\end{subfigure}
\caption{\textcolor{black}{Radiative diagrams for scalar field that contributes upto $N^{(4)}LO$.}}
\label{Figure10}
\end{figure}
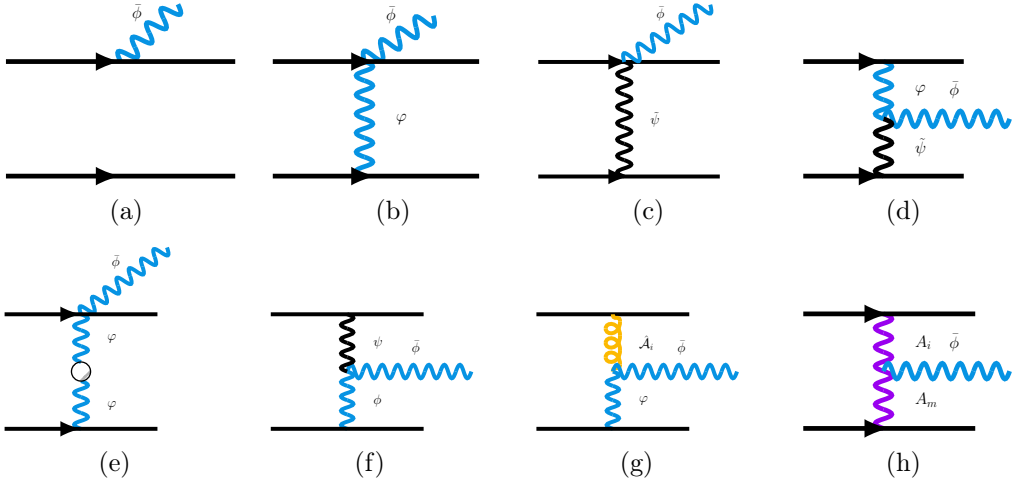
\item The $N^{(1)}LO$ contribution comes from the worldline vertex $\rmint dt\, \frac{1}{2}\,v_a^2\, m_as_a \frac{\phi}{m_p}$ and it's contribution to the effective action has the following form,
\begin{align}
    \begin{split}
        \mathcal{S}_{\text{eff}}=\frac{1}{2}\sum_{a}m_a s_a v_a^2\,\Bar{\phi}(\Vec{x}_a(t_a))\sim \mathcal{O}(L^{1/2}v^{3/2}).
    \end{split}
\end{align}

\item The $N^{(2)}LO$ contribution comes from the vertex: $-\rmint dt\, g_a \varphi\, \Bar{\phi}$ and $-\rmint dt\,s_a \varphi$. The amplitude corresponds to the diagram in Fig.~(\ref{10b}):
\begin{align}
    \begin{split}
          \mathcal{S}_{\text{eff}}\Big|_{\text{fig}(\ref{10b})}&=\frac{m_1 m_2}{m_p^2}\rmint\Bar{\mathcal{D}}\Hat{\xi}\rmint dt_1\,g_1 \varphi(\Vec{x}_1(t_1))\frac{\Bar{\phi}(\Vec{x}_1(t_1))}{m_p}\rmint dt_2\,s_2\,\varphi(\Vec{x}_2(t_2))\,,\\ &
        =\frac{4 \pi  g_1 s_2 m_1 m_2}{m_p^2} \rmint dt\rmint_{\Vec{k}}\frac{e^{i\Vec{k}\cdot \Vec{r}}}{\Vec{k^2}+m^2}\frac{\Bar{\phi}(\Vec{x}_1(t_1))}{m_p}\,,\\ &
        =\frac{g_1 s_2 m_1 m_2}{m_p^2}\rmint dt \frac{e^{-mr}}{r}\frac{\Bar{\phi}(\Vec{x}_1(t))}{m_p}+(1\leftrightarrow 2)\,\sim\mathcal{O}(L^{1/2}v^{5/2}).
    \end{split}
\end{align}

\item \textcolor{black}{Another $N^{(2)}LO$ contribution for the source comes from the worldline vertex:
-$\rmint dts_a m_a\bar{\phi}\tilde{\psi}\,.$}
The amplitude corresponds to the diagram in Fig.~(\ref{10c}):

\begin{align}
    \begin{split}
        \mathcal{S}_{\text{eff}}^{0}\Big|_{\text{fig}(\ref{10c})}&=\frac{s_1m_1m_2}{m_p^3}\rmint\Bar{\mathcal{D}}\Hat{\xi}\rmint dt_1 \,\frac{\Bar{\phi}(\Vec{x}_1(t_1))}{m_p}\tilde{\psi}(\Vec{x}_1(t_1))\rmint dt_2 \tilde{\psi}(\Vec{x}_2(t_2))\,,\\ &
        =\frac{s_1m_1m_2}{m_p^3}\rmint dt_1\,dt_2\,\Bar{\phi}(\Vec{x}_1(t_1))\Big\langle\tilde{\psi}(\Vec{x}_1(t_1))\tilde{\psi}(\Vec{x}_2(t_2))\Big\rangle\,, \\ &
        =\frac{s_1m_1m_2}{8\pi m_p^2}\rmint dt\, \frac{1}{r} \frac{\Bar{\phi}(\Vec{x}_1(t))}{m_p}+(1\leftrightarrow 2)\,\,\sim\mathcal{O}({L^{1/2}v^{5/2}}).
    \end{split}
\end{align}
\item  The $N^{(3)}LO$ contribution coming from vertex $m^2 \rmint d^4x \Tilde{\psi}\varphi\Bar{\phi}(x)$ corresponding to the diagram in Fig.~(\ref{10h}), contributes to the effective action in the following way \footnote{We further comment on the radiation coming out from the middle vertices in Appendix~(\ref{D}).}:
\begin{align}
    \begin{split}
\mathcal{S}_{\text{eff}}\Big |_{\text{fig}(\ref{10h})}&=\frac{m^2 s_1m_1m_2}{m_p^3}\rmint dt_1dt_2\rmint d^4x \Big\langle \psi(\Vec{x}_2(t_2))\psi(x)\Big\rangle \Big\langle\varphi(\Vec{x}_1(t_1))\varphi(x) \Big\rangle \bar{\phi}(x_0)\,,\\&
=\frac{s_1m_1m_2}{8\pi m_p^2}\rmint dt \frac{(1-e^{-mr})}{r}\frac{\bar{\phi}(\vec x_0(t))}{m_p}+1\leftrightarrow 2\,\sim\mathcal{O}(L^{1/2}v^{7/2}).
    \end{split}
\end{align}
\item \textcolor{black}{One of the diagrams that contribute at $N^{(4)}LO$ term comes from the relativistic time correction, as shown in Fig.~(\ref{10d}), and the corresponding amplitude has the following form}:

\vspace{-0.6cm}

\begin{align}
    \begin{split}
 \mathcal{S}_{\text{eff}}\Big|_{\text{fig}(\ref{10d})} = & \rmint\Bar{\mathcal{D}}\Hat{\xi} \rmint dt_1\, \frac{m_1g_1}{m_p^2}\,\varphi(\Vec{x}_1(t_1))\,\bar{\phi}(\Vec{x}_1(t_1))\rmint dt_2\,\frac{s_2m_2}{m_p}\varphi(\Vec{x}_2(t_2)) \rmint d^4 x\,\partial_0\varphi(x)\partial_0\varphi(x)\,,\\&
 =\frac{m_1m_2s_2g_1}{m_p^3}\rmint dt dt_1 dt_2 \rmint d^3x\, \partial_0 \Big\langle \,\varphi(\vec{x}_1(t_1))\varphi(x)\Big\rangle\,\partial_0 \Big\langle\varphi(\vec{x}_2(t_2)) \varphi(x)\Big\rangle \,\bar{\phi}(\Vec{x}_1,t)\,,\\&
=\frac{m_1 m_2 g_1 s_2}{16 \pi m_p^2}\rmint dt \,e^{-mr}\Bigg\{\Big(\frac{\Vec{v}_1\cdot\Vec{v}_2}{r}-\frac{(\Vec{v}_1\cdot \hat{n})(\Vec{v}_2\cdot \hat{n})}{r}\Big)+\frac{m}{2}\,(\Vec{v}_2\cdot \Hat{n})(\Vec{v}_1\cdot \hat{n})\Bigg\}\,\frac{\Bar{\phi}(\Vec{x}_1(t))}{m_p} \\& \hspace{2.5cm}+ T(\Dot{\Bar{\phi}})+ (1 \leftrightarrow 2)\,\sim\mathcal{O}(L^{1/2}v^{9/2}).
\label{5.26}
 \end{split}
\end{align}
Here, 
\begin{align}
    \begin{split}
        T(\Dot{\Bar{\phi}})&=-\frac{m_1 m_2 g_1 s_2}{16\pi m_p^2}\rmint dt\, e^{-mr}(\Vec{v}_2\cdot \Hat{n})\frac{ \Dot{\Bar{\phi}}(\Vec{x}_1(t))}{m_p},\\ &
        =\frac{m_1 m_2 g_1 s_2}{16\pi m_p^2}\rmint dt \frac{e^{-mr}}{r}(\Vec{v}_2\cdot \Dot{\Vec{r}}+\Vec{a}_2\cdot \Vec{r})\,\frac{\Bar{\phi}(\Vec{x}_1(t))}{m_p}\,.
    \end{split}
\end{align}
We have ignored the total derivative term by assuming the time variation of $\Bar{\phi}$ is very small and $|\Dot{\Vec{r}}|=0$ for the circular orbit. \newpage
\item Amplitude of another term \textcolor{black}{at $N^{(4)}LO$} as shown in the Fig.~(\ref{10e}) with the interaction vertex $\rmint d^4x \psi\partial_0\phi\partial_0\bar{\phi}:$ 
\begin{align}
    \begin{split}
\mathcal{S}_{\text{eff}}\Big |_{\text{fig}(\ref{10e})}&=\frac{m_1m_2s_2}{m_p^2}\rmint\Bar{\mathcal{D}}\Hat{\xi}\rmint dt_1dt_2\tilde{\psi}(\Vec{x}_1(t_1))\phi(\Vec{x}_2(t_2))\rmint d^4x \,\frac{\psi}{m_p}\partial_0\phi\partial_0\bar{\phi}\,,
\\&
=\frac{-m_1m_2s_2}{m_p^2}\rmint dt dt_1 dt_2 \rmint d^3x \Big\langle\Tilde{\psi}(\vec{x}_1(t_1))\Tilde{\psi}(x)\Big\rangle\,\partial_0 \Big\langle\varphi(\vec{x}_2(t_2)) \varphi(x)\Big\rangle \,\partial_0\frac{\bar{\phi}}{m_p}\,,\\&
=\frac{-s_2m_1m_2}{8\pi m^2 m_p^2} \rmint dt\,  \frac{\bar\phi(\Vec{x}_0(t))}{m_p}\, \partial_0\Big[\frac{e^{-m r} \left(m r-e^{m r}+1\right)}{ r^2}(\vec{v}_2-\vec{v}_1)\cdot\hat{n}\Big]\,,\\&
=\frac{-s_2m_1m_2}{8\pi m^2 m_p^2}\rmint dt \,  \Big[\partial_{0}f(r)(\Vec{v}_2-\Vec{v}_1)\cdot \Hat{n}+f(r)\Big\{(\Vec{a}_2-\Vec{a}_1)\cdot \Hat{n}\\ &
\,\,\,\,\,\,\,\,\,\,\,\,\,\,\,\,\,\,\,\,\,\,\,\,\,\,\,\,\,\,\,\,\,\,\,\,\,\,\,\,\,\,\,\,\,\,\,+(\Vec{v}_2-\Vec{v}_1)\cdot \Big(\Hat{n}\frac{(\Vec{v}_1-\Vec{v}_2)\cdot \Hat{n}}{r}+\frac{(\Vec{v}_2-\Vec{v}_1)}{r}\Big)\Big\}\Big]\frac{\bar\phi(\Vec{x}_0(t))}{m_p}\,+(1\leftrightarrow 2),\\&\sim\mathcal{O}(L^{1/2}v^{9/2})\,,
\label{5.29}
    \end{split}
\end{align}
where in (\ref{5.29}) the function $f(r)$ is given by,
\begin{align}
    \begin{split}
        f(r)=\frac{e^{-mr}(mr-e^{mr}+1)}{r^2}.
        \label{5.46}
    \end{split}
\end{align}
\vspace{-0.925cm}
\item A third diagram that contributes at  $N^{(4)}LO$ with the bulk interaction vertex $\rmint d^4x\,\hat{\mathcal{A}}_i\,\,\partial_i\varphi\,\partial_0\bar{\phi}(x)$, is shown in Fig.~(\ref{10f}). The corresponding amplitude is given by: 
\begin{align}
\begin{split}
S_{eff}\Big |_{\text{fig}(\ref{10f})}=&\frac{m_1m_2s_2}{m_p^3} \rmint\Bar{\mathcal{D}}\Hat{\xi}\rmint dt_1 dt_2\hat{\mathcal{A}}_j\,v_1^j\, \varphi(\Vec{x}_2(t_2))\rmint d^4x\,\hat{\mathcal{A}}_i\,\,\partial_i\varphi\,\partial_0\bar{\phi}(x)\,,\\&
=\frac{m_1m_2s_2}{m_p^3}\rmint dt dt_1 dt_2 \rmint d^3x \Big\langle\hat{\mathcal{A}}_j(\vec{x}_1(t_1))\hat{\mathcal{A}}_i(x)\Big\rangle\,\partial_i \Big\langle\varphi(\vec{x}_2(t_2)) \varphi(x)\Big\rangle \,\partial_0\bar{\phi}\,,\\&
=\frac{1}{2\pi }\frac{m_1m_2s_2}{m_p^3m^2}\rmint dt \partial_0\bar{\phi}v_1^i\Bigg\{\partial_i\frac{(1-e^{-mr})}{r}\Bigg\}\,,\\&
=\textstyle{\frac{-1}{2\pi }\frac{m_1m_2s_2}{m_p^3m^2}\rmint dt\,\,\bar{\phi}\,\Bigg[(\vec{v}_1\cdot\hat{ n}) \partial_0\Big[\frac{e^{-m r} \left(m r-e^{m r}+1\right)}{r^2}\Big]+\partial_0(\vec{v}_1\cdot n)\Bigg\{\frac{e^{-m r} \left(m r-e^{m r}+1\right)}{r^2}\Bigg\}\Bigg] }\,,\\&
=\frac{-1}{2\pi }\frac{m_1m_2s_2}{m_p^2m^2}\rmint dt\,\,\frac{\Bar{\phi}(\vec{x}_0(t))}{m_p}\,\Bigg[(\vec{v}_1\cdot \hat{n}) \partial_0f(r)+\\&\,\,\,\,\,\,\, \Bigg\{(\vec{a}_1\cdot n)+  \vec{v}_1\cdot \Big[
\frac{(\vec{v}_1-\vec{v}_2)\cdot \Hat{n}}{r}\,\Hat{n}+\frac{(\vec{v}_2-\vec{v}_1)}{r} \Big]\Bigg\}f(r)\Bigg] +1\leftrightarrow 2\, 
\sim\mathcal{O}(L^{1/2}v^{9/2}).
\end{split}
\end{align}
Here $f(r)$ is defined in Eq.~(\ref{5.46}).
\newpage
\item \textcolor{black}{A bulk vertex term causing scalar radiation at $N^{(5)}LO$ is the following: \\$g_{a\gamma\gamma}\,\epsilon^{0ikm}\rmint d^4x \partial_{0}\mathcal{A}_{i}\partial_{k}\mathcal{A}_{m}\Bar{\phi}\,.$ This comes from the \textit{Theta} term.} The amplitude corresponds to the diagram in Fig.~(\ref{10g})\,,
\begin{align}
    \begin{split}
          \mathcal{S}_{\text{eff}}\Big|_{\text{fig}(\ref{10g})}&=Q_1Q_2g_{a\gamma\gamma}\epsilon^{0ikm}\rmint\Bar{\mathcal{D}}\Hat{\xi}\rmint dt_1 v_{1}^{j}(t_1)\mathcal{A}_{j}(\Vec{x}_1(t_1))\rmint dt_2 \,v_2^{l}\,\mathcal{A}_{l}(\Vec{x}_2(t_2))\\&\hspace{5cm}\rmint d^4 x\,\partial_{0}\mathcal{A}_{i}\partial_{k}\mathcal{A}_{m}(x)\frac{\Bar{\phi}(\Vec{x}_0(t))}{m_p}\,,\\ &
=Q_1Q_2g_{a\gamma\gamma}\epsilon^{0ikm}\rmint dt dt_1 dt_2 v_{1}^{j}(t_1)\,v_2^{l}(t_2)\,\rmint d^3 x
\,\partial_{0}\Big\langle\mathcal{A}_{j}(\Vec{x}_1(t_1))\mathcal{A}_{i}(x)\Big\rangle\,\\&\hspace{8cm}\partial_{k}\Big\langle\mathcal{A}_{l}(\Vec{x}_2(t_2))\mathcal{A}_{m}(x)\Big\rangle\frac{\Bar{\phi}(\Vec{x}_0(t))}{m_p}\,,\\ &
=Q_1 Q_2 g_{a\gamma\gamma}\epsilon^{0ikm}\rmint dt v_2^{m}(t)\Big[a_1^{i}(t)\rmint_{k}\frac{ik^{k}e^{-i\Vec{k}\cdot\Vec{r}}}{\Vec{k}^4}-\underbrace{v_1^{i}v_1^{a}\rmint_{k}\frac{k^a\,k^k\,e^{-i\Vec{k}\cdot\Vec{r}}}{\Vec{k}^4}}_{\text{No contribution}}\Big]\frac{\Bar{\phi}(\Vec{x}_0(t))}{m_p}\,,\\ &
=\frac{Q_1 Q_2 g_{a\gamma\gamma}}{8\pi}\rmint dt \,[\Vec{a}_1\cdot (\Vec{v}_2\times \Hat{n})+1\leftrightarrow 2]\,\frac{\Bar{\phi}(\Vec{x}_0(t))}{m_p}\,\sim\mathcal{O}(L^{1/2}v^{11/2}).\label{5.32m}
    \end{split}
\end{align}
\end{itemize}
Although we get a non-trivial contribution to the radiative effective action from the \textit{theta} term in (\ref{5.32m}), it vanishes for any orbit confined to a \textit{plane}. 
\textcolor{black}{Apart from the above-mentioned terms, we can have some more 3-point bulk interaction vertices with a radiating field of the following form,
\begin{equation}
\rmint d^4x \, \boldsymbol{\zeta}_{qr}\partial_{i}\varphi\partial_{j}\Bar{\phi}. \label{5.36n}
\end{equation}
But they would not contribute to the source term because of the following reason. To identify the source term, the effective action should have the following form $\rmint d^4x J(x)\Bar{\phi}(\Vec{x}_0,t)$. 
 Now to write down the interaction action in the specified form, we need to  integrate by parts:
 \begin{equation}
    \rmint dt \rmint d^3x \, \boldsymbol{\zeta}_{qr}\partial_{i}\varphi\partial_{j}\Bar{\phi}=-\rmint dt\Bar{\phi}(\Vec{x}_0(t))\rmint d^3x \partial_{j}\{\boldsymbol{\zeta}_{qr}\partial_{i}\varphi\}\,.
 \end{equation} 
 After that, demanding that the potential fields vanish at spatial infinity, we get the contribution to zero $\implies \text{the source term}\,J=0$ for that specific interaction.}
\\
Now we will calculate the power radiation in the centre of mass frame defined by, $$\Vec{x}_1=\frac{m_2}{m_1+m_2}\Vec{r} \textrm{ and } \Vec{x}_2=\frac{-m_1}{m_1+m_2}\Vec{r}\,.$$
It is evident that the scalar monopole moment does not contribute. We are considering a circular orbit parameterized by, $$x=r \cos(\Omega t),\quad y=r \sin(\Omega t), \quad z=0\,,$$ with the orbital frequency $\Omega$.
The source term can be calculated from the effective action by omitting the radiative fields as defined in (\ref{5.10}). The source term contributing to scalar radiation has the form
\begin{align}
    \begin{split}
        J_{\phi}=& -\sum_{a=1}^{2}m_a\,s_a\delta(\Vec{x}-\Vec{x}_a)+\frac{m_1m_2}{8\pi m_p^2 r}\sum_{a=1}^{2}s_a \delta(\Vec{x}-\Vec{x}_a)+\frac{1}{2}\sum_{a}m_as_av_a^2\delta(\Vec{x}-\Vec{x}_a)\\&
       +\underbrace{ \frac{m_1^2 m_2^2}{16 \pi \,m_p^2(m_1+m_2)^2}}_{\frac{\mu^2}{16\pi}}e^{-mr}r^2\Omega^2\sum_{a\ne b}^{2}g_a s_{b}\,\delta(\Vec{x}-\Vec{x}_a)\\ &+\frac{m_1m_2}{m_p^2} \frac{e^{-mr}}{r}\Big[1+\frac{1}{16}\Big(\frac{m_1}{M}-\nu\Big)\Omega^2\,r^2\Big]\sum_{a\ne b}g_a s_b \delta(\Vec{x}-\Vec{x}_a)\\&+\underbrace{\Big\{[\tilde{c}_1+\tilde{c}_2]r^2\Omega^2+\Big[\frac{\gamma_2m_1-\gamma_1m_2}{M}\Big]r^2\Omega^2+[\tilde{G}_1+\tilde{G}_2]+(\tilde{d}_1-\tilde{d}_2)\Omega^2+\Big[\frac{\Gamma_2m_1-\Gamma_1m_2}{M}\Big]\Omega^2\Big\}}_{\eta(r,\Omega)}\\& \,\,\,\,\,\,\,\,\delta(x-\Vec{x}_0)\,,
    \end{split}
\end{align}
where the functions are defined below,
\begin{align}
\begin{split}
&\tilde{c}_{1,2}=\frac{-s_{2,1}\,\mu\, M f(r)}{8\pi m^2m_p^2r}\,\,\,,
    \tilde{G}_{1,2}=\frac{s_{1,2}\,\mu\, M}{8\pi m_p^2}\frac{(1-e^{-mr})}{r}\,\, ,\,\gamma_{1,2}(r)=-\frac{1}{2\pi}\frac{\mu \,M\, s_{2,1}\,f(r)}{m_p^2 m^2r},\,\\& \tilde{d}_{1,2}=\tilde{c}_{1,2}\, r,\,\,\,\,\,
    \Gamma_{1,2}=\gamma_{1,2}\, r.
    \end{split}
\end{align}
and  $f(r)$ is defined in equation (\ref{5.29}). \par
 Now the dipole moment is essentially given by,
\begin{align}
    \begin{split}
        I_{\phi}^{i}(t)&=\rmint d^3 x\,x^{i} J_{\phi}+\textcolor{black}{\frac{m^2}{10} \rmint d^3 x\,  x^i x^2\,J_{\phi} }\,,\\ &
        =\sum_{a\ne b}\underbrace{[-m_a s_a(1-\frac{v_a^2}{2})+\frac{ m_1 m_2}{8\pi m_p^2r}s_a +\frac{\mu^2 (g_a s_b)}{16\pi m_p^2}e^{-mr}r^2 \Omega^2+\frac{m_1m_2}{m_p^2}\frac{e^{-mr}}{r}\Big[1+\frac{1}{16}\Big(\frac{m_1}{M}-\nu\Big)\Omega^2\,r^2\Big]g_a s_b]}_{\mathcal{C}_{a}(r)}x_a^{i}\\ &
        +\textcolor{black}{\frac{m^2}{10}\sum_{a\ne b}\mathcal{C}_a x_a^2 x_a^{i}}+\frac{m^2}{10}\eta(r,\Omega){x}_0^i\,\Vec{x}_0^2+\eta(r,\Omega){x}_0^i\,.
    \end{split}
\end{align}
In the frequency domain, the moments are given by,
\begin{align}
    \begin{split}
      &  I_{\phi}^{i}(\omega)=\rmint dt e^{i\omega t}I_{\phi}^{i}(t)\,,\\ &
        \implies I_{\phi}^{x}(\omega)=\Big[ \frac{\mathcal{C}_1 m_2-\mathcal{C}_2 m_1}{m_1+m_2}+\textcolor{black}{\frac{m^2r^2}{10}\frac{\mathcal{C}_1m_2^3-\mathcal{C}_2 m_1^3}{(m_1+m_2)^3} }+\frac{\eta(r,\Omega)}{2}+\frac{m^2}{80}\eta(r,\Omega)r^2\Big]r\,\sqrt{\frac{\pi}{2}}\delta(\omega-\Omega)\,,\\&\quad \text{and}\quad \,I_{\phi}^{y}(\omega)=i I_{\phi}^{x}(\omega)\,.
    \end{split}
\end{align}
Then the radiated power is given by \footnote{The formula for scalar power radiation matches up to some numerical factor with the result of \cite{Huang:2018pbu} due to the different choices of overall normalization of the gravitational action. Also, the mass-dependent second term, i.e., the $\frac{m^2r^2}{10}(...)$ is not there. This is because of the  mass dependence of the multipole moment, coming from the equation of motion, in (\ref{5.12mn}). },\par
\begin{equation}
\boxed{
\begin{aligned}
        P_{\phi}&=\frac{1}{12\pi m_p^2 }\Big(\frac{\mathcal{C}_1 m_2-\mathcal{C}_2 m_1}{m_1+m_2}+\textcolor{black}{\frac{m^2 r^2}{10}\frac{\mathcal{C}_1m_2^3-\mathcal{C}_2 m_1^3}{(m_1+m_2)^3} }+\frac{\eta(r,\Omega)}{2}+\frac{m^2}{80}\eta(r,\Omega)r^2\Big)^2\\&\,\,\,\,\,\,\,\,\,\,\,\,\,\,\,\,\,\,\,\,\,\,\,\,\,\,\,\,\,\,\,\,\,\,\,\,\,\,\,\,\,\,\,\,\,\,\,\,\,\,\,\,\,\,\,\,\,\,\,\,\,\,\,\,\,\,\,\,\,\,\,\,\,\,\,\,\,\,\,\,\,\,\,\,\,\,\,\,\,\,\,\,\,\,\,\,\,\,\,\,\,\,\,\,\,\,\,\,\,\,\,\,\,\,\,\,\,\,\,\,\,\,\,\,\,\,\,\,\,\,\,\,\rmint d\omega \,\omega (\omega^2-m^2)^{3/2} r^2\delta(\omega-\Omega)\,,\\ &
        =\frac{1}{12\pi m_p^2 }\Big(\frac{\mathcal{C}_1 m_2-\mathcal{C}_2 m_1}{m_1+m_2}+\textcolor{black}{\frac{m^2r^2}{10}\frac{\mathcal{C}_1m_2^3-\mathcal{C}_2 m_1^3}{(m_1+m_2)^3} }+\frac{\eta(r,\Omega)}{2}+\frac{m^2}{80}\eta(r,\Omega)r^2\Big)^2\,r^2 \, \Omega^4 (1-\frac{m^2}{\Omega^2})^{3/2}\,.
\end{aligned}
}
\end{equation}
Before we end this section, some comments are in order. 
\begin{itemize}
   \item  \textit{To the best of our knowledge, it is the first calculation of scalar radiation upto $N^{(4)}LO$. Leading order computations are done in \cite{Huang:2018pbu, Kuntz:2019zef}}.
    \item The contribution coming from the $g_{a\gamma\gamma}$ term drops for the orbits which are confined to a plane (e.g. for the circular orbit that we have considered in this paper), but it would have contributed otherwise for orbits embedded in 3-dimensions (e.g. some helical orbits embedded in 3d \cite{2021EPJC...81.1048L,Chen:2022qvg,Liu:2020vsy}).
\end{itemize}
\subsection{Electromagnetic interaction and the multipole decomposition}\label{emRad}
Now we consider the radiative photons with the source.  
The effective action can be written as
\begin{align}
    \begin{split}
        \mathcal{S}_{\text{eff}}=\rmint d^4 x \Big(-\frac{1}{4}F_{\mu\nu}F^{\mu\nu}+J^{\mu}\Bar{a}_{\mu}\Big)\,.
    \end{split}
\end{align}
We impose the Lorenz gauge: $$\partial_{\mu}\Bar{a}^{\mu}=0.$$ The equation of motion takes the following form,
\begin{eqnarray}
    \Box\Bar{a}_{\mu}=-J_{\mu}\,.
\end{eqnarray}

To compensate for the extra term we need to add an extra term $\frac{1}{6}\rmint d^3 x\, x^2 J^{\mu}\nabla^2\Bar{a}_{\mu}$ with the monopole term. Now use $\Box \Bar{a}_{\mu}=0$ (as we are measuring outside the source) we can see the term can be written as
\begin{align}
    \begin{split}
        S_{int}'&=\frac{1}{6}\rmint dt \rmint d^3 x\, x^2\,J^\mu (\Box+\partial_{t}^2)\Bar{a}_{\mu}\,,\\ &
        =\frac{1}{6}\rmint dt\rmint d^3x\,\Bar{a}_{\mu}\, \partial_{t}^2 J^{\mu} x^2\,.
    \end{split}
\end{align}
We compute the effective action for the source following the same approach (in Sec.~(\ref{subsec:radiated_scalars})) as the scalar field case. By doing a similar computation like the scalar field case, the imaginary part of the effective action for the source in terms of multipole moments can be found as  \cite{Kuntz:2019zef}, 
\begin{align}
    \begin{split}
       \text{Im} [\gamma_\mathrm{eff}^{(\bar a_\mu)}]&=\frac{1}{8\pi^2}\rmint dk^0\rmint d|\Vec{k}|\,|\Vec{k}|^2 \, \delta(k_0^2-\Vec{k}^2)\Big[|\mathcal{I}^{\mu}(k_0)|^2+\frac{1}{3}|\mathcal{I}^{(\mu)ij}(k_0)|^2|\Vec{k}|^2+\frac{1}{30}|\mathcal{I}^{(\mu)ij}(k_0)|^2|\Vec{k}|^4\Big]\,,\\ &
    =\frac{1}{8\pi^2}\rmint dk^0\rmint d|\Vec{k}|\,|\Vec{k}|^2 \frac{1}{2|k^0|}[\delta(k^0+|\Vec{k}|)+\delta(k^0-|\Vec{k}|)]\Big[|\mathcal{I}^{\mu}(k_0)|^2+\frac{1}{3}|\mathcal{I}^{(\mu)ij}(k_0)|^2|\Vec{k}|^2\\ &\hspace{3cm}
    \,\,\,\,\,\,\,\,\,\,+\frac{1}{30}|\mathcal{I}^{(\mu)ij}(k_0)|^2|\Vec{k}|^4\Big]\,,\\ &
    =\frac{1}{8\pi^2}\rmint_0^{\infty} d\omega\,\Big[|\mathcal{I}^{\mu}(\omega)|^2\omega+\frac{1}{3}|\mathcal{I}^{(\mu)i}(\omega)|^2\omega^3
    +\frac{1}{30}|\mathcal{I}^{(\mu)ij}(\omega)|^2\omega^5\Big]\,.
    \label{5.42}
    \end{split}
\end{align}
 We next compute this effective action upto LO.  The diagrams that contribute in this order are shown in Fig.~(\ref{Fig:12}).
\begin{itemize}
\item The $LO$ contribution comes from a diagram with the worldline vertex $\rmint dt Q_a \Bar{a}_{0}$ as shown in Fig.~(\ref{fig12a}) has the following form,
\begin{align}
    \begin{split}
        \mathcal{S}_{\text{eff}}\Big |_{\text{fig}(\ref{fig12a})}=\sum_{a}\rmint dt \,Q_a \,\Bar{a}_{0}\sim \mathcal{O}(L^{1/2}v^{1/2}).
    \end{split}
\end{align}
\begin{figure}
    \centering
    \begin{subfigure}{0.22\textwidth}
      \scalebox{0.3}{\begin{feynman}
    \fermion[lineWidth=6]{4.00, 5.80}{7.40, 5.80}
    \fermion[lineWidth=6]{4.00, 4.00}{7.40, 4.00}
    \dashed[showArrow=false, lineWidth=6, label=$\Bar{a_0}$]{5.60, 5.80}{7.40, 7.20}
\end{feynman}
}
    \caption{}
    \label{fig12a}
    \end{subfigure}
\begin{subfigure}{0.22\textwidth}
    \centering
\scalebox{0.3}{\begin{feynman}
    \fermion[lineWidth=6]{4.00, 5.80}{7.40, 5.80}
    \fermion[lineWidth=6]{4.00, 4.00}{7.40, 4.00}
    \electroweak[color=9900ef, lineWidth=6, label=$\Bar{a_i}$]{5.60, 5.80}{7.40, 7.20}
\end{feynman}
}
    \caption{}
    \label{fig12b}
\end{subfigure}
\begin{subfigure}{0.22\textwidth}
    \centering
 \scalebox{0.25}{
\begin{feynman}
    \fermion[lineWidth=6]{4.00, 6.20}{7.40, 6.20}
    \electroweak[lineWidth=6, label=$\Tilde{\psi}$]{5.60, 6.20}{5.60, 4.00}
    \dashed[showArrow=false, lineWidth=6, label=$\Bar{a_0}$]{5.60, 6.20}{7.60, 7.80}
    \fermion[lineWidth=6]{4.00, 4.00}{7.40, 4.00}
\end{feynman}}
 \caption{}
    \label{fig12d}
\end{subfigure}
\begin{subfigure}{0.22\textwidth}
    \centering
 \scalebox{0.3}{\begin{feynman}
    \electroweak[label=$\tilde{\psi}$, lineWidth=6]{5.60, 5.80}{5.60, 4.00}
    \fermion[lineWidth=6]{4.00, 5.80}{7.40, 5.80}
    \fermion[lineWidth=6]{4.00, 4.00}{7.40, 4.00}
    \electroweak[color=9900ef, label=$\Bar{a}_i$, lineWidth=6]{5.60, 5.80}{7.40, 7.20}
\end{feynman}
}
    \caption{}
    \label{fig12c}
\end{subfigure}
\caption{Radiative diagrams for electromagnetic field up to $N^{(3)}LO$. }
\label{Fig:12}
\end{figure}
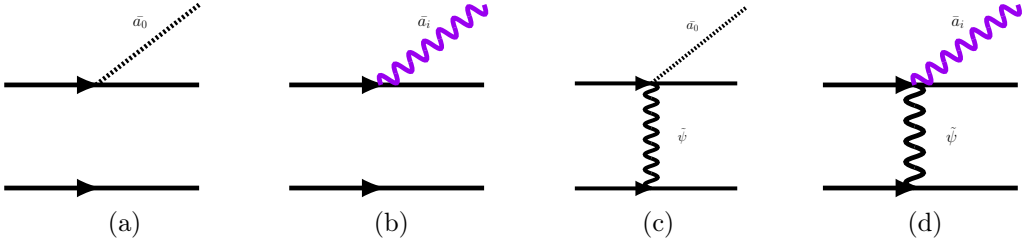

\item The diagram shown in the Fig.~(\ref{fig12b}) contributes at $N^{(1)}LO.$ It consists of a worldline vertex $\rmint dt\,Q_av_a^{i}\Bar{a}_{i}$ and the corresponding amplitude is given by,
\begin{align}
    \begin{split}
          \mathcal{S}_{\text{eff}}\Big |_{\text{fig}(\ref{fig12b})}=\sum_{a}\rmint dt \, Q_a\,v_a^{i}\, \Bar{a}_{i}(\Vec{x}_a(t))\,\sim\mathcal{O}(L^{1/2}v^{3/2}).
    \end{split}
\end{align}
\item \textcolor{black}{The $N^{(2)}LO$ contribution for the source of $J^{0}$ comes from the diagram shown in Fig.~(\ref{fig12d}) with the worldline vertex:
$-\rmint dt Q_a\,\Bar{a}_{0}\tilde{\psi}$\,.} The amplitude corresponds to the diagram is: 
\begin{align}
    \begin{split} \mathcal{S}_{\text{eff}}\Big |_{\text{fig}(\ref{fig12d})}&=\frac{Q_1m_2}{m_p^2}\rmint\Bar{\mathcal{D}}\Hat{\xi}\rmint dt_1 \Bar{a}_{0}(x_1,t_1)\tilde{\psi}(x_1,t_1)\rmint dt_2 \tilde{\psi}(x_2,t_2)\,,\\ &
        =\frac{Q_1m_2}{m_p^2}\rmint dt_1\,dt_2\,\Bar{a}_{0}(x_1,t_1)\Big\langle\tilde{\psi}(x_1,t_1)\tilde{\psi}(x_2,t_2)\Big\rangle\,, \\ &
        =\frac{1}{8\pi m_p^2}\,\sum_{a\ne b}Q_a m_b\rmint dt\, \frac{1}{r}\,\Bar{a}_{0}(x_a,t)\,\sim\mathcal{O}(L^{1/2}v^{5/2}).
    \end{split}
\end{align}
\item Other radiation diagram consisting of the following world line vertices, $-\rmint dt Q_a v_a^{i}\tilde{\psi}\Bar{a}^{i}$ and $\rmint  dt \,\tilde{\psi}$ as shown in Fig.~(\ref{fig12c}) contributes at $N^{(3)}LO\,.$ The corresponding amplitude is given by,
\begin{align}
    \begin{split}
          \mathcal{S}_{\text{eff}}\Big |_{\text{fig}(\ref{fig12c})}&=-Q_1m_2\rmint\Bar{\mathcal{D}}\Hat{\xi}\rmint dt_1\, v_1^{i}(t_1)\tilde{\psi}(\Vec{x}_1(t_1))\Bar{a}_{i}(\Vec{x}_1(t_1))\rmint dt_2 \, \tilde{\psi}(\Vec{x}_2(t_2))\,,\\ &
    =-\frac{Q_1m_2}{2m_p^2}\rmint dt v_1^{i}\rmint_{\Vec{k}}\frac{e^{i\Vec{k}\cdot\Vec{r}}}{\Vec{k}^2}\,\Bar{a}_{i}(\Vec{x}_1(t))\,,\\ &
    =-\frac{Q_1m_2}{2m_p^2}\rmint dt v_1^{i}\, \frac{1}{4\pi r}\,\Bar{a}_{i}(\Vec{x}_1(t))+(1\leftrightarrow 2)\,\sim\mathcal{O}(L^{1/2}v^{7/2}).
    \end{split}
\end{align}
\end{itemize}
By adding all these contributions we get the total effective action for the radiation sector. Then we can calculate the source term and it is given by,
\begin{align}
    \begin{split}
      &  J_{a}^{0}=\frac{1}{8\pi m_p^2\,r}\sum_{a\ne b}Q_{a}m_b\delta(\Vec{x}-\Vec{x_a})+\sum_{a}Q_a\delta(\Vec{x}-\Vec{x}_a)\,,\\ &
J_{a}^{i}=\sum_{a}Q_{a}\,v_a^{i}\delta(\Vec{x}-\Vec{x_a})-\frac{1}{8\pi m_p^2\,r}\sum_{a\ne b}Q_a\, m_b \,v_a^{i}\delta(\Vec{x}-\Vec{x}_a)\,.
    \end{split}
\end{align}
Finally, we compute the multipole moment using this source term and then using (\ref{5.42}) (as well using the optical theorem), we get the following power radiation formula. Here we have confined ourselves to the dipole term (i.e. $l=1$).\\
\begin{equation}
    \boxed{
 \begin{aligned}
        P_{em}=&\frac{1}{12\pi}\Bigg\{\Bigg[\frac{(Q_1m_2-Q_2 m_1)}{M}r+\frac{(\mathcal{C}_1m_2^2-\mathcal{C}_2m_1^2)}{M}\Bigg]^2\Omega^4\\ &
        +\Bigg[-\frac{(\mathcal{D}_1m_2^2+\mathcal{D}_2m_1^2)}{M^2}r
+\frac{(\tilde{d_1}m_2^4+\tilde{d_2}m_1^4)}{M^4}r^3\Omega^2
+\frac{\Delta_1m_2^3+\Delta_2m_1^3}{M^2}\\ &
+\frac{(G\tilde{d_1}m_2^5+G\tilde{d_2}m_1^5)}{M^4}r^2\Omega^2
 \Bigg]^2 16r^2\Omega^6\Bigg\}\,.
\end{aligned}
}
\end{equation}
 
 Here, 
 $$ \mathcal{C}_a=\frac{Q_a}{8\pi m_p^2},\mathcal{D}_a=\frac{Q_a}{2},\tilde{d_a}=\frac{Q_a}{20},\Delta_a=\frac{\mathcal{D}_a}{8\pi m_p^2}, $$ with $a=1,2.$\par 
\textit{Again, to the best of our knowledge, the electromagnetic dipole radiation has been computed previously only up to $LO$ \cite{Cardoso:2020iji}. Here we have computed the same up to  $N^{(3)}LO$.}

\subsection{Proca interaction and multipole decomposition}
Now we will focus on the Proca sector. In order to derive the power radiation formula, we need to calculate the imaginary part of the effective action of the source. Following the procedure outlined in Sec.~(\ref{subsec:radiated_scalars}) we get, 
\begin{align}
    \begin{split}
        i \gamma_\mathrm{eff}^{(\bar b_\mu)}&\sim \rmint dt_1\rmint d t_2 \Big[\mathcal{I}^{(\mu)}_{b}\mathcal{I}^{(\nu)}_{b}\Big\langle T \Bar{b}_{\mu}(t_1,\boldsymbol{0})\Bar{b}_{\nu}(t_2,\boldsymbol{0})\Big\rangle+\mathcal{I}^{(\mu)i}_{b}\mathcal{I}^{(\nu)j}_{b}\Big\langle T \partial_i \Bar{b}_{\mu}(t_1,\boldsymbol{0})\partial_j \Bar{b}_{\nu}(t_2,\boldsymbol{0})\Big\rangle+\\ &
        \,\,\,\,\,\,\,\,\,\,\,\,\,\,\,\,\,\,\,\,\,\,\,\,\,\,\,\,\,\,\,\,\,\,\,\,\,\,\,\,\frac{1}{4}\mathcal{I}^{(\mu)ij}_{a}\mathcal{I}^{(\nu)kl}_{b}\Big\langle T \partial_i\partial_j\Bar{b}_{\mu}(t_1,\boldsymbol{0})\partial_k\partial_l\Bar{b}_{\nu}(t_2,\boldsymbol{0})\Big\rangle\Big]\,.\label{5.1}
    \end{split}
\end{align}
The main difference between Proca and an electromagnetic propagator is that, in the case of a Proca field, the propagator is a massive propagator. The Feynman propagator for this case is given below,
\begin{align}
    \begin{split}
        \Big\langle T \Bar{b}_{\mu}(t_1,\boldsymbol{x}_1)\Bar{b}_{\nu}(t_2,\boldsymbol{x}_2)\Big\rangle=\rmint \frac{d^4k}{(2\pi)^4}\frac{-i\,\left(g_{\mu\nu}-\frac{k_\mu k_\nu}{\mu_\gamma ^2}\right)}{-k^2-\mu_{\gamma}^2+i\epsilon}e^{-ik\cdot(x_1-x_2)}\,.\label{5.2}
    \end{split}
\end{align}
Therefore, following the same method discussed in Sec.~(\ref{subsec:radiated_scalars}), we get the imaginary part of the effective action.
\begin{align}
    \begin{split}
        \text{Im}[\gamma^{\bar{b}_{\mu}}_{\text{eff}}]\Big|_{(1)}&=\frac{1}{8\pi^2}\rmint dk_{0}\rmint d|\boldsymbol{k}|\,|\boldsymbol{k}|^2\delta(k_0^2-\boldsymbol{k}^2-\mu_{\gamma}^2)\Big[|\mathcal{I}^{\mu}(k_0)|^2+\frac{1}{3}|\mathcal{I}^{(\mu)i}(k_0)|^2|\boldsymbol{k}|^2\\ &
   \hspace{7cm}+\frac{1}{30}|\mathcal{I}^{(\mu)ij}(k_0)|^2|\boldsymbol{k}|^4\Big]
   \,,\\ &
    =\frac{1}{8\pi^2}\rmint dk_0\rmint d|\boldsymbol{k}|\,|\boldsymbol{k}|^2\frac{1}{2\sqrt{k_0^2-\mu_{\gamma}^2}}\Big[\delta(-|\boldsymbol{k}|+\sqrt{k_0^2-\mu_{\gamma}^2})+\delta(|\boldsymbol{k}|+\sqrt{k_0^2-\mu_{\gamma}^2})\Big]\Big[|\mathcal{I}^{\mu}(k_0)|^2\\ &
   \hspace{7cm}+\frac{1}{3}|\mathcal{I}^{(\mu)i}(k_0)|^2|\boldsymbol{k}|^2+\frac{1}{30}|\mathcal{I}^{(\mu)ij}(k_0)|^2|\boldsymbol{k}|^4\Big]\,,\\ &
    =\frac{1}{8\pi^2}\rmint d\omega \, \sum_{L}\frac{1}{l!(2l+1)!!}\,(\omega^2-\mu_{\gamma}^2)^{1/2+l}\,|\mathcal{I}^{(\mu)}_{L}(\omega)|^2\,.
    \label{5.50}
    \end{split}
\end{align}
The other part comes from the second part of the propagator,
\begin{align}
    \begin{split}
\text{Im}[\gamma^{\bar{b}_{\mu}}_{\text{eff}}]\Big|_{(2)}&=-\frac{1}{8\pi^2}\int dt_1 dt_2 I^{\mu}(t_1)I^{\mu}(t_2)\int dk_0\frac{e^{ik_0\,(t_1-t_2)}}{2\sqrt{k_0^2-\mu_{\gamma}^2}}\int d^3k \,\delta\left(-|\boldsymbol{k}|+\sqrt{k_0^2-\mu_{\gamma}^2}\right)\frac{k_\mu k_\nu}{\mu_{\gamma}^2}\\ &
=-\frac{1}{8\pi^2}\int \frac{d\omega}{2\sqrt{\omega^2-\mu_\gamma^2}} I^{\mu}(\omega)I^{*\nu}(\omega)\int d^3k \,\delta\left(-|\boldsymbol{k}|+\sqrt{k_0^2-\mu_{\gamma}^2}\right)\frac{k_\mu k_\nu}{\mu_{\gamma}^2}\\ &
=-\frac{1}{16\pi^2\mu_{\gamma^2}}\int d\omega \,\sqrt{\omega^2-\mu_\gamma^2}\left(\omega^2|I^{0}(\omega)|^2-\frac{1}{3}(\omega^2-\mu_\gamma^2)|I^{i}(\omega)|^2\right)
\end{split}
\end{align}
However, in our following computations, we ignore the second part of the radiative effective action as we are only looking for the leading contribution for this massive mode. Although the correction can be straightforwardly done.
Therefore, the power radiation is given by,
\begin{align}
    \begin{split}
        P_{pr}=\frac{1}{4\pi^2\mathcal{T}}\sum_{l}\frac{1}{l!(2l+1)!!}\rmint_0^\infty d\omega\,\omega\,(\omega^2-\mu_{\gamma}^2)^{l+1/2}|\mathcal{I}^{\mu}_{(L)}(\omega)|^2\,.
    \end{split}
\end{align}
 We need to calculate the two source terms, $J_{b}^{0}$ and $J_{b}^{i}$, and the computation is exactly analogous to the electromagnetic case discussed in Sec.~(\ref{emRad}). Again, we restrict ourselves to \textcolor{black}{$N^{(3)}LO$}. Below, we quote the results for the source terms. 
\begin{align}
    \begin{split}
      &  J_{b}^{0}=\frac{1}{8\pi m_p^2 r}\sum_{a\neq b}Q_{a}'m_b\delta(\boldsymbol{x}-\boldsymbol{x}_a)+\sum_{a}Q_a'\delta(\boldsymbol{x}-\boldsymbol{x}_a)\,\\ &
      J_{b}^{i}=\sum_{a}Q_{a}'\,v_a^{i}\delta(\boldsymbol{x}-\boldsymbol{x}_a)-\frac{1}{8\pi m_p^2 r}\sum_{a\ne b}Q_a'\, m_b \,v_a^{i}\delta(\boldsymbol{x}-\boldsymbol{x}_a)\,.
    \end{split}
\end{align}

Now that we have the source terms, we can use (\ref{5.50}) to calculate the multipole moments up to $N^{(3)}LO$ as follows,
 \footnote{ We use $\delta(x-a)^2=\delta(x-a)\delta(0)$.}
\begin{align}
\begin{split}
 |\mathcal{I}^{\mu}_{(L)}(\omega)|^2= &
\Bigg[\frac{({Q}_1'm_2-{Q}_2'm_1)r}{M}+\frac{\mu_{\gamma}^2}{10}\frac{({Q}_1'm_2^3-{Q}_2'm_1^3)}{M^3}r^2+\frac{(\mathcal{C}_1'm_2^2-\mathcal{C}_2'm_1^2)}{M}+\frac{\mu_{\gamma}^2}{10}\frac{(\mathcal{C}_1'm_2^3-\mathcal{C}_2'm_1^3)}{M^3}r^2\Bigg]^2\\&{\frac{\pi\delta(\omega-\Omega)\delta(0)}{2}}+\Bigg[- \frac{(\mathcal{D}_1'm_2^2+\mathcal{D}_2'm_1^2)}{M^2}r
+\frac{(\tilde{d_1}'m_2^4+\tilde{d_2}'m_1^4)}{M^4}r^3\Omega^2\textcolor{black}{(1-\frac{\mu_{\gamma}^2}{\Omega^2}})
\\&+\frac{\Delta_1'm_2^3+\Delta_2'm_1^3}{M^2} +\frac{(\tilde{d_1}'m_2^5+\tilde{d_2}'m_1^5)}{8\pi m_p^2 M^4}r^2\Omega^2\textcolor{black}{(1-\frac{\mu_{\gamma}^2}{\Omega^2}})
 \Bigg]^2\frac{\pi\, r^2\Omega^2}{2}\delta(\omega-2\Omega)\delta(0).
 \end{split}
 \end{align}

Now we have the source terms from which we can calculate using (\ref{5.50}) the multipole moments upto $N^{(3)}LO$ as follows,
 \footnote{ We use $\delta(x-a)^2=\delta(x-a)\delta(0)$.}
\begin{align}
\begin{split}
 |\mathcal{I}^{\mu}_{(L)}(\omega)|^2= &
\Bigg[\frac{({Q}_1'm_2-{Q}_2'm_1)r}{M}+\frac{\mu_{\gamma}^2}{10}\frac{({Q}_1'm_2^3-{Q}_2'm_1^3)}{M^3}r^2+\frac{(\mathcal{C}_1'm_2^2-\mathcal{C}_2'm_1^2)}{M}+\frac{\mu_{\gamma}^2}{10}\frac{(\mathcal{C}_1'm_2^3-\mathcal{C}_2'm_1^3)}{M^3}r^2\Bigg]^2\\&{\frac{\pi\delta(\omega-\Omega)\delta(0)}{2}}+\Bigg[- \frac{(\mathcal{D}_1'm_2^2+\mathcal{D}_2'm_1^2)}{M^2}r
+\frac{(\tilde{d_1}'m_2^4+\tilde{d_2}'m_1^4)}{M^4}r^3\Omega^2\textcolor{black}{(1-\frac{\mu_{\gamma}^2}{\Omega^2}})
\\&+\frac{\Delta_1'm_2^3+\Delta_2'm_1^3}{M^2} +\frac{(\tilde{d_1}'m_2^5+\tilde{d_2}'m_1^5)}{8\pi m_p^2 M^4}r^2\Omega^2\textcolor{black}{(1-\frac{\mu_{\gamma}^2}{\Omega^2}})
 \Bigg]^2\frac{\pi\, r^2\Omega^2}{2}\delta(\omega-2\Omega)\delta(0).
 \end{split}
 \end{align}

\textcolor{black}{Finally we get the expression for the power radiation  for the Proca sector from the dipole part ($l=1$)}
 \begin{equation}
     \boxed{
 \begin{aligned}
      P_{\text{pr}}=&\frac{1}{12\pi} \Bigg\{\Bigg[\frac{({Q}_1'm_2-{Q}_2'm_1)r}{M}+\frac{\mu_{\gamma}^2}{10}\frac{({Q}_1'm_2^3-{Q}_2'm_1^3)}{M^3}r^2+\frac{(\mathcal{C}_1'm_2^2-\mathcal{C}_2'm_1^2)}{M}+\frac{\mu_{\gamma}^2}{10}\frac{(\mathcal{C}_1'm_2^4-\mathcal{C}_2'm_1^4)}{M^3}r^2\Bigg]^2\\ &
       \Omega^4(1-\frac{\mu_{\gamma}^2}{\Omega^2})^{\frac{3}{2}} +\Bigg[-\frac{(\mathcal{D}_1' m_2^2+\mathcal{D}_2' m_1^2)}{M^2}r
+\frac{(\tilde{d_1}'m_2^4+\tilde{d_2}'m_1^4)}{M^4}r^3\Omega^2\textcolor{black}{(1-\frac{\mu_{\gamma}^2}{\Omega^2}})
+\frac{\Delta'_1m_2^3+\Delta'_2m_1^3}{M^2}\\ &
+\frac{(\tilde{d_1}'m_2^5+\tilde{d_2}' m_1^5)}{8\pi m_p^2 M^4}r^2\Omega^2\textcolor{black}{(1-\frac{\mu_{\gamma}^2}{\Omega^2}})
 \Bigg]^2 16r^2\Omega^6(1-\frac{\mu_{\gamma}^2}{4\Omega^2})^{\frac{3}{2}}\Bigg\}\,,
\end{aligned}
}
 \end{equation}
 
 where $$ \mathcal{C}_a'=\frac{Q_a'}{8\pi m_p^2},\mathcal{D}_a'=\frac{Q_a'}{2},\tilde{d_a}'=\frac{Q_a'}{20},\Delta_a'=\frac{\mathcal{D}_a}{8\pi m_p^2}. $$\par 
\textit{Again, to the best of our knowledge, we have computed for the first time the Proca dipole radiation up to  $N^{(3)}LO$.}
\subsection{Gravitational radiation and Multipole decomposition}\label{sec5.5}
Finally, we focus on the gravitational sector. This is already studied in the \cite{Kuntz:2019zef, Levi:2018nxp}. We provide the computation for the gravitational power radiation for completeness, and also compute the corrections coming from different field vertices due to their coupling with the gravitational field at different orders of perturbation, which are explicitly mentioned. 
First, we will discuss the EFT  power counting similar to the one discussed in Sec.~(\ref{EFTcounting}) to understand which diagrams contribute to a \textcolor{black}{specific order}.\par
To see how the multipole moments scale, we need to know the scaling of the pseudo stress-energy tensor $T^{\mu\nu}.$

\begin{equation}\rmint d^3xT^{\mu\nu}\sim \begin{cases}
\begin{array}{ll}
M & (\mu=0,\nu=0),\\
Mv & (\mu=0,\nu=i),\\
Mv^2 & (\mu=i,\nu=j),\\
\end{array}
\end{cases}
\end{equation}
where $M$ denotes the mass-scale. From this we get,

\begin{equation}\rmint d^3x T^{\mu\nu}x^L\sim \begin{cases}
\begin{array}{ll}
L^{\frac{1}{2}}r^lv^{\frac{1}{2}} & (\mu=0,\nu=0),\\
L^{\frac{1}{2}}r^lv^{\frac{3}{2}} & (\mu=0,\nu=i),\\
L^{\frac{1}{2}}r^lv^{\frac{5}{2}} & (\mu=i,\nu=j),\\
\end{array}
\end{cases}
\end{equation}

where $L$ denotes the length-scale. \\
The formula for power radiation can be obtained in the same way as discussed in (\ref{5.18n}). First, we write down the imaginary part of the effective action.
\begin{align}
    \begin{split}
        i\gamma_{\text{eff}}^{\text{g}}=-\frac{1}{32 m_p^2}\rmint dt_1 dt_2\, I^{ij}_{g}(t_1)I^{kl}_{g}(t_2)\,\Big\langle T\ddot{\Bar{h}}_{ij}^{\text{TT}}(t_1,\Vec{0})\ddot{\Bar{h}}_{kl}^{\text{TT}}(t_2,\Vec{0})\Big\rangle+\cdots,\label{5.56mm}
    \end{split}
\end{align}
where ${\Bar{h}}_{ij}^{\text{TT}}$ is the radiative gravitational field in \textit{transverse-traceless (TT)} gauge and ($\cdots$) denote the higher order moments. Now use the following propagator for ${\Bar{h}}_{ij}^{\text{TT}}$ \cite{Goldberger:2004jt} to evaluate the effective action.
\begin{align}
    \begin{split}
        \Big\langle T\ddot{\Bar{h}}_{ij}^{\text{TT}}(t_1,\Vec{0})\ddot{\Bar{h}}_{kl}^{\text{TT}}(t_2,\Vec{0})\Big\rangle=\frac{8}{5}\Big[\frac{1}{2}(\delta_{ik}\delta_{jl}+\delta_{ik}\delta_{jl}-\frac{1}{3}\delta_{ij}\delta_{kl})\Big]\rmint \frac{d^4k}{(2\pi)^4}\frac{k_0^4}{-k^2-i\epsilon}e^{ik_0(t_1-t_2)},\label{5.57mm}
    \end{split}
\end{align}
Then using (\ref{5.57mm}) into (\ref{5.56mm}) we can write the effective action as \cite{Goldberger:2004jt,Kuntz:2019zef},
\begin{align}
    \begin{split}
        \gamma_{\text{eff}}^{\text{g}}=\frac{1}{20m_p^2}\rmint \frac{d^4 k}{(2\pi)^4}\frac{k_0^4}{-k^2-i\epsilon}|I^{ij}(k_0)|^2+\cdots.
    \end{split}
\end{align}
From this, we can write down the  power radiation in terms of \textit{quadrupole} moment,
\begin{align}
    \begin{split}
        P_{g}=\frac{2G}{5\mathcal{T}}\rmint \frac{d\omega}{2\pi}\omega^6\,|I^{ij}_{g}(\omega)|^2.
    \end{split}
\end{align}
In the rest of the section, we will mainly use this to compute the power radiation. This can be systematically extended by including contributions from higher multipole moments. Interested readers are referred to \cite{Goldberger:2004jt} for more details. For the sake of completeness, we also write down the expression for power radiation upto \textit{octupole} moment in the following way \cite{Goldberger:2004jt},
\begin{align}
    \begin{split}
        \mathcal{P}_{g}=\frac{2G}{5\mathcal{T}}\rmint \frac{d\omega}{2\pi}\, \Bigg\{\omega^6\, |{I}_{h}^{ij}(\omega)|^2+\frac{16\,\omega^6}{45}|J^{ij}(\omega)|^2+\frac{\omega^8}{189}|I^{ijk}(\omega)|^2+.........\Bigg\}\,.
        \label{5.65}
    \end{split}
\end{align}
 \textcolor{black}{ The explicit forms of ${I}^{ij}_{h}, \,J^{ij}$ and $I^{ijk}$ are given by:
\begin{align}
    \begin{split}
      &   {I}^{ij}_{h}(t)=\rmint d^3x \,T^{00}Q^{ij}.\\ &
      J^{ij}=-\frac{1}{2}\rmint d^3x (\epsilon^{ikl}T^{0k}x^ix^j+\epsilon^{jkl}T^{0k}x^ix^l)+.........\\ &
      I^{ijk}=\rmint d^3x (T^{00}+T^{ll})[x^ix^jx^k]^{STF}\,.
    \end{split}
\end{align}
}
\begin{figure}[b!]
\centering
\scalebox{0.3}{
\begin{feynman}
    \electroweak[lineWidth=4, label=$\Bar{\psi}$]{5.60, 6.00}{7.60, 7.20}
    \fermion[lineWidth=4]{4.00, 4.00}{7.40, 4.00}
    \fermion[lineWidth=4]{4.00, 6.00}{7.40, 6.00}
\end{feynman}}
\caption{LO radiation coming from the pure gravitational sector.}
\label{figpq}
\end{figure}
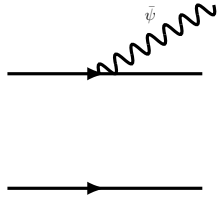

\underline{\textit{{$LO$ result for pure gravity}}}:\\\\
First, we will briefly discuss the $LO$ power radiation from the pure gravitational sector (i.e. no other field couplings are present).  Only one diagram consists of the vertices $-\sum_a m_a\rmint dt\,\psi$ as shown in Fig.~(\ref{figpq}) contributes to this case. The corresponding amplitude is given by: 
\begin{align}
    \begin{split}
S_{\text{eff}}\Big|_{{\text{fig}(\ref{figpq})}}=-\sum_{a}m_a\rmint dt\, \frac{\Bar{\psi}(\Vec{x}_a(t))}{m_p}\sim\mathcal{O}(L^{1/2}v^{1/2})\,.
    \end{split}
\end{align}

Now, to compute power radiation we first define the centre of mass velocities (at leading order) in the following way: $$\Vec{v}_1=\frac{m_2}{m_1+m_2} \Vec{v}\quad \& \quad \Vec{v}_2=-\frac{m_1}{m_1+m_2}  \Vec{v}.$$\\

Then we compute the quadrupole moment,
\begin{align}
\begin{split}
   \mathcal{I}^{ij}_h&=-\rmint d^3 x\rmint dt\underbrace{\sum_a m_a\delta(\vec{x}-\vec{x_a})}_{T^{00}}[x^ix^j]_{STF}\,,\\&
    =-\rmint dt{\sum_a m_a}[x_a^ix_a^j]_{STF}\,.\label{5.59m}
\end{split}
\end{align}
\vspace{-0.1cm}

Using (\ref{5.59m}) one can obtain the formula for $LO$ power radiation as ,
\begin{align}
\begin{split}
{P}_g&=\frac{2G}{5}\rmint \frac{d\omega}{2\pi}\omega^6\underbrace{(\frac{\pi}{2}\mu^2r^4\delta(\omega-2\Omega))}_{|\mathcal{I}^{ij}_h|^2}\,,\\&
=\frac{32G}{5}\mu^2r^4\Omega^6.
\end{split}
\end{align}
which agrees with our known GR radiation formula in leading order \cite{2009GReGr..41.1667H}. As there are couplings of the gravitational field with other fields for our case, we will also have contributions from them to the gravitational power radiation. Next, we calculate those non-trivial corrections of the gravitational power radiation to their respective leading orders.\\

\begin{figure}
    \centering
    \begin{subfigure}{0.2\textwidth}
        \centering
        \scalebox{0.25}{\begin{feynman}
    \fermion[lineWidth=6]{4.00, 6.40}{7.20, 6.40}
    \fermion[lineWidth=6]{4.00, 4.00}{7.20, 4.00}
    \electroweak[lineWidth=6, label=$\tilde{\psi}$]{5.60, 6.40}{5.60, 4.00}
\electroweak[label=$\Bar{\psi}$, lineWidth=6]{5.60, 6.40}{7.20, 8.00}
\end{feynman}
}
\caption{}
\label{fig13e}
    \end{subfigure}
    \begin{subfigure}{0.2\textwidth}
        \centering
        \scalebox{0.25}{
    \begin{feynman}
    \fermion[lineWidth=6]{4.00, 6.40}{7.20, 6.40}
    \fermion[lineWidth=6]{4.00, 4.00}{7.20, 4.00}
    \dashed[showArrow=false, lineWidth=6, color=9900ef, label=$\mathcal{B}_0$]{5.60, 6.40}{5.60, 4.00}
\electroweak[label=$\Bar{\psi}$, lineWidth=6]{5.60, 6.40}{7.20, 8.00}
\end{feynman}
}\caption{}
\label{fig13f}
    \end{subfigure}
      \begin{subfigure}{0.2\textwidth}
        \centering
        \scalebox{0.25}{
\begin{feynman}
    \dashed[showArrow=false, lineWidth=6, color=9900ef, label=$\mathcal{B}_0$]{5.60, 5.40}{5.60, 4.00}
    \fermion[lineWidth=6]{4.00, 6.40}{7.20, 6.40}
    \electroweak[lineWidth=6, label=$\Bar{\psi}$]{5.60, 6.40}{7.60, 7.80}
    \fermion[lineWidth=6]{4.00, 4.00}{7.20, 4.00}
    \dashed[lineWidth=6, showArrow=false, label=$\mathcal{A}_0$]{5.60, 6.40}{5.60, 5.40}
\end{feynman}
}
\caption{}
\label{fig13l}
\end{subfigure}
\begin{subfigure}{0.2\textwidth}
        \centering
        \scalebox{0.25}{
\begin{feynman}
    \fermion[lineWidth=6]{4.00, 6.40}{7.20, 6.40}
    \dashed[showArrow=false, lineWidth=6, label=$\mathcal{A}_0$]{5.60, 6.40}{5.60, 4.00}
    \fermion[lineWidth=6]{4.00, 4.00}{7.20, 4.00}
\electroweak[label=$\Bar{\psi}$, lineWidth=6]{5.60, 6.40}{7.20, 8.00}
\end{feynman}
}
\caption{}
\label{fig13g}
    \end{subfigure}
    \begin{subfigure}{0.25\textwidth}
    \centering
\scalebox{0.22}{\begin{feynman}
\electroweak[label=$\varphi$, lineWidth=6, color=0693e3]{6.20, 5.10}{6.20, 4.00}
    \electroweak[lineWidth=5, label=$\Bar{\psi}$]{6.50, 5.40}{8.00, 5.40}
    \fermion[lineWidth=6]{4.00, 4.00}{8.40, 4.00}
    \electroweak[lineWidth=6, color=0693e3, label=$\varphi$]{6.20, 5.70}{6.20, 6.80}
    \fermion[lineWidth=6]{4.00, 6.80}{8.40, 6.80}
    \parton[color=eb144c]{6.20,5.40}{0.30}
\end{feynman}}
    \caption{}
\label{fig13n}
\end{subfigure}
    \begin{subfigure}{0.2\textwidth}
        \centering
        \scalebox{0.25}{
        \begin{feynman}
    \gluon[color=eb144c, label=$\mathcal{B}_i$, lineWidth=6, flip=true]{5.20, 6.40}{5.20, 4.00}
    \fermion[lineWidth=6]{4.00, 4.00}{6.60, 4.00}
    \fermion[lineWidth=6]{4.00, 6.40}{6.60, 6.40}
    \electroweak[lineWidth=6, label=$\bar{\psi}$]{5.20, 6.40}{6.80, 8.00}
\end{feynman}
}
\caption{}
\label{fig13a}
    \end{subfigure}
    \begin{subfigure}{0.2\textwidth}
        \centering
        \scalebox{0.25}{
        \begin{feynman}
    \fermion[lineWidth=6]{4.00, 4.00}{7.00, 4.00}
    \electroweak[lineWidth=6, color=9900ef, label=$\mathcal{A}_i$]{5.40, 6.40}{5.40, 4.00}
    \electroweak[lineWidth=6, label=$\Bar{\psi}$]{5.40, 6.40}{7.60, 8.00}
    \fermion[lineWidth=6]{4.00, 6.40}{7.00, 6.40}
\end{feynman}
}\caption{}
\label{fig13b}
    \end{subfigure}
\begin{subfigure}{0.2\textwidth}
        \centering
        \scalebox{0.25}{
\begin{feynman}
    \fermion[lineWidth=6]{4.00, 6.40}{7.20, 6.40}
    \dashed[showArrow=false, label=$\mathcal{B}_0$, color=9900ef, lineWidth=6]{4.40, 4.00}{5.60, 6.40}
    \fermion[lineWidth=6]{4.00, 4.00}{7.20, 4.00}
    \electroweak[lineWidth=6]{5.60, 6.40}{6.80, 4.00}
\electroweak[label=$\Bar{\psi}$, lineWidth=6]{5.60, 6.40}{7.20, 8.00}
\end{feynman}
}
\caption{}
\label{fig13c}
    \end{subfigure}
\begin{subfigure}{0.2\textwidth}
        \centering
        \scalebox{0.25}{
    \begin{feynman}
    \fermion[lineWidth=6]{4.00, 6.40}{7.20, 6.40}
    \dashed[showArrow=false, lineWidth=6, label=$\mathcal{A}_0$]{4.40, 4.00}{5.60, 6.40}
    \fermion[lineWidth=6]{4.00, 4.00}{7.20, 4.00}
    \electroweak[lineWidth=6]{5.60, 6.40}{6.80, 4.00}
\electroweak[label=$\Bar{\psi}$, lineWidth=6]{5.60, 6.40}{7.20, 8.00}
\end{feynman}
}\caption{}
\label{fig13d}
    \end{subfigure}
\begin{subfigure}{0.2\textwidth}
        \centering
        \scalebox{0.25}{
    \begin{feynman}
    \fermion[lineWidth=6]{4.00, 6.40}{7.20, 6.40}
    \electroweak[lineWidth=6, color=0693e3, label=$\varphi$]{5.60, 6.40}{5.60, 4.00}
    \fermion[lineWidth=6]{4.00, 4.00}{7.20, 4.00}
\electroweak[label=$\Bar{\psi}$, lineWidth=6]{5.60, 6.40}{7.20, 8.00}
\end{feynman}
}\caption{}
\label{fig13h}
    \end{subfigure}
\begin{subfigure}{0.2\textwidth}
        \centering
        \scalebox{0.25}{
\begin{feynman}
    \fermion[lineWidth=6]{4.00, 6.40}{7.20, 6.40}
    \electroweak[lineWidth=6, color=0693e3, label=$\varphi$]{5.60, 6.40}{6.80, 4.00}
    \fermion[lineWidth=6]{4.00, 4.00}{7.20, 4.00}
    \electroweak[lineWidth=6, label=$\varphi$, color=0693e3]{4.40, 4.00}{5.60, 6.40}
\electroweak[label=$\Bar{\psi}$, lineWidth=6]{5.60, 6.40}{7.20, 8.00}
\end{feynman}
}
\caption{}
\label{fig13i}
    \end{subfigure}
\begin{subfigure}{0.2\textwidth}
        \centering
        \scalebox{0.25}{
\begin{feynman}
    \gluon[flip=true, lineWidth=6, label=$\mathcal{B}_i$, color=eb144c]{5.60, 5.20}{5.60, 4.00}
    \fermion[lineWidth=6]{4.00, 6.40}{7.20, 6.40}
    \electroweak[lineWidth=6, label=$\Bar{\psi}$]{5.60, 5.20}{8.20, 5.20}
    \fermion[lineWidth=6]{4.00, 4.00}{7.20, 4.00}
    \electroweak[lineWidth=6, color=9900ef, label=$\mathcal{A}_i$]{5.60, 6.40}{5.60, 5.20}
\end{feynman}
} 
\caption{}
\label{fig13j}
\end{subfigure}
\begin{subfigure}{0.2\textwidth}
        \centering
        \scalebox{0.25}{\begin{feynman}
    \gluon[lineWidth=6, label=$\mathcal{B}_k$, color=eb144c, flip=true]{5.60, 5.40}{5.60, 4.00}
    \fermion[lineWidth=6]{4.00, 6.40}{7.20, 6.40}
    \dashed[lineWidth=6, label=$\mathcal{A}_0$, showArrow=false]{5.60, 6.40}{5.60, 5.40}
    \electroweak[lineWidth=6, label=$\Bar{\psi}$]{5.60, 6.40}{7.60, 7.80}
    \fermion[lineWidth=6]{4.00, 4.00}{7.20, 4.00}
\end{feynman}
}
\caption{}
\label{fig13k}
\end{subfigure}
\begin{subfigure}{0.2\textwidth}
        \centering
        \scalebox{0.25}{
\begin{feynman}
    \dashed[showArrow=false, lineWidth=6, color=9900ef, label=$\mathcal{B}_0$]{5.60, 5.20}{5.60, 4.00}
    \electroweak[lineWidth=6, label=$\mathcal{A}_i$, color=9900ef]{5.60, 5.20}{5.60, 6.40}
    \fermion[lineWidth=6]{4.00, 6.40}{7.20, 6.40}
    \electroweak[lineWidth=6, label=$\Bar{\psi}$]{5.60, 6.40}{7.60, 7.80}
    \fermion[lineWidth=6]{4.00, 4.00}{7.20, 4.00}
\end{feynman}
}
\caption{}
\label{fig13m}
\end{subfigure}
\begin{subfigure}{0.22\textwidth}
    \centering
\scalebox{0.25}{
\begin{feynman}
    \electroweak[lineWidth=6, label=$\bar{\psi}$]{5.60, 6.40}{7.00, 7.60}
    \fermion[lineWidth=6]{4.00, 4.00}{7.20, 4.00}
    \electroweak[color=9900ef, lineWidth=6, label=$\mathcal{A}_l$]{5.60, 5.20}{6.80, 4.00}
    \electroweak[color=0693e3, lineWidth=6, label=$\varphi$]{5.60, 6.40}{5.60, 5.20}
    \electroweak[color=9900ef, lineWidth=6, label=$\mathcal{A}_j$]{4.40, 4.00}{5.60, 5.20}
    \fermion[lineWidth=6]{4.00, 6.40}{7.20, 6.40}
\end{feynman}}
 \caption{}
\label{fig13o}
\end{subfigure}
\begin{subfigure}{0.22\textwidth}
        \centering
        \scalebox{0.25}{\begin{feynman}
    \electroweak[flip=true, lineWidth=6, label=$\varphi$, color=0693e3]{5.80, 5.80}{6.80, 7.00}
    \electroweak[lineWidth=6, color=9900ef, label=$\mathcal{A}_l$]{5.80, 5.80}{5.80, 4.60}
    \electroweak[color=9900ef, lineWidth=6, label=$\mathcal{A}_j$]{4.80, 7.00}{5.80, 5.80}
    \fermion[lineWidth=6]{4.00, 4.60}{7.40, 4.60}
    \electroweak[lineWidth=6, label=$\bar{\psi}$, flip=true]{5.80, 4.60}{7.20, 4.00}
    \fermion[lineWidth=6]{4.00, 7.00}{7.60, 7.00}
\end{feynman}
}
 \caption{}
\label{fig13p}
\end{subfigure}
\caption{Radiative diagrams for gravitational fields.}
\label{Fig15}
\end{figure}
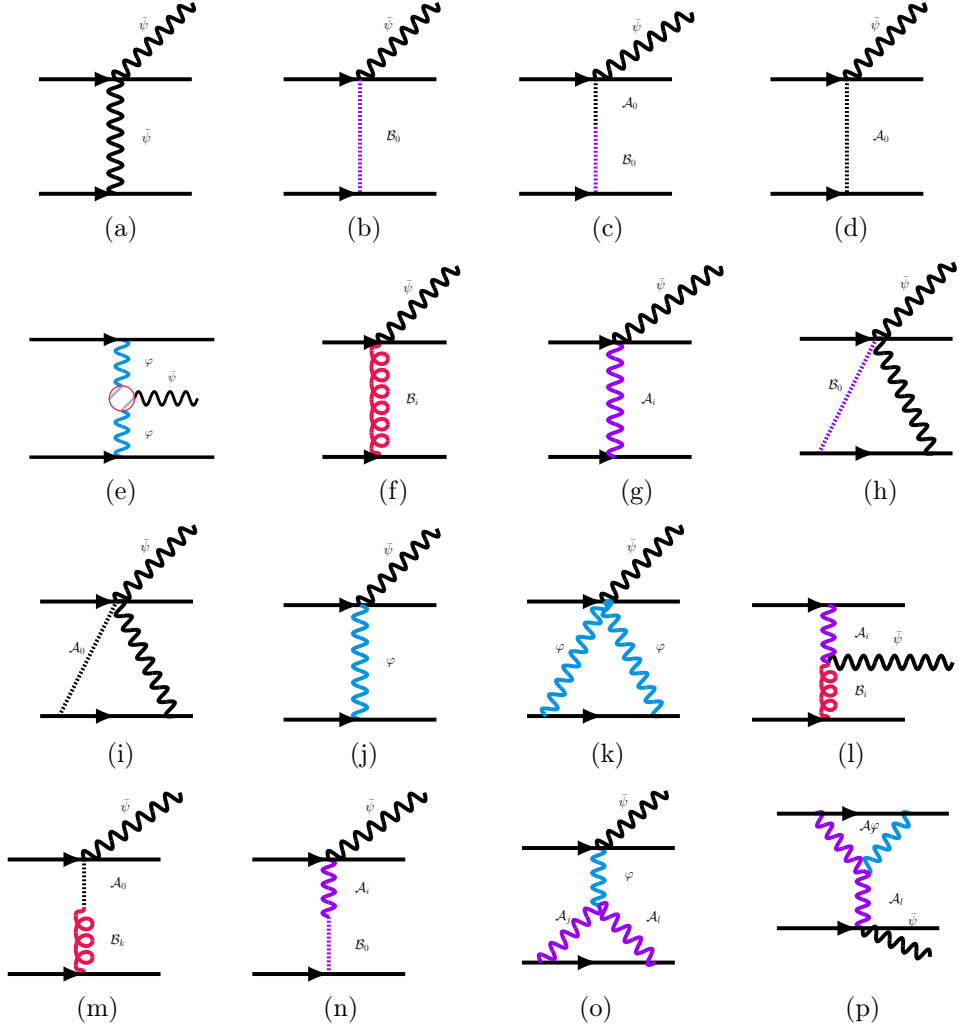

\vspace{0.2 cm}
\underline{\textit{{Radiation from Pure  gravity at higher order ($N^{(2)}LO$)}}}:
\begin{itemize}
\item \textcolor{black}{The diagram that contributes at $N^{(2)}LO$ is shown Fig.~($\ref{figpq}$). The corresponding amplitude is given by,
\begin{align}
\begin{split}
\mathcal{S}_\text{eff}\Big |_{\text{fig}(\ref{figpq})}=-\sum_a m_a \rmint dt \,\frac{3}{2}v_a^2\,\frac{\Bar{\psi}(\vec{x}_a(t))}{m_p}\sim\mathcal{O}(L^{1/2}v^{5/2})\,.
\end{split}
\end{align}}
\item The amplitude corresponding to the diagram in Fig.~(\ref{fig13e}) contributing at $N^{(2)}LO$ with worldline coupling $\frac{m_a}{2}\rmint dt\,\tilde{\psi}\Bar{\psi} :$
\begin{align}
    \begin{split}
         \mathcal{S}_{\text{eff}}\Big |_{\text{fig}(\ref{fig13e})}&=\frac{m_1m_2}{2m_p^2}\rmint \Bar{\mathcal{D}}\hat{\xi}\rmint dt_1 \Tilde{\psi}(\Vec{x}_1(t_1))\rmint dt_2\Tilde{\psi}(\Vec{x}_2(t_2))\frac{\Bar{\psi}(\Vec{x}_1(t_1))}{m_p}\,\\&
         =\frac{m_1m_2}{2m_p^2}\rmint dt_1 dt_2 \Big{\langle}\tilde{\psi}(\Vec{x}_1(t_1))\Tilde{\psi}(\Vec{x}_2(t_2))\Big{\rangle}\frac{\Bar{\psi}(\Vec{x}_1(t_1))}{m_p}\,,\\&
      =\frac{m_1m_2}{16m_p^2\pi}\rmint dt\frac{1}{r}\frac{\Bar{\psi}(\Vec{x}_1(t))}{m_p}+(1\leftrightarrow 2)\,,\sim\mathcal{O}(L^{1/2}v^{5/2}).
    \end{split}
\end{align}
\end{itemize}
\underline{\textit{Radiation due to other field couplings with gravity}}\footnote{There are many diagrams contributing to the pure gravitational radiation at different $N^{(n)}LO$. But we don't calculate all of them explicitly as the results are already there in literature \cite{Blanchet:2013haa}.}:
\begin{itemize}
 \item The amplitude corresponding to the diagram in Fig.~(\ref{fig13f}) with worldline interaction vertex $-Q_a'\rmint dt\,\mathcal{B}_0\Bar{\psi}$ at $N^{(2)}LO :$
\begin{align}
    \begin{split}
         \mathcal{S}_{\text{eff}}\Big |_{\text{fig}(\ref{fig13f})}&=-Q_1'Q_2'\rmint\Bar{\mathcal{D}}\Hat{\xi}\rmint dt_1\mathcal{B}_0(\vec{x_1}(t_1))\rmint dt_2\,\mathcal{B}_0(\vec{x_2}(t_2))\frac{\Bar{\psi}(\Vec{x}_1(t_1))}{m_p}\\&
       =-Q_1'Q_2'\rmint dt_1 dt_2\Big{\langle}\,\mathcal{B}_0(\vec{x}_1(t_1))\mathcal{B}_0(\vec{x}_2(t_2))\Big{\rangle}\frac{\Bar{\psi}(\Vec{x}_1(t_1))}{m_p}\,, \\&
       =Q_1'Q_2'\rmint dt\frac{e^{-\mu_{\gamma} r}}{4\pi r}\frac{\Bar{\psi}(\Vec{x}_1(t))}{m_p}+(1\leftrightarrow 2)\,\,\sim\mathcal{O}(L^{1/2}v^{5/2})\,.
    \end{split}
\end{align}
\end{itemize}

\begin{itemize}

\item  The amplitude corresponding to the diagram in Fig.~(\ref{fig13l}) contributing at $N^{(2)}LO$ with the following bulk interaction vertex $\rmint d^4x\,\partial_k \mathcal{A}_0\partial_k \mathcal{B}_0 :$ 
\vspace{-0.5cm}
 \begin{align}
    \begin{split}
\mathcal{S}_{\text{eff}}\Big|_{\text{fig}(\ref{fig13l})}&=-\gamma Q_1 Q_2'\rmint \Bar{\mathcal{D}}\hat{\xi}\rmint dt_1\mathcal{A}_0(\Vec{x}_1(t_1))\rmint dt_2\mathcal{B}_0(\Vec{x}_2(t_2))\rmint d^4x\,\partial_k\mathcal{A}_0(x)\partial_k\mathcal{B}_0(x)\frac{\Bar{\psi}}{m_p}\,,\\&=-\gamma Q_1 Q_2'\rmint dt_1dt_2\rmint d^4x\partial_k\Big{\langle}\mathcal{A}_0(x)\mathcal{A}_0(\Vec{x}_1(t_1))\Big{\rangle}\partial_k\Big{\langle}\mathcal{B}_0(x)\mathcal{B}_0(\Vec{x}_2(t_2))\Big{\rangle}\frac{\Bar{\psi}}{m_p}\,,\\ &
=-\gamma Q_1Q_2'\rmint dt\,\frac{e^{-\mu_{\gamma}r}}{4\pi r}\Big(\frac{\Bar{\psi}(\Vec{x}_1(t))}{m_p}+\frac{\Bar{\psi}(\Vec{x}_2(t))}{m_p}\Big)+(1\leftrightarrow 2)\,\,\sim \mathcal{O}(L^{1/2}v^{5/2}).
    \end{split}
\end{align}
\item The amplitude corresponding to the diagram in Fig.~(\ref{fig13g})   contributing at $N^{(2)}LO$ with the worldline interaction vertex $-Q_a\rmint dt\,\mathcal{A}_0\Bar{\psi} :$
\begin{align}
    \begin{split}
         \mathcal{S}_{\text{eff}}\Big|_{\text{fig}(\ref{fig13g})}&=-Q_1Q_2\rmint \Bar{\mathcal{D}}\hat{\xi}\rmint dt_1dt_2\mathcal{A}_0(\vec{x}_1(t_1))\mathcal{A}_0(\vec{x}_2(t_2))\frac{\Bar{\psi}(\Vec{x}_1(t_1))}{m_p}\,,\\&
          =-Q_1Q_2\rmint dt_1 dt_2\Big{
       \langle}\mathcal{A}_0(\vec{x_1}(t_1))\mathcal{A}_0(\vec{x_2}(t_2))\Big{\rangle}\frac{\Bar{\psi}(\Vec{x}_1(t_1))}{m_p} \,,\\&
       =Q_1Q_2\rmint dt\frac{1}{4\pi r}\frac{\Bar{\psi}(\Vec{x}_1(t))}{m_p}\,+(1\leftrightarrow 2)\,\sim\mathcal{O}(L^{1/2}v^{5/2})\,.
    \end{split}
\end{align}
\item Diagram as shown in Fig.~(\ref{fig13n}) consists of a scalar propagator with relativistic correction, contributing to the gravitational radiation at $N^{(4)}LO.$ The amplitude is given by, 
\begin{align}
    \begin{split}
  \mathcal{S}_{\text{eff}}^{ikm}\Big |_{{\text{fig}(\ref{fig13n})}} &=\frac{m_1m_2 s_1 s_2}{m_p^3}\rmint \Bar{\mathcal{D}}\hat\xi\rmint dt_1 \varphi(\Vec{x}_1(t_1))\bar{\psi}(x_0) \rmint d^4 x\partial_0\varphi(x)\partial_0\varphi(x)\rmint dt_2 \varphi(\Vec{x}_2(t_2))\,, \\&
=\frac{m_1 m_2 s_1 s_2}{16 \pi m_p^2}\rmint dt \,e^{-mr}\Bigg\{\Big(\frac{\Vec{v}_1\cdot\Vec{v}_2}{r}-\frac{(\Vec{v}_1\cdot \hat{n})(\Vec{v}_2\cdot \hat{n})}{r}\Big)+\frac{m}{2}\,(\Vec{v}_2\cdot \Hat{n})(\Vec{v}_1\cdot \hat{n})\Bigg\}\,\frac{\Bar{\psi}(\Vec{x}_0(t)}{m_p}\\&+T(\Dot{\Bar{\psi}})+ (1 \leftrightarrow 2)
\sim \mathcal{O}({L^{1/2}v^{9/2}}),
\end{split}
\end{align}
where $T(\dot{\bar{\psi}})$ is given in  equation (\ref{5.26}) and we just need to replace $\phi$ by $\psi$. 
\begin{align}
    \begin{split}
        T(\Dot{\Bar{\psi}})&=-\frac{m_1 m_2 s_1 s_2}{16\pi m_p^2}\rmint dt\, e^{-mr}(\Vec{v}_2\cdot \Hat{n})\frac{ \Dot{\Bar{\psi}}(\Vec{x}_0(t))}{m_p},\\ &
        =\frac{m_1 m_2 s_1 s_2}{16\pi m_p^2}\rmint dt \frac{e^{-mr}}{r}(\Vec{v}_2\cdot \Dot{\Vec{r}}+\Vec{a}_2\cdot \Vec{r})\,\frac{\Bar{\psi}(\Vec{x}_0(t))}{m_p}\,.
    \end{split}
\end{align}
\item \textcolor{black}{The amplitude corresponding to the diagram with worldline interaction vertex $Q_a'\rmint dt\,v_a\Bar{\psi}\mathcal{B}_i$ in Fig.~(\ref{fig13a}) which contributes at $N^{(4)}LO :$ }
\begin{align}
    \begin{split}
         \mathcal{S}_{\text{eff}}\Big |_{\text{fig}(\ref{fig13a})}&=-Q_1' Q_2'\rmint\Bar{\mathcal{D}}\Hat{\xi}\rmint dt_1\,v_1^i(t_1)\mathcal{B}_i(\vec{x_1}(t_1))\rmint dt_2 \,v_2^j(t_2)\mathcal{B}_j(\vec{x}_2(t_2))\frac{\Bar{\psi}(\Vec{x}_1(t_1))}{m_p}\,,\\ &
         =-Q_1'Q_2'\rmint dt_1 dt_2 v_1^i(t_1)v_2^{j}(t_2)\Big\langle\mathcal{B}_i(\vec{x_1}(t_1))\mathcal{B}_j(\vec{x}_2(t_2))\Big\rangle\,\frac{\bar\psi(\Vec{x}_1(t_1))}{m_p},\\ &
       =\frac{-Q_1'Q_2'}{4\pi }\rmint dt \,(\Vec{v}_1 \cdot \Vec{v}_2)\frac{e^{-\mu r}}{r}\frac{\Bar{\psi}(\Vec{x}_1(t))}{m_p}+(1\leftrightarrow 2)\,\sim \mathcal{O}(L^{1/2}v^{9/2}).
    \end{split}
\end{align}
\item The amplitude corresponding to the diagram with the worldline vertex $Q_a\rmint dt\,v_a\Bar{\psi}\mathcal{A}_i$ in Fig.~(\ref{fig13b}) which contributes at $N^{(4)}LO :$ 
\begin{align}
    \begin{split}
         \mathcal{S}_{\text{eff}}\Big |_{\text{fig}(\ref{fig13b})}&=-Q_1 Q_2\rmint\Bar{\mathcal{D}}\Hat{\xi}\rmint dt_1\,v_1^i(t_1)\mathcal{A}_i(\vec{x_1}(t_1))\rmint dt_2 \,v_2^j(t_2)\mathcal{A}_j(\vec{x}_2(t_2))\frac{\Bar{\psi}(\Vec{x}_1(t_1))}{m_p}\,,\\ &
         =-Q_1Q_2\rmint dt_1 dt_2 v_1^i(t_1)v_2^{j}(t_2)\Big\langle\mathcal{A}_i(\vec{x_1}(t_1))\mathcal{A}_j(\vec{x}_2(t_2))\Big\rangle\,\frac{\bar\psi(\Vec{x}_1(t_1))}{m_p},\\ &
       =\frac{-Q_1Q_2}{4\pi }\rmint dt \,\frac{\Vec{v}_1 \cdot \Vec{v}_2}{r}\frac{\Bar{\psi}(\Vec{x}_1(t))}{m_p}+(1\leftrightarrow 2)\,\sim \mathcal{O}(L^{1/2}v^{9/2}).
    \end{split}
\end{align}
\item The amplitude corresponding to the diagram in Fig.~(\ref{fig13c}) with worldline interaction vertex $Q_a'\rmint dt\,\mathcal{B}_0\Tilde{\psi}\Bar{\psi}$ which contributes at $N^{(4)}LO :$ 
\begin{align}
    \begin{split}
         \mathcal{S}_{\text{eff}}\Big |_{\text{fig}(\ref{fig13c})}&=-\frac{m_2Q_1'Q_2'}{m_p^2}\rmint\Bar{\mathcal{D}}\Hat{\xi}\rmint dt_1\mathcal{B}_0(\Vec{x}_1(t_1))\Tilde{\psi}(\Vec{x}_1(t_1))\rmint dt_2\Tilde{\psi}(\Vec{x}_2(t_2)\rmint dt_3\mathcal{B}_0(\Vec{x}_2(t_3))\frac{\Bar{\psi}(\Vec{x}_1(t_1))}{m_p}\\&
         =-\frac{m_2Q_1'Q_2'}{m_p^2}'\rmint dt_1dt_2dt_3\Big{\langle}\Tilde{\psi}(\Vec{x}_2(t_2)\Tilde{\psi}(\Vec{x}_1(t_1))\Big{\rangle}\Big{\langle}\mathcal{B}_0(\Vec{x}_2(t_3))\mathcal{B}_0(\Vec{x}_1(t_1))\Big{\rangle}\frac{\Bar{\psi}(\Vec{x}_1(t_1))}{m_p}\,,\\&
       =-\frac{m_2Q_1'Q_2'}{32m_p^2\pi^2}\rmint dt \frac{e^{-\mu r}}{r^2}\frac{\Bar{\psi}(\Vec{x}_1(t))}{m_p}\,+(1\leftrightarrow 2)\,\sim\mathcal{O}(L^{1/2}v^{9/2}).
    \end{split}
\end{align}
\item The amplitude corresponding to the diagram in Fig.~(\ref{fig13d}) with worldline coupling $Q_a\rmint dt\,\mathcal{A}_0\tilde\psi\Bar{\psi}$ which contributes at $N^{(4)}LO :$ 
\begin{align}
    \begin{split}
         \mathcal{S}_{\text{eff}}\Big |_{\text{fig}(\ref{fig13d})}&=-\frac{m_2Q_1Q_2}{m_p^2}\rmint\Bar{\mathcal{D}}\Hat{\xi}\rmint dt_1\mathcal{A}_0(\Vec{x}_1(t_1))\Tilde{\psi}(\Vec{x}_1(t_1))\rmint dt_2\Tilde{\psi}(\Vec{x}_2(t_2)\rmint dt_3\mathcal{A}_0(\Vec{x}_2(t_3))\frac{\Bar{\psi}(\Vec{x}_1(t_1))}{m_p}\\&=-\frac{m_2Q_1Q_2}{m_p^2}\rmint dt_1dt_2dt_3\Big{\langle}\tilde{\psi}(\Vec{x}_2(t_2)\tilde{\psi}(\Vec{x}_1(t_1))\Big{\rangle}\Big{\langle}\mathcal{A}_0(\Vec{x}_2(t_3)\mathcal{A}_0(\Vec{x}_1(t_1))\Big{\rangle}\frac{\Bar{\psi}(\Vec{x}_1(t_1))}{m_p}\,,\\&
       =-\frac{m_2Q_1Q_2}{32m_p^2\pi^2}\rmint dt \frac{1}{r^2}\frac{\Bar{\psi}(\Vec{x}_1(t))}{m_p}+(1\leftrightarrow 2)\,\,\sim\mathcal{O}(L^{1/2}v^{9/2})\,.
    \end{split}
\end{align}
 \item The amplitude corresponding to the diagram in Fig.~(\ref{fig13h}) is given by: 
\begin{align}
    \begin{split}
         \mathcal{S}_{\text{eff}}\Big |_{\text{fig}(\ref{fig13h})}&=\frac{m_1m_2s_1s_2}{m_p^2}\rmint \Bar{\mathcal{D}}\hat{\xi}\rmint dt_1  \Big(\frac{1}{m_p}+\frac{3v_1^2}{2m_p}\Big)\varphi(\Vec{x}_1(t_1)){\Bar{\psi}(\Vec{x}_1(t_1))}\rmint dt_2\varphi(\vec{x}_1(t_2))\,,\\&
       =\frac{m_1m_2s_1s_2}{m_p^2}\rmint dt_1 dt_2 \Big(\frac{1}{m_p}+\frac{3v_1^2}{2m_p}\Big) \Big{\langle}\varphi(\Vec{x_1}(t_1)) \varphi(\Vec{x_2}(t_2))\Big{\rangle}{\Bar{\psi}(\Vec{x}_1(t_1))}\,,\\&
       =\frac{m_1m_2s_1s_2}{m_p^2} \rmint dt_1 dt_2 \Big(\frac{1}{m_p}+\frac{3v_1^2}{2m_p}\Big)\delta(t_1-t_2)\rmint \frac{d^3k}{(2\pi)^3}\frac{e^{i\Vec{k}\cdot\Vec{r}}}{\Vec{k}^2+m^2}{\Bar{\psi}(\Vec{x}_1(t_1))}\,,\\ &
       =\frac{m_1m_2s_1s_2}{m_p^2} \rmint dt \Big(1+\frac{3v_1^2}{2}\Big) \frac{1}{4\pi r}e^{-mr}\frac{\Bar{\psi}(\Vec{x}_1(t))}{m_p}+(1\leftrightarrow 2)\,,\\&\sim \mathcal{O}(L^{1/2}v^{5/2})+\mathcal{O}(L^{1/2}v^{9/2})\,.
    \end{split}
\end{align}
Note that, a part of it contributes at $N^{(2)}LO$ and the other part at $N^{(4)}LO.$ 
 \item Amplitude corresponding to the diagram in Fig.~(\ref{fig13i})  with the radiation coming from a triangle vertex $\phi\phi\Bar{\psi}$  is given by: 
\begin{align}
    \begin{split}
         \mathcal{S}_{\text{eff}}\Big |_{\text{fig}(\ref{fig13i})}&=\frac{m_1m_2^2s_2^2g_1}{m_p^4}\rmint \Bar{\mathcal{D}}\hat{\xi}\rmint dt_1\Big(1+\frac{3}{2}v_1^2\Big)\varphi(\Vec{x}_1(t_1))\varphi(\Vec{x}_1(t_1))\rmint dt_2 \varphi(\Vec{x_2}(t_2)) \,\, \\& \,\,\,\,\,\,\,\,\,\,\,\,\,\,\,\,\,\,\,\,\,\,\,\,\,\,\,\,\,\,\,\,\,\,\,\,\,\,\,\,\,\,\,\,\,\,\,\,\,\,\,\,\,\,\,\,\,\,\,\,\,\,\,\,\,\,\,\,\,\,\,\,\,\,\,\,\,\,\,\,\,\,\,\,\,\,\,\,\,\,\,\,\,\,\,\,\,\,\,\,\,\,\,\,\,\,\,\,\,\,\,\,\,\,\,\,\,\,\,\,\,\,\,\,\,\,\rmint dt_3\varphi(\Vec{x}_2(t_3)) \frac{\Bar{\psi}(\Vec{x}_1(t_1))}{m_p}\,,\\&
         =\frac{m_1m_2^2s_2^2g_1}{m_p^4}\rmint dt_1 dt_2 dt_3\Big(1+\frac{3}{2}v_1^2\Big)\Big{\langle}\varphi(\Vec{x}_1(t_1)) \varphi(\Vec{x_2}(t_2))\Big{\rangle}\Big{\langle}\varphi(\Vec{x}_1(t_1)) \varphi(\Vec{x}_2(t_3))\Big{\rangle}\\& \,\,\,\,\,\,\,\,\,\,\,\,\,\,\,\,\,\,\,\,\,\,\,\,\,\,\,\,\,\,\,\,\,\,\,\,\,\,\,\,\,\,\,\,\,\,\,\,\,\,\,\,\,\,\,\,\,\,\,\,\,\,\,\,\,\,\,\,\,\,\,\,\,\,\,\,\,\,\,\,\,\,\,\,\,\,\,\,\,\,\,\,\,\,\,\,\,\,\,\,\,\,\,\,\,\,\,\,\,\,\,\,\,\,\,\,\,\,\,\,\,\,\,\,\,\, \frac{\Bar{\psi}(\Vec{x}_1(t_1))}{m_p}\,,\\&
       =\frac{m_1m_2^2s_2^2g_1}{16\pi^2m_p^4}\rmint dt \Big(1+\frac{3}{2}v_1^2\Big)\frac{e^{-2mr}}{r^2}\frac{\Bar{\psi}(\Vec{x}_1(t))}{m_p}+(1\leftrightarrow 2)\,,\\&\sim \mathcal{O}(L^{1/2}v^{9/2})+\mathcal{O}(L^{1/2}v^{13/2}).
    \end{split}
\end{align}
Again note that a part of it is contributing at $N^{(4)}LO$ and the other part is contributing at $N^{(6)}LO.$
\item The amplitude corresponds to the diagram in Fig.~(\ref{fig13j}) which contributes at $N^{(4)}LO$ with the bulk vertex $\rmint d^4x\,\psi \partial_l\mathcal{A}_i\partial_l\mathcal{B}_i :$ 
\vspace{-0.3cm}
\begin{align}
    \begin{split}
   \mathcal{S}_{\text{eff}}\Big |_{\text{fig}(\ref{fig13j})}&=-4\gamma Q_1Q_2'\rmint \Bar{\mathcal{D}}\hat{\xi}\rmint dt_1\mathcal{A}_j(\Vec{x}_1(t_1))v_1^j\rmint dt_2\mathcal{B}_m(\Vec{x}_2(t_2))v_2^m(t_2)\,\\&  \,\,\,\,\,\,\,\,\,\,\,\,\,\,\,\,\,\,\,\,\,\,\,\,\,\,\,\,\,\, \,\,\,\,\,\,\,\,\,\,\,\,\,\,\,\,\,\,\,\,\,\,\,\,\,\,\,\,\,\, \,\,\,\,\,\,\,\,\,\,\,\,\,\,\,\,\,\,\,\,\,\,\,\,\,\,\,\,\,\, \,\,\,\,\,\,\,\,\,\,\,\,\,\,\,\,\,\,\,\,\,\,\,\, \rmint d^4x\partial_l\mathcal{A}_i\partial_l\mathcal{B}_i \frac{ \Bar{\psi}(\Vec{x}_0(t))}{m_p}\\&=-4\gamma Q_1Q_2'\rmint \prod_{i=1}^2dt_iv_1^j(t_1)v_2^m(t_2)\rmint d^4x\partial_l\Big{\langle}\mathcal{A}_i(x)\mathcal{A}_j(\Vec{x}_1(t_1))\Big{\rangle}\\&\hspace{6cm}\partial_l\Big{\langle}\mathcal{B}_i(x)\mathcal{B}_m(\Vec{x}_2(t_2))\Big{\rangle}\frac{\Bar{\psi}(\Vec{x}_0(t))}{m_p}\,,\\ &
=-4\gamma Q_1Q_2'\rmint dt\, \,\Bigg[\Vec{v}_1(t)\cdot \Vec{v}_2(t)\rmint_k \frac{e^{-i\vec k\cdot \vec{r}}}{(\vec{k}^2+\mu^2)}\Bigg]                \frac{\Bar{\psi}(\Vec{x}_0(t))}{m_p}\,,\\ &
=-\frac{\gamma Q_1 Q_2'}{\pi}\rmint dt\, \vec{v}_1\cdot \vec{v}_2\, \frac{e^{-\mu_{\gamma}r}}{r}\Big(\frac{\Bar{\psi}(\vec{x}_{0}(t))}{m_p}\Big)+(1\leftrightarrow 2)\,\sim \mathcal{O}(L^{1/2}v^{9/2}).
\end{split}
\end{align}
\item \textcolor{black}{ The amplitude that governs the effective action corresponding to the bulk interaction vertex $\rmint d^4x\,\psi \partial_l\mathcal{A}_i\partial_i\mathcal{B}_l$ at $N^{(4)}LO$,} 
\begin{align}
    \begin{split}
   \mathcal{S}_{\text{eff}}\Big |_{\text{fig}(\ref{fig13j})}&=-4\gamma Q_1Q_2'\rmint \Bar{\mathcal{D}}\hat{\xi}\rmint dt_1\mathcal{A}_j(\Vec{x}_1(t_1))v_1^j\rmint dt_2\mathcal{B}_m(\Vec{x}_2(t_2))v_2^m(t_2)\\& \,\,\,\,\,\,\,\,\,\,\,\,\,\,\,\,\,\,\,\,\,\,\,\,\,\,\,\,\,\, \,\,\,\,\,\,\,\,\,\,\,\,\,\,\,\,\,\,\,\,\,\,\,\,\,\,\,\,\,\, \,\,\,\,\,\,\,\,\,\,\,\,\,\,\,\,\,\,\,\,\,\,\,\,\,\,\,\,\,\,\,\,\,\,\,\,\,\,\,\,\,\,\,\,\,\,\,\,\,\,\,\,\,\,\,\,\,\,\,\,  \rmint d^4x\,\partial_l\mathcal{A}_i\partial_i\mathcal{B}_l \frac{ \Bar{\psi}(\Vec{x}_0(t))}{m_p}\\&
   =-4\gamma\textstyle{Q_1 Q_2'\rmint dt_1dt_2v_1^j(t_1)v_2^m(t_2)\rmint d^4x\partial_l\Big{\langle}\mathcal{A}_i(x)\mathcal{A}_j(\Vec{x}_1(t_1))\Big{\rangle}}\\&\hspace{6cm}\textstyle{\partial_i\Big{\langle}\mathcal{B}_l(x)\mathcal{B}_m(\Vec{x}_2(t_2))\Big{\rangle}\frac{\Bar{\psi}(\Vec{x}_0(t))}{m_p}}\,,\\ &
=-4\gamma Q_1 Q_2'\rmint dt \, \Big[\Vec{v_1}\cdot \Vec{v}_2\frac{e^{-r \mu_{\gamma }} \left(-r \mu _{\gamma }+e^{r \mu _{\gamma }}-1\right)}{2 \pi  r^3 \mu_{\gamma }^2}\\ & \hspace{1cm}
+(\Vec{v}_1\cdot \Hat{n})(\Vec{v}_2\cdot \Hat{n})\frac{e^{-r \mu _{\gamma }} \Big(r \mu _{\gamma } \left(r \mu _{\gamma }+3\right)-3 e^{r \mu _{\gamma }}+3\Big)}{2 \pi  r^3 \mu _{\gamma }^2}\Big]\Big(\frac{\Bar{\psi}(\Vec{x}_0(t))}{m_p}\Big)+(1\leftrightarrow 2)\,\\&\sim\mathcal{O}(L^{1/2}v^{9/2}).
\end{split} 
\end{align}
\item The amplitude corresponds to the diagram in Fig.~(\ref{fig13k}) that contributes at $N^{(4)}LO$ with the bulk interaction vertex $\rmint d^4x\,\partial_k\mathcal{A}_0\partial_0 \mathcal{B}_k :$
 \begin{align}
    \begin{split}
   \mathcal{S}_{\text{eff}}\Big|_{\text{fig}(\ref{fig13k})}&=-\gamma Q_1Q_2'\rmint \Bar{\mathcal{D}}\hat{\xi}\rmint dt_1\mathcal{A}_0(\Vec{x}_1(t_1))\rmint dt_2\mathcal{B}_j(\Vec{x}_2(t_2))v_2^j(t_2)\rmint d^4x \partial_k\mathcal{A}_0\partial_0\mathcal{B}_k(x) \frac{ \Bar{\psi}}{m_p}\,,\\&
   =-\gamma Q_1Q_2'\rmint dt_1dt_2v_2^j(t_2)\rmint d^4x\partial_k\Big{\langle}\mathcal{A}_0(x)\mathcal{A}_0(\Vec{x}_1(t_1))\Big{\rangle}\partial_0\Big{\langle}\mathcal{B}_k(x) \mathcal{B}_j(\Vec{x}_2(t_2))\Big{\rangle}\frac{\Bar{\psi}}{m_p}\,,\\ &
=\frac{\gamma Q_1Q_2'}{2\pi}\rmint dt \Bigg[(\Vec{a}_2\cdot \Hat{n})\Big\{\frac{  e^{-\mu_{\gamma} r}}{2 \mu_{\gamma} \,r}-\frac{1 -  e^{-\mu_{\gamma} r}}{2\, \mu_{\gamma}^2\, r^2}\Big\}+v_2^2 \,\frac{e^{-r \mu _{\gamma }} \left(-r \mu _{\gamma }+e^{r \mu _{\gamma }}-1\right)}{  r^3 \mu _{\gamma }^2}\\ &
\,\,\,\,\,\,\,\,\,\,\,\,\,\,\,\,\,\,\,\,\,\,\,\,\,\,\,\,\,
(\Vec{v}_2\cdot \Hat{n})^2\,\frac{e^{-r \mu _{\gamma }} \Big(r \mu _{\gamma } \left(r \mu _{\gamma }+3\right)-3 e^{r \mu _{\gamma }}+3\Big)}{  r^3 \mu _{\gamma }^2}
\Bigg]\Big(\frac{\Bar{\psi}(\Vec{x}_1(t))}{m_p}+\frac{\Bar{\psi}(\Vec{x}_2(t))}{m_p}\Big)\\& +(1\leftrightarrow 2)\,,\sim\mathcal{O}(L^{1/2}v^{9/2}).
\end{split}
\end{align}
\vspace{-0.7cm}
\item  The amplitude corresponds to the diagram in Fig.~(\ref{fig13m}) that contributes at $N^{(4)}LO$ with bulk interaction vertex $\rmint d^4x\,\partial_0\mathcal{A}_i\partial_i\mathcal{B}_0$ :\footnote{It is important to remember that we have diagrams with asymmetric vertices, i.e., vertices connected with two different fields. Hence, along with shuffling the worldlines, we also have to consider the radiation from the other field vertex (e.g., if we only shuffle the worldlines indices, it will generate a diagram producing radiation from the same field but from different worldlines). But for asymmetric vertices, it demands considering the same diagram with radiation from the other field coupled to one of the worldlines before shuffling the indices).}
\begin{align}
    \begin{split}
   \mathcal{S}_{\text{eff}}\Big |_{{\text{fig}(\ref{fig13m})}}&=-\gamma Q_1Q_2'\rmint \Bar{\mathcal{D}}\hat{\xi}\rmint dt_1v_1^j(t_1)\mathcal{A}_j(\Vec{x}_1(t_1))\rmint dt_2\mathcal{B}_0(\Vec{x}_2(t_2))\rmint d^4x\,\partial_0\mathcal{A}_i(x)\partial_i\mathcal{B}_0(x)\frac{\Bar{\psi}}{m_p}\,,\\ &=-\gamma Q_1Q_2'\rmint dt_1dt_2v_1^j(t_1)\rmint d^4x\,\partial_0\Big{\langle}\mathcal{A}_i(x)\mathcal{A}_j(\Vec{x}_1(t_1))\Big{\rangle}\partial_i\Big{\langle}\mathcal{B}_0(x)\mathcal{B}_0(\Vec{x}_2(t_2))\Big{\rangle}\frac{\Bar{\psi}}{m_p}\,,\\ &
=\frac{\gamma Q_1Q_2'}{2\pi}\rmint dt \Bigg[(\Vec{a}_1\cdot \Hat{n})\Big\{\frac{  e^{-\mu_{\gamma} r}}{2 \mu_{\gamma} \,r}-\frac{1 -  e^{-\mu_{\gamma} r}}{2\, \mu_{\gamma}^2\, r^2}\Big\}+v_1^2 \,\frac{e^{-r \mu _{\gamma }} \left(-r \mu _{\gamma }+e^{r \mu _{\gamma }}-1\right)}{  r^3 \mu _{\gamma }^2}\\ &
\,\,\,\,\,\,\,\,\,\,\,\,\,\,\,\,\,\,\,\,\,\,\,\,\,\,\,\,\,
(\Vec{v}_1\cdot \Hat{n})^2\,\frac{e^{-r \mu _{\gamma }} \Big(r \mu _{\gamma } \left(r \mu _{\gamma }+3\right)-3 e^{r \mu _{\gamma }}+3\Big)}{  r^3 \mu _{\gamma }^2}
\Bigg]\Big(\frac{\Bar{\psi}(\Vec{x}_1(t))}{m_p}+\frac{\Bar{\psi}(\Vec{x}_2(t))}{m_p}\Big)+\\&(1\leftrightarrow 2)\,,\sim\mathcal{O}(L^{1/2}v^{9/2})\,.
\end{split}
\end{align}
\newpage
\item The amplitude corresponds to the diagram in Fig.~(\ref{fig13o}) that contributes at $N^{(7)}LO$ with the  3-point bulk interaction vertex $\rmint d^4x\,\varphi\,\partial_{0}\mathcal{A}_{i}\partial_{k}\mathcal{A}_{m}(x) :$ 
\begin{align}
    \begin{split}
          \mathcal{S}_{\text{eff}}^{ikm}\Big |_{{\text{fig}(\ref{fig13o})}}&=-\frac{Q_2^2m_1s_1}{m_p^2}\rmint\Bar{\mathcal{D}}\Hat{\xi}\rmint dt_2\, v_2^{j}(t_2)\mathcal{A}_{j}(\Vec{x}_2(t_2))\rmint dt_3 \,v_2^{l}(t_3)\mathcal{A}_{l}(\Vec{x}_2(t_3))\\& \rmint dt_1\,\varphi(\Vec{x}_1(t_1))\frac{\Bar{\psi}}{m_p}(\Vec{x}_1(t_1))\rmint d^4x \,\varphi\,\partial_0 \mathcal{A}_{i}\partial_{k} \mathcal{A}_{m}(x)\,,\\&    
        =-\frac{Q_2^2m_1s_1}{m_p^2}\rmint dt_1dt_2dt_3dt\,v_2^{j}(t_2)v_2^{l}(t_3)\frac{\Bar{\psi}(\Vec{x}_1(t_1))}{m_p}\rmint d^3 x\, \partial_{k}\Big\langle\mathcal{A}_{j}(\Vec{x}_2(t_2))\mathcal{A}_{m}(x)\Big\rangle \\&\hspace{5cm} \partial_{0}\Big\langle \mathcal{A}_{l}(\Vec{x}_2(t_3))\mathcal{A}_{i}(x)\Big\rangle \Big\langle \varphi(\Vec{x}_1(t_1))\varphi(\Vec{x}_2(t_2))\Big\rangle\,,\\ &
    =-\frac{Q_2^2m_1s_1}{m_p^2}\rmint dt\, v_2^{m}(t)a_2^{i}(t)\Bigg[\frac{1}{16} \pi^{3/2}\, m \,G_{1,3}^{2,1}\left(\frac{m^2 r^2}{4}\Big|
\begin{array}{c}
 -\frac{1}{2} \\
 -\frac{1}{2},-\frac{1}{2},0 \\
\end{array}
\right)\\&\hspace{5cm}-\frac{\,\pi^{3/2}\, G_{1,3}^{2,1}\left(\frac{m^2 r^2}{4}\Big|
\begin{array}{c}
 \frac{1}{2} \\
 \frac{1}{2},\frac{1}{2},0 \\
\end{array}
\right)}{8 \,m\, r^2}\Bigg]n^k\,
\frac{\Bar{\psi}(\Vec{x}_1(t))}{m_p} \\& +\text{term symmetric in $i,m$}+1\leftrightarrow 2\,\sim \mathcal{O}(L^{1/2} v^{15/2}).
    \end{split}
\end{align}
 \item \textcolor{black}{A similar type of interaction arises from two different electromagnetic worldline couplings: $\rmint dt Q_1v_1^{i}\mathcal{A}_{i}$ and $-\rmint dt Q_2v_2^{i}\mathcal{A}_{i}\Bar{\psi}$. This is shown in Fig.~(\ref{fig13p}) and it contributes at $N^{(7)}LO.$
Corresponding amplitude is given by (to calculate this amplitude, we need to invoke the one-loop massive scalar integral \cite{Anastasiou:1999ui,2022arXiv220103593W})}: 
\begin{align}
\begin{split}
    \mathcal{S}_{\text{eff}}^{ikm}\Big |_{{\text{fig}(\ref{fig13p})}}&=-\frac{1}{m_p^2}\rmint \Bar{\mathcal{D}}\hat{\xi}\rmint dt_1 Q_1v_1^{j}(t_1)\mathcal{A}_{j}(\Vec{x}_1(t_1))\rmint dt_2 m_1s_1 \varphi(\Vec{x}_1(t_2))\rmint dt_3 Q_2 v_2^{l}(t_3)\mathcal{A}_{l}(\Vec{x}_2(t_3))\\&\,\,\,\,\,\,\,\,\,\,\,\,\,\,\,\,\,\,\,\,\,\,\,\,\,\,\frac{\Bar{\psi}(\Vec{x}_2(t_2))}{m_p}\rmint d^4x \varphi\partial_0 \mathcal{A}_{i}\partial_{k} \mathcal{A}_{m}(x)\,,\\ &
        =-\frac{Q_1Q_2m_1s_1}{m_p^2}\rmint dt_1dt_2dt_3dt\,v_1^{j}(t_1)v_2^{l}(t_3)\frac{\Bar{\psi}(\Vec{x}_2(t_2))}{m_p}\rmint d^3 x \,\partial_{t}\Big\langle\mathcal{A}_{j}(\Vec{x}_1(t_1))\mathcal{A}_{i}(x)\Big\rangle\\ &
\hspace{6cm}\partial_{k}\Big\langle\mathcal{A}_{l}(\Vec{x}_2(t_3))  \mathcal{A}_{m}(x)  \Big\rangle\Big\langle \varphi(x_1(t_2))\varphi(x)\Big\rangle\,,\\ &
    =-\frac{Q_1Q_2m_1s_1}{m_p^2}\rmint dt\, dt_1 v_1^{i}(t_1)v_2^{m}(t)\frac{\Bar{\psi}(\Vec{x}_2(t_2))}{m_p}\partial_{t}\delta(t-t_1)\,\\&\hspace{2cm} \rmint_{\Vec{x}} \rmint_{\Vec{k_1},\Vec{k_2},\Vec{k_3}}\frac{-ik_2^{k}}{\Vec{k_1}^2\Vec{k_2}^2(\Vec{k_3}^2+m^2)}e^{i\Vec{k_1}\cdot[\Vec{x_1}(t_1)-\Vec{x}(t)]}e^{i\Vec{k_2}\cdot[\Vec{x_2}(t)-\Vec{x}(t)]} 
       e^{i\Vec{k_3}\cdot[\Vec{x_1}(t)-\Vec{x}(t)]}\,,\\ &
  =\frac{Q_1 Q_2 m_1 s_1}{4\pi\,m_p^2}\rmint dt \, v_2^m (t)a_1^i(t)\,\partial_r\alpha(m;r)n^k \frac{\bar{\psi}}{m_p}(\vec{x_2}(t))\,\sim \mathcal{O}(L^{1/2}v^{15/2})+(1\leftrightarrow 2)\,.\label{5.85n}
  \end{split}
\end{align}
The details of computations are given in Appendix~(\ref{app3}). Also the function  $\alpha(m;r)$ is defined (\ref{C3m}). One can obtain the contribution to the effective action by contracting with $\epsilon^{0ikm}$.
\end{itemize}
\color{black}
Then, from this, as discussed in the previous subsections, we can read out the source term, i.e., $T^{00}$, and from that, we can compute the total power radiation. Before writing that, we make the following comment about the contribution coming from axion-photon coupling ($g_
{a\gamma\gamma}$) to the gravitational radiation.  
\vspace{0.2cm}

\underline{{\textit{ The $g_{a\gamma\gamma}$ contribution}}}:\par
The source term coming from the diagrams (\ref{fig13o}) and (\ref{fig13p}), captures the effect of the \textit{theta} term.  
\begin{align}
\begin{split}
T^{00}&\sim
\frac{2Q_1Q_2m_1s_1g_{a\gamma \gamma}}{4\pi\,m_p^2}\, [(\hat n\times \vec{v}_2)\cdot \Vec{a}_1]
\partial_r\alpha(m;r) \delta(\vec{x}-\vec{x}_2)\\&
   + \frac{\pi^{3/2}}{4\,m_p^2}g_{a\gamma\gamma}Q_2^2\,m_1\,s_1\,[\Vec{a_2}\cdot(\hat{n}\times\Vec{v_2})]\Bigg[\frac{1}{2}\, m \,G_{1,3}^{2,1}\left(\frac{m^2 r^2}{4}\Big|
\begin{array}{c}
 -\frac{1}{2} \\
 -\frac{1}{2},-\frac{1}{2},0 \\
\end{array}
\right)-\frac{ G_{1,3}^{2,1}\left(\frac{m^2 r^2}{4}\Big|
\begin{array}{c}
 \frac{1}{2} \\
 \frac{1}{2},\frac{1}{2},0 \\
\end{array}
\right)}{ \,m\, r^2}\Bigg]\\&\delta(\vec{x}-\vec{x}_1)+(1\leftrightarrow 2)\label{5.85}
\end{split}
\end{align}
But the source term in (\ref{5.85}) is zero for any orbit that lies in two dimensions. Hence the term does not contribute to the gravitational power radiation. This term will only contribute to the power radiation expression for orbits which lie in three dimensions instead.\par

\begin{table}[t!]
 \centering
 \scalebox{0.91}{\begin{tabular}{|c|c|c|c|c|c|c|c|c|}
 \hline
  Fields & $LO$ & $N^{(1)}LO$ &  $N^{(2)}LO$ &  $N^{(3)}LO$ &   $N^{(4)}LO$&   $N^{(5)}LO$&   $N^{(6)}LO$ &   $N^{(7)}LO$\\
  \hline
 \textit {scalar} & 1 & 1 & 2 & 1 & 3 & 1 $\sim g_{a\gamma\gamma}$ & - & -\\
  \hline
 \textit{electromagnetic} & 1 & 1 & 1 & 1 & - &-& -& -\\
 \hline 
 \textit{proca} & 1 & 1 & 1 & 1 & - &-& -& -\\
 \hline 
 \textit{gravitational} & 1 & - & 1($\sim \gamma$)+5=6 & - & 4($\sim \gamma$)+7=11 & - & 1 & 2 $\sim g_{a\gamma\gamma}$\\
 \hline
 \end{tabular}}
 \caption{Table showing the number of diagrams contributing to a specific $N^{(n)}LO$ for different radiative fields. We have excluded the pure gravitational sector as that is already studied extensively in the literature (e.g. readers are referred to \cite{Blanchet:2013haa}). We have only shown the contribution due to the interaction of the other fields with the gravitational field.}
 \label{table2m}
 \end{table}
 
\textit{Before we end this section, we summarize our results in Table~(\ref{table2m}). We show the number of diagrams that contribute to gravitational radiation and arise due to the (minimal) coupling of different fields with the gravitational field. We also highlight the order in which the two couplings constant $g_{a\gamma\gamma}$ and $\gamma$ contribute to the gravitational radiation. Again we emphasize that the pure gravitational sector is already well-studied \cite{Blanchet:2013haa} as mentioned before. Here we mainly focus on the leading order contributions to the gravitational radiation due to the coupling of scalar, em and Proca fields with the gravitational field. To the best of our knowledge, these are some of the new results.} 
Finally, we write down the total gravitational power for \textit{circular orbit} radiation considering the term which arises at $LO$  for pure gravity and terms which arise at their respective leading orders due to the interaction between other fields (scalar, em, and Proca) with the gravitational field.
\begin{equation}
\boxed{
\begin{aligned}
     P_g=&\frac{32G}{5}\Omega^6\Bigg[\mu r^2+\frac{3}{2}\Bigg\{\frac{m_2^3+m_1^3}{M^2}\Bigg\}\nu r^4\Omega^2+\\&\mathcal{D}\Bigg\{\frac{m_1^2+m_2^2}{M^2}\Bigg\}\nu r^4\Omega^2+\Bigg\{\frac{\mathcal{E}_2m_1^2+\mathcal{E}_1m_2^2}{M^2}\Bigg\}r^2+\mathcal{D}(r)r^2\frac{(m_2^2+m_1^2)}{M^2}+\\&\nu r^2\Bigg\{G_1(r)m_2^3(1+\frac{3m_2^2r^2\Omega^2}{2M^2})+G_2(r)m_1^3(1+\frac{3m_1^2r^2\Omega^2}{2M^2})\Bigg\}+\\&\nu r^4\Omega^2\Bigg\{A_1(r)(\frac{m_2+m_1}{2M})^2+A_2(r)(\frac{m_2+m_1}{2M})^2\Bigg\}+\Bigg\{(\frac{H_1(r)m_2^2}{ M^2}+\frac{H_2(r)m_1^2}{ M^2})\frac{(m_1^2+m_2^2)}{M^2}(r^4\Omega^2)\Bigg\}+\\&\Bigg\{\Big(\frac{H_1(r)m_1^2}{ M^2}+\frac{H_2(r)m_2^2}{ M^2}\Big)\frac{(m_1^2+m_2^2)}{M^2}(r^4\Omega^2)\Bigg\}+\Bigg\{\frac{\mathcal{P}_2m_1^2+\mathcal{P}_1m_2^2}{M^2}r^2\Bigg\}+\\&
       \Bigg\{\frac{\mathcal{P}_1m_1^2+\mathcal{P}_2m_2^2}{M^2}r^2\Bigg\}+\Bigg\{\frac{V_2m_1^3}{M^3}-\frac{V_1m_2^3}{M^3}\Bigg\}r^3\Omega^2-
       \Bigg\{\frac{V_1m_1-V_2m_2}{M}\Bigg\}(\nu r^3 \Omega^2)\\&+\Bigg\{\frac{{V}_1m_1^3}{M^3}-\frac{{V}_2m_2^3}{M^3}\Bigg\}r^3\Omega^2+
       \Bigg\{\frac{{V}_1m_2-{V}_2m_1}{M}\Bigg\}(\nu r^3 \Omega^2)+R(r)\Bigg\{\frac{m_1-m_2}{M}\Bigg\}r^4\Omega^2\Bigg]^2
\end{aligned}
}\label{5.81n}
\end{equation}

where, 
\begin{align}
\begin{split}
  & \mathcal{D}(r)= \Bigg\{\frac{Q_1Q_2}{4\pi } \frac{1}{r}+\frac{Q_1'Q_2'}{4\pi} \frac{e^{-\mu_{\gamma} r}}{r}\Bigg\}\,,\quad
   G_1(r)=\frac{s_2^2g_1}{16\pi^2m_p^4}\frac{e^{-2mr}}{r^2}\,,\\& 
\mathcal{E}_1(r)=\Bigg[\frac{m_1m_2s_1s_2}{m_p}  \Big(\frac{1}{m_p}+\frac{3\frac{m_2^2}{M^2}r^2\Omega^2}{2m_p}\Big) \frac{e^{-mr}}{4\pi r}
+\frac{m_1m_2}{16\pi m_p^2 r}-\frac{m_2 Q_1Q_2}{32\pi^2m_p^2} \frac{1}{r^2}-\frac{m_2 Q_1'Q_2'}{32\pi^2m_p^2} \frac{e^{-\mu_{\gamma} r}}{r^2}\Bigg]\,,\\&
A_1(r)=\Bigg\{\frac{-\gamma Q_1Q_2'}{2\pi}\Big[  \left(\frac{  e^{-\mu_{\gamma} r}}{ r}+\frac{4e^{-r\mu_{\gamma}}\left(-1-r\mu_{\gamma}+e^{\mu_{\gamma} r}\right)}{\mu_{\gamma}^2 r^3}\right) \Big]+\frac{m_1 m_2 s_1 s_2}{16 \pi m_p^2}\,\frac{e^{-m\,r}}{r}\Bigg\}\,,\\&
\mathcal{P}_1(r)=-\frac{\gamma Q_1Q_2'}{4\pi r}e^{-\mu_{\gamma} r}\,,\quad 
H_1(r)=\frac{\gamma Q_1Q_2'}{2\pi}\Bigg[\,\frac{e^{-r \mu _{\gamma }} \left(-r \mu _{\gamma }+e^{r \mu _{\gamma }}-1\right)}{ r^3 \mu _{\gamma }^2}
\Bigg]\,,\\&
V_1(r)=\gamma Q_1Q_2'\frac{1}{2\pi} \Bigg[\Big\{\frac{  e^{-\mu_{\gamma} r}}{2 \mu_{\gamma} \,r}-\frac{1 -  e^{-\mu_{\gamma} r}}{2\, \mu_{\gamma}^2\, r^2}\Big\}\Bigg]\,,\quad 
R(r)=\frac{m_1m_2s_1s_2e^{-mr}}{64\pi m_p^2r}\,.\label{5.81m}
\end{split}
\end{align}
We have used an index subscript for the functions which are not symmetric in two indices (denoting the two worldlines) i.e. one and two, and no indices for the symmetric ones. Also, $\mathcal{E}_2,G_2,A_2,\mathcal{P}_2,H_2,V_2$ can be defined by just replacing the index one by two. There are a few comments that are in order. 

\begin{itemize}
    \item As expected, the total power radiation is a function of the binary separation.
\item In (\ref{5.81m})  Also, we have a term like $\frac{m_1m_2}{16\pi m_p^2 r}$, which doesn't go away even when we set the em charges, dark charges and screening terms to zero. So it is a pure gravitational radiation correction at $N^{(2)}LO$. We can easily check that it matches the $X^2$ term in  \cite{Huang:2018pbu}. This also serves as a consistency check of our computation. 
\end{itemize}

\section{Discussions and outlooks}\label{disc}
Motivated by recent prospects of searching dark matter using GW observations \cite{Zhang:2021mks, Coogan:2021uqv, Singh:2022wvw, Becker:2021ivq, Yue:2019ozq, Cardoso:2020iji,Bhattacharya:2023stq}, in this paper, we consider a black hole binary in a dark matter environment modelled by certain \textit{ultra-light particle} (in the form of a massive scalar field with an anomalous coupling with EM field via a Chern-Simons type term) and the \textit{ultra-light vector field} (in the form of a massive vector field (Proca field) which also couples with EM field by kinetic mixing term). Eventually, we like to see whether, apart from the masses of the scalar and Proca field,  we can put constraints on the axion-photon coupling parameters $g_{a\gamma\gamma}$ as well as the Proca-electromagnetic coupling constant $\gamma$ (coming from the kinetic mixing term) using GW observations and compare them with those coming from other phenomenological considerations as mentioned earlier. To do that, we took a primary step forward in this paper. We initiated a computation of the power radiation from a binary system in this dark matter environment, which is the key to computing the phase of the gravitational waveform. We resort to the WEFT formalism to carry out the computation. The study done in this paper is mainly focused on the regime where gravity is perturbative. This is approximately true for the adiabatic binary inspiral phase, which is important for gravitational wave detection. There is a set of length scales that are relevant for such arrangements. All of them play a crucial role in the dynamics, ranging from the gravitational radius to the length scale of the objects, to their typical orbital separation, and finally, the wavelength of the radiation emitted by the inspiralling binaries. Remembering these length scales and integrating over different potential fields, we achieve the orbital and dissipative dynamics, truncating at a specific PN order. Then we compute the power radiation from the binary moving in a \textit{circular orbit}, which is the first step in the computation of the phase of the GW waveform. Below, we summarize our results. 
\begin{itemize}
\item We calculated electromagnetic and proca bound sector upto $1PN \sim \mathcal{O}(Lv^2)$. 

\item We also investigated the \textit{em-axion} bound sector, which contributes to the bound sector at  $2.5PN\sim \mathcal{O}(Lv^5)$ and the pure scalar bound sector upto $2PN \sim \mathcal{O}(Lv^4)$ order. \textit{We have provided the details of the 2PN scalar bound sector and upto $N^{(4)}LO$ radiative sector, which is, to the best of our knowledge, a new result.}

\item \textit{Electromagnetic and Proca radiation has been investigated up to $N^{(3)}LO$. We restricted ourselves to the dipole radiation only. Again, this is a new result to the best of our knowledge.} 

\item We also calculated the gravitational radiation due to different field couplings with the gravitational field (results for the pure gravitational sector are well known, and we do not reproduce them here beyond $LO$). \textit{We show that the effect of kinetic term mixing (between photon and dark photon) coupling,  $\gamma$ appears at $N^{(2)}LO (L^{1/2}v^{5/2})$ and $N^{(4)}LO (L^{1/2}v^{9/2}) $ order and
axion-photon, $g_{a\gamma\gamma}$ appears at $\sim N^{(7)}LO(L^{1/2}v^{15/2})$}.  
        
\item \textit{We obtained a novel result that the parity-violating axionic term ($g_{a\gamma\gamma}$)  would lead to detected power radiation only for orbits in three dimensions for non-spinning binaries. However, it vanishes for the orbits confined to a two-dimensional plane.} However, it's easy to see that if two binaries have a spin and are not parallel, then that may lead to orbital oscillation leading to the novel possibility of detecting axions in radiation at infinity.

\end{itemize}

We now end here by stating some future directions. First and foremost, we like to compute the phase of the GW waveform using our results. This will enable one to estimate various parameters of the theory along the line of \cite{Zhang:2021mks} for possible constraints on the couplings $g_{a\gamma\gamma}$ and $\gamma.$ This will help us to get some constraints on this dark matter model and the parameter space of ultra-light scalar and vector fields. We hope to report on this soon. Furthermore, it will be interesting to generalize our computations for spinning binaries. This will help us to build a complete waveform model. \par 

Another extension of the computation presented in this paper will be to generalize the results for non-Abelian gauge fields and supersymmetric dark matter models. By computing the conservative and radiative dynamics, one will be able to comment on the dependence of our main results on the \textit{color} indices of the gauge fields. One can also introduce finite size effects (tidal deformation etc.)  and investigate the effects of those in our setup. \par 
Last but the least, it will be interesting to investigate the binary dynamics in Post-Minkowskian (PM) regime. One can use Schwinger-Keyldysh `in-in' formalism to study the conservative as well as the dissipative (radiation-reaction) dynamics in a unified framework \cite{Dlapa:2021npj,Kalin:2022hph,Dlapa:2022lmu,Dlapa:2021vgp,Dlapa:2023hsl}. This will enable us to do a full relativistic computation. In recent times, a method known as \textit{Worldline Quantum Field Theory} (WQFT) has been developed to perform such full relativistic computation \cite{Mogull:2020sak,Jakobsen:2022fcj,Jakobsen:2021zvh,Jakobsen:2021lvp,Jakobsen:2021smu,Jakobsen:2022psy} where not only the fields are quantized but also the worldline deflections are quantized. 
Using the WQFT method, it would be interesting to compute the full relativistic waveform, i.e., $\langle h_{\mu\nu}\rangle, \langle A_{\mu}\rangle$, etc., and then taking the non-relativistic result, one can compare it to our result. It would be a consistency check of the two different formalisms. Another interesting formalism based on scattering amplitude, namely the KMOC formalism \cite{Kosower:2018adc}, has been proposed to calculate the gravitational radiation from a binary system. Again in \cite{Mogull:2020sak}, it is shown that the gravitational waveform calculated from the KMOC formalism is related to the one-point function in WQFT. Hence, it is interesting to find the correspondence between three completely different `new' formalisms using different examples. We hope to report on some of these issues in the near future. 


\section*{Acknowledgements}
  The authors would like to thank Alec Aivazis for developing a graphical interface for Feynman diagram drawing \cite{fdraw}. The authors also thank Koustubh Guha for drawing Fig.~(\ref{newfig}). Research of S.G is supported by the Prime Minister's Research Fellowship (ID:1702711) by the Ministry of Education, Government of India. A.B is supported by the Mathematical Research Impact Centric Support Grant (MTR/2021/000490) by the Department of Science and Technology, Science and Engineering Research Board (India) and the Relevant Research Project grant (202011BRE03RP06633-BRNS) by the Board Of Research In Nuclear Sciences (BRNS), Department of Atomic Energy (DAE), India. We also thank the speakers of the (virtual) workshop ``Testing Aspects of  General Relativity-II" for useful discussions. This work was presented by S.G at the conference ``New insights into particle physics from quantum information and gravitational waves" (12-13th June, 2023) at Lethbridge University, Canada, funded by McDonald Research Partnership-Building Workshop grant by McDonald Institute. Authors also thank Robin Fynn Diedrichs and Daniel Schmitt for their useful correspondence regarding the radiation vertices.
\appendix 
\section{Details of computation used in Section~(\ref{Sec4.2})}\label{app2}
In order to evaluate the amplitude in (\ref{Sec4.2}) we need to do two  Feynman loop integrals (\ref{2.15}). Although the integrals are not in the form of Feynman master integrals, one can reduce the integrals by Fourier transforming them into the known master integral form. We have the following two integrals,
\begin{align}
    \begin{split}
       & \mathcal{I}(r)=\rmint_{\Vec{k_1}\Vec{{k_2}}} \frac{k_2^k}{\Vec{k_1}^2\Vec{k_2}^2[(\Vec{k_1}+\Vec{k_2})^2+\textcolor{black}{m^2}]}e^{i(\Vec{k_1}+\Vec{k_2}).\Vec{r}(t)},\,\Tilde{\mathcal{I}}(r)=\rmint_{\Vec{k_1}\Vec{k_2}}\frac{\,k_2^k\, k_1^a\,e^{i(\Vec{k_1}+\Vec{k_2}).\Vec{r}(t)}}{\Vec{k_1}^2\Vec{k_2}^2[(\Vec{k_1}+\Vec{k_2})^2+\textcolor{black}{m^2}]}\,.
    \end{split}
\end{align}
We start with the first one,
\begin{align}
    \begin{split}
      &  \mathcal{I}(r)=\rmint_{\Vec{k_1}\Vec{{k_2}}} \frac{k_2^k}{\Vec{k_1}^2\Vec{k_2}^2[(\Vec{k_1}+\Vec{k_2})^2+\textcolor{black}{m^2}]}e^{i(\Vec{k_1}+\Vec{k_2}).\Vec{r}(t)}\,.
    \end{split}
\end{align}
After doing the Fourier transform we get,
\begin{align}
    \begin{split}
        \hat{I}(\Vec{k})&=\rmint d^3r e^{-i\Vec{k}.\Vec{r}}\,\mathcal{I}(r)\,,\\ &
        =\rmint_{\Vec{k_1}}\frac{k^k-k_1^k}{\Vec{k_1}^2\,(\Vec{k}^2+\textcolor{black}{m^2})\,(\Vec{k_1}-\Vec{k})^2}\,.
    \end{split}
\end{align}
Now do the inverse Fourier transform,
\begin{align}
    \begin{split}
        \mathcal{I}(r)&=\iint_{\Vec{k},\Vec{k_1}}e^{i\Vec{k}.\Vec{r}}\,\Big(\frac{k^k-k_1^k}{\Vec{k_1}^2\,(\Vec{k}^2+\textcolor{black}{m^2})\,(\Vec{k_1}-\Vec{k})^2}\Big)\,,\\ &
    =\iint_{\Vec{k},\Vec{k_1}} e^{i\Vec{k}.\Vec{r}}\frac{k^k}{\Vec{k_1}^2\,(\Vec{k}^2+\textcolor{black}{m^2})\,(\Vec{k_1}-\Vec{k})^2}-\iint_{\Vec{k},\Vec{k_1}} e^{i\Vec{k}.\Vec{r}}\frac{k_1^k}{\Vec{k_1}^2\,(\Vec{k}^2+\textcolor{black}{m^2})\,(\Vec{k_1}-\Vec{k})^2}\,,\\ &
=\frac{1}{8}\rmint_{\Vec{k_1}}\frac{k_1^k \,e^{i\Vec{k_1}.\Vec{r}}}{|\Vec{k_1}|(\Vec{k_1}^2+\textcolor{black}{m^2})}-\frac{1}{16}\rmint_{\Vec{k_1}}\frac{k_1^k \,e^{i\Vec{k_1}.\Vec{r}}}{|\Vec{k_1}|(\Vec{k_1}^2+\textcolor{black}{m^2})}\,,\\ &
=\textstyle{\Bigg[\frac{i}{16} \pi^{3/2}\, m \,G_{1,3}^{2,1}\left(\frac{m^2 r^2}{4}\Big|
\begin{array}{c}
 -\frac{1}{2} \\
 -\frac{1}{2},-\frac{1}{2},0 \\
\end{array}
\right)-\frac{i\,\pi^{3/2}\, G_{1,3}^{2,1}\left(\frac{m^2 r^2}{4}\Big|
\begin{array}{c}
 \frac{1}{2} \\
 \frac{1}{2},\frac{1}{2},0 \\
\end{array}
\right)}{8 \,m\, r^2}\Bigg]n^k}\,.
    \end{split}
\end{align}
If we expand the integral in a small mass limit, we will get,
\begin{align}
    \begin{split}
        -i\mathcal{I}(r)\approx\frac{1}{r}+\frac{1}{3} m^2 r [\log \left(mr\right)+ \gamma_E -\frac{4}{3}]\,.\label{50}
    \end{split}
\end{align}
We can see there is a smooth massless limit of the expression: $-i\mathcal{I}(r)\approx\frac{1}{r}$ as expected. Another interesting point to note is that the integral in (\ref{50}) in the small mass limit has the same form as a scalar one-loop integral without the divergent term. That implies that although we are doing some complicated loop integral, the results are convergent. That ensures they are not quantum loops; rather, we can call them ``Classical Loops''.
Now do to the second integral,
\begin{equation}
    \Tilde{\mathcal{I}}(r)=\rmint_{\Vec{k_1}\Vec{k_2}}\frac{\,k_2^k\, k_1^a\,e^{i(\Vec{k_1}+\Vec{k_2}).\Vec{r}(t)}}{\Vec{k_1}^2\Vec{k_2}^2[(\Vec{k_1}+\Vec{k_2})^2+\textcolor{black}{m^2}]}\,.
\end{equation}
Proceeding as before we first Fourier transform it.
\begin{align}
    \begin{split}
        \hat{\Tilde{\mathcal{I}}}(\Vec{k})&=\rmint d^3 r\, e^{-i\Vec{k}.\Vec{r}}\, \Tilde{\mathcal{I}}(r)\,,\\ &
        =\rmint_{k_1}\frac{k_1^a(k^k-k_1^k)}{\Vec{k_1}^2\,(\Vec{k}^2+\textcolor{black}{m^2})\,(\Vec{k_1}-\Vec{k})^2}\,.\\ &
    \end{split}
\end{align}
Now do the inverse Fourier transform and get,
\begin{align}
    \begin{split}
    \Tilde{\mathcal{I}}(r)&=\iint_{\Vec{k},\Vec{k_1}}\frac{k_1^a\,k^ke^{i\Vec{k}.\Vec{r}}}{\Vec{k_1}^2\,(\Vec{k}^2+\textcolor{black}{m^2})\,(\Vec{k_1}-\Vec{k})^2}-\iint_{\Vec{k},\Vec{k_1}}\frac{k_1^a\,k_1^ke^{i\Vec{k}.\Vec{r}}}{\Vec{k_1}^2\,(\Vec{k}^2+\textcolor{black}{m^2})\,(\Vec{k_1}-\Vec{k})^2}\,,\\  &
         =\rmint_{\Vec{k_1}}\frac{k_1^k \,e^{i\Vec{k_1}.\Vec{r}}}{\Vec{k_1}^2+\textcolor{black}{m^2}}\rmint_{\Vec{k}}\frac{k^a}{\Vec{k}^2(\Vec{k}-\Vec{k_1})^2}-\rmint_{\Vec{k_1}}\frac{e^{i\Vec{k_1}.\Vec{r}}}{\Vec{k_1}^2+\textcolor{black}{m^2}}\rmint_{\Vec{k}}\frac{k^a\,k^k}{\Vec{k}^2(\Vec{k}-\Vec{k_1})^2}\,,\\ &
         =\frac{1}{16}\rmint_{\Vec{k_1}}\frac{k_1^k\,k_1^a\,e^{i\Vec{k_1}.\Vec{r}}}{|\Vec{k_1}|(\Vec{k_1}^2+\textcolor{black}{m^2})}+\frac{1}{16}\rmint_{\Vec{k_1}}\frac{e^{i\Vec{k_1}.\Vec{r}}}{|\Vec{k_1}|(\Vec{k_1}^2+\textcolor{black}{m^2})}\Bigg[\frac{1}{4}\Vec{k_1}^2\delta^{ak}-\frac{3}{4}k_1^a\,k_1^k\Bigg]\,,\\ &
         =\frac{1}{64}\rmint_{\Vec{k_1}}\frac{k_1^k\,k_1^a\,e^{i\Vec{k_1}.\Vec{r}}}{|\Vec{k_1}|(\Vec{k_1}^2+\textcolor{black}{m^2})}+\mathcal{F}\,(|\Vec{r}|)\delta^{ak}\,,\\ &
         =-\frac{1}{64}\frac{\partial^2}{\partial x^{k}\partial x^{a}}\rmint_{\Vec{k_1}}\frac{e^{i\Vec{k_1}.\Vec{r}}}{|\Vec{k_1}|(\Vec{k_1}^2+\textcolor{black}{m^2})}+\mathcal{F}\,(|\Vec{r}|)\delta^{ak}\,,\\ &
         =-\frac{1}{64}\Bigg[ \frac{1}{r}(\delta^{ak}-n^a\,n^k)\Bigg\{\pi ^{3/2} m G_{1,3}^{2,1}\left(\frac{m^2 r^2}{4}|
\begin{array}{c}
 -\frac{1}{2} \\
 -\frac{1}{2},-\frac{1}{2},0 \\
\end{array}
\right)-\frac{2 \pi ^{3/2} G_{1,3}^{2,1}\left(\frac{m^2 r^2}{4}|
\begin{array}{c}
 \frac{1}{2} \\
 \frac{1}{2},\frac{1}{2},0 \\
\end{array}
\right)}{m r^2}\Bigg\}+\\ &
n^a n^k\Bigg\{\frac{\pi ^{3/2} \left(-2 m^2 r^2 G_{1,3}^{2,1}\left(\frac{m^2 r^2}{4}|
\begin{array}{c}
 -\frac{1}{2} \\
 -\frac{1}{2},-\frac{1}{2},0 \\
\end{array}
\right)+8 G_{1,3}^{2,1}\left(\frac{m^2 r^2}{4}|
\begin{array}{c}
 \frac{1}{2} \\
 \frac{1}{2},\frac{1}{2},0 \\
\end{array}
\right)+ m^4 r^4 G_{1,3}^{2,1}\left(\frac{m^2 r^2}{4}|
\begin{array}{c}
 -\frac{3}{2} \\
 -\frac{3}{2},-\frac{3}{2},0 \\
\end{array}
\right)\right)}{2 m r^3}\Bigg\}\\ &
\,\,\,\,\,\,\,+\mathcal{F}(|\Vec{r}|)\delta^{ak}\Bigg]\,.
    \end{split}
\end{align}
Finally using these results we can write down the amplitude mentioned in (\ref{4.31}).
\section{ Details of the Feynman integral computation  used in Section~(\ref{Sec4.3})}\label{App1}
The integral in (\ref{4.38}) can be evaluated as follows:
\begin{align}
    \begin{split}
         \mathcal{I}_1&= \rmint_{k,k_1}\frac{k_{1i}k_{1j}\,e^{i\Vec{k}\cdot \Vec{r}}}{\Vec{k}^2\,[(\Vec{k}-\Vec{k}_1)^2+m^2](\Vec{k}_{1}^2+m^2)}\,,\\ &
         =\rmint_{k}\frac{e^{i\Vec{k}\cdot \Vec{r}}}{\Vec{k}^2}\underbrace{\rmint_{k_1}\frac{k_{1i}k_{1j}}{(\Vec{k}_1^2+m^2)[(\Vec{k}_1-\Vec{k})^2+m^2]}}_{k_i k_j B_{21}+\delta_{ij}B_{22}}\,.\label{A.1}
    \end{split}
\end{align}
Then the tensor integral in (\ref{A.1}) can be reduced to a scalar integral using Passarino-Veltman reduction \cite{Passarino:1978jh,2022arXiv220103593W} where $B_{21,22}$ is given by,
 \begin{align}
     \begin{split}
          & B_{21}=\frac{1}{2\Vec{k}^2}\Big[\frac{3}{2}\,\Vec{k}^2\,B_{1}+m^2\,B_0-\frac{1}{2}A_0(im)\Big]\,,\\ &
          B_{22}=\frac{1}{4}\Big[-B_1\Vec{k}^2-2m^2B_{0}-A_{0}(im)\Big]\,,
     \end{split}
 \end{align}
 where
 \begin{align}
     \begin{split}
      B_{0}&=\rmint_{k_1}\frac{1}{(\Vec{k}_1^2+m^2)[(\Vec{k}_1-\Vec{k})^2+m^2]}\\ &
      =\frac{1}{4\pi|\Vec{k}|}\arctan(|\Vec{k}|/2m),
    \label{A.3}
     \end{split}
 \end{align}
 and $B_1,A_0(im)$ are given by,
 \begin{align}
     \begin{split}
           &  B_1=\frac{1}{2}B_0=\frac{1}{8\pi|\Vec{k}|}\arctan(|\Vec{k}|/2m)\,,\\ &
       A_0(im)=\frac{m}{4\pi}\,.
     \end{split}
 \end{align}
Therefore the integral in (\ref{A.1}) can be written as,
\begin{align}
    \begin{split}
    \mathcal{I}_{1}&=\rmint_{k}\frac{e^{i\Vec{k}\cdot\Vec{r}}}{\Vec{k}^2}\Big[\frac{3}{4}\frac{1}{8\pi|\Vec{k}|}\arctan(|\Vec{k}|/2m)+\frac{m^2}{8\pi|\Vec{k}|^3}\arctan(|\Vec{k}|/2m)-\frac{m}{16\pi|\Vec{k}|^2}\Big]k_ik_j\\ &
   \,\,\,\, +\rmint_{k}\frac{e^{i\Vec{k}\cdot \Vec{r}}}{\Vec{k}^2}\Big[-\frac{1}{32\pi}|\Vec{k}|\arctan(|\Vec{k}|/2m)-\frac{m^2}{8\pi |\Vec{k}|}\arctan(|\Vec{k}|/2m)-\frac{m}{16\pi}\Big]\delta_{ij}\,,
   \\ & =\frac{3}{32\pi}\Big[\delta_{ij}\frac{1-e^{-2 m r}}{8 \pi  m r^3}+n_in_j\frac{e^{-2m r} \left(2m r-3 e^{2m r}+3\right)}{8 \pi  m r^3}\Big]+\frac{m^2}{8\pi}\Big[\Big\{-\frac{2 m^2 r^2+e^{-2 m r} (2 m r+1)-1}{32 \pi  m^3 r^3}\,n_in_j\Big\}\\ &
  \,\,\,\,\,\, +\Big\{\frac{e^{-2 m r} \left(e^{2 m r} \left(-8 m^3 r^3 \text{Ei}(-2 m r)+6 m^2 r^2-1\right)-4 m^2 r^2+2 m r+1\right)}{96 \pi  m^3 r^3}\delta_{ij}\Big\}\Big]
   \\&-\frac{m}{16\pi}\Big[\frac{1}{8\pi r}\Big(\delta_{ij}-n_i n_j\Big)\Big]
  \,
  +\Big[\frac{e^{-2mr}}{4\pi r^2}-\frac{m}{64 \pi^2 r}\Big(1-e^{-2mr}-2mr \text{Ei}(-2mr)\Big)-\frac{m}{128\pi^2 r}\Big]\delta_{ij}\,,\\ &
   =\lambda_1(r)\delta_{ij}+\lambda_2(r)n_i n_j\,.
    \end{split}\label{B.5m}
\end{align}
The explicit forms of the functions $\lambda_1(r)$ and $\lambda_2(r)$ are given by,
\begin{align}
    \begin{split}
        \lambda_1(r)=&\frac{3-3e^{-2mr}}{256\pi^2m r^3}+\frac{e^{-2 m r} \left(e^{2 m r} \left(-8 m^3 r^3 \text{Ei}(-2 m r)+6 m^2 r^2-1\right)-4 m^2 r^2+2 m r+1\right)}{768\pi^2m r^3}\\ &
        -\frac{m}{128\pi^2 r}+\frac{e^{-2mr}}{4\pi r^2}-\frac{m}{64 \pi^2 r}\Big(1-e^{-2mr}-2mr \text{Ei}(-2mr)\Big)-\frac{m}{128\pi^2 r}\,,
    \end{split}
\end{align}
and,

\begin{align}
    \begin{split}
        \lambda_2(r)=&\frac{e^{-2m r} \left(2m r-3 e^{2m r}+3\right)}{256\pi^2 m r^3}-\frac{2m^2r^2+e^{-2mr}(2mr+1)-1}{256\pi^2m r^3}+\frac{m}{128\pi^2 r}\,.
    \end{split}
\end{align}
The integral in (\ref{B.5m}) can be done as follows,
\begin{align}
    \begin{split}
        \mathcal{I}_{c}(x;r)&=\rmint_{k}\frac{e^{i\Vec{k}\cdot \Vec{r}}}{\Vec{k}^5}\arctan(|\Vec{k}|/x)\,,\\ &
        =\frac{1}{2\pi^2 r x^3}\underbrace{\rmint_{0}^{\infty}dk \, \frac{\sin(\gamma k)}{k^4}\arctan(k)}_{\Hat{\delta}(\gamma)}\,\,,\gamma=x\,r\,.\\ &
    \end{split}\label{B.6m}
\end{align}
The second integral in (\ref{B.5m}) is IR finite. But direct computation is a hard task. So, we first compute the integral in (\ref{B.6m}) and then take darivatives two times w.r.t $x_i.$  But the integral $\Hat{\delta}(\gamma)$ has \textcolor{black}{IR divergence} but one can extract out the finite part by the Feynman trick. For that, we first take double derivative with respect to $\gamma$, and then integrate twice to find out the mother integral with some infinite constant.
\begin{align}
    \begin{split}
         \frac{d^2 \Hat{\delta}(\gamma)}{d\gamma^2}&=-\rmint_{0}^{\infty}dk \, \frac{\sin(\gamma k)}{k^2}\arctan(k)\,,\\ &
         =\frac{1}{2} \pi  \left(\gamma\,  \text{Ei}(-\gamma )+e^{-\gamma }-1\right)\,.
    \end{split}\label{B.7m}
\end{align}
Integrating (\ref{B.7m}) we will have,
\begin{align}
    \begin{split}
\Hat{\delta}(\gamma)&=\rmint^{\gamma}d\gamma'\rmint^{\gamma'}d\gamma''\,\frac{d^2 \Hat{\delta}(\gamma'')}{d\gamma''^2}+C_1 \gamma+C_{2},\,(\text{$C_1, C_2$ are $\gamma$ independent constant})\,,\\ &
        =\frac{\pi}{12}  e^{-\gamma } \Big(\left(1-3 e^{\gamma }\right) \gamma ^2-\gamma +e^{\gamma } \gamma ^3 \text{Ei}(-\gamma )+2\Big)+C_1\,\gamma+C_2.
    \end{split}\label{B.8m}
\end{align}
In order to find the constants, one need to use the following initial conditions: $\Hat{\delta}(0)=0\,\text{and}\,|\Hat{\delta}'(0)|\rightarrow \infty$. Hence using equation (\ref{B.8m}) one can find that, 
\begin{align}
    \begin{split}
      &  0=\Hat{\delta}(0)=\frac{\pi}{6}+C_2\implies C_2=-\frac{\pi}{6}.\\ &
      \text{and},\,\infty=|\Hat{\delta}'(0)|=|-\frac{\pi}{4}+C_1|\implies |C_1|\rightarrow \infty\,.
    \end{split}
\end{align}
Therefore,
\begin{align}
    \begin{split}
        \mathcal{I}_c=\frac{e^{- x\,r} \Big[(r^2 x^2\, e^{x\,r}\, (x\,r\, \text{Ei}(-x\,r)-3)+r^2 x^2-x\,r+2\Big]}{24 \pi  r x^3}-\frac{1}{12\pi r x^3}+ \text{$`r$' independent infinite constant.}\label{B.9m}
    \end{split}
 \end{align}
Eventually, using the result of (\ref{B.9m}), one can compute the following integral.
\begin{align}
    \begin{split}
     \mathcal{I}_m &= \rmint_{k}\frac{e^{i\Vec{k}\cdot \Vec{r}}}{\Vec{k}^5}\arctan(|\Vec{k}|/x)k_ik_j,\,\,x=2m\\ &
     =-\partial^{x}_{i}\partial^{x}_{j}\,\mathcal{I}_c(2m;r)\,,\\ &
     =\frac{e^{-2 m r} \left(e^{2 m r} \left(-8 m^3 r^3 \text{Ei}(-2 m r)+6 m^2 r^2-1\right)-4 m^2 r^2+2 m r+1\right)}{96 \pi  m^3 r^3}\delta_{ij}\\ &
     \,\,\,\,\,\,\,\,\,\,-\frac{2 m^2 r^2+e^{-2 m r} (2 m r+1)-1}{32 \pi  m^3 r^3}\,n_in_j\,.
    \end{split}
\end{align}
One can obtain the two coefficients by matching the coefficients of $\delta_{ij}$ and $n_i n_j$. Furthermore, We can omit the infinite constant because it does not contribute to the equation of motion obtained from the variation of the effective action.\par 

Another integral relevant for evaluating (\ref{4.38}) is:
\begin{align}
    \begin{split}
         \mathcal{I}_2&= \rmint_{k,k_1}\frac{k_{1i}k_{j}\,e^{i\Vec{k}\cdot \Vec{r}}}{\Vec{k}^2\,[(\Vec{k}-\Vec{k}_1)^2+m^2](\Vec{k}_{1}^2+m^2)}\,,\\ &
         =\rmint_{k}\frac{e^{i\Vec{k}\cdot \Vec{r}}k_j}{\Vec{k}^2}\underbrace{\rmint_{k_1}\frac{k_{1i}}{(\Vec{k}_1^2+m^2)[(\Vec{k}_1-\Vec{k})^2+m^2]}}_{k_{i}B_{1}}\,,\label{A.5}
    \end{split}
\end{align}
where $B_1$ is given in (\ref{A.3}). Therefore the integral in (\ref{A.5}) can be written as,
\begin{align}
    \begin{split} \mathcal{I}_2&=\rmint_{k}\frac{e^{i\Vec{k}\cdot \Vec{r}}k_i k_j}{\Vec{k}^2}\frac{1}{16|\Vec{k}|}\arctan(|\Vec{k}|/2m)\,,\\ &
    =\delta_{ij}\frac{1-e^{-2 m r}}{128 \pi  m r^3}+n_in_j\frac{e^{-2m r} \left(2m r-3 e^{2m r}+3\right)}{128 \pi  m r^3}\,,\\ &
    =\chi_1(r)\delta_{ij}+\chi_2(r)n_i n_j\,.
        \label{A.6}
    \end{split}
\end{align}
Again the two coefficients $\chi_{1,2}$ can be obtained by matching the coefficients of $\delta_{ij}$ and $n_i n_j$ in (\ref{A.6}) as,
\begin{align}
    \begin{split}
        &\chi_1(r)=\frac{1-e^{-2mr}}{128\pi m r^3}\,,\\ &
        \text{and,}\\ &
        \chi_2(r)=\frac{e^{-2mr}(2mr-3e^{2mr}+3)}{128\pi m r^3}\,.
    \end{split}
\end{align}
\section{ Details of computation of Feynman integral used in (\ref{Sec4}) and (\ref{Sec5})}\label{app3}
In (\ref{4.33mm}) we have the following integral to evaluate:
\begin{align}
    \begin{split}
        I&=\rmint_{k}\frac{k^a e^{i\Vec{k}\cdot \Vec{r}}}{\Vec{k}^2}\underbrace{\rmint_{k_3}\frac{1}{(\Vec{k}-\Vec{k}_3)^2(\Vec{k}_3^2+m^2)}}\\ &
        =\rmint_{k}\frac{k^a e^{i\Vec{k}\cdot \Vec{r}}}{\Vec{k}^2}
        \begin{cases}
        \frac{1}{8|\Vec{k}|}\Big[1-\frac{2}{\pi}\arctan(m/|\Vec{k}|)\Big] & ,k^2> m^2\\
        \frac{1}{4\pi|\Vec{k}|}\arctan(|\Vec{k}|/m) &, k^2<m^2
        \end{cases}\\ &
        =\frac{1}{4\pi}\rmint_{k}\frac{k^a e^{i\Vec{k}\cdot\Vec{r}}}{\Vec{k}^3}\arctan(|\Vec{k}|/m)\,,\\ &
        =-\frac{i}{4\pi}n^{a}\partial_{r}\,\underbrace{\rmint_{k}\frac{e^{i\Vec{k}\cdot\Vec{r}}}{\Vec{k}^3}\arctan(|\Vec{k}|/m)}_{\alpha(m;r)}\,.\label{C.1}
    \end{split}
\end{align}
In (\ref{C.1}) we have used the identity, $\arctan(x)+\arctan(1/x)=\pi/2, \forall x>0$ and $\alpha(m;r)$ has the following form,
\begin{align}
    \begin{split}
        \alpha(m;r)&=\frac{4\pi}{mr}\rmint_{0}^{\infty}dk \frac{\sin(mrk)}{k^2}\arctan(k)\,,\\ &
        =\frac{2 \pi ^2 r \left(-m r \text{Ei}(-m r)-e^{-m r}+1\right)}{m}\,.
    \end{split}
\end{align}
\begin{figure}
    \centering
    \includegraphics[scale=0.5]{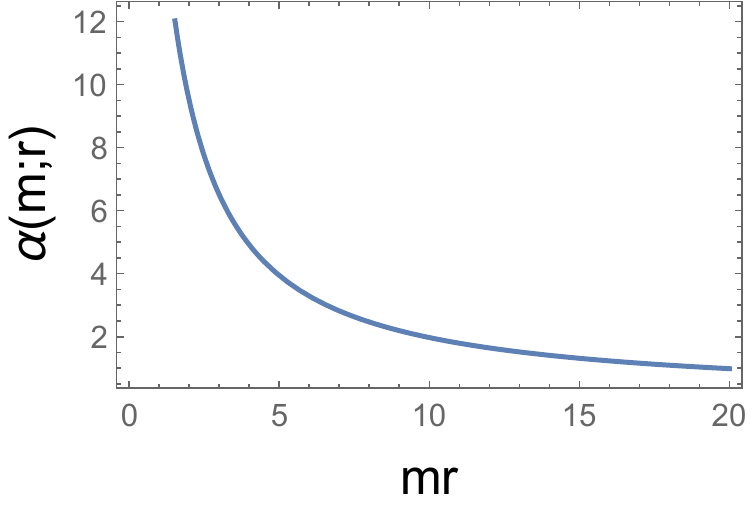}
    \caption{$\alpha(m;r)$ vs $mr$ plot}
    \label{fig12}
\end{figure}
 We numerically integrate and show the nature of the integral in Fig~(\ref{fig12}).\par
\begin{figure}[b!]
    \centering
    \includegraphics[scale=0.4]{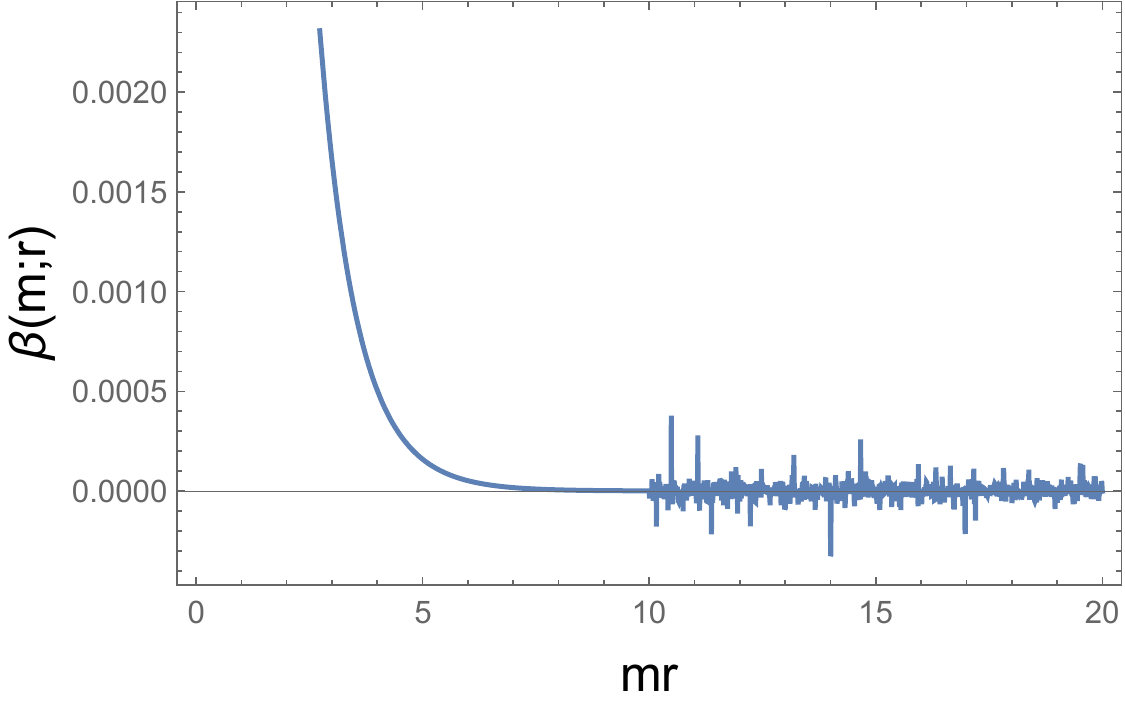}
    \caption{$\beta(m;r)$ vs $mr$ plot}
    \label{lastfig}
\end{figure}

Another useful relevant integral for (\ref{4.26}) has the following form,
\begin{align}
    \begin{split}
        \beta(m;r)&=\rmint_{k}\frac{e^{i\vec k \cdot \vec r}}{\vec k^2+m^2}\rmint_{k_1}\frac{1}{\vec k_1^2[(\vec k_1-\vec k)^2+m^2]}\\ &
        = \rmint_{k}\frac{e^{i\vec k \cdot \vec r}}{\vec k^2+m^2}
        \begin{cases}
                \frac{1}{8|\Vec{k}|}[1-\frac{2}{\pi}\arctan(m/|\Vec{k}|)]\, & \Vec{k}^2>m^2\\
                \frac{1}{4\pi|\Vec{k}|}\arctan(|\Vec{k}|/m)\, & \Vec{k}^2<m^2
            \end{cases}\,,\\ &
 =\rmint_{k}\frac{e^{i\Vec{k}\cdot\Vec{r}}}{(\Vec{k}^2+m^2)|\Vec{k}|}\arctan(|\Vec{k}|/m)\\ &
 =\frac{1}{2\pi^2 mr}\rmint_{0}^{\infty}dk\,\frac{\sin(mrk)}{k^2+1}\arctan(k)\,.\label{C3m}
 \end{split}
\end{align}
The integral in (\ref{C3m}) does not have any closed form expression. One can numerically integrate it and the nature of the function $\beta(m;r)$ is shown in Fig.~(\ref{lastfig}).

\begin{figure}[t!]
    \centering
    \scalebox{0.4}{\begin{feynman}
    \fermion[label=$$, lineWidth=4]{10.40, 4.80}{11.40, 4.80}
    \fermion[showArrow=false, lineWidth=4]{8.00, 4.40}{7.60, 4.80}
    \fermion[lineWidth=4]{12.60, 4.80}{13.60, 4.80}
    \dashed[showArrow=false, color=eb144c]{4.80, 4.00}{6.00, 4.80}
    \dashed[showArrow=false, color=eb144c]{5.20, 4.00}{6.00, 4.40}
    \dashed[showArrow=false, color=eb144c]{4.80, 4.80}{6.00, 5.60}
    \dashed[showArrow=false, color=eb144c]{4.80, 4.40}{6.00, 5.20}
    \fermion[showArrow=false]{6.00, 4.00}{6.00, 5.60}
    \fermion[showArrow=false, lineWidth=4]{9.20, 5.20}{9.60, 4.80}
    \fermion[lineWidth=4, label=$$]{4.00, 4.00}{6.80, 4.00}
    \fermion[showArrow=false, lineWidth=4]{9.20, 4.40}{9.60, 4.80}
    \fermion[lineWidth=4]{4.00, 5.60}{6.80, 5.60}
    \dashed[showArrow=false, color=eb144c]{4.80, 5.20}{5.60, 5.60}
    \fermion[showArrow=false]{4.80, 4.00}{4.80, 5.60}
    \fermion[showArrow=false, lineWidth=4]{7.60, 4.80}{8.00, 5.20}
    \fermion[lineWidth=5, showArrow=false]{7.60, 4.80}{9.60, 4.80}
    \parton[color=eb144c]{12.00,4.80}{0.60}
\end{feynman}

}
    \caption{Correspondence between PN diagrams (left) and one-loop scalar Feynman diagram (right).}
    \label{fig16}
\end{figure}
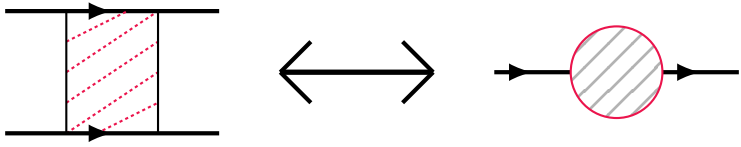

\section{Comments on middle vertex contribution to the radiation}\label{D}
In the radiative diagrams, where one radiative field is coming out from the vertex, we implicitly assume that the radiation is coming out from an average position $\boldsymbol{x}_0$. But in general, the radiative field is integrated over all the spacetime i.e, the radiation could be from anywhere in the spacetime. \\

\textbf{Scalar diagrams consisting of radiative bulk vertices:}

1)\,Fig.~(\ref{10h}):

\begin{align}
    \begin{split}
\mathcal{S}_{\text{eff}}\Big |_{\text{fig}(\ref{10h})}&=\frac{m^2 s_1m_1m_2}{m_p^3}\rmint dt_1dt_2\rmint d^4x \Big\langle \psi(\boldsymbol{x}_2(t_2))\psi(x)\Big\rangle \Big\langle\varphi(\boldsymbol{x}_1(t_1))\varphi(x) \Big\rangle \Big\{\bar{\phi}(\vec 0,t)+x^i\partial_{i}\bar\phi(\vec 0,t)+\cdots\Big\}\,.\label{E.1}
    \end{split}
\end{align}
The first and second terms in (\ref{E.1}) give the monopole and dipole terms, respectively. First, compute the dipole term. Here, it is difficult to isolate the source term $J$, but one can directly compute the dipole moment by isolating the coefficient of the $\partial_i \bar \phi(\vec 0,t)$. Hence, the effective action has the following, 
\begin{align}
    \begin{split}
        \mathcal{S}_{\text{eff}}\Big |_{\text{fig}(\ref{10h})}&=\frac{m^2 s_1m_1m_2}{2 m_p^3}\rmint dt\,\partial_i\bar\phi(\vec 0,t)\rmint d^3x\,x^i \rmint_{\boldsymbol{k}_1,\boldsymbol{k}_2}\frac{e^{i\boldsymbol{k}_1\cdot (\boldsymbol{x}_2-\boldsymbol{x})}e^{i\boldsymbol{k}_2\cdot (\boldsymbol{x}_1-\boldsymbol{x})}}{\boldsymbol{k}_1^2(\boldsymbol{k}_2^2+m^2)}\,.\label{E.2}
    \end{split}
\end{align}
Now, from \eqref{E.2}, it is clear that the dipole moment has the following form,
\begin{align}
    \begin{split}
        I_{\phi}^{i}\Big|_{\text{fig.} (\ref{10h})}&=\frac{m^2 s_1m_1m_2}{2 m_p^3}\rmint d^3x\,x^i \rmint_{\boldsymbol{k}_1,\boldsymbol{k}_2}\frac{e^{i\boldsymbol{k}_1\cdot (\boldsymbol{x}_2-\boldsymbol{x})}e^{i\boldsymbol{k}_2\cdot (\boldsymbol{x}_1-\boldsymbol{x})}}{\boldsymbol{k}_1^2(\boldsymbol{k}_2^2+m^2)}\,,\\ &
        =\frac{m^2 s_1m_1m_2}{2 m_p^3}\Big[r^i\frac{(m^2r^2-2)+2e^{-mr}(mr)}{4\pi m^4 r^3}+x_2^i\frac{1-e^{-mr}}{4\pi m^2 r}\big]\,.
    \end{split}
\end{align}
2) Fig.~(\ref{10e}): Here, the computation is more involved as we have a time derivative over the radiation field.
\begin{align}
    \begin{split}
        \mathcal{S}_{\text{eff}}\Big |_{\text{fig}(\ref{10e})}&=\frac{m_1m_2s_2}{m_p^2}\rmint\Bar{\mathcal{D}}\Hat{\xi}\rmint dt_1dt_2\tilde{\psi}(\boldsymbol{x}_1(t_1))\phi(\boldsymbol{x}_2(t_2))\rmint d^4x \,\frac{\psi}{m_p}\partial_0\phi\partial_0\bar{\phi}\,,\\ &
        =-\frac{m_1m_2s_2}{m_p^2}\rmint \Bar{\mathcal{D}}\Hat{\xi}\rmint dt_1dt_2\tilde{\psi}(\boldsymbol{x}_1(t_1))\phi(\boldsymbol{x}_2(t_2))\rmint d^4x \,\Big[\partial_0\tilde\psi\partial_0 \varphi+\tilde\psi\partial_0^2 \varphi\Big]\Big\{\bar\phi(\vec 0,t)+x^i \partial_i \bar\phi(\vec 0,t)\Big\}\,.
    \end{split}
\end{align}
Again, focus on the dipole term only,
\begin{align}
    \begin{split}
    \mathcal{S}_{\text{eff}}\Big |_{\text{fig}(\ref{10e})}&=\frac{m_1m_2s_2}{m_p^2}\rmint dt_1 dt_2 \rmint d^4x \Big[\partial_0\Big\langle\tilde\psi(\boldsymbol{x}_1(t_1))\tilde\psi(x)\Big\rangle \partial_0\Big\langle\varphi(\boldsymbol{x}_2(t_2))\varphi(x)\Big\rangle\\ &
    \hspace{3 cm}+\Big\langle\tilde\psi(\boldsymbol{x}_1(t_1))\tilde\psi(x)\Big\rangle \partial_0^2\Big\langle\varphi(\boldsymbol{x}_2(t_2))\varphi(x)\Big\rangle\Big]x^{i}\partial_i\bar\phi(\vec 0,t)\,.
    \end{split}
\end{align}
Considering the first term only,
\begin{align}
    \begin{split}
         \mathcal{S}_{\text{eff}}\Big |_{\text{fig}(\ref{10e})}^{(1)}&=\frac{m_1m_2s_2}{m_p^2}\rmint dt_1 dt_2 \rmint d^4x \partial_0\Big\langle\tilde\psi(\boldsymbol{x}_1(t_1))\tilde\psi(x)\Big\rangle \partial_0\Big\langle\varphi(\boldsymbol{x}_2(t_2))\varphi(x)\Big\rangle x^{i}\partial_i \bar\phi(\vec 0,t)\,,\\ &
         =\frac{m_1m_2s_2}{2m_p^2}\rmint dt \,\partial_{i}\bar\phi(\vec 0,t) \rmint d^3x \rmint_{k_1}\frac{(i\boldsymbol{k}_1\cdot \boldsymbol{v}_1) }{\boldsymbol{k}_1^2}\Big[x_1^{i}+i\partial_{k_1^i}\Big]e^{i\boldsymbol{k}_1\cdot (\boldsymbol{x}_1-\boldsymbol{x})}\rmint_{k_2}(i\boldsymbol{k}_2\cdot \boldsymbol{v}_2)\frac{e^{ik_2\cdot (\boldsymbol{x}_2-\boldsymbol{x})}}{\boldsymbol{k}_2^2+m^2}\,.\\ &
    \end{split}
\end{align}
Therefore, the dipole moment can be isolated as,
\begin{align}
    \begin{split}
        I_{\phi}^{i}(t)\Big|_{\text{fig.}(\ref{10e})}&=-v_1^a v_2^b\rmint d^3x \rmint_{k_1}\Big[\frac{k_1^a\,x_1^i}{\boldsymbol{k}_1^2} -i\Big(\frac{\delta_{ai}}{\boldsymbol{k}_1^2}-\frac{2k_1^ik_1^a}{\boldsymbol{k}_1^4}\Big)\Big] e^{i\boldsymbol{k}_1\cdot (\boldsymbol{x}_1-\boldsymbol{x})}\rmint_{k_2}k_2^b\,\frac{e^{i\boldsymbol{k}_2\cdot(\boldsymbol{x}_2-\boldsymbol{x})}}{\boldsymbol{k}_2^2+m^2}\,,\\ &
        =v_1^a v_2^b\rmint_{k_1}\Big[\frac{k_1^a\,x_1^i}{\boldsymbol{k}_1^2} -i\Big(\frac{\delta_{ai}}{\boldsymbol{k}_1^2}-\frac{2k_1^ik_1^a}{\boldsymbol{k}_1^4}\Big)\Big]\frac{k_1^b}{\boldsymbol{k}_1^2+m^2}e^{i\boldsymbol{k}_1\cdot \vec r}\,,\\ &
        =v_1^a v_2^b\Bigg[x_1^i\rmint_{\boldsymbol{k}_1}\frac{k_1^a k_1^b}{\boldsymbol{k}_1^2(\boldsymbol{k}_1^2+m^2)}-i\delta_{ai}\rmint_{k_1}\frac{k_1^b}{\boldsymbol{k}_1^2(\boldsymbol{k}_1^2+m^2)}+2i\rmint_{k_1}\frac{k_1^i k_1^a k_1^b}{\boldsymbol{k}_1^4(\boldsymbol{k}_1^2+m^2)}\Bigg]e^{i\boldsymbol{k}\cdot \vec r}\,.
    \end{split}
\end{align}
The integrals can be done by investigating two separate families of Feynman-Fourier integrals that we discuss in the subsequent subsection. However, we are not showing the explicit results of these integrals, as they can be straightforwardly evaluated using the results we provide in the subsequent section.\\\\
\textbf{IBP with exponential and systematic computation of the Fourier integrals.}\\ 
Let's take a moment to systematically compute the Fourier Feynman Integrals.\\\\
\textbf{\texttt{One-loop Fourier family}:}\,Let us define a family of integrals,
\begin{align}
    \begin{split}
        {J}_{a_1,a_2}= \int d^dk\, \frac{e^{i\boldsymbol{k}\cdot \boldsymbol r}}{(i \boldsymbol{k}\cdot \boldsymbol r)^{a_1}(\boldsymbol{k}^2+m^2)^{a_2}}.
    \end{split}
\end{align}
After solving the IBP relations (with a simple modified version of \textbf{\texttt{LiteRed}} \cite{Lee:2013mka}) we have three master integrals for the first family: ${J}^{}_{0,1},{J}^{}_{0,2}, {J}^{}_{1,1}$. First do the following rescaling, $\boldsymbol{k}\to \boldsymbol{\ell}=\,\boldsymbol{k}/m$ and $\boldsymbol{r}\to \boldsymbol{s}=\boldsymbol{r}m$. We left with the integral,
\begin{align}
    J_{a_1,a_2}=(m^2)^{d/2-a_2}\int d^d\boldsymbol{\ell} \frac{e^{i\boldsymbol{\ell}\cdot\boldsymbol{s}}}{(i\boldsymbol{\ell}\cdot \boldsymbol{s}-i\varepsilon)^{a_1}(\boldsymbol{\ell}^2+1)^{a_2}}\,.
\end{align}
The differential equation satisfied by the master integral is given by (with $x=s^2$),
\begin{align}
\partial_{x}\boldsymbol{\mathcal {J}}=\boldsymbol{\Omega}_{x}\cdot \boldsymbol{\mathcal {J}},\qquad   \boldsymbol{\Omega}_{x}=\left(
\begin{array}{ccc}
 -\frac{d-2}{2 x} & -\frac{1}{x} & 0 \\
 -\frac{1}{4} & 0 & 0 \\
 \frac{1}{2 x} & 0 & -\frac{1}{2 x} \\
\end{array}
\right)\,.
\end{align}
Therefore, we have three coupled differential equations to be solved,
\begin{align}
    \begin{split}&\partial_{x}J_{0,1}=\frac{2-d}{2x}J_{0,1}-\frac{1}{x}J_{0,2},\\&
    \partial_{x}J_{0,2}=-\frac{1}{4}J_{0,1},\\ &
    \partial_{x}J_{1,1}=\frac{1}{2x}J_{0,1}-\frac{1}{2x}J_{1,1}.
    \end{split}
\end{align}
The solution is given by,
\begin{align}
\begin{split}
&    J_{0,1}=x^{\frac{2-d}{4}}\left(C_1 I_{\frac{d-2}{2}}(\sqrt{x})+C_2 K_{\frac{d-2}{2}}(\sqrt{x})\right),\\ &
J_{0,2}
=
\frac{1}{2}\,x^{\frac{4-d}{4}}
\left(
-\,C_1 I_{\frac{d-4}{2}}(\sqrt{x})
+
C_2 K_{\frac{d-4}{2}}(\sqrt{x})
\right),\\ &
J_{1,1}(x)
=
\frac{1}{\sqrt{x}}
\left[
C_3
+
\int^{\sqrt{x}}
t^{-\frac{d-2}{2}}
\left(
C_1 I_{\frac{d-2}{2}}(t)
+
C_2 K_{\frac{d-2}{2}}(t)
\right)\,dt
\right].\label{2.201j}
    \end{split}
\end{align}
The constants $C_1, C_2$ can be fixed using the asymptotic expansion around $x\to 0$ of the integral. In this asymptotic, the integral has two distinct momentum scalings,
\begin{align}
   \texttt{hard:} |\boldsymbol{\ell}|\sim\frac{1}{s},\,\,  \texttt{soft:} |\boldsymbol{\ell}|\sim1\,.
\end{align}
There are therefore two regions: from the method of regions \cite{Beneke:1997zp},\\\\
\textbf{\texttt{hard region}.} In this region the propagator expanded as massless, $\frac{1}{\boldsymbol{\ell}^2+1}\sim \frac{1}{\boldsymbol{\ell}^2}$ and hence,
\begin{align}
    J_{0,1}^{(h)}(x)=2^{d-2}\pi^{d/2} \Gamma\left(\frac{d}{2}-1\right)x^{1-\frac{d}{2}}\,.
\end{align}\\
\textbf{\texttt{soft region}.} In this region, the exponent has a soft expansion, and considering the leading order, one has,
\begin{align}
    J_{0,1}^{(s)}(x)=\int d^d\boldsymbol{\ell} \frac{1}{\boldsymbol{\ell}^2+1}=\pi^{d/2}\Gamma\left(1-\frac{d}{2}\right)\,.
\end{align}
Collecting all the leading contributions from the asymptotic region,
\begin{align}
    J_{0,1}\overset{x\to 0}{\sim} 2^{d-2}\pi^{d/2}\Gamma\left(\frac{d}{2}-1\right) x^{d/2-1}+\pi^{d/2}\Gamma\left(1-\frac{d}{2}\right)+\cdots\,.\label{2.205a}
\end{align}
Now, expanding $J_{0,1}$ in \eqref{2.201j} we get,
\begin{align}
   J_{0,1}\overset{x\to 0}{\sim}  \frac{2^{-\frac{d}{2}} \left(2 C_1+C_2 \Gamma \left(\frac{d}{2}\right) \Gamma \left(1-\frac{d}{2}\right)\right)}{\Gamma \left(\frac{d}{2}\right)}+C_2 2^{\frac{d}{2}-2} x^{1-\frac{d}{2}} \Gamma \left(\frac{d-2}{2}\right)\,.\label{2.206a}
\end{align}
Now, matching \eqref{2.205a} and \eqref{2.206a} we get,
\begin{align}
    C_1=0,\,C_2=(2\pi)^{d/2}\,.
\end{align}
In a similar way, $C_3$ can be fixed from the asymptotic expansion of  $J_{1,1}$, in soft region,
\begin{align}
\begin{split}
    J_{1,1}^{(s)}&\overset{x\to0}{\sim} \int d^d\ell \frac{1-i\boldsymbol{\ell}\cdot \boldsymbol{s}}{(i\boldsymbol{\ell\cdot s-i\varepsilon})(\boldsymbol{\ell}^2+1)}+\cdots\\ &
    \sim \int d^d\ell \frac{1}{(i\boldsymbol{\ell\cdot s-i\varepsilon})(\boldsymbol{\ell}^2+1)}+\cdots=\frac{\pi^{\frac{d+1}{2}}\Gamma\left(\frac{3-d}{2}\right)}{\sqrt{x}\,}+\cdots
    \end{split}
\end{align}
Comparing we have, $C_3={\pi^{\frac{d+1}{2}}}\Gamma\left(\frac{3-d}{2}\right)$
Therefore, the integrals become,
\begin{align}
\begin{split}
 &   J_{0,1}=(2\pi)^{d/2} \left(\frac{m}{r}\right)^{d/2-1}K_{\frac{d-2}{2}}(mr)\,,\\ &
    J_{0,2}=\frac{1}{2}(m^2)^{d/2}\left(mr\right)^{\frac{4-d}{4}} K_{\frac{d-4}{2}}(mr)\,,\\ &
    J_{1,1}=\frac{1}{\sqrt{m r}}\Bigg[2^{\frac{d}{2}-3} \pi ^d \csc \left(\frac{\pi  d}{2}\right) \sqrt[4]{m r} \Big(4 \pi ^{3/2} \, _1\tilde{F}_2\left(\frac{1}{2};\frac{d}{2},\frac{3}{2};\frac{\sqrt{m r}}{4}\right)\\ &-\pi  2^d \Gamma \left(\frac{3}{2}-\frac{d}{2}\right) (m r)^{\frac{2-d}{4}} \, _1\tilde{F}_2\left(\frac{3}{2}-\frac{d}{2};2-\frac{d}{2},\frac{5}{2}-\frac{d}{2};\frac{\sqrt{m r}}{4}\right)\Big)+\pi ^{\frac{d+1}{2}} \Gamma \left(\frac{3}{2}-\frac{d}{2}\right)\Bigg]\,.
    \end{split}
\end{align}
\paragraph{Two-loop Fourier family.}
We consider the two-loop family
\begin{align}
G_{a_1,a_2,a_3,a_4,a_5}(\boldsymbol{r})
=
\int d^d k\, d^d k_1\,
\frac{e^{D_1}}{D_1^{a_1} D_2^{a_2} D_3^{a_3} D_4^{a_4} D_5^{a_5}},
\end{align}
where
\begin{align}
D_1 = i\,\boldsymbol{k}\!\cdot\!\boldsymbol{r},
\qquad
D_2 = i\,\boldsymbol{k}_1\!\cdot\!\boldsymbol{r},
\qquad
D_3 = \boldsymbol{k}^2,
\qquad
D_4 = \boldsymbol{k}_1^2 + m^2,
\qquad
D_5 = (\boldsymbol{k}_1-\boldsymbol{k})^2 + m^2.
\end{align}
After solving the IBP relations, the family reduces to $17$ master integrals. For our purposes, it is sufficient to consider the subset
\begin{align}
\mathcal{G}(x)
=
\bigl(
G_1(x),G_2(x),G_3(x),G_4(x),G_5(x),G_6(x)
\bigr)^T,
\end{align}
with
\begin{align}
G_1 &= G_{0,0,0,1,1},
\qquad
G_2 = G_{0,0,0,2,1},
\qquad
G_3 = G_{0,0,0,1,2},
\\
G_4 &= G_{0,0,-1,1,1},
\qquad
G_5 = G_{0,0,1,0,1},
\qquad
G_6 = G_{0,0,1,1,0}.
\end{align}
and we define the dimensionless variable
\begin{align}
x = (m r)^2,
\qquad r = |\boldsymbol{r}|.
\end{align}
The differential system satisfied by these master integrals is
\begin{align}
\partial_x \mathcal G(x) = \Delta_x\, \mathcal{G}(x),\quad \Delta_x =
\begin{pmatrix}
\dfrac{2-d}{x} & -\dfrac{1}{x} & -\dfrac{1}{x} & 0 & 0 & 0 \\[6pt]
\dfrac{(d-2)^2}{4x} & 0 & \dfrac{d-2}{2x} & \dfrac{1}{8} & 0 & 0 \\[6pt]
\dfrac{(d-2)^2}{4x} & \dfrac{d-2}{2x} & 0 & \dfrac{1}{8} & 0 & 0 \\[6pt]
\dfrac{2(d-3)}{x} & \dfrac{4}{x} & \dfrac{4}{x} & \dfrac{1-d}{x} & 0 & 0 \\[6pt]
0 & 0 & 0 & 0 & -\dfrac{d-2}{2x} & 0 \\[6pt]
0 & 0 & 0 & 0 & 0 & -\dfrac{d-2}{2x}
\end{pmatrix}.
\end{align}
By symmetry, one has
\begin{align}
G_2(x)=G_3(x),
\qquad
G_5(x)=G_6(x),
\end{align}
and therefore the system reduces to four coupled equations:
\begin{align}
\begin{split}
G_1'(x) &= \frac{2-d}{x}\,G_1(x) - \frac{2}{x}\,G_2(x), \\
G_2'(x) &= \frac{(d-2)^2}{4x}\,G_1(x) + \frac{d-2}{2x}\,G_2(x) + \frac{1}{8}\,G_4(x), \\
G_4'(x) &= \frac{2(d-3)}{x}\,G_1(x) + \frac{8}{x}\,G_2(x) + \frac{1-d}{x}\,G_4(x), \\
G_5'(x) &= \frac{2-d}{2x}\,G_5(x).
\end{split}
\end{align}
Eliminating $G_2$ and $G_4$ from the first three equations, one finds that $G_1$ satisfies the third-order differential equation
\begin{align}
2x^2 G_1'''(x) + 3d\,x\,G_1''(x) + \bigl(d^2-d-2x\bigr) G_1'(x) + (1-d) G_1(x) = 0.
\end{align}
Its general solution can be written as
\begin{align}
G_1(x) = c_1 u_1(x) + c_2 u_2(x) + c_3 u_3(x),
\end{align}
where
\begin{align}
\begin{split}
u_1(x) &= {}_1F_2\!\left(\frac{d-1}{2};\, d-1,\frac{d}{2};\, x\right), \\
u_2(x) &= x^{1-\frac{d}{2}}\,
{}_1F_2\!\left(\frac{1}{2};\, 2-\frac{d}{2},\frac{d}{2};\, x\right), \\
u_3(x) &= x^{2-d}\,
{}_1F_2\!\left(\frac{3-d}{2};\, 3-d,2-\frac{d}{2};\, x\right).
\end{split}
\end{align}
Consequently,
\begin{align}
\begin{split}
G_2(x) &= (2-d)\,G_1(x) - x\,G_1'(x), \\
G_4(x) &= 8\,G_2'(x) - \frac{2(d-2)^2}{x}\,G_1(x) - \frac{4(d-2)}{x}\,G_2(x), \\
G_5(x) &= c_4\,x^{1-\frac{d}{2}}.
\end{split}
\end{align}
Although $G_1$ can be identified with the square of the one-loop master integral discussed previously, let us assume that this relation is not known a priori and determine the constants $c_1,c_2,c_3$ directly from boundary data. To this end, we introduce the dimensionless variables
\begin{align}
\boldsymbol{s} = m \boldsymbol{r},
\qquad
x = \boldsymbol{s}^2,
\end{align}
and rescale the loop momenta according to
\begin{align}
\boldsymbol{q} = \frac{\boldsymbol{k}_1}{m},
\qquad
\boldsymbol{p} = \frac{\boldsymbol{k}_1-\boldsymbol{k}}{m}.
\end{align}
Then the dimensionless master integral can be written as
\begin{align}
G_1(x)
=
m^{4-2d} G_{0,0,0,1,1}(\boldsymbol{r})
=
\int d^d q\, d^d p\,
\frac{e^{i(\boldsymbol{q}-\boldsymbol{p})\cdot \boldsymbol{s}}}{(\boldsymbol{q}^2+1)(\boldsymbol{p}^2+1)}.
\end{align}
To determine the small-$x$ behavior of $G_1$, we use the method of regions in the limit $x \to 0$, or equivalently $s \to 0$. The relevant momentum scalings are
\begin{align}
\texttt{hard:}\quad |\boldsymbol{q}|,|\boldsymbol{p}| \sim \frac{1}{s},
\qquad
\texttt{soft:}\quad |\boldsymbol{q}|,|\boldsymbol{p}| \sim 1.
\end{align}
There are therefore four asymptotic regions.

\paragraph{\texttt{hard-hard} region.}
In this region, both propagators may be expanded as massless,
\begin{align}
\frac{1}{\boldsymbol{q}^2+1} \sim \frac{1}{\boldsymbol{q}^2},
\qquad
\frac{1}{\boldsymbol{p}^2+1} \sim \frac{1}{\boldsymbol{p}^2},
\end{align}
and hence
\begin{align}
G_1^{(hh)}(x)
&=
\int d^d q\,\frac{e^{i\boldsymbol{q}\cdot \boldsymbol{s}}}{\boldsymbol{q}^2}
\int d^d p\,\frac{e^{-i\boldsymbol{p}\cdot \boldsymbol{s}}}{\boldsymbol{p}^2}
\nonumber\\
&=
2^{2d-4}\pi^d
\Gamma\!\left(\frac{d}{2}-1\right)^2
x^{2-d}.
\end{align}

\paragraph{\texttt{hard-soft} and \texttt{soft-hard} regions.}
Consider, for instance, the region $|\boldsymbol{q}| \sim s^{-1}$ and $|\boldsymbol{p}| \sim 1$. The soft exponential is expanded as
\begin{align}
e^{-i\boldsymbol{p}\cdot \boldsymbol{s}}
=
1 - i\,\boldsymbol{p}\cdot \boldsymbol{s} - \frac{1}{2}(\boldsymbol{p}\cdot \boldsymbol{s})^2 + \cdots.
\end{align}
The odd terms vanish upon angular integration, and at leading order, we obtain
\begin{align}
G_1^{(hs)}(x)
&=
\int d^d q\,\frac{e^{i\boldsymbol{q}\cdot \boldsymbol{s}}}{\boldsymbol{q}^2}
\int d^d p\,\frac{1}{\boldsymbol{p}^2+1}.
\end{align}
Including the symmetric \texttt{soft-hard} contribution, one finds
\begin{align}
G_1^{(hs)}(x) + G_1^{(sh)}(x)
=
2^{d-1}\pi^d
\Gamma\!\left(\frac{d}{2}-1\right)
\Gamma\!\left(1-\frac{d}{2}\right)
x^{1-\frac{d}{2}}.
\end{align}

\paragraph{\texttt{soft-soft} region.}
In this case both loop momenta are of order unity, and the exponential can be expanded as
\begin{align}
e^{i(\boldsymbol{q}-\boldsymbol{p})\cdot \boldsymbol{s}} = 1 + \mathcal O(s)\,.
\end{align}
The leading term is therefore
\begin{align}
G_1^{(ss)}(x)
&=
\left(
\int d^d p\,\frac{1}{\boldsymbol{p}^2+1}
\right)^2
=
\pi^d \Gamma\!\left(1-\frac{d}{2}\right)^2.
\end{align}
Collecting the leading contributions from all regions, we obtain the asymptotic expansion
\begin{align}
G_1(x)
\sim
2^{2d-4}\pi^d
\Gamma\!\left(\frac{d}{2}-1\right)^2
x^{2-d}
+
2^{d-1}\pi^d
\Gamma\!\left(\frac{d}{2}-1\right)
\Gamma\!\left(1-\frac{d}{2}\right)
x^{1-\frac{d}{2}}
+
\pi^d
\Gamma\!\left(1-\frac{d}{2}\right)^2
+\cdots.
\end{align}
Matching this expansion against the Frobenius basis $\{u_1,u_2,u_3\}$ immediately yields
\begin{align}
\begin{split}
c_1 &= \pi^d \Gamma\!\left(1-\frac{d}{2}\right)^2, \\
c_2 &= 2^{d-1}\pi^d
\Gamma\!\left(\frac{d}{2}-1\right)
\Gamma\!\left(1-\frac{d}{2}\right), \\
c_3 &= 2^{2d-4}\pi^d
\Gamma\!\left(\frac{d}{2}-1\right)^2.
\end{split}
\end{align}
We consider another two-loop family that arises in our previous computations,
\begin{align}
Q_{a_1,a_2,a_3,a_4,a_5}(\boldsymbol{r})
=
\int d^d k\, d^d k_1\,
\frac{e^{D_1}}{D_1^{a_1} D_2^{a_2} D_3^{a_3} D_4^{a_4} D_5^{a_5}},
\end{align}
where
\begin{align}
D_1 = i\,\boldsymbol{k}\!\cdot\!\boldsymbol{r},
\qquad
D_2 = i\,\boldsymbol{k}_1\!\cdot\!\boldsymbol{r},
\qquad
D_3 = \boldsymbol{k}^2+m^2,
\qquad
D_4 = \boldsymbol{k}_1^2 ,
\qquad
D_5 = (\boldsymbol{k}_1-\boldsymbol{k})^2.
\end{align}
Solving IBP identities, we were left with 10 master integrals. However, for our purpose, we need three of them,
\begin{align}
    \mathcal {Q}:Q_{0,0,0,1,1},\,Q_{0,0,1,1,1},\, Q_{0,0,2,1,1}.
\end{align}
The differential equation satisfied by them is given by,
\begin{align}
    \partial_{x} \mathcal{Q}(x)=\Lambda_x \mathcal{Q}(x), \qquad \Lambda_x =\left(
\begin{array}{ccc}
 \frac{2-d}{x} & 0 & 0 \\
 0 & \frac{3-d}{x} & -\frac{1}{x} \\
 \frac{1}{4} & \frac{2 (d-7) d-x+24}{4 x} & \frac{d-4}{2 x} \\
\end{array}
\right)\,.
\end{align}
Therefore, the system reduces to three coupled differential equations,
\begin{align}
    \begin{split}
        &Q_1'(x)=\frac{2-d}{x}Q_1,\\ &
        Q_2'(x)=\frac{3-d}{x}Q_1-\frac{1}{x}Q_2,\\ &
        Q_3'(x)=\frac{1}{4}Q_1(x)+\frac{2 (d-7) d-x+24}{4 x}Q_2+\frac{d-4}{2x}Q_3.
    \end{split}
\end{align}
As we see, $Q_1$ is decoupled and can be integrated to,
\begin{align}
    Q_1(x)=c_1 x^{2-d}.
\end{align}
After massaging the equations a bit, we find that $Q_2$ satisfies the following differential equation,
\begin{align}
    4 x^2 Q_2'''(x)+(4-6 d)\, x\, Q_2''(x)+(x-2 (d-2) d) Q_2'(x)+(d-2) Q_2(x)=0\,.
\end{align}
 The general solution takes the form,
 \begin{align}
     Q_2(x)=x^{\frac{2-d}{4}}\left(c_1 I_{\frac{d-2}{2}}\left(\sqrt{x}\right)+c_2 K_{\frac{d-2}{2}}(\sqrt{x})\right)+c_3 x^{3-d}\, _1F_2\left(1;4-d,3-\frac{d}{2};\frac{x}{4}\right)\,.
 \end{align}
 Similarly, $Q_3=(3-d)Q_1-x\, Q_2'(x)$. The constants $c_i$ can be fixed by studying the asymptotic expansion of the integrals, as we showed previously.
\section{Useful Feynman integrals} \label{ch1:app:E}
We have seen that there is a direct connection between PN diagrams and the one-loop scalar Feynman diagrams, Fig.~(\ref{fig16}). Below, we list master integrals that will be useful in our context. We take the integrals from \cite{Levi:2011eq}.
\begingroup
\small
\begin{eqnarray}
\label{ch1:tensorfourierindentity}
I^i\equiv\rmint\frac{d^d\bf{k}}{(2\pi)^d}\frac{k^ie^{i\bf{k}\cdot\bf{r}}}{({\bf{k}}^2)^\alpha}&=&\frac{i}{(4\pi)^{d/2}}\frac{\Gamma(d/2-\alpha+1)}{\Gamma(\alpha)}\left(\frac{{\bf{r}}^2}{4}\right)^{\alpha-d/2-1/2}n^i,\\
I^{ij}\equiv\rmint\frac{d^d\bf{k}}{(2\pi)^d}\frac{k^ik^je^{i\bf{k}\cdot\bf{r}}}{({\bf{k}}^2)^\alpha}&=&\frac{1}{(4\pi)^{d/2}}\frac{\Gamma(d/2-\alpha+1)}{\Gamma(\alpha)}\left(\frac{{\bf{r}}^2}{4}\right)^{\alpha-d/2-1}\left(\frac{1}{2}\delta^{ij}+(\alpha-1-d/2)n^in^j\right),\\
I^{ijl}\equiv\rmint\frac{d^d\bf{k}}{(2\pi)^d}\frac{k^ik^jk^le^{i\bf{k}\cdot\bf{r}}}{({\bf{k}}^2)^\alpha}&=&\frac{i}{(4\pi)^{d/2}}\frac{\Gamma(d/2-\alpha+2)}{\Gamma(\alpha)}\left(\frac{{\bf{r}}^2}{4}\right)^{\alpha-d/2-3/2}\nonumber\\
&&\quad
\times\left(\frac{1}{2}\left(\delta^{ij}n^l+\delta^{il}n^j+\delta^{jl}n^i\right)+(\alpha-d/2-2)n^in^jn^l\right),\\
I^{ijlm}\textstyle{\equiv\rmint\frac{d^d\bf{k}}{(2\pi)^d}\frac{k^ik^jk^lk^me^{i\bf{k}\cdot\bf{r}}}{({\bf{k}}^2)^\alpha}}&=&\textstyle{\frac{1}{(4\pi)^{d/2}}\frac{\Gamma(d/2-\alpha+2)}{\Gamma(\alpha)}\left(\frac{{\bf{r}}^2}{4}\right)^{\alpha-d/2-2}\left(\frac{1}{4}\left(\delta^{ij}\delta^{lm}+\delta^{il}\delta^{jm}+\delta^{im}\delta^{jl}\right)\right.} \\
&&
\textstyle{+\frac{\alpha-d/2-2}{2}\left(\delta^{ij}n^ln^m+\delta^{il}n^jn^m+\delta^{im}n^jn^l+\delta^{jl}n^in^m+\delta^{jm}n^in^l+\delta^{lm}n^in^j\right) }\nonumber\\
&&\left.\quad
+(\alpha-d/2-2)(\alpha-d/2-3)n^in^jn^ln^m\right).\nonumber
\end{eqnarray}
\endgroup
We use the d-dimensional master formula for one-loop scalar integrals given by,
\begin{align}
\begin{split}
   & J\equiv\rmint \frac{d^d\bf{k}}{(2\pi)^d}\frac{1}{\left[{\bf{k}}^2\right]^\alpha\left[({\bf{k}-\bf{q}})^2\right]^\beta}=  \frac{1}{(4\pi)^{d/2}}\frac{\Gamma(\alpha+\beta-d/2)}{\Gamma(\alpha)\Gamma(\beta)}\frac{\Gamma(d/2-\alpha)\Gamma(d/2-\beta)}{\Gamma(d-\alpha-\beta)}\left(q^2\right)^{d/2-\alpha-\beta}.\label{ch1:eq:1loop}
    \end{split}
\end{align}
The d-dimensional master formula for one-loop tensor integrals is taken from \cite{Levi:2011eq}. 
Similarly, one can also derive the following d-dimensional formulas for the one-loop tensor integrals: 
\begin{align}
\begin{split}
J^i\equiv\rmint \frac{d^d\bf{k}}{(2\pi)^d}\frac{k^i}{\left[{\bf{k}}^2\right]^\alpha\left[({\bf{k}-\bf{q}})^2\right]^\beta}=\frac{1}{(4\pi)^{d/2}}\frac{\Gamma(\alpha+\beta-d/2)}{\Gamma(\alpha)\Gamma(\beta)}\frac{\Gamma(d/2-\alpha+1)\Gamma(d/2-\beta)}{\Gamma(d-\alpha-\beta+1)}\\ \left(q^2\right)^{d/2-\alpha-\beta}q^i\end{split}
\end{align}
\begin{align}
\begin{split}
J^{ij}\equiv\rmint \frac{d^d\bf{k}}{(2\pi)^d}\frac{k^ik^j}{\left[{\bf{k}}^2\right]^\alpha\left[({\bf{k}-\bf{q}})^2\right]^\beta}=\frac{1}{(4\pi)^{d/2}}\frac{\Gamma(\alpha+\beta-d/2-1)}{\Gamma(\alpha)\Gamma(\beta)}\frac{\Gamma(d/2-\alpha+1)\Gamma(d/2-\beta)}{\Gamma(d-\alpha-\beta+2)}\left(q^2\right)^{d/2-\alpha-\beta}\\
\times\left(\frac{d/2-\beta}{2}q^2\delta^{ij}+(\alpha+\beta-d/2-1)(d/2-\alpha+1)q^iq^j\right),\end{split}
\end{align}
\begin{align}
\begin{split}
J^{ijl}\equiv\rmint \frac{d^d\bf{k}}{(2\pi)^d}\frac{k^ik^jk^l}{\left[{\bf{k}}^2\right]^\alpha\left[({\bf{k}-\bf{q}})^2\right]^\beta}=\frac{1}{(4\pi)^{d/2}}\frac{\Gamma(\alpha+\beta-d/2-1)}{\Gamma(\alpha)\Gamma(\beta)}\frac{\Gamma(d/2-\alpha+2)\Gamma(d/2-\beta)}{\Gamma(d-\alpha-\beta+3)}\left(q^2\right)^{d/2-\alpha-\beta} \\ 
\times\left(\frac{d/2-\beta}{2}q^2\left(\delta^{ij}q^l+\delta^{il}q^j+\delta^{jl}q^i\right)\right.  \left. 
+(\alpha+\beta-d/2-1)(d/2-\alpha+2)q^iq^jq^l\right),
\end{split}
\end{align}

\begin{align}
\begin{split}
&J^{ijlm}\equiv\rmint \frac{d^d\bf{k}}{(2\pi)^d}\frac{k^ik^jk^lk^m}{\left[{\bf{k}}^2\right]^\alpha\left[({\bf{k}-\bf{q}})^2\right]^\beta}=\frac{1}{(4\pi)^{d/2}}\frac{\Gamma(\alpha+\beta-d/2-2)}{\Gamma(\alpha)\Gamma(\beta)}\\& \frac{\Gamma(d/2-\alpha+2)\Gamma(d/2-\beta)}{\Gamma(d-\alpha-\beta+4)}\left(q^2\right)^{d/2-\alpha-\beta}\,\,\,\,
\times\left(\frac{(d/2-\beta)(d/2-\beta+1)}{4}q^4\left(\delta^{ij}\delta^{lm}+\delta^{il}\delta^{jm}+\delta^{jl}\delta^{im}\right)\right.\\
& \textstyle{
+(\alpha+\beta-d/2-2)(d/2-\alpha+2)\frac{d/2-\beta}{2}q^2 
\times\left(\delta^{ij}q^lq^m+\delta^{il}q^jq^m+\delta^{im}q^jq^l+\delta^{jl}q^iq^m+\delta^{jm}q^iq^l+\delta^{lm}q^iq^j\right)}\\
&\left.\,\,\,\,\,\,\,\,\,
+(\alpha+\beta-d/2-2)(\alpha+\beta-d/2-1)(d/2-\alpha+2)(d/2-\alpha+3)\right.
\left.
\times q^iq^jq^lq^m\right).
\end{split}
\end{align}
For more details on the Feynman integral, especially the massive integrals, can be found in \cite{smirnov}.

\bibliographystyle{utphysmodb}
\bibliography{ref}

@article{Anastasiou:1999ui,
    author = "Anastasiou, C. and Glover, E. W. Nigel and Oleari, C.",
    title = "{Scalar one loop integrals using the negative dimension approach}",
    eprint = "hep-ph/9907494",
    archivePrefix = "arXiv",
    reportNumber = "DTP-99-80",
    doi = "10.1016/S0550-3213(99)00637-9",
    journal = "Nucl. Phys. B",
    volume = "572",
    pages = "307--360",
    year = "2000"
}

@article{Goldberger:2004jt,
    author = "Goldberger, Walter D. and Rothstein, Ira Z.",
    title = "{An Effective field theory of gravity for extended objects}",
    eprint = "hep-th/0409156",
    archivePrefix = "arXiv",
    reportNumber = "UCSD-PTH-04-17, CMU-HEP-04-06",
    doi = "10.1103/PhysRevD.73.104029",
    journal = "Phys. Rev. D",
    volume = "73",
    pages = "104029",
    year = "2006"
}

@article{Levi:2018nxp,
    author = "Levi, Mich\`ele",
    title = "{Effective Field Theories of Post-Newtonian Gravity: A comprehensive review}",
    eprint = "1807.01699",
    archivePrefix = "arXiv",
    primaryClass = "hep-th",
    doi = "10.1088/1361-6633/ab12bc",
    journal = "Rept. Prog. Phys.",
    volume = "83",
    number = "7",
    pages = "075901",
    year = "2020"
}

@article{Mogull:2020sak,
    author = "Mogull, Gustav and Plefka, Jan and Steinhoff, Jan",
    title = "{Classical black hole scattering from a worldline quantum field theory}",
    eprint = "2010.02865",
    archivePrefix = "arXiv",
    primaryClass = "hep-th",
    reportNumber = "UUITP-37/20, HU-EP-20/22-RTG",
    doi = "10.1007/JHEP02(2021)048",
    journal = "JHEP",
    volume = "02",
    pages = "048",
    year = "2021"
}

@article{Kol:2007bc,
    author = "Kol, Barak and Smolkin, Michael",
    title = "{Non-Relativistic Gravitation: From Newton to Einstein and Back}",
    eprint = "0712.4116",
    archivePrefix = "arXiv",
    primaryClass = "hep-th",
    doi = "10.1088/0264-9381/25/14/145011",
    journal = "Class. Quant. Grav.",
    volume = "25",
    pages = "145011",
    year = "2008"
}

@article{Kuntz:2019zef,
    author = "Kuntz, Adrien and Piazza, Federico and Vernizzi, Filippo",
    title = "{Effective field theory for gravitational radiation in scalar-tensor gravity}",
    eprint = "1902.04941",
    archivePrefix = "arXiv",
    primaryClass = "gr-qc",
    doi = "10.1088/1475-7516/2019/05/052",
    journal = "JCAP",
    volume = "05",
    pages = "052",
    year = "2019"
}

@book{Burgess:2020tbq,
    author = "Burgess, C. P.",
    title = "{Introduction to Effective Field Theory}",
    doi = "10.1017/9781139048040",
    isbn = "978-1-139-04804-0, 978-0-521-19547-8",
    publisher = "Cambridge University Press",
    month = "12",
    year = "2020"
}

@article{Cardoso:2018tly,
    author = "Cardoso, Vitor and Dias, \'Oscar J. C. and Hartnett, Gavin S. and Middleton, Matthew and Pani, Paolo and Santos, Jorge E.",
    title = "{Constraining the mass of dark photons and axion-like particles through black-hole superradiance}",
    eprint = "1801.01420",
    archivePrefix = "arXiv",
    primaryClass = "gr-qc",
    doi = "10.1088/1475-7516/2018/03/043",
    journal = "JCAP",
    volume = "03",
    pages = "043",
    year = "2018"
}

@article{LIGOScientific:2016aoc,
    author = "Abbott, B. P. and others",
    collaboration = "LIGO Scientific, Virgo",
    title = "{Observation of Gravitational Waves from a Binary Black Hole Merger}",
    eprint = "1602.03837",
    archivePrefix = "arXiv",
    primaryClass = "gr-qc",
    reportNumber = "LIGO-P150914",
    doi = "10.1103/PhysRevLett.116.061102",
    journal = "Phys. Rev. Lett.",
    volume = "116",
    number = "6",
    pages = "061102",
    year = "2016"
}

@article{LIGOScientific:2016vlm,
    author = "Abbott, B. P. and others",
    collaboration = "LIGO Scientific, Virgo",
    title = "{Properties of the Binary Black Hole Merger GW150914}",
    eprint = "1602.03840",
    archivePrefix = "arXiv",
    primaryClass = "gr-qc",
    reportNumber = "LIGO-P1500218",
    doi = "10.1103/PhysRevLett.116.241102",
    journal = "Phys. Rev. Lett.",
    volume = "116",
    number = "24",
    pages = "241102",
    year = "2016"
}

@article{LIGOScientific:2016sjg,
    author = "Abbott, B. P. and others",
    collaboration = "LIGO Scientific, Virgo",
    title = "{GW151226: Observation of Gravitational Waves from a 22-Solar-Mass Binary Black Hole Coalescence}",
    eprint = "1606.04855",
    archivePrefix = "arXiv",
    primaryClass = "gr-qc",
    reportNumber = "LIGO-P151226",
    doi = "10.1103/PhysRevLett.116.241103",
    journal = "Phys. Rev. Lett.",
    volume = "116",
    number = "24",
    pages = "241103",
    year = "2016"
}

@article{LIGOScientific:2017bnn,
    author = "Abbott, Benjamin P. and others",
    collaboration = "LIGO Scientific, VIRGO",
    title = "{GW170104: Observation of a 50-Solar-Mass Binary Black Hole Coalescence at Redshift 0.2}",
    eprint = "1706.01812",
    archivePrefix = "arXiv",
    primaryClass = "gr-qc",
    reportNumber = "LIGO-P170104",
    doi = "10.1103/PhysRevLett.118.221101",
    journal = "Phys. Rev. Lett.",
    volume = "118",
    number = "22",
    pages = "221101",
    year = "2017",
    note = "[Erratum: Phys.Rev.Lett. 121, 129901 (2018)]"
}

@article{Porto:2016pyg,
    author = "Porto, Rafael A.",
    title = "{The effective field theorist\textquoteright{}s approach to gravitational dynamics}",
    eprint = "1601.04914",
    archivePrefix = "arXiv",
    primaryClass = "hep-th",
    doi = "10.1016/j.physrep.2016.04.003",
    journal = "Phys. Rept.",
    volume = "633",
    pages = "1--104",
    year = "2016"
}

@article{Goldberger:2006bd,
    author = "Goldberger, Walter D. and Rothstein, Ira Z.",
    title = "{Towers of Gravitational Theories}",
    eprint = "hep-th/0605238",
    archivePrefix = "arXiv",
    doi = "10.1142/S0218271806009698",
    journal = "Gen. Rel. Grav.",
    volume = "38",
    pages = "1537--1546",
    year = "2006"
}

@article{Blanchet:2004ek,
    author = "Blanchet, Luc and Damour, Thibault and Esposito-Farese, Gilles and Iyer, Bala R.",
    title = "{Gravitational radiation from inspiralling compact binaries completed at the third post-Newtonian order}",
    eprint = "gr-qc/0406012",
    archivePrefix = "arXiv",
    doi = "10.1103/PhysRevLett.93.091101",
    journal = "Phys. Rev. Lett.",
    volume = "93",
    pages = "091101",
    year = "2004"
}

@article{Damour:2000ni,
    author = "Damour, Thibault and Jaranowski, Piotr and Schaefer, Gerhard",
    title = "{Equivalence between the ADM-Hamiltonian and the harmonic coordinates approaches to the third postNewtonian dynamics of compact binaries}",
    eprint = "gr-qc/0010040",
    archivePrefix = "arXiv",
    doi = "10.1103/PhysRevD.63.044021",
    journal = "Phys. Rev. D",
    volume = "63",
    pages = "044021",
    year = "2001",
    note = "[Erratum: Phys.Rev.D 66, 029901 (2002)]"
}

@article{Itoh:2003fy,
    author = "Itoh, Yousuke and Futamase, Toshifumi",
    title = "{New derivation of a third postNewtonian equation of motion for relativistic compact binaries without ambiguity}",
    eprint = "gr-qc/0310028",
    archivePrefix = "arXiv",
    doi = "10.1103/PhysRevD.68.121501",
    journal = "Phys. Rev. D",
    volume = "68",
    pages = "121501",
    year = "2003"
}

@article{PhysRevD.12.329,
  title = {Gravitational two-body problem with arbitrary masses, spins, and quadrupole moments},
  author = {Barker, B. M. and O'Connell, R. F.},
  journal = {Phys. Rev. D},
  volume = {12},
  issue = {2},
  pages = {329--335},
  numpages = {0},
  year = {1975},
  month = {Jul},
  publisher = {American Physical Society},
  doi = {10.1103/PhysRevD.12.329},
  url = {https://link.aps.org/doi/10.1103/PhysRevD.12.329}
}

@article{PhysRevD.2.1428,
  title = {Derivation of the Equations of Motion of a Gyroscope from the Quantum Theory of Gravitation},
  author = {Barker, B. M. and O'Connell, R. F.},
  journal = {Phys. Rev. D},
  volume = {2},
  issue = {8},
  pages = {1428--1435},
  numpages = {0},
  year = {1970},
  month = {Oct},
  publisher = {American Physical Society},
  doi = {10.1103/PhysRevD.2.1428},
  url = {https://link.aps.org/doi/10.1103/PhysRevD.2.1428}
}

@article{Kidder:1992fr,
    author = "Kidder, Lawrence E. and Will, Clifford M. and Wiseman, Alan G.",
    title = "{Spin effects in the inspiral of coalescing compact binaries}",
    eprint = "gr-qc/9211025",
    archivePrefix = "arXiv",
    reportNumber = "PRINT-92-0520",
    doi = "10.1103/PhysRevD.47.R4183",
    journal = "Phys. Rev. D",
    volume = "47",
    number = "10",
    pages = "R4183--R4187",
    year = "1993"
}

@article{Tagoshi:2000zg,
    author = "Tagoshi, Hideyuki and Ohashi, Akira and Owen, Benjamin J.",
    title = "{Gravitational field and equations of motion of spinning compact binaries to 2.5 postNewtonian order}",
    eprint = "gr-qc/0010014",
    archivePrefix = "arXiv",
    doi = "10.1103/PhysRevD.63.044006",
    journal = "Phys. Rev. D",
    volume = "63",
    pages = "044006",
    year = "2001"
}

@article{Faye:2006gx,
    author = "Faye, Guillaume and Blanchet, Luc and Buonanno, Alessandra",
    title = "{Higher-order spin effects in the dynamics of compact binaries. I. Equations of motion}",
    eprint = "gr-qc/0605139",
    archivePrefix = "arXiv",
    doi = "10.1103/PhysRevD.74.104033",
    journal = "Phys. Rev. D",
    volume = "74",
    pages = "104033",
    year = "2006"
}

@article{Blanchet:2006gy,
    author = "Blanchet, Luc and Buonanno, Alessandra and Faye, Guillaume",
    title = "{Higher-order spin effects in the dynamics of compact binaries. II. Radiation field}",
    eprint = "gr-qc/0605140",
    archivePrefix = "arXiv",
    reportNumber = "APC-06-25",
    doi = "10.1103/PhysRevD.81.089901",
    journal = "Phys. Rev. D",
    volume = "74",
    pages = "104034",
    year = "2006",
    note = "[Erratum: Phys.Rev.D 75, 049903 (2007), Erratum: Phys.Rev.D 81, 089901 (2010)]"
}

@article{Steinhoff:2007mb,
    author = "Steinhoff, Jan and Hergt, Steven and Schaefer, Gerhard",
    title = "{On the next-to-leading order gravitational spin(1)-spin(2) dynamics}",
    eprint = "0712.1716",
    archivePrefix = "arXiv",
    primaryClass = "gr-qc",
    doi = "10.1103/PhysRevD.77.081501",
    journal = "Phys. Rev. D",
    volume = "77",
    pages = "081501",
    year = "2008"
}

@article{Steinhoff:2008ji,
    author = "Steinhoff, Jan and Hergt, Steven and Schaefer, Gerhard",
    title = "{Spin-squared Hamiltonian of next-to-leading order gravitational interaction}",
    eprint = "0809.2200",
    archivePrefix = "arXiv",
    primaryClass = "gr-qc",
    doi = "10.1103/PhysRevD.78.101503",
    journal = "Phys. Rev. D",
    volume = "78",
    pages = "101503",
    year = "2008"
}

@article{Hergt:2008jn,
    author = "Hergt, Steven and Schaefer, Gerhard",
    title = "{Higher-order-in-spin interaction Hamiltonians for binary black holes from Poincare invariance}",
    eprint = "0809.2208",
    archivePrefix = "arXiv",
    primaryClass = "gr-qc",
    doi = "10.1103/PhysRevD.78.124004",
    journal = "Phys. Rev. D",
    volume = "78",
    pages = "124004",
    year = "2008"
}

@article{Hergt:2010pa,
    author = "Hergt, Steven and Steinhoff, Jan and Schaefer, Gerhard",
    title = "{Reduced Hamiltonian for next-to-leading order Spin-Squared Dynamics of General Compact Binaries}",
    eprint = "1002.2093",
    archivePrefix = "arXiv",
    primaryClass = "gr-qc",
    doi = "10.1088/0264-9381/27/13/135007",
    journal = "Class. Quant. Grav.",
    volume = "27",
    pages = "135007",
    year = "2010"
}

@article{Porto:2010zg,
    author = "Porto, Rafael A. and Ross, Andreas and Rothstein, Ira Z.",
    title = "{Spin induced multipole moments for the gravitational wave flux from binary inspirals to third Post-Newtonian order}",
    eprint = "1007.1312",
    archivePrefix = "arXiv",
    primaryClass = "gr-qc",
    doi = "10.1088/1475-7516/2011/03/009",
    journal = "JCAP",
    volume = "03",
    pages = "009",
    year = "2011"
}

@article{Levi:2011eq,
    author = "Levi, Michele",
    title = "{Binary dynamics from spin1-spin2 coupling at fourth post-Newtonian order}",
    eprint = "1107.4322",
    archivePrefix = "arXiv",
    primaryClass = "gr-qc",
    doi = "10.1103/PhysRevD.85.064043",
    journal = "Phys. Rev. D",
    volume = "85",
    pages = "064043",
    year = "2012"
}

@article{Levi:2015uxa,
    author = "Levi, Michele and Steinhoff, Jan",
    title = "{Next-to-next-to-leading order gravitational spin-orbit coupling via the effective field theory for spinning objects in the post-Newtonian scheme}",
    eprint = "1506.05056",
    archivePrefix = "arXiv",
    primaryClass = "gr-qc",
    doi = "10.1088/1475-7516/2016/01/011",
    journal = "JCAP",
    volume = "01",
    pages = "011",
    year = "2016"
}

@article{Levi:2015ixa,
    author = "Levi, Michele and Steinhoff, Jan",
    title = "{Next-to-next-to-leading order gravitational spin-squared potential via the effective field theory for spinning objects in the post-Newtonian scheme}",
    eprint = "1506.05794",
    archivePrefix = "arXiv",
    primaryClass = "gr-qc",
    doi = "10.1088/1475-7516/2016/01/008",
    journal = "JCAP",
    volume = "01",
    pages = "008",
    year = "2016"
}

@article{Levi:2014sba,
    author = "Levi, Michele and Steinhoff, Jan",
    title = "{Equivalence of ADM Hamiltonian and Effective Field Theory approaches at next-to-next-to-leading order spin1-spin2 coupling of binary inspirals}",
    eprint = "1408.5762",
    archivePrefix = "arXiv",
    primaryClass = "gr-qc",
    doi = "10.1088/1475-7516/2014/12/003",
    journal = "JCAP",
    volume = "12",
    pages = "003",
    year = "2014"
}

@article{Lehner:2014asa,
    author = "Lehner, Luis and Pretorius, Frans",
    title = "{Numerical Relativity and Astrophysics}",
    eprint = "1405.4840",
    archivePrefix = "arXiv",
    primaryClass = "astro-ph.HE",
    doi = "10.1146/annurev-astro-081913-040031",
    journal = "Ann. Rev. Astron. Astrophys.",
    volume = "52",
    pages = "661--694",
    year = "2014"
}

@article{LIGOScientific:2014oec,
    author = "Aasi, J. and others",
    collaboration = "LIGO Scientific, VIRGO, NINJA-2",
    title = "{The NINJA-2 project: Detecting and characterizing gravitational waveforms modelled using numerical binary black hole simulations}",
    eprint = "1401.0939",
    archivePrefix = "arXiv",
    primaryClass = "gr-qc",
    reportNumber = "LIGO-P1300199",
    doi = "10.1088/0264-9381/31/11/115004",
    journal = "Class. Quant. Grav.",
    volume = "31",
    pages = "115004",
    year = "2014"
}

@ARTICLE{2022arXiv220103593W,
       author = {{Weinzierl}, Stefan},
        title = "{Feynman Integrals}",
      journal = {arXiv e-prints},
         year = 2022,
        month = jan,
          eid = {arXiv:2201.03593},
        pages = {arXiv:2201.03593},
          doi = {10.48550/arXiv.2201.03593},
archivePrefix = {arXiv},
       eprint = {2201.03593},
 primaryClass = {hep-th},
       adsurl = {https://ui.adsabs.harvard.edu/abs/2022arXiv220103593W}
}

@article{Dlapa:2021npj,
    author = {Dlapa, Christoph and K\"alin, Gregor and Liu, Zhengwen and Porto, Rafael A.},
    title = "{Dynamics of binary systems to fourth Post-Minkowskian order from the effective field theory approach}",
    eprint = "2106.08276",
    archivePrefix = "arXiv",
    primaryClass = "hep-th",
    reportNumber = "DESY 21-093, DESY-21-093, MPP-2021-83",
    doi = "10.1016/j.physletb.2022.137203",
    journal = "Phys. Lett. B",
    volume = "831",
    pages = "137203",
    year = "2022"
}

@article{Huang:2018pbu,
    author = "Huang, Junwu and Johnson, Matthew C. and Sagunski, Laura and Sakellariadou, Mairi and Zhang, Jun",
    title = "{Prospects for axion searches with Advanced LIGO through binary mergers}",
    eprint = "1807.02133",
    archivePrefix = "arXiv",
    primaryClass = "hep-ph",
    doi = "10.1103/PhysRevD.99.063013",
    journal = "Phys. Rev. D",
    volume = "99",
    number = "6",
    pages = "063013",
    year = "2019"
}

@inproceedings{Adams:2022pbo,
    author = "Adams, C. B. and others",
    title = "{Axion Dark Matter}",
    booktitle = "{Snowmass 2021}",
    eprint = "2203.14923",
    archivePrefix = "arXiv",
    primaryClass = "hep-ex",
    month = "3",
    year = "2022"
}

@article{Fukuda:2021drn,
    author = "Fukuda, Hajime and Shirai, Satoshi",
    title = "{Detection of QCD axion dark matter by coherent scattering}",
    eprint = "2112.13536",
    archivePrefix = "arXiv",
    primaryClass = "hep-ph",
    reportNumber = "IPMU21-0089",
    doi = "10.1103/PhysRevD.105.095030",
    journal = "Phys. Rev. D",
    volume = "105",
    number = "9",
    pages = "095030",
    year = "2022"
}

@article{CAPP:2020utb,
    author = "Kwon, Ohjoon and others",
    collaboration = "CAPP",
    title = "{First Results from an Axion Haloscope at CAPP around 10.7  $\mu$eV}",
    eprint = "2012.10764",
    archivePrefix = "arXiv",
    primaryClass = "hep-ex",
    doi = "10.1103/PhysRevLett.126.191802",
    journal = "Phys. Rev. Lett.",
    volume = "126",
    number = "19",
    pages = "191802",
    year = "2021"
}

@article{Sakhelashvili:2021eid,
    author = "Sakhelashvili, Otari",
    title = "{Consistency of the dual formulation of axion solutions to the strong CP problem}",
    eprint = "2110.03386",
    archivePrefix = "arXiv",
    primaryClass = "hep-th",
    doi = "10.1103/PhysRevD.105.085020",
    journal = "Phys. Rev. D",
    volume = "105",
    number = "8",
    pages = "085020",
    year = "2022"
}

@article{Peccei:2006as,
    author = "Peccei, R. D.",
    editor = "Kuster, Markus and Raffelt, Georg and Beltran, Berta",
    title = "{The Strong CP problem and axions}",
    eprint = "hep-ph/0607268",
    archivePrefix = "arXiv",
    doi = "10.1007/978-3-540-73518-2_1",
    journal = "Lect. Notes Phys.",
    volume = "741",
    pages = "3--17",
    year = "2008"
}

@article{RevModPhys.82.557,
  title = {Axions and the strong $CP$ problem},
  author = {Kim, Jihn E. and Carosi, Gianpaolo},
  journal = {Rev. Mod. Phys.},
  volume = {82},
  issue = {1},
  pages = {557--601},
  numpages = {0},
  year = {2010},
  month = {Mar},
  publisher = {American Physical Society},
  doi = {10.1103/RevModPhys.82.557},
  url = {https://link.aps.org/doi/10.1103/RevModPhys.82.557}
}

@article{Zhang:2021mks,
    author = "Zhang, Jun and Lyu, Zhenwei and Huang, Junwu and Johnson, Matthew C. and Sagunski, Laura and Sakellariadou, Mairi and Yang, Huan",
    title = "{First Constraints on Nuclear Coupling of Axionlike Particles from the Binary Neutron Star Gravitational Wave Event GW170817}",
    eprint = "2105.13963",
    archivePrefix = "arXiv",
    primaryClass = "hep-ph",
    reportNumber = "Imperial/TP/2021/JZ/01, KCL-PH-TH-2021-25, CERN-TH-2021-061,
  LIGO-P2100161, CERN-TH-2021-061, LIGO-P2100161",
    doi = "10.1103/PhysRevLett.127.161101",
    journal = "Phys. Rev. Lett.",
    volume = "127",
    number = "16",
    pages = "161101",
    year = "2021"
}

@inproceedings{Cicoli:2013ana,
    author = "Cicoli, Michele",
    title = "{Axion-like Particles from String Compactifications}",
    booktitle = "{9th Patras Workshop on Axions, WIMPs and WISPs}",
    eprint = "1309.6988",
    archivePrefix = "arXiv",
    primaryClass = "hep-th",
    doi = "10.3204/DESY-PROC-2013-04/cicoli_michele",
    pages = "235--242",
    year = "2013"
}

@article{Svrcek:2006hf,
    author = "Svrcek, Peter",
    title = "{Cosmological Constant and Axions in String Theory}",
    eprint = "hep-th/0607086",
    archivePrefix = "arXiv",
    reportNumber = "SLAC-PUB-11957",
    month = "7",
    year = "2006"
}

@article{Jakobsen:2022fcj,
    author = "Jakobsen, Gustav Uhre and Mogull, Gustav",
    title = "{Conservative and Radiative Dynamics of Spinning Bodies at Third Post-Minkowskian Order Using Worldline Quantum Field Theory}",
    eprint = "2201.07778",
    archivePrefix = "arXiv",
    primaryClass = "hep-th",
    reportNumber = "HU-EP-22/03-RTG",
    doi = "10.1103/PhysRevLett.128.141102",
    journal = "Phys. Rev. Lett.",
    volume = "128",
    number = "14",
    pages = "141102",
    year = "2022"
}

@article{Jakobsen:2021zvh,
    author = "Jakobsen, Gustav Uhre and Mogull, Gustav and Plefka, Jan and Steinhoff, Jan",
    title = "{SUSY in the sky with gravitons}",
    eprint = "2109.04465",
    archivePrefix = "arXiv",
    primaryClass = "hep-th",
    reportNumber = "HU-EP-21/28-RTG",
    doi = "10.1007/JHEP01(2022)027",
    journal = "JHEP",
    volume = "01",
    pages = "027",
    year = "2022"
}

@article{Jakobsen:2021lvp,
    author = "Jakobsen, Gustav Uhre and Mogull, Gustav and Plefka, Jan and Steinhoff, Jan",
    title = "{Gravitational Bremsstrahlung and Hidden Supersymmetry of Spinning Bodies}",
    eprint = "2106.10256",
    archivePrefix = "arXiv",
    primaryClass = "hep-th",
    reportNumber = "HU-EP-21/15-RTG",
    doi = "10.1103/PhysRevLett.128.011101",
    journal = "Phys. Rev. Lett.",
    volume = "128",
    number = "1",
    pages = "011101",
    year = "2022"
}

@article{Jakobsen:2021smu,
    author = "Jakobsen, Gustav Uhre and Mogull, Gustav and Plefka, Jan and Steinhoff, Jan",
    title = "{Classical Gravitational Bremsstrahlung from a Worldline Quantum Field Theory}",
    eprint = "2101.12688",
    archivePrefix = "arXiv",
    primaryClass = "gr-qc",
    reportNumber = "HU-EP-21/03-RTG",
    doi = "10.1103/PhysRevLett.126.201103",
    journal = "Phys. Rev. Lett.",
    volume = "126",
    number = "20",
    pages = "201103",
    year = "2021"
}

@article{Jakobsen:2022psy,
    author = "Jakobsen, Gustav Uhre and Mogull, Gustav and Plefka, Jan and Sauer, Benjamin",
    title = "{All things retarded: radiation-reaction in worldline quantum field theory}",
    eprint = "2207.00569",
    archivePrefix = "arXiv",
    primaryClass = "hep-th",
    reportNumber = "HU-EP-22/24-RTG",
    doi = "10.1007/JHEP10(2022)128",
    journal = "JHEP",
    volume = "10",
    pages = "128",
    year = "2022"
}

@article{PhysRevLett.38.1440,
  title = {$\mathrm{CP}$ Conservation in the Presence of Pseudoparticles},
  author = {Peccei, R. D. and Quinn, Helen R.},
  journal = {Phys. Rev. Lett.},
  volume = {38},
  issue = {25},
  pages = {1440--1443},
  numpages = {0},
  year = {1977},
  month = {Jun},
  publisher = {American Physical Society},
  doi = {10.1103/PhysRevLett.38.1440},
  url = {https://link.aps.org/doi/10.1103/PhysRevLett.38.1440}
}

@article{PhysRevD.16.1791,
  title = {Constraints imposed by $\mathrm{CP}$ conservation in the presence of pseudoparticles},
  author = {Peccei, R. D. and Quinn, Helen R.},
  journal = {Phys. Rev. D},
  volume = {16},
  issue = {6},
  pages = {1791--1797},
  numpages = {0},
  year = {1977},
  month = {Sep},
  publisher = {American Physical Society},
  doi = {10.1103/PhysRevD.16.1791},
  url = {https://link.aps.org/doi/10.1103/PhysRevD.16.1791}
}

@article{PhysRevLett.40.279,
  title = {Problem of Strong $P$ and $T$ Invariance in the Presence of Instantons},
  author = {Wilczek, F.},
  journal = {Phys. Rev. Lett.},
  volume = {40},
  issue = {5},
  pages = {279--282},
  numpages = {0},
  year = {1978},
  month = {Jan},
  publisher = {American Physical Society},
  doi = {10.1103/PhysRevLett.40.279},
  url = {https://link.aps.org/doi/10.1103/PhysRevLett.40.279}
}

@article{Baker:2006ts,
    author = "Baker, C. A. and others",
    title = "{An Improved experimental limit on the electric dipole moment of the neutron}",
    eprint = "hep-ex/0602020",
    archivePrefix = "arXiv",
    doi = "10.1103/PhysRevLett.97.131801",
    journal = "Phys. Rev. Lett.",
    volume = "97",
    pages = "131801",
    year = "2006"
}

@article{GrillidiCortona:2015jxo,
    author = "Grilli di Cortona, Giovanni and Hardy, Edward and Pardo Vega, Javier and Villadoro, Giovanni",
    title = "{The QCD axion, precisely}",
    eprint = "1511.02867",
    archivePrefix = "arXiv",
    primaryClass = "hep-ph",
    doi = "10.1007/JHEP01(2016)034",
    journal = "JHEP",
    volume = "01",
    pages = "034",
    year = "2016"
}

@article{Kamionkowski:2014zda,
    author = "Kamionkowski, Marc and Pradler, Josef and Walker, Devin G. E.",
    title = "{Dark energy from the string axiverse}",
    eprint = "1409.0549",
    archivePrefix = "arXiv",
    primaryClass = "hep-ph",
    reportNumber = "SLAC-PUB-16085",
    doi = "10.1103/PhysRevLett.113.251302",
    journal = "Phys. Rev. Lett.",
    volume = "113",
    number = "25",
    pages = "251302",
    year = "2014"
}

@article{PhysRevLett.120.151301,
  title = {Search for Invisible Axion Dark Matter with the Axion Dark Matter Experiment},
  author = {Du, N. and Force, N. and Khatiwada, R. and Lentz, E. and Ottens, R. and Rosenberg, L. J and Rybka, G. and Carosi, G. and Woollett, N. and Bowring, D. and Chou, A. S. and Sonnenschein, A. and Wester, W. and Boutan, C. and Oblath, N. S. and Bradley, R. and Daw, E. J. and Dixit, A. V. and Clarke, J. and O'Kelley, S. R. and Crisosto, N. and Gleason, J. R. and Jois, S. and Sikivie, P. and Stern, I. and Sullivan, N. S. and Tanner, D. B and Hilton, G. C.},
  collaboration = {ADMX Collaboration},
  journal = {Phys. Rev. Lett.},
  volume = {120},
  issue = {15},
  pages = {151301},
  numpages = {5},
  year = {2018},
  month = {Apr},
  publisher = {American Physical Society},
  doi = {10.1103/PhysRevLett.120.151301},
  url = {https://link.aps.org/doi/10.1103/PhysRevLett.120.151301}
}

@article{Dlapa:2022lmu,
    author = {Dlapa, Christoph and K\"alin, Gregor and Liu, Zhengwen and Neef, Jakob and Porto, Rafael A.},
    title = "{Radiation Reaction and Gravitational Waves at Fourth Post-Minkowskian Order}",
    eprint = "2210.05541",
    archivePrefix = "arXiv",
    primaryClass = "hep-th",
    doi = "10.1103/PhysRevLett.130.101401",
    journal = "Phys. Rev. Lett.",
    volume = "130",
    number = "10",
    pages = "101401",
    year = "2023"
}

@article{Dlapa:2021vgp,
    author = {Dlapa, Christoph and K\"alin, Gregor and Liu, Zhengwen and Porto, Rafael A.},
    title = "{Conservative Dynamics of Binary Systems at Fourth Post-Minkowskian Order in the Large-Eccentricity Expansion}",
    eprint = "2112.11296",
    archivePrefix = "arXiv",
    primaryClass = "hep-th",
    reportNumber = "DESY 21-226",
    doi = "10.1103/PhysRevLett.128.161104",
    journal = "Phys. Rev. Lett.",
    volume = "128",
    number = "16",
    pages = "161104",
    year = "2022"
}

@article{Passarino:1978jh,
    author = "Passarino, G. and Veltman, M. J. G.",
    title = "{One Loop Corrections for e+ e- Annihilation Into mu+ mu- in the Weinberg Model}",
    reportNumber = "Print-79-0284 (UTRECHT)",
    doi = "10.1016/0550-3213(79)90234-7",
    journal = "Nucl. Phys. B",
    volume = "160",
    pages = "151--207",
    year = "1979"
}

@inbook{nastase_2019, place={Cambridge}, title={The Optical Theorem and the Cutting Rules}, DOI={10.1017/9781108624992.023}, booktitle={Introduction to Quantum Field Theory}, publisher={Cambridge University Press}, author={Nastase, Horatiu}, year={2019}, pages={188–196}}

@inbook{smirnov, title={Feynman Integral Calculus}, DOI={https://doi.org/10.1007/3-540-30611-0}, booktitle={Feynman Integral Calculus}, publisher={Springer Berlin, Heidelberg}, author={Vladimir A. Smirnov}, year={2006}}

@article{Porto:2017dgs,
    author = "Porto, Rafael A. and Rothstein, Ira Z.",
    title = "{Apparent ambiguities in the post-Newtonian expansion for binary systems}",
    eprint = "1703.06433",
    archivePrefix = "arXiv",
    primaryClass = "gr-qc",
    doi = "10.1103/PhysRevD.96.024062",
    journal = "Phys. Rev. D",
    volume = "96",
    number = "2",
    pages = "024062",
    year = "2017"
}

@article{Maia:2017gxn,
    author = "Maia, Natalia T. and Galley, Chad R. and Leibovich, Adam K. and Porto, Rafael A.",
    title = "{Radiation reaction for spinning bodies in effective field theory I: Spin-orbit effects}",
    eprint = "1705.07934",
    archivePrefix = "arXiv",
    primaryClass = "gr-qc",
    doi = "10.1103/PhysRevD.96.084064",
    journal = "Phys. Rev. D",
    volume = "96",
    number = "8",
    pages = "084064",
    year = "2017"
}

@article{Maia:2017yok,
    author = "Maia, Natalia T. and Galley, Chad R. and Leibovich, Adam K. and Porto, Rafael A.",
    title = "{Radiation reaction for spinning bodies in effective field theory II: Spin-spin effects}",
    eprint = "1705.07938",
    archivePrefix = "arXiv",
    primaryClass = "gr-qc",
    doi = "10.1103/PhysRevD.96.084065",
    journal = "Phys. Rev. D",
    volume = "96",
    number = "8",
    pages = "084065",
    year = "2017"
}

@article{Foffa:2019yfl,
    author = "Foffa, Stefano and Porto, Rafael A. and Rothstein, Ira and Sturani, Riccardo",
    title = "{Conservative dynamics of binary systems to fourth Post-Newtonian order in the EFT approach II: Renormalized Lagrangian}",
    eprint = "1903.05118",
    archivePrefix = "arXiv",
    primaryClass = "gr-qc",
    doi = "10.1103/PhysRevD.100.024048",
    journal = "Phys. Rev. D",
    volume = "100",
    number = "2",
    pages = "024048",
    year = "2019"
}

@article{Kalin:2022hph,
    author = {K\"alin, Gregor and Neef, Jakob and Porto, Rafael A.},
    title = "{Radiation-reaction in the Effective Field Theory approach to Post-Minkowskian dynamics}",
    eprint = "2207.00580",
    archivePrefix = "arXiv",
    primaryClass = "hep-th",
    reportNumber = "DESY-22-109, DESY 22-109",
    doi = "10.1007/JHEP01(2023)140",
    journal = "JHEP",
    volume = "01",
    pages = "140",
    year = "2023"
}

@article{Dlapa:2023hsl,
    author = {Dlapa, Christoph and K\"alin, Gregor and Liu, Zhengwen and Porto, Rafael A.},
    title = "{Bootstrapping the relativistic two-body problem}",
    eprint = "2304.01275",
    archivePrefix = "arXiv",
    primaryClass = "hep-th",
    reportNumber = "DESY 23-041",
    month = "4",
    year = "2023"
}

@article{Porto:2008jj,
    author = "Porto, Rafael A and Rothstein, Ira Z.",
    title = "{Next to Leading Order Spin(1)Spin(1) Effects in the Motion of Inspiralling Compact Binaries}",
    eprint = "0804.0260",
    archivePrefix = "arXiv",
    primaryClass = "gr-qc",
    doi = "10.1103/PhysRevD.78.044013",
    journal = "Phys. Rev. D",
    volume = "78",
    pages = "044013",
    year = "2008",
    note = "[Erratum: Phys.Rev.D 81, 029905 (2010)]"
}

@article{Porto:2008tb,
    author = "Porto, Rafael A. and Rothstein, Ira Z.",
    title = "{Spin(1)Spin(2) Effects in the Motion of Inspiralling Compact Binaries at Third Order in the Post-Newtonian Expansion}",
    eprint = "0802.0720",
    archivePrefix = "arXiv",
    primaryClass = "gr-qc",
    doi = "10.1103/PhysRevD.78.044012",
    journal = "Phys. Rev. D",
    volume = "78",
    pages = "044012",
    year = "2008",
    note = "[Erratum: Phys.Rev.D 81, 029904 (2010)]"
}

@inproceedings{Porto:2007pw,
    author = "Porto, Rafael A. and Sturani, Riccardo",
    title = "{Scalar gravity: Post-Newtonian corrections via an effective field theory approach}",
    booktitle = "{Les Houches Summer School - Session 86: Particle Physics and Cosmology: The Fabric of Spacetime}",
    eprint = "gr-qc/0701105",
    archivePrefix = "arXiv",
    month = "1",
    year = "2007"
}

@inproceedings{Porto:2007px,
    author = "Porto, Rafael A.",
    title = "{New results at 3PN via an effective field theory of gravity}",
    booktitle = "{11th Marcel Grossmann Meeting on General Relativity}",
    eprint = "gr-qc/0701106",
    archivePrefix = "arXiv",
    doi = "10.1142/9789812834300_0442",
    pages = "2493--2496",
    month = "1",
    year = "2007"
}

@article{Mandal:2022nty,
    author = "Mandal, Manoj K. and Mastrolia, Pierpaolo and Patil, Raj and Steinhoff, Jan",
    title = "{Gravitational spin-orbit Hamiltonian at NNNLO in the post-Newtonian framework}",
    eprint = "2209.00611",
    archivePrefix = "arXiv",
    primaryClass = "hep-th",
    reportNumber = "HU-EP-22/28-RTG",
    doi = "10.1007/JHEP03(2023)130",
    journal = "JHEP",
    volume = "03",
    pages = "130",
    year = "2023"
}

@article{Mandal:2022ufb,
    author = "Mandal, Manoj K. and Mastrolia, Pierpaolo and Patil, Raj and Steinhoff, Jan",
    title = "{Gravitational Quadratic-in-Spin Hamiltonian at NNNLO in the post-Newtonian framework}",
    eprint = "2210.09176",
    archivePrefix = "arXiv",
    primaryClass = "hep-th",
    reportNumber = "HU-EP-22/33-RTG",
    month = "10",
    year = "2022"
}

@article{Patil:2020dme,
    author = "Patil, Raj",
    title = "{EFT approach to general relativity: correction to EIH Lagrangian due to electromagnetic charge}",
    eprint = "2009.11107",
    archivePrefix = "arXiv",
    primaryClass = "gr-qc",
    doi = "10.1007/s10714-020-02748-1",
    journal = "Gen. Rel. Grav.",
    volume = "52",
    number = "9",
    pages = "95",
    year = "2020"
}

@article{Gupta:2022spq,
    author = "Gupta, Pawan Kumar",
    title = "{Binary dynamics from Einstein-Maxwell theory at second post-Newtonian order using effective field theory}",
    eprint = "2205.11591",
    archivePrefix = "arXiv",
    primaryClass = "gr-qc",
    month = "5",
    year = "2022"
}

@article{Blanchet:2023sbv,
    author = "Blanchet, Luc and Faye, Guillaume and Henry, Quentin and Larrouturou, Fran\c{c}ois and Trestini, David",
    title = "{Gravitational Wave Flux and Quadrupole Modes from Quasi-Circular Non-Spinning Compact Binaries to the Fourth Post-Newtonian Order}",
    eprint = "2304.11186",
    archivePrefix = "arXiv",
    primaryClass = "gr-qc",
    reportNumber = "DESY-23-044",
    month = "4",
    year = "2023"
}

@article{Blanchet:2023bwj,
    author = "Blanchet, Luc and Faye, Guillaume and Henry, Quentin and Larrouturou, Fran\c{c}ois and Trestini, David",
    title = "{Gravitational-Wave Phasing of Compact Binary Systems to the Fourth-and-a-Half post-Newtonian Order}",
    eprint = "2304.11185",
    archivePrefix = "arXiv",
    primaryClass = "gr-qc",
    reportNumber = "DESY-23-043",
    month = "4",
    year = "2023"
}

@article{Larrouturou:2021gqo,
    author = "Larrouturou, Fran\c{c}ois and Blanchet, Luc and Henry, Quentin and Faye, Guillaume",
    title = "{The quadrupole moment of compact binaries to the fourth post-Newtonian order: II. Dimensional regularization and renormalization}",
    eprint = "2110.02243",
    archivePrefix = "arXiv",
    primaryClass = "gr-qc",
    reportNumber = "DESY-22-006",
    doi = "10.1088/1361-6382/ac5ba0",
    journal = "Class. Quant. Grav.",
    volume = "39",
    number = "11",
    pages = "115008",
    year = "2022"
}

@article{Blanchet:2004bb,
    author = "Blanchet, Luc and Iyer, Bala R.",
    title = "{Hadamard regularization of the third post-Newtonian gravitational wave generation of two point masses}",
    eprint = "gr-qc/0409094",
    archivePrefix = "arXiv",
    doi = "10.1103/PhysRevD.71.024004",
    journal = "Phys. Rev. D",
    volume = "71",
    pages = "024004",
    year = "2005"
}

@article{PhysRevLett.40.223,
  title = {A New Light Boson?},
  author = {Weinberg, Steven},
  journal = {Phys. Rev. Lett.},
  volume = {40},
  issue = {4},
  pages = {223--226},
  numpages = {0},
  year = {1978},
  month = {Jan},
  publisher = {American Physical Society},
  doi = {10.1103/PhysRevLett.40.223},
  url = {https://link.aps.org/doi/10.1103/PhysRevLett.40.223}
}

@article{Clowe:2006eq,
    author = "Clowe, Douglas and Bradac, Marusa and Gonzalez, Anthony H. and Markevitch, Maxim and Randall, Scott W. and Jones, Christine and Zaritsky, Dennis",
    title = "{A direct empirical proof of the existence of dark matter}",
    eprint = "astro-ph/0608407",
    archivePrefix = "arXiv",
    reportNumber = "SLAC-PUB-12078",
    doi = "10.1086/508162",
    journal = "Astrophys. J. Lett.",
    volume = "648",
    pages = "L109--L113",
    year = "2006"
}

@article{Preskill:1982cy,
    author = "Preskill, John and Wise, Mark B. and Wilczek, Frank",
    editor = "Srednicki, M. A.",
    title = "{Cosmology of the Invisible Axion}",
    reportNumber = "HUTP-82-A048, NSF-ITP-82-103",
    doi = "10.1016/0370-2693(83)90637-8",
    journal = "Phys. Lett. B",
    volume = "120",
    pages = "127--132",
    year = "1983"
}

@article{Gorghetto:2018ocs,
    author = "Gorghetto, Marco and Villadoro, Giovanni",
    title = "{Topological Susceptibility and QCD Axion Mass: QED and NNLO corrections}",
    eprint = "1812.01008",
    archivePrefix = "arXiv",
    primaryClass = "hep-ph",
    doi = "10.1007/JHEP03(2019)033",
    journal = "JHEP",
    volume = "03",
    pages = "033",
    year = "2019"
}

@article{DiLuzio:2021pxd,
    author = "Di Luzio, Luca and Gavela, Belen and Quilez, Pablo and Ringwald, Andreas",
    title = "{An even lighter QCD axion}",
    eprint = "2102.00012",
    archivePrefix = "arXiv",
    primaryClass = "hep-ph",
    reportNumber = "DESY-21-010, DESY 21-010, IFT-UAM/CSIC-20-143, FTUAM-20-21",
    doi = "10.1007/JHEP05(2021)184",
    journal = "JHEP",
    volume = "05",
    pages = "184",
    year = "2021"
}

@article{Berezhiani:2000gh,
    author = "Berezhiani, Zurab and Gianfagna, Leonida and Giannotti, Maurizio",
    title = "{Strong CP problem and mirror world: The Weinberg-Wilczek axion revisited}",
    eprint = "hep-ph/0009290",
    archivePrefix = "arXiv",
    reportNumber = "DFAQ-TH-2000-04",
    doi = "10.1016/S0370-2693(00)01392-7",
    journal = "Phys. Lett. B",
    volume = "500",
    pages = "286--296",
    year = "2001"
}

@article{Hsu:2004mf,
    author = "Hsu, S. D. H. and Sannino, F.",
    title = "{New solutions to the strong CP problem}",
    eprint = "hep-ph/0408319",
    archivePrefix = "arXiv",
    doi = "10.1016/j.physletb.2004.11.040",
    journal = "Phys. Lett. B",
    volume = "605",
    pages = "369--375",
    year = "2005"
}

@article{Fukuda:2015ana,
    author = "Fukuda, Hajime and Harigaya, Keisuke and Ibe, Masahiro and Yanagida, Tsutomu T.",
    title = "{Model of visible QCD axion}",
    eprint = "1504.06084",
    archivePrefix = "arXiv",
    primaryClass = "hep-ph",
    reportNumber = "IPMU15-0050",
    doi = "10.1103/PhysRevD.92.015021",
    journal = "Phys. Rev. D",
    volume = "92",
    number = "1",
    pages = "015021",
    year = "2015"
}

@article{Dimopoulos:2016lvn,
    author = "Dimopoulos, Savas and Hook, Anson and Huang, Junwu and Marques-Tavares, Gustavo",
    title = "{A collider observable QCD axion}",
    eprint = "1606.03097",
    archivePrefix = "arXiv",
    primaryClass = "hep-ph",
    doi = "10.1007/JHEP11(2016)052",
    journal = "JHEP",
    volume = "11",
    pages = "052",
    year = "2016"
}

@article{Gherghetta:2016fhp,
    author = "Gherghetta, Tony and Nagata, Natsumi and Shifman, Mikhail",
    title = "{A Visible QCD Axion from an Enlarged Color Group}",
    eprint = "1604.01127",
    archivePrefix = "arXiv",
    primaryClass = "hep-ph",
    reportNumber = "UMN-TH-3522-16, FTPI-MINN-16-11",
    doi = "10.1103/PhysRevD.93.115010",
    journal = "Phys. Rev. D",
    volume = "93",
    number = "11",
    pages = "115010",
    year = "2016"
}

@article{Agrawal:2017ksf,
    author = "Agrawal, Prateek and Howe, Kiel",
    title = "{Factoring the Strong CP Problem}",
    eprint = "1710.04213",
    archivePrefix = "arXiv",
    primaryClass = "hep-ph",
    reportNumber = "FERMILAB-PUB-17-500-T",
    doi = "10.1007/JHEP12(2018)029",
    journal = "JHEP",
    volume = "12",
    pages = "029",
    year = "2018"
}

@article{Gupta:2020vxb,
    author = "Gupta, R. S. and Khoze, V. V. and Spannowsky, M.",
    title = "{Small instantons and the strong CP problem in composite Higgs models}",
    eprint = "2012.00017",
    archivePrefix = "arXiv",
    primaryClass = "hep-ph",
    doi = "10.1103/PhysRevD.104.075011",
    journal = "Phys. Rev. D",
    volume = "104",
    number = "7",
    pages = "075011",
    year = "2021"
}

@article{PhysRevLett.43.103,
  title = {Weak-Interaction Singlet and Strong $\mathrm{CP}$ Invariance},
  author = {Kim, Jihn E.},
  journal = {Phys. Rev. Lett.},
  volume = {43},
  issue = {2},
  pages = {103--107},
  numpages = {0},
  year = {1979},
  month = {Jul},
  publisher = {American Physical Society},
  doi = {10.1103/PhysRevLett.43.103},
  url = {https://link.aps.org/doi/10.1103/PhysRevLett.43.103}
}

@article{Kim:1998va,
    author = "Kim, Jihn E.",
    title = "{Constraints on very light axions from cavity experiments}",
    eprint = "hep-ph/9802220",
    archivePrefix = "arXiv",
    reportNumber = "HUTP-98-A009",
    doi = "10.1103/PhysRevD.58.055006",
    journal = "Phys. Rev. D",
    volume = "58",
    pages = "055006",
    year = "1998"
}

@article{DiLuzio:2020wdo,
    author = "Di Luzio, Luca and Giannotti, Maurizio and Nardi, Enrico and Visinelli, Luca",
    title = "{The landscape of QCD axion models}",
    eprint = "2003.01100",
    archivePrefix = "arXiv",
    primaryClass = "hep-ph",
    reportNumber = "DESY 20-036, DESY-20-036",
    doi = "10.1016/j.physrep.2020.06.002",
    journal = "Phys. Rept.",
    volume = "870",
    pages = "1--117",
    year = "2020"
}

@article{Conlon:2006tq,
    author = "Conlon, Joseph P.",
    title = "{The QCD axion and moduli stabilisation}",
    eprint = "hep-th/0602233",
    archivePrefix = "arXiv",
    reportNumber = "DAMTP-2006-17",
    doi = "10.1088/1126-6708/2006/05/078",
    journal = "JHEP",
    volume = "05",
    pages = "078",
    year = "2006"
}

@article{Chakraborty:2021fkp,
    author = "Chakraborty, Sabyasachi and Jung, Tae Hyun and Okui, Takemichi",
    title = "{Composite neutrinos and the QCD axion: Baryogenesis, dark matter, small Dirac neutrino masses, and vanishing neutron electric dipole moment}",
    eprint = "2108.04293",
    archivePrefix = "arXiv",
    primaryClass = "hep-ph",
    reportNumber = "KEK-TH-2341",
    doi = "10.1103/PhysRevD.105.015024",
    journal = "Phys. Rev. D",
    volume = "105",
    number = "1",
    pages = "015024",
    year = "2022"
}

@article{Harigaya:2019qnl,
    author = "Harigaya, Keisuke and Leedom, Jacob M.",
    title = "{QCD Axion Dark Matter from a Late Time Phase Transition}",
    eprint = "1910.04163",
    archivePrefix = "arXiv",
    primaryClass = "hep-ph",
    doi = "10.1007/JHEP06(2020)034",
    journal = "JHEP",
    volume = "06",
    pages = "034",
    year = "2020"
}

@article{Hiramatsu:2010yu,
    author = "Hiramatsu, Takashi and Kawasaki, Masahiro and Sekiguchi, Toyokazu and Yamaguchi, Masahide and Yokoyama, Jun'ichi",
    title = "{Improved estimation of radiated axions from cosmological axionic strings}",
    eprint = "1012.5502",
    archivePrefix = "arXiv",
    primaryClass = "hep-ph",
    reportNumber = "IPMU10-0229, RESCEU-29-10, YITP-10-111",
    doi = "10.1103/PhysRevD.83.123531",
    journal = "Phys. Rev. D",
    volume = "83",
    pages = "123531",
    year = "2011"
}

@inproceedings{Kokeyama:2020dkg,
    author = "Kokeyama, Keiko",
    collaboration = "KAGRA",
    title = "{Observing the Universe from Underground Gravitational Wave Telescope KAGRA}",
    booktitle = "{3rd World Summit on Exploring the Dark Side of the Universe}",
    pages = "41--48",
    year = "2020"
}

@article{AbhishekChowdhuri:2022ora,
    author = "Chowdhuri, Abhishek and Bhattacharyya, Arpan",
    title = "{Study of eccentric binaries in Horndeski gravity}",
    eprint = "2203.09917",
    archivePrefix = "arXiv",
    primaryClass = "gr-qc",
    doi = "10.1103/PhysRevD.106.064046",
    journal = "Phys. Rev. D",
    volume = "106",
    number = "6",
    pages = "064046",
    year = "2022"
}

@article{Kosower:2018adc,
    author = "Kosower, David A. and Maybee, Ben and O'Connell, Donal",
    title = "{Amplitudes, Observables, and Classical Scattering}",
    eprint = "1811.10950",
    archivePrefix = "arXiv",
    primaryClass = "hep-th",
    doi = "10.1007/JHEP02(2019)137",
    journal = "JHEP",
    volume = "02",
    pages = "137",
    year = "2019"
}

@article{Goldberger:2009qd,
    author = "Goldberger, Walter D. and Ross, Andreas",
    title = "{Gravitational radiative corrections from effective field theory}",
    eprint = "0912.4254",
    archivePrefix = "arXiv",
    primaryClass = "gr-qc",
    doi = "10.1103/PhysRevD.81.124015",
    journal = "Phys. Rev. D",
    volume = "81",
    pages = "124015",
    year = "2010"
}

@article{Ross:2012fc,
    author = "Ross, Andreas",
    title = "{Multipole expansion at the level of the action}",
    eprint = "1202.4750",
    archivePrefix = "arXiv",
    primaryClass = "gr-qc",
    doi = "10.1103/PhysRevD.85.125033",
    journal = "Phys. Rev. D",
    volume = "85",
    pages = "125033",
    year = "2012"
}

@article{Porto:2012as,
    author = "Porto, Rafael A. and Ross, Andreas and Rothstein, Ira Z.",
    title = "{Spin induced multipole moments for the gravitational wave amplitude from binary inspirals to 2.5 Post-Newtonian order}",
    eprint = "1203.2962",
    archivePrefix = "arXiv",
    primaryClass = "gr-qc",
    doi = "10.1088/1475-7516/2012/09/028",
    journal = "JCAP",
    volume = "09",
    pages = "028",
    year = "2012"
}

@article{Goldberger:2012kf,
    author = "Goldberger, Walter D. and Ross, Andreas and Rothstein, Ira Z.",
    title = "{Black hole mass dynamics and renormalization group evolution}",
    eprint = "1211.6095",
    archivePrefix = "arXiv",
    primaryClass = "hep-th",
    doi = "10.1103/PhysRevD.89.124033",
    journal = "Phys. Rev. D",
    volume = "89",
    number = "12",
    pages = "124033",
    year = "2014"
}

@article{Galley:2015kus,
    author = "Galley, Chad R. and Leibovich, Adam K. and Porto, Rafael A. and Ross, Andreas",
    title = "{Tail effect in gravitational radiation reaction: Time nonlocality and renormalization group evolution}",
    eprint = "1511.07379",
    archivePrefix = "arXiv",
    primaryClass = "gr-qc",
    doi = "10.1103/PhysRevD.93.124010",
    journal = "Phys. Rev. D",
    volume = "93",
    pages = "124010",
    year = "2016"
}

@article{Goldberger:2022rqf,
    author = "Goldberger, Walter D.",
    title = "{Effective Field Theory for Compact Binary Dynamics}",
    eprint = "2212.06677",
    archivePrefix = "arXiv",
    primaryClass = "hep-th",
    month = "12",
    year = "2022"
}

@inproceedings{Goldberger:2022ebt,
    author = "Goldberger, Walter D.",
    title = "{Effective field theories of gravity and compact binary dynamics: A Snowmass 2021 whitepaper}",
    booktitle = "{Snowmass 2021}",
    eprint = "2206.14249",
    archivePrefix = "arXiv",
    primaryClass = "hep-th",
    month = "6",
    year = "2022"
}

@article{Goldberger:2020fot,
    author = "Goldberger, Walter D. and Li, Jingping and Rothstein, Ira Z.",
    title = "{Non-conservative effects on spinning black holes from world-line effective field theory}",
    eprint = "2012.14869",
    archivePrefix = "arXiv",
    primaryClass = "hep-th",
    doi = "10.1007/JHEP06(2021)053",
    journal = "JHEP",
    volume = "06",
    pages = "053",
    year = "2021"
}

@article{Goldberger:2017ogt,
    author = "Goldberger, Walter D. and Li, Jingping and Prabhu, Siddharth G.",
    title = "{Spinning particles, axion radiation, and the classical double copy}",
    eprint = "1712.09250",
    archivePrefix = "arXiv",
    primaryClass = "hep-th",
    doi = "10.1103/PhysRevD.97.105018",
    journal = "Phys. Rev. D",
    volume = "97",
    number = "10",
    pages = "105018",
    year = "2018"
}

@article{Goldberger:2005cd,
    author = "Goldberger, Walter D. and Rothstein, Ira Z.",
    title = "{Dissipative effects in the worldline approach to black hole dynamics}",
    eprint = "hep-th/0511133",
    archivePrefix = "arXiv",
    doi = "10.1103/PhysRevD.73.104030",
    journal = "Phys. Rev. D",
    volume = "73",
    pages = "104030",
    year = "2006"
}

@misc{fdraw,
  author = {Alec Aivazis},
  howpublished = "\url{https://feynman.aivazis.com/}",
}

@article{Yoshida:2017cjl,
    author = "Yoshida, Daiske and Soda, Jiro",
    title = "{Exploring the string axiverse and parity violation in gravity with gravitational waves}",
    eprint = "1708.09592",
    archivePrefix = "arXiv",
    primaryClass = "gr-qc",
    reportNumber = "KOBE-COSMO-17-10",
    doi = "10.1142/S0218271818500967",
    journal = "Int. J. Mod. Phys. D",
    volume = "27",
    number = "09",
    pages = "1850096",
    year = "2018"
}

@article{Machado:2018nqk,
    author = "Machado, Camila S. and Ratzinger, Wolfram and Schwaller, Pedro and Stefanek, Ben A.",
    title = "{Audible Axions}",
    eprint = "1811.01950",
    archivePrefix = "arXiv",
    primaryClass = "hep-ph",
    reportNumber = "MITP/18-107",
    doi = "10.1007/JHEP01(2019)053",
    journal = "JHEP",
    volume = "01",
    pages = "053",
    year = "2019"
}

@article{Machado:2019xuc,
    author = "Machado, Camila S. and Ratzinger, Wolfram and Schwaller, Pedro and Stefanek, Ben A.",
    title = "{Gravitational wave probes of axionlike particles}",
    eprint = "1912.01007",
    archivePrefix = "arXiv",
    primaryClass = "hep-ph",
    reportNumber = "MITP/19-083",
    doi = "10.1103/PhysRevD.102.075033",
    journal = "Phys. Rev. D",
    volume = "102",
    number = "7",
    pages = "075033",
    year = "2020"
}

@article{Zhang:2017srh,
    author = "Zhang, Xing and Liu, Tan and Zhao, Wen",
    title = "{Gravitational radiation from compact binary systems in screened modified gravity}",
    eprint = "1702.08752",
    archivePrefix = "arXiv",
    primaryClass = "gr-qc",
    doi = "10.1103/PhysRevD.95.104027",
    journal = "Phys. Rev. D",
    volume = "95",
    number = "10",
    pages = "104027",
    year = "2017"
}

@article{Ishii:2022lwc,
    author = "Ishii, Takaaki and Kaku, Youka and Murata, Keiju",
    title = "{Energy extraction from AdS black holes via superradiance}",
    eprint = "2207.03123",
    archivePrefix = "arXiv",
    primaryClass = "hep-th",
    reportNumber = "RUP-22-14",
    doi = "10.1007/JHEP10(2022)024",
    journal = "JHEP",
    volume = "10",
    pages = "024",
    year = "2022"
}

@article{Hill:2015vma,
    author = "Hill, Christopher T.",
    title = "{Axion Induced Oscillating Electric Dipole Moment of the Electron}",
    eprint = "1508.04083",
    archivePrefix = "arXiv",
    primaryClass = "hep-ph",
    reportNumber = "FERMILAB-PUB-15-350-T",
    doi = "10.1103/PhysRevD.93.025007",
    journal = "Phys. Rev. D",
    volume = "93",
    number = "2",
    pages = "025007",
    year = "2016"
}

@article{Balkin:2022qer,
    author = "Balkin, Reuven and Serra, Javi and Springmann, Konstantin and Stelzl, Stefan and Weiler, Andreas",
    title = "{White dwarfs as a probe of light QCD axions}",
    eprint = "2211.02661",
    archivePrefix = "arXiv",
    primaryClass = "hep-ph",
    reportNumber = "IFT-UAM/CSIC-22-136",
    month = "11",
    year = "2022"
}

@article{Zhang:2019eid,
    author = "Zhang, Jun and Yang, Huan",
    title = "{Dynamic Signatures of Black Hole Binaries with Superradiant Clouds}",
    eprint = "1907.13582",
    archivePrefix = "arXiv",
    primaryClass = "gr-qc",
    reportNumber = "Imperial/TP/2019/JZ/02",
    doi = "10.1103/PhysRevD.101.043020",
    journal = "Phys. Rev. D",
    volume = "101",
    number = "4",
    pages = "043020",
    year = "2020"
}

@article{Dar:2018dra,
    author = "Dar, Furqan and De Rham, Claudia and Deskins, J. Tate and Giblin, John T. and Tolley, Andrew J.",
    title = "{Scalar Gravitational Radiation from Binaries: Vainshtein Mechanism in Time-dependent Systems}",
    eprint = "1808.02165",
    archivePrefix = "arXiv",
    primaryClass = "hep-th",
    reportNumber = "Imperial/TP/2018/CdR/04",
    doi = "10.1088/1361-6382/aaf5e8",
    journal = "Class. Quant. Grav.",
    volume = "36",
    number = "2",
    pages = "025008",
    year = "2019"
}

@book{poissonwill,
    author = "Eric Poisson, Clifford M. Will",
    title = "{Gravity:
Newtonian, Post-Newtonian, Relativistic}",
    publisher = "Cambridge University Press",
    doi = "https://doi.org/10.1063/PT.3.2951",
    ISBN = "9781107032866",
    month = "5",
    year = "2014"
}

@article{Dyadina:2018ryl,
    author = "Dyadina, P. I. and Avdeev, N. A. and Alexeyev, S. O.",
    title = "{Horndeski gravity without screening in binary pulsars}",
    eprint = "1811.05393",
    archivePrefix = "arXiv",
    primaryClass = "astro-ph.HE",
    doi = "10.1093/mnras/sty3094",
    journal = "Mon. Not. Roy. Astron. Soc.",
    volume = "483",
    number = "1",
    pages = "947--963",
    year = "2019"
}

@article{Zhang:2018prg,
    author = "Zhang, Xing and Zhao, Wen and Liu, Tan and Lin, Kai and Zhang, Chao and Zhao, Xiang and Zhang, Shaojun and Zhu, Tao and Wang, Anzhong",
    title = "{Angular momentum loss for eccentric compact binary in screened modified gravity}",
    eprint = "1811.00339",
    archivePrefix = "arXiv",
    primaryClass = "gr-qc",
    doi = "10.1088/1475-7516/2019/01/019",
    journal = "JCAP",
    volume = "01",
    pages = "019",
    year = "2019"
}

@article{Saffer:2018jmx,
    author = "Saffer, Alexander and Yunes, Nicolas",
    title = "{Angular momentum loss for a binary system in Einstein-\AE{}ther theory}",
    eprint = "1807.08049",
    archivePrefix = "arXiv",
    primaryClass = "gr-qc",
    doi = "10.1103/PhysRevD.98.124015",
    journal = "Phys. Rev. D",
    volume = "98",
    number = "12",
    pages = "124015",
    year = "2018"
}

@article{Lin:2018ken,
    author = "Lin, Kai and Zhao, Xiang and Zhang, Chao and Liu, Tan and Wang, Bin and Zhang, Shaojun and Zhang, Xing and Zhao, Wen and Zhu, Tao and Wang, Anzhong",
    title = "{Gravitational waveforms, polarizations, response functions, and energy losses of triple systems in Einstein-aether theory}",
    eprint = "1810.07707",
    archivePrefix = "arXiv",
    primaryClass = "astro-ph.GA",
    doi = "10.1103/PhysRevD.99.023010",
    journal = "Phys. Rev. D",
    volume = "99",
    number = "2",
    pages = "023010",
    year = "2019"
}

@article{Enoki:2006kj,
    author = "Enoki, Motohiro and Nagashima, Masahiro",
    title = "{The Effect of Orbital Eccentricity on Gravitational Wave Background Radiation from Cosmological Binaries}",
    eprint = "astro-ph/0609377",
    archivePrefix = "arXiv",
    doi = "10.1143/PTP.117.241",
    journal = "Prog. Theor. Phys.",
    volume = "117",
    pages = "241",
    year = "2007"
}

@article{Favata:2011qi,
    author = "Favata, Marc",
    title = "{The Gravitational-wave memory from eccentric binaries}",
    eprint = "1108.3121",
    archivePrefix = "arXiv",
    primaryClass = "gr-qc",
    doi = "10.1103/PhysRevD.84.124013",
    journal = "Phys. Rev. D",
    volume = "84",
    pages = "124013",
    year = "2011"
}

@article{Munna:2019fjz,
    author = "Munna, Christopher and Evans, Charles R.",
    title = "{Eccentric-orbit extreme-mass-ratio-inspiral radiation: Analytic forms of leading-logarithm and subleading-logarithm flux terms at high PN orders}",
    eprint = "1909.05877",
    archivePrefix = "arXiv",
    primaryClass = "gr-qc",
    doi = "10.1103/PhysRevD.100.104060",
    journal = "Phys. Rev. D",
    volume = "100",
    number = "10",
    pages = "104060",
    year = "2019"
}

@inproceedings{Blanchet:2023soy,
    author = "Blanchet, L. and Faye, G. and Henry, Q. and Larrouturou, F. and Trestini, D.",
    title = "{Gravitational waves from compact binaries to the fourth post-Newtonian order}",
    booktitle = "{57th Rencontres de Moriond on Gravitation}",
    eprint = "2304.13647",
    archivePrefix = "arXiv",
    primaryClass = "gr-qc",
    month = "4",
    year = "2023"
}

@article{Arun:2004hn,
    author = "Arun, K. G. and Iyer, Bala R. and Sathyaprakash, B. S. and Sundararajan, Pranesh A.",
    title = "{Parameter estimation of inspiralling compact binaries using 3.5 post-Newtonian gravitational wave phasing: The Non-spinning case}",
    eprint = "gr-qc/0411146",
    archivePrefix = "arXiv",
    doi = "10.1103/PhysRevD.71.084008",
    journal = "Phys. Rev. D",
    volume = "71",
    pages = "084008",
    year = "2005",
    note = "[Erratum: Phys.Rev.D 72, 069903 (2005)]"
}

@article{Boetzel:2019nfw,
    author = "Boetzel, Yannick and Mishra, Chandra Kant and Faye, Guillaume and Gopakumar, Achamveedu and Iyer, Bala R.",
    title = "{Gravitational-wave amplitudes for compact binaries in eccentric orbits at the third post-Newtonian order: Tail contributions and postadiabatic corrections}",
    eprint = "1904.11814",
    archivePrefix = "arXiv",
    primaryClass = "gr-qc",
    doi = "10.1103/PhysRevD.100.044018",
    journal = "Phys. Rev. D",
    volume = "100",
    number = "4",
    pages = "044018",
    year = "2019"
}

@article{Fujita:2010xj,
    author = "Fujita, Ryuichi and Iyer, Bala R.",
    title = "{Spherical harmonic modes of 5.5 post-Newtonian gravitational wave polarisations and associated factorised resummed waveforms for a particle in circular orbit around a Schwarzschild black hole}",
    eprint = "1005.2266",
    archivePrefix = "arXiv",
    primaryClass = "gr-qc",
    doi = "10.1103/PhysRevD.82.044051",
    journal = "Phys. Rev. D",
    volume = "82",
    pages = "044051",
    year = "2010"
}

@article{Faye:2012we,
    author = "Faye, Guillaume and Marsat, Sylvain and Blanchet, Luc and Iyer, Bala R.",
    title = "{The third and a half post-Newtonian gravitational wave quadrupole mode for quasi-circular inspiralling compact binaries}",
    eprint = "1204.1043",
    archivePrefix = "arXiv",
    primaryClass = "gr-qc",
    doi = "10.1088/0264-9381/29/17/175004",
    journal = "Class. Quant. Grav.",
    volume = "29",
    pages = "175004",
    year = "2012"
}

@article{Mishra:2013rna,
    author = "Mishra, Chandra Kant and Arun, K. G. and Iyer, Bala R.",
    editor = "Bi\v{c}\'ak, Ji\v{r}\'\i{} and Ledvinka, Tom\'a\v{s}",
    title = "{2.5PN kick from black-hole binaries in circular orbit: Nonspinning case}",
    eprint = "1304.5915",
    archivePrefix = "arXiv",
    primaryClass = "gr-qc",
    doi = "10.1007/978-3-319-06761-2_21",
    journal = "Springer Proc. Phys.",
    volume = "157",
    pages = "169--175",
    year = "2014"
}

@article{Faye:2014fra,
    author = "Faye, Guillaume and Blanchet, Luc and Iyer, Bala R.",
    title = "{Non-linear multipole interactions and gravitational-wave octupole modes for inspiralling compact binaries to third-and-a-half post-Newtonian order}",
    eprint = "1409.3546",
    archivePrefix = "arXiv",
    primaryClass = "gr-qc",
    doi = "10.1088/0264-9381/32/4/045016",
    journal = "Class. Quant. Grav.",
    volume = "32",
    number = "4",
    pages = "045016",
    year = "2015"
}

@article{Cho:2022syn,
    author = "Cho, Gihyuk and Porto, Rafael A. and Yang, Zixin",
    title = "{Gravitational radiation from inspiralling compact objects: Spin effects to the fourth post-Newtonian order}",
    eprint = "2201.05138",
    archivePrefix = "arXiv",
    primaryClass = "gr-qc",
    reportNumber = "DESY-22-004, ET-0001A-22, DESY-22-004; ET-0001A-22",
    doi = "10.1103/PhysRevD.106.L101501",
    journal = "Phys. Rev. D",
    volume = "106",
    number = "10",
    pages = "L101501",
    year = "2022"
}

@article{Li:2022grj,
    author = "Li, Zhao and Qiao, Jin and Liu, Tan and Zhu, Tao and Zhao, Wen",
    title = "{Gravitational waveform and polarization from binary black hole inspiral in dynamical Chern-Simons gravity: from generation to propagation}",
    eprint = "2211.12188",
    archivePrefix = "arXiv",
    primaryClass = "gr-qc",
    doi = "10.1088/1475-7516/2023/04/006",
    journal = "JCAP",
    volume = "04",
    pages = "006",
    year = "2023"
}

@article{Shiralilou:2021mfl,
    author = "Shiralilou, Banafsheh and Hinderer, Tanja and Nissanke, Samaya M. and Ortiz, N\'estor and Witek, Helvi",
    title = "{Post-Newtonian gravitational and scalar waves in scalar-Gauss\textendash{}Bonnet gravity}",
    eprint = "2105.13972",
    archivePrefix = "arXiv",
    primaryClass = "gr-qc",
    doi = "10.1088/1361-6382/ac4196",
    journal = "Class. Quant. Grav.",
    volume = "39",
    number = "3",
    pages = "035002",
    year = "2022"
}

@article{Julie:2019sab,
    author = "Juli\'e, F\'elix-Louis and Berti, Emanuele",
    title = "{Post-Newtonian dynamics and black hole thermodynamics in Einstein-scalar-Gauss-Bonnet gravity}",
    eprint = "1909.05258",
    archivePrefix = "arXiv",
    primaryClass = "gr-qc",
    doi = "10.1103/PhysRevD.100.104061",
    journal = "Phys. Rev. D",
    volume = "100",
    number = "10",
    pages = "104061",
    year = "2019"
}

@article{Arias:2012az,
    author = "Arias, Paola and Cadamuro, Davide and Goodsell, Mark and Jaeckel, Joerg and Redondo, Javier and Ringwald, Andreas",
    title = "{WISPy Cold Dark Matter}",
    eprint = "1201.5902",
    archivePrefix = "arXiv",
    primaryClass = "hep-ph",
    reportNumber = "DESY-11-226, MPP-2011-140, CERN-PH-TH-2011-323, IPPP-11-80, DCPT-11-160",
    doi = "10.1088/1475-7516/2012/06/013",
    journal = "JCAP",
    volume = "06",
    pages = "013",
    year = "2012"
}

@article{QuilezLasanta:2021yzt,
    author = "Quilez Lasanta, Pablo and Di Luzio, Luca and Gavela, Belen and Ringwald, Andreas",
    title = "{An exceptionally light axion: Strong CP and Dark Matter}",
    eprint = "2111.03149",
    archivePrefix = "arXiv",
    primaryClass = "hep-ph",
    reportNumber = "DESY 21-180, DESY 21-180 DESY 21-180 DESY 21-180 DESY 21-180 DESY 21-180",
    doi = "10.22323/1.398.0177",
    journal = "PoS",
    volume = "EPS-HEP2021",
    pages = "177",
    year = "2022"
}

@article{Flambaum:2019cih,
    author = "Flambaum, V. V. and Samsonov, I. B.",
    title = "{Ultralight dark photon as a model for early universe dark matter}",
    eprint = "1908.09432",
    archivePrefix = "arXiv",
    primaryClass = "astro-ph.CO",
    doi = "10.1103/PhysRevD.100.063541",
    journal = "Phys. Rev. D",
    volume = "100",
    number = "6",
    pages = "063541",
    year = "2019"
}

@Article{10.21468/SciPostPhys.12.5.171,
	title={{Audible axions with a booster: Stochastic gravitational waves from  rotating ALPs}},
	author={Eric Madge and Wolfram Ratzinger and Daniel Schmitt and Pedro Schwaller},
	journal={SciPost Phys.},
	volume={12},
	pages={171},
	year={2022},
	publisher={SciPost},
	doi={10.21468/SciPostPhys.12.5.171},
	url={https://scipost.org/10.21468/SciPostPhys.12.5.171},
}

@article{Ejlli:2022zah,
    author = "Ejlli, Aldo and Vermeulen, Sander M. and Schwartz, Eyal and Aiello, Lorenzo and Grote, Hartmut",
    title = "{Probing dark matter with polarimetry techniques}",
    eprint = "2211.09922",
    archivePrefix = "arXiv",
    primaryClass = "hep-ph",
    doi = "10.1103/PhysRevD.107.083035",
    journal = "Phys. Rev. D",
    volume = "107",
    number = "8",
    pages = "083035",
    year = "2023"
}

@article{Nagano:2021kwx,
    author = "Nagano, Koji and Nakatsuka, Hiromasa and Morisaki, Soichiro and Fujita, Tomohiro and Michimura, Yuta and Obata, Ippei",
    title = "{Axion dark matter search using arm cavity transmitted beams of gravitational wave detectors}",
    eprint = "2106.06800",
    archivePrefix = "arXiv",
    primaryClass = "hep-ph",
    doi = "10.1103/PhysRevD.104.062008",
    journal = "Phys. Rev. D",
    volume = "104",
    number = "6",
    pages = "062008",
    year = "2021"
}

@article{Blanchet:2013haa,
    author = "Blanchet, Luc",
    title = "{Gravitational Radiation from Post-Newtonian Sources and Inspiralling Compact Binaries}",
    eprint = "1310.1528",
    archivePrefix = "arXiv",
    primaryClass = "gr-qc",
    doi = "10.12942/lrr-2014-2",
    journal = "Living Rev. Rel.",
    volume = "17",
    pages = "2",
    year = "2014"
}

@article{Brito:2015oca,
    author = "Brito, Richard and Cardoso, Vitor and Pani, Paolo",
    title = "{Superradiance}: {New Frontiers in Black Hole
Physics}",
    eprint = "1501.06570",
    archivePrefix = "arXiv",
    primaryClass = "gr-qc",
    doi = "10.1007/978-3-319-19000-6",
    journal = "Lect. Notes Phys.",
    volume = "906",
    pages = "pp.1--237",
    year = "2015"
}

@article{Sanchis-Gual:2022ooi,
    author = "Sanchis-Gual, Nicolas and Izquierdo, Paula",
    title = "{Ultralight bosonic dark matter in white dwarfs and potential observational consequences}",
    eprint = "2202.00434",
    archivePrefix = "arXiv",
    primaryClass = "gr-qc",
    doi = "10.1103/PhysRevD.105.084023",
    journal = "Phys. Rev. D",
    volume = "105",
    number = "8",
    pages = "084023",
    year = "2022"
}

@article{Goldstein:2022pxu,
    author = "Goldstein, Isabelle S. and Koushiappas, Savvas M. and Walker, Matthew G.",
    title = "{Viability of ultralight bosonic dark matter in dwarf galaxies}",
    eprint = "2206.05244",
    archivePrefix = "arXiv",
    primaryClass = "astro-ph.GA",
    doi = "10.1103/PhysRevD.106.063010",
    journal = "Phys. Rev. D",
    volume = "106",
    number = "6",
    pages = "063010",
    year = "2022"
}

@article{Schutz:2020jox,
    author = "Schutz, Katelin",
    title = "{Subhalo mass function and ultralight bosonic dark matter}",
    eprint = "2001.05503",
    archivePrefix = "arXiv",
    primaryClass = "astro-ph.CO",
    reportNumber = "MIT-CTP/5173",
    doi = "10.1103/PhysRevD.101.123026",
    journal = "Phys. Rev. D",
    volume = "101",
    number = "12",
    pages = "123026",
    year = "2020"
}

@article{LIGOScientific:2019hgc,
    author = "Abbott, Benjamin P and others",
    collaboration = "LIGO Scientific, Virgo",
    title = "{A guide to LIGO\textendash{}Virgo detector noise and extraction of transient gravitational-wave signals}",
    eprint = "1908.11170",
    archivePrefix = "arXiv",
    primaryClass = "gr-qc",
    doi = "10.1088/1361-6382/ab685e",
    journal = "Class. Quant. Grav.",
    volume = "37",
    number = "5",
    pages = "055002",
    year = "2020"
}

@article{Cardoso:2020iji,
    author = "Cardoso, Vitor and Macedo, Caio F. B. and Vicente, Rodrigo",
    title = "{Eccentricity evolution of compact binaries and applications to gravitational-wave physics}",
    eprint = "2010.15151",
    archivePrefix = "arXiv",
    primaryClass = "gr-qc",
    doi = "10.1103/PhysRevD.103.023015",
    journal = "Phys. Rev. D",
    volume = "103",
    number = "2",
    pages = "023015",
    year = "2021"
}

@ARTICLE{2021EPJC...81.1048L,
       author = {{Liu}, Lang and {Christiansen}, {\O}yvind and {Ruan}, Wen-Hong and {Guo}, Zong-Kuan and {Cai}, Rong-Gen and {Kim}, Sang Pyo},
        title = "{Gravitational and electromagnetic radiation from binary black holes with electric and magnetic charges: elliptical orbits on a cone}",
      journal = {European Physical Journal C},
         year = 2021,
        month = nov,
       volume = {81},
       number = {11},
          eid = {1048},
        pages = {1048},
          doi = {10.1140/epjc/s10052-021-09849-4},
archivePrefix = {arXiv},
       eprint = {2011.13586},
 primaryClass = {gr-qc},
       adsurl = {https://ui.adsabs.harvard.edu/abs/2021EPJC...81.1048L}
}

@ARTICLE{2009GReGr..41.1667H,
       author = {{Husa}, Sascha},
        title = "{Michele Maggiore: Gravitational waves. Volume 1: theory and experiments. Oxford University Press, 2007, 576p., GBP47.00, ISBN13: 978-0-19-857074-5}",
      journal = {General Relativity and Gravitation},
         year = 2009,
        month = jul,
       volume = {41},
       number = {7},
        pages = {1667-1669},
          doi = {10.1007/s10714-009-0762-5},
       adsurl = {https://ui.adsabs.harvard.edu/abs/2009GReGr..41.1667H}
}

@article{Coogan:2021uqv,
    author = "Coogan, Adam and Bertone, Gianfranco and Gaggero, Daniele and Kavanagh, Bradley J. and Nichols, David A.",
    title = "{Measuring the dark matter environments of black hole binaries with gravitational waves}",
    eprint = "2108.04154",
    archivePrefix = "arXiv",
    primaryClass = "gr-qc",
    doi = "10.1103/PhysRevD.105.043009",
    journal = "Phys. Rev. D",
    volume = "105",
    number = "4",
    pages = "043009",
    year = "2022"
}

@article{Singh:2022wvw,
    author = "Singh, Divya and Gupta, Anuradha and Berti, Emanuele and Reddy, Sanjay and Sathyaprakash, B. S.",
    title = "{Constraining properties of asymmetric dark matter candidates from gravitational-wave observations}",
    eprint = "2210.15739",
    archivePrefix = "arXiv",
    primaryClass = "gr-qc",
    doi = "10.1103/PhysRevD.107.083037",
    journal = "Phys. Rev. D",
    volume = "107",
    number = "8",
    pages = "083037",
    year = "2023"
}

@article{Becker:2021ivq,
    author = "Becker, Niklas and Sagunski, Laura and Prinz, Lukas and Rastgoo, Saeed",
    title = "{Circularization versus eccentrification in intermediate mass ratio inspirals inside dark matter spikes}",
    eprint = "2112.09586",
    archivePrefix = "arXiv",
    primaryClass = "gr-qc",
    doi = "10.1103/PhysRevD.105.063029",
    journal = "Phys. Rev. D",
    volume = "105",
    number = "6",
    pages = "063029",
    year = "2022"
}

@article{Yue:2019ozq,
    author = "Yue, Xiao-Jun and Cao, Zhoujian",
    title = "{Dark matter minispike: A significant enhancement of eccentricity for intermediate-mass-ratio inspirals}",
    eprint = "1908.10241",
    archivePrefix = "arXiv",
    primaryClass = "astro-ph.HE",
    doi = "10.1103/PhysRevD.100.043013",
    journal = "Phys. Rev. D",
    volume = "100",
    number = "4",
    pages = "043013",
    year = "2019"
}

@article{Chen:2022qvg,
    author = "Chen, Zu-Cheng and Kim, Sang Pyo and Liu, Lang",
    title = "{Gravitational and electromagnetic radiation from binary black holes with electric and magnetic charges: Hyperbolic orbits on a cone}",
    eprint = "2210.15564",
    archivePrefix = "arXiv",
    primaryClass = "gr-qc",
    month = "10",
    year = "2022"
}

@article{Ghosh:2023tyz,
    author = "Ghosh, Dilip Kumar and Ghoshal, Anish and Jeesun, Sk",
    title = "{Axion-like particle (ALP) portal freeze-in dark matter confronting ALP search experiments}",
    eprint = "2305.09188",
    archivePrefix = "arXiv",
    primaryClass = "hep-ph",
    month = "5",
    year = "2023"
}

@article{Bhattacharya:2023stq,
    author = "Bhattacharya, Sulagna and Dasgupta, Basudeb and Laha, Ranjan and Ray, Anupam",
    title = "{Can LIGO Detect Asymmetric Dark Matter?}",
    eprint = "2302.07898",
    archivePrefix = "arXiv",
    primaryClass = "hep-ph",
    reportNumber = "TIFR/TH/23-1, N3AS-23-006",
    month = "2",
    year = "2023"
}

@article{Liu:2020vsy,
    author = "Liu, Lang and Christiansen, \O{}yvind and Guo, Zong-Kuan and Cai, Rong-Gen and Kim, Sang Pyo",
    title = "{Gravitational and electromagnetic radiation from binary black holes with electric and magnetic charges: Circular orbits on a cone}",
    eprint = "2008.02326",
    archivePrefix = "arXiv",
    primaryClass = "gr-qc",
    doi = "10.1103/PhysRevD.102.103520",
    journal = "Phys. Rev. D",
    volume = "102",
    number = "10",
    pages = "103520",
    year = "2020"
}

@article{Beneke:1997zp,
    author = "Beneke, M. and Smirnov, Vladimir A.",
    title = "{Asymptotic expansion of Feynman integrals near threshold}",
    eprint = "hep-ph/9711391",
    archivePrefix = "arXiv",
    reportNumber = "CERN-TH-97-315",
    doi = "10.1016/S0550-3213(98)00138-2",
    journal = "Nucl. Phys. B",
    volume = "522",
    pages = "321--344",
    year = "1998"
}

@article{Lee:2013mka,
    author = "Lee, Roman N.",
    editor = "Wang, Jianxiong",
    title = "{LiteRed 1.4: a powerful tool for reduction of multiloop integrals}",
    eprint = "1310.1145",
    archivePrefix = "arXiv",
    primaryClass = "hep-ph",
    doi = "10.1088/1742-6596/523/1/012059",
    journal = "J. Phys. Conf. Ser.",
    volume = "523",
    pages = "012059",
    year = "2014"
}
\end{document}